\documentclass[twoside,12pt]{article}
\usepackage{epsfig}
\usepackage[latin2]{inputenc}
\usepackage[T1]{fontenc}
\usepackage{amsthm}
\usepackage{amssymb}
\usepackage{amsmath}
\usepackage{graphicx}
\usepackage{threeparttable}
\begin{document}

\title{ \vspace{1cm} Two-proton emission and related phenomena}
\author{M.\ Pf\"utzner,$^{1}$  I.\ Mukha,$^2$ S.M.\ Wang,$^{3,4}$ \\
\\
$^1$Faculty of Physics, University of Warsaw, Pasteura 5, PL-02-093 Warszawa, Poland\\
$^2$GSI Helmholtzzentrum f\"ur Schwerionenforschung, 64291 Darmstadt, Germany\\
$^3$Key Laboratory of Nuclear Physics and Ion-beam Application (MOE), \\
Institute of Modern Physics, Fudan University, Shanghai 200433, China\\
$^4$Shanghai Research Center for Theoretical Nuclear Physics, \\NSFC and Fudan University, Shanghai 200438, China
}
\maketitle

\begin{abstract}
One of characteristic phenomena for nuclei beyond the proton dripline is the simultaneous
emission of two protons (2\emph{p}). The current status of our knowledge of this most recently observed and the least known decay mode is presented. First, different approaches to theoretical description
of this process, ranging from effective approximations to advanced three-body models are
overviewed. Then, after a brief survey of main experimental methods to produce 2\emph{p}-emitting
nuclei and techniques to study their decays, experimental findings in this research field
are presented and discussed. This review covers decays of short-lived resonances and excited states of unbound nuclei
as well as longer-lived, ground-state radioactive decays. In addition, more exotic decays
like three- and four-proton emission are addressed. Finally, related few-body topics,
like two-neutron and four-neutron radioactivity, and the problem of the tetraneutron are shortly discussed.
\end{abstract}


\section{Introduction}

One of the current frontiers in low-energy nuclear-physics research
is the study of nuclei at the limits of nuclear stability. These limits are
determined by nuclear binding energies and are characterized by
the drip lines which separate bound systems from the unbound ones on the
chart of nuclei. This article is focused on the very neutron-deficient
nuclei, beyond the proton drip line. This region is very interesting,
mainly due to unique phenomena resulting from interplay between nuclear
forces, in particular pairing interaction, and the Coulomb force.
The emerging Coulomb barrier hampers emission of unbound protons
which brings into competition beta decays governed by weak interactions.
In addition, proton unbound nuclei are examples of open quantum systems
in which coupling to the continuum of scattering states affects
various features like Thomas-Ehrman shift or clustering phenomena.

Beyond the proton drip line, unbound protons can be emitted from a nucleus
leading to proton radioactivity. Single-proton emission from the nuclear
ground-state, occurring for nuclei with odd number of protons ($Z$),
has been studied since 40 years
and developed into a powerful spectroscopic tool providing a wealth of
information on properties and structure of exotic nuclei \cite{Pfutzner:2012}.
On the other hand, a characteristic decay of even-$Z$ unbound
nuclei is the two-proton (2\emph{p})
radioactivity in which two protons are ejected simultaneously from a
 nucleus. This is the most recently discovered decay mode and
still the least known. This phenomenon is the central topic of the present
article in which we make a summary of the current status of the experimental
and theoretical studies on the 2\emph{p} emission.

In the past, developments in this research field were presented in
review articles by Blank and Borge~\cite{Blank:2008}, Blank and
P\l{}oszajczak~\cite{Blank:2008c}, and Grigorenko~\cite{Grigorenko:2009d}.
Recently, a status of two-proton radioactivity was summarized by
Zhou et al.~\cite{Zhou:2022}. A broader overview of radioactive decays at
the limits of nuclear stability was given in Ref.~\cite{Pfutzner:2012}.
A pedagogical introduction to nuclear physics at the proton dripline
and to radioactive decays with emission of charged particles can be found
in~\cite{Pfutzner:2022} and \cite{Blank:2022}, respectively.

\subsection{\it Basic concepts}

The limit of nuclear stability for neutron-deficient even-$Z$ nucleus
is determined by the two-proton separation energy:
\begin{equation}\label{eq:1_S2p}
  S_{2p} = B(N,Z) - B(N, Z-2) \, ,
\end{equation}
where $B(N,Z)$ is the binding energy of the nuclide with $N$ neutrons
 and $Z$ protons.
The (two)proton drip-line is defined as the border between the last
bound isotope and the first one with the negative value of the $S_{2p}$.
Thus, the exact location of the drip-line requires precise knowledge
of nuclear masses in the region of interest.

\begin{figure}[tb]
\begin{center}
\begin{minipage}[t]{10 cm}
\includegraphics[width = \columnwidth]{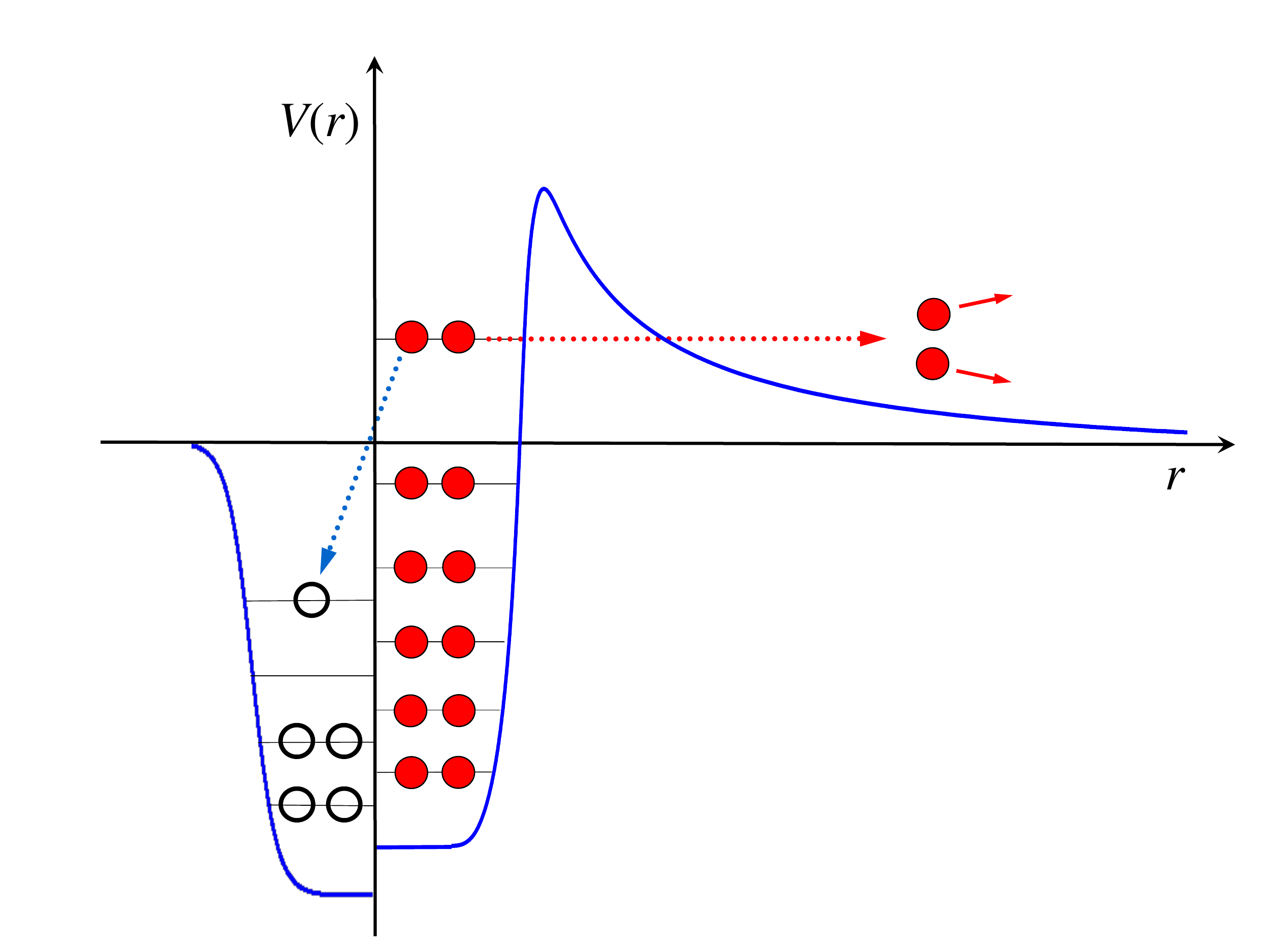}
\end{minipage}
\begin{minipage}[t]{17 cm}
\caption{(Color online) Schematic picture of a 2\emph{p}-unbound nucleus. The solid line
illustrates the radial part of nuclear potential (including Coulomb interaction)
as a function of the distance
from the nuclear center ($r$). The potential felt by protons, represented by full
circles, is shown on the right side. The sum of the repulsive, long range Coulomb
interaction and the attractive, short range nuclear force creates the Coulomb barrier.
The potential felt by neutrons (open circles) is shown on the left. The dotted
lines indicate alternative decay modes: 2\emph{p} emission and $\beta^+$ decay.
 }
 \label{fig:1_Model_Potential}
\end{minipage}
\end{center}
\end{figure}

\begin{figure}[tb]
\begin{center}
\begin{minipage}[t]{12 cm}
\includegraphics[width = \columnwidth]{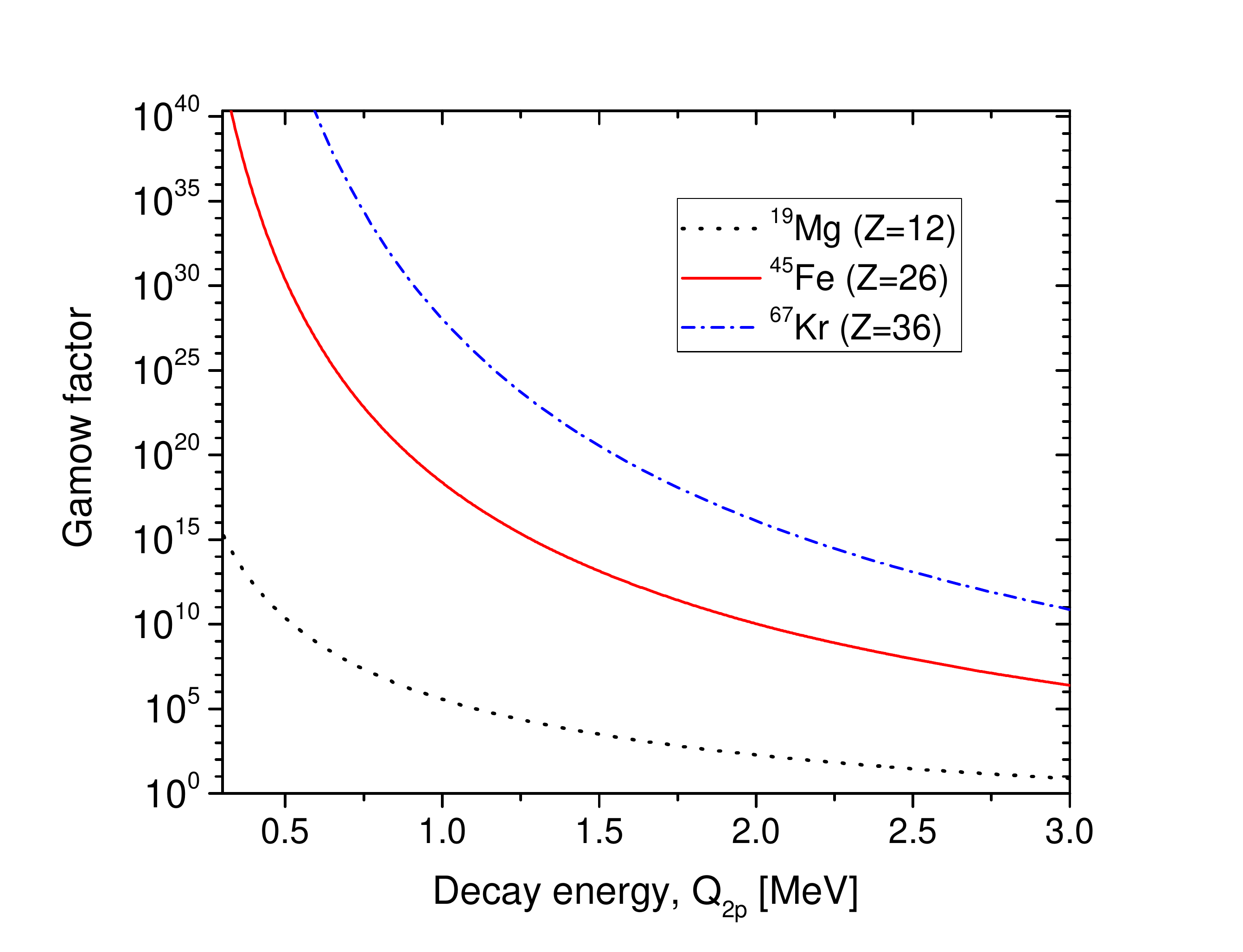}
\end{minipage}
\begin{minipage}[t]{17 cm}
\caption{(Color online) Gamow factor for the diproton penetration through the potential barrier
 as a function of the 2\emph{p} decay energy $Q_{2p}$ for the three 2\emph{p} emitters. }
\label{fig:1_Gamow}
\end{minipage}
\end{center}
\end{figure}

The nuclear potential for protons and neutrons in a 2\emph{p}-unbound nucleus
is schematically shown in Fig. \ref{fig:1_Model_Potential}. Although the last
two protons are unbound, they are confined by the Coulomb barrier which
prevents them form the rapid emission. The partial half-life of
the 2\emph{p} decay is strongly affected by the tunneling probability through
this potential barrier, which in turn is extremely sensitive to the
decay energy $Q_{2p} = - S_{2p}$. To illustrate this sensitivity we
consider the WKB approximation \cite{Gurvitz:1987} in which the partial
half-life for the emission of a charged particle from a nucleus is
proportional to the Gamow factor describing the tunneling process:
\begin{equation}\label{eq:1_Gamow}
  G = \exp \left[ 2 \, \int_{r_{\rm in}}^{r_{\rm out}} dr \, \sqrt{\frac{2 \, \mu}{\hbar^2} \, |Q-V(r)|} \right] \, ,
\end{equation}
where $\mu$ is the reduced mass of the particle and the daughter nucleus,
$Q$ is the decay energy, and the $V(r)$ is the radial part of the
particle-nucleus potential. The integral runs over the classically
forbidden region, where particle remains under the barrier.
Taking the Woods-Saxon potential with the Coulomb term and
assuming the tunneling of a diproton ($^2$He)
with the orbital angular momentum of $\ell=0$, we have calculated
the $G$-factor for the three cases of 2\emph{p} emission as a function
of the decay energy, see Fig. \ref{fig:1_Gamow}. It can be seen that
a change of the decay energy by a few hundred keV changes the $G$-factor
by a few orders of magnitude.
Thus, the reliable prediction of 2\emph{p} decaying candidates requires
very accurate knowledge of the decay energy, in addition to a model of the 2\emph{p} decay mechanism.

On the other hand, the $\beta^+$ decay
channel is opened with the partial decay constant proportional to $Q_{\beta}^5$.
Since the beta decay energy $Q_{\beta}$ is large for nuclei far from stability,
the resulting $\beta$ half-lives can be as short as a few milliseconds.
This has important consequences for experimental observation of the
2\emph{p} decay. To win the competition with $\beta^+$ decay, it has to proceed
faster, with the half-life of the order of milliseconds and shorter.
Hence, production and separation of such a nucleus must be fast and
calls for special experimental techniques. We note that when the Coulomb
barrier is sufficiently high, the $\beta^+$ decay of an unbound nucleus
will be faster that the proton emission and will dominate the decay.
Thus, an observation of the proton emission proves that the nucleus is
beyond the drip-line but does not allow to locate this line.

\begin{figure}[tb]
\begin{center}
\begin{minipage}[t]{14 cm}
\includegraphics[width = \columnwidth]{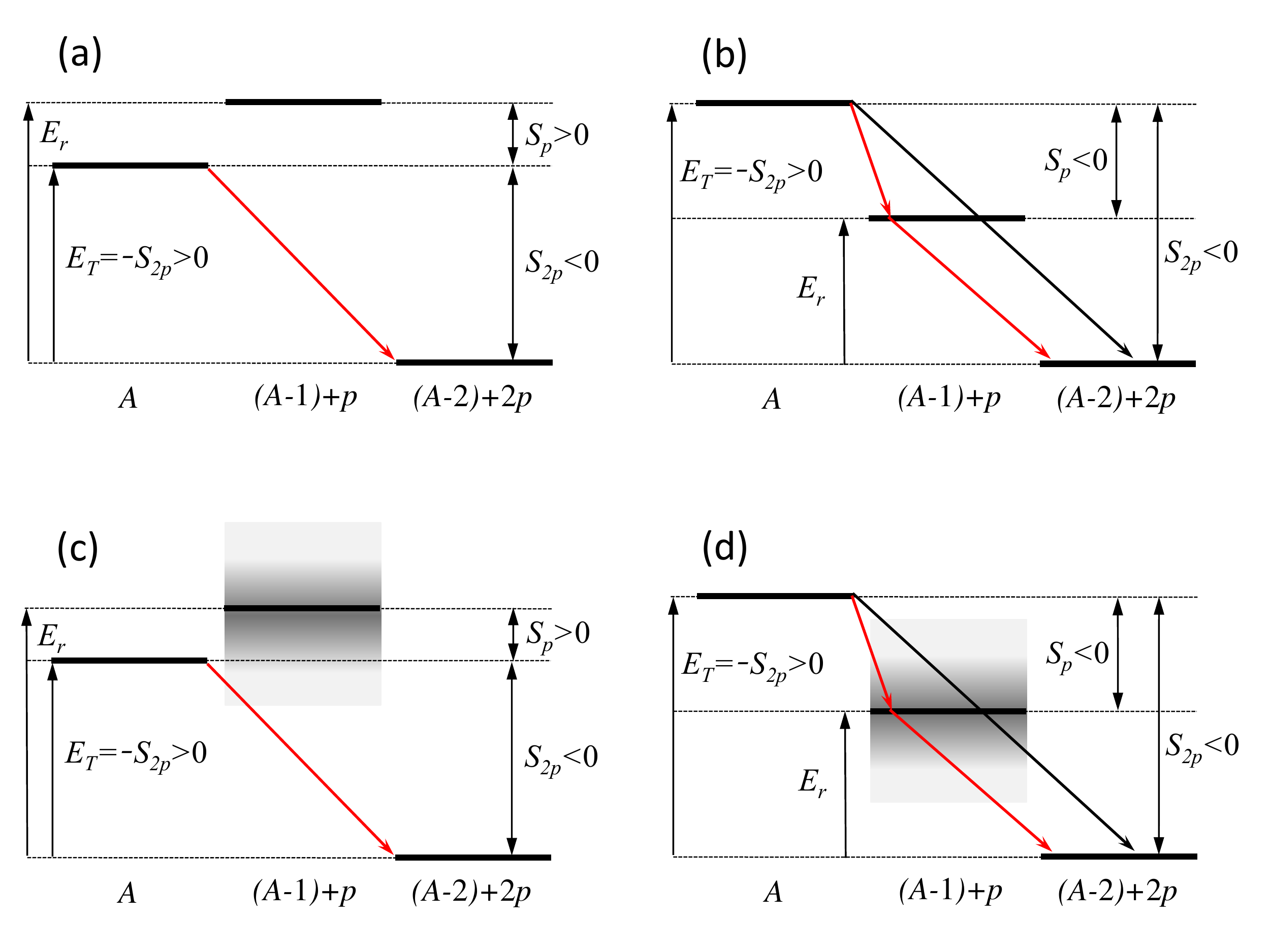}
\end{minipage}
\begin{minipage}[t]{17 cm}
\caption{Energy conditions for different modes of two-proton emission:
(a) true 2\emph{p} decay, (b) sequential emission via a narrow intermediate
state. The cases (c) and (d) represent ``democratic'' decays. }
\label{fig:1_EnergyConditions}
\end{minipage}
\end{center}
\end{figure}

Another important aspect of the 2\emph{p} emission is the competition with the
single proton emission. Possible scenarios are illustrated by respective decay
energy schemes in Fig. \ref{fig:1_EnergyConditions}.
In this Figure, in addition to the
separation energies, we indicated two energy values for future reference:
$E_T$ is the energy of the three-body system, relative to the break-up
threshold, and $E_{r}$ is the lowest two-body resonance relative to this
threshold.
The typical situation for the ground state
of an even-$Z$ medium mass nucleus beyond the proton drip-line, and the most favorable
for the 2\emph{p} radioactivity, is shown in the panel (a). Due to the pairing interaction,
the $S_{2p}$ energy is negative, while the single proton separation energy,
$S_p$ is positive, or close to zero. Then, the emission of a single proton,
and thus also sequential emission of two protons, is energetically
forbidden or strongly suppressed. This type of decay has essentially
three-body character and we refer to it as the \emph{true 2p decay}.

In the situation shown in the panel (b), the proton separation energy $S_p$ is
negative so that the initial nucleus can decay by sequential emission of two
protons. Such a scenario may occur in particular when the initial nucleus
is located further beyond the proton drip-line (has less neutrons) than
the isotope decaying by the true 2\emph{p} emission (a) or when it is in an
excited state. In this situation the simultaneous emission of two protons
is also possible but it is difficult to experimentally isolate such a
channel from the dominant sequential contribution.
The panels (c) and (d) represent scenarios analogous to
(a) and (b), respectively, but with the broad intermediate state.
Such cases, when the energies of emitted protons are of the same order than
the width of the intermediate state, are called ``democratic'' decays.
The distinction between simultaneous and sequential emission losses its
meaning and no strong correlations between outgoing fragments are formed.

\begin{figure}[tb]
\begin{center}
\begin{minipage}[t]{18 cm}
\includegraphics[width = \columnwidth]{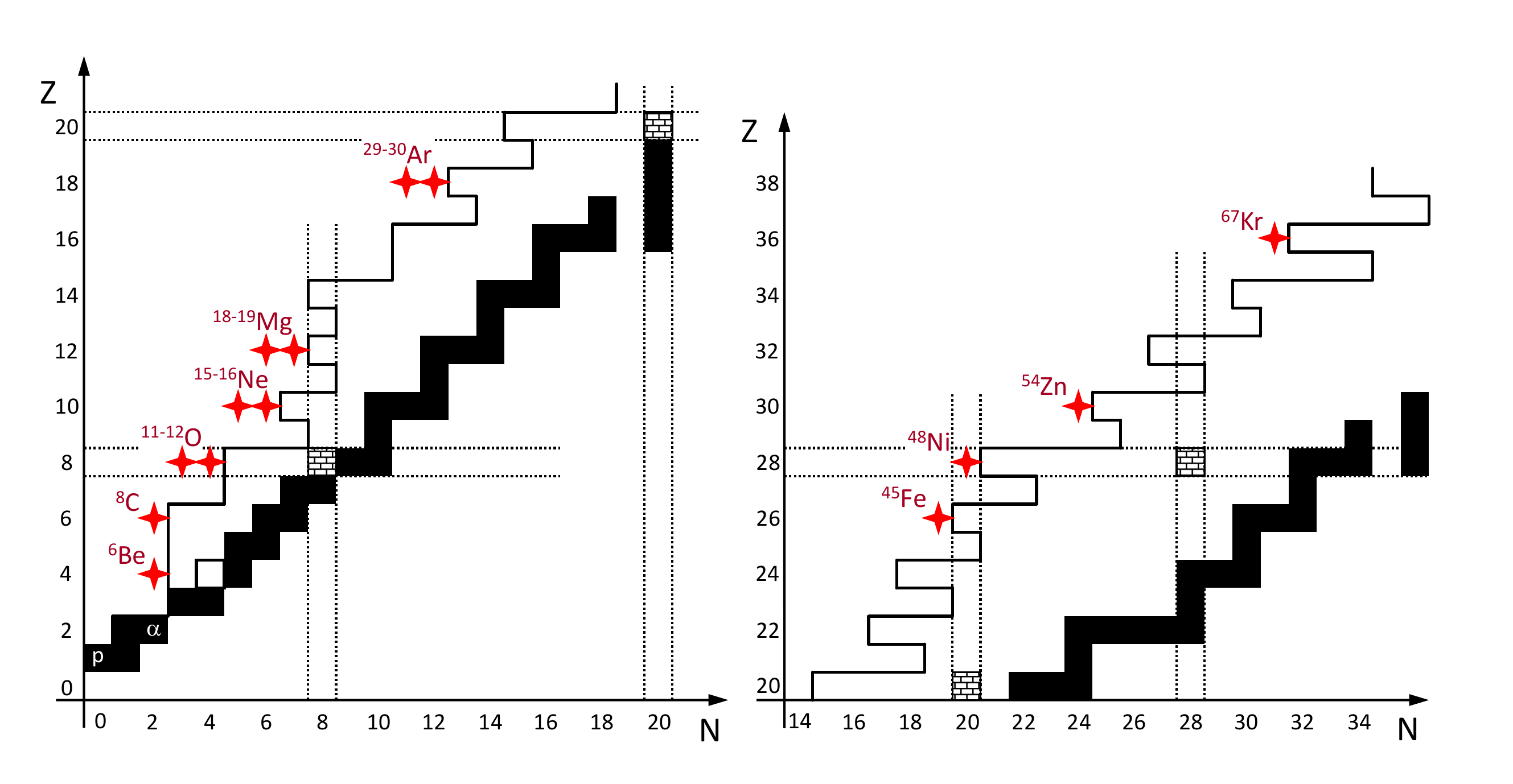}
\end{minipage}
\begin{minipage}[t]{17 cm}
\caption{(Color online) Region of the chart of nuclei discussed in this article.
Solid line denotes the proton drip-line, doubly magic nuclei are
indicated by a square filled with a pattern, other stable nuclei
are marked with black squares. Stars represent known ground-state 2\emph{p} emitters.  }
\label{fig:1_Chart}
\end{minipage}
\end{center}
\end{figure}

Experimental work devoted to 2\emph{p} emission has been focused on light and
medium mass nuclei. Presently, these studies reached the $Z=36$ (krypton).
The relevant part of the chart of nuclei is shown in Fig. \ref{fig:1_Chart}.
The proton drip-line is shown by the solid line and the nuclei discussed in
this review are indicated.

\subsection{\it Historical note}

The classical era of radioactivity studies was completed with the discovery of
spontaneous fission by Flerov and Petrzhak in 1940.
By then, all most common radioactive processes, including $\alpha$ decay, $\beta^-$, $\beta^+$, and electron capture (EC) decays, as well as $\gamma$ radiation
were known. A detailed account of this fascinating period in the subatomic
physics was given by Pais \cite{Pais:1986}. The modern times in research on radioactivity started with
investigations of nuclei very far from stability, at the limits of nuclear
binding. In 1960 Zeldovich was the first to consider the proton drip-line
for light nuclei and to notice the possibility of simultaneous
2\emph{p} emission \cite{Zeldovich:1960}. It was, however, Goldansky who made the
first systematical study of properties of very neutron-deficient nuclei, and
considered their possible decay modes, including one-proton and two-proton
radioactivity \cite{Goldansky:1960,Goldansky:1961}.
The first theoretical description of 2\emph{p} emission was attempted by
Galitsky and Cheltsov \cite{Galitsky:1964}. The candidates for the experimental
observation of 2\emph{p} radioactivity were considered by Goldansky \cite{Goldansky:1961}
and by J\"anecke \cite{Janecke:1965}. Precision of their mass models, however,
was not sufficient for accurate predictions and
experimental techniques at that time were not developed enough to reach
the considered candidates.

\begin{figure}[tb]
\begin{center}
\begin{minipage}[t]{13 cm}
\includegraphics[width = \columnwidth]{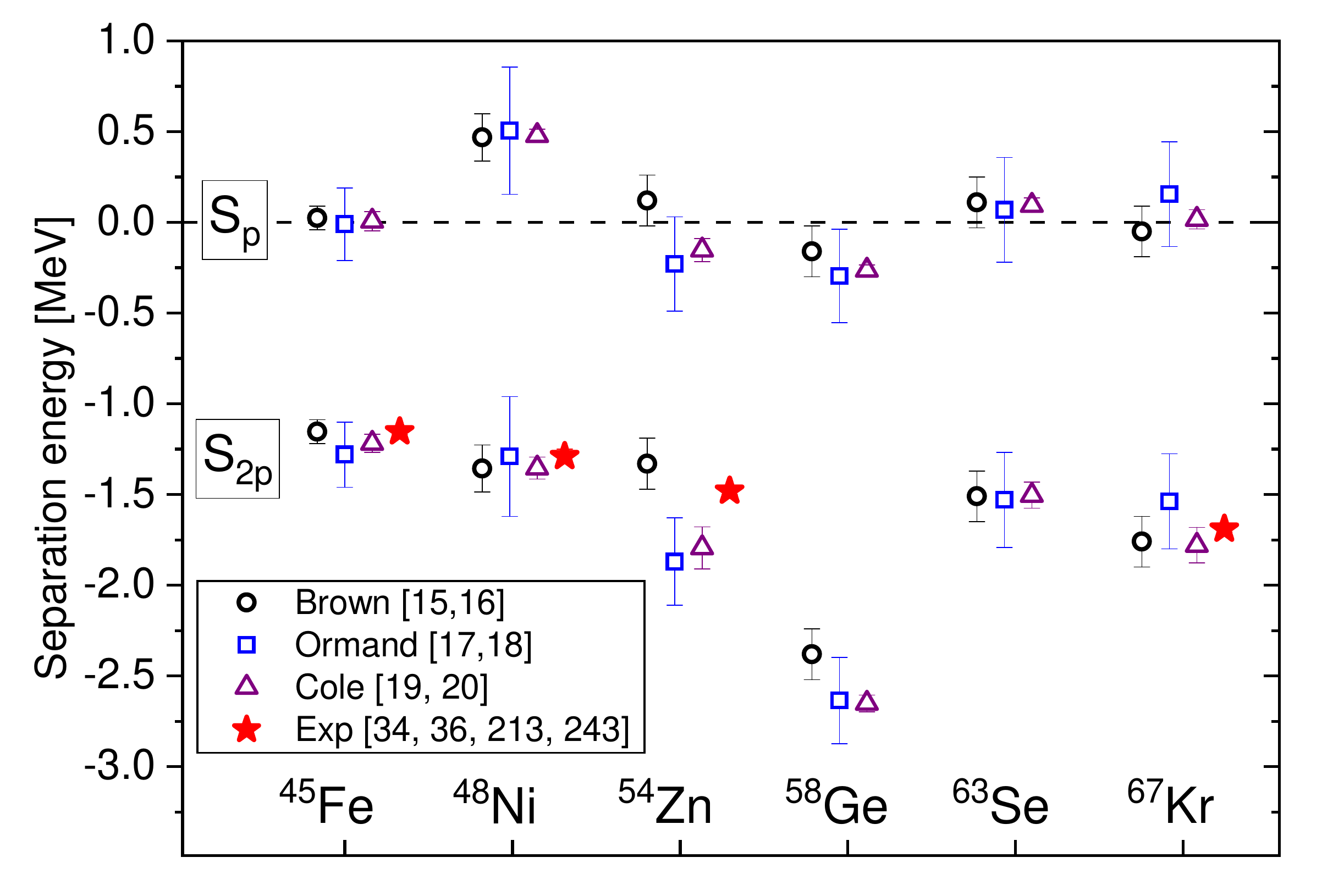}
\end{minipage}
\begin{minipage}[t]{17 cm}
\caption{(Color online) Theoretical predictions of one-proton ($S_p$) and two-proton
($S_{2p}$) separation energies for candidates of true 2\emph{p} decay, together
with experimental results (stars).}
\label{fig:1_S1pS2p}
\end{minipage}
\end{center}
\end{figure}

Most exact predictions of masses for 2\emph{p} radioactivity candidates were based
on Coulomb energy differences between partners of isospin multiplets.
This difference can be calculated, either using microscopic models or
Coulomb energy systematics. Then, using experimentally measured mass
of the neutron-rich member of the multiplet, the mass
of the exotic proton-rich member can be determined \cite{Brown:1991,Brown:2002,Ormand:1996,Ormand:1997,Cole:1996,Cole:1999}.
This approach proved to be very successful and played an important role
motivating experimental efforts to search for the 2\emph{p} radioactivity.
Selected results of these predictions are shown in Fig. \ref{fig:1_S1pS2p}.

Experimental studies on 2\emph{p} emission started in the end of seventies.
Naturally, the first cases tried were light nuclei which are
relatively easier to produce: $^6$Be \cite{Geesaman:1977,Bochkarev:1984},
$^{12}$O \cite{KeKelis:1978,Kryger:1995}, and $^{16}$Ne \cite{KeKelis:1978}.
Because of the low Coulomb barrier these nuclei have very short half-lives
and intermediate states are also broad, so they fall into category of
democratic decays. Subsequent studies of these cases, in particular on $^6$Be, provided interesting insights into three-body dynamics as will be discussed
later in more detail. Two-proton emission from excited nuclear states,
populated in $\beta$ decay or in inelastic nuclear reactions, can
provide a way to study the 2\emph{p} decay mechanism for less exotic nuclei.
In 1983 the first case of $\beta$-delayed two-proton emission ($\beta$2\emph{p})
was discovered in $^{22}$Al \cite{Cable:1983}. Later many more cases were
observed and some of them, like $^{31}$Ar, were studied in more detail \cite{Blank:2008}.
In all of them, however, sequential emission was found to dominate.
Nuclear reactions were used to excite states in a few nuclei to investigate
2\emph{p} emission. This approach started in 1996 with the $2^+$ state at 7.77 MeV in $^{14}$O produced by 2\emph{p} transfer on $^{12}$C in 1996 \cite{Bain:1996}. Later
excited states in $^{17}$Ne, $^{18}$Ne, and $^{19}$Ne were populated by various methods \cite{Chromik:2002,Zerguerras:2004,delCampo:2001,Raciti:2008,Oliveira:2005}.
A question whether the simultaneous emission of two protons does contribute
to these decays remains unsettled and will be discussed in the following.

The observation of the 2\emph{p} radioactivity from a long-lived nuclear ground-state
had to wait more than 40 years from the first insight of Goldansky.
This breakthrough was achieved in 2002 for $^{45}$Fe in two independent
experiments at GSI Darmstadt \cite{Pfutzner:2002} and GANIL \cite{Giovinazzo:2002}.
It was possible thanks to a new experimental technique based on the
reaction of projectile fragmentation and the in-flight identification
of single ions arriving to the detection setup where they decay at rest.
The same method allowed discovery of 2\emph{p} radioactivity of $^{54}$Zn at
GANIL \cite{Blank:2005}, $^{48}$Ni at the NSCL MSU laboratory \cite{Pomorski:2011}, and $^{67}$Kr at RIKEN Nishina Center \cite{Goigoux:2016}. As can be
seen in the Fig. \ref{fig:1_S1pS2p} the measured decay energies for these
four 2\emph{p} emitters were found to agree very well with the predictions.
A different technique based on tracking the products of the decay in-flight,
developed for studies of 2\emph{p} decays with very short half-life, was
used at GSI Darmstadt to identify 2\emph{p} radioactivity of $^{19}$Mg in 2007 \cite{Mukha:2007}.

\subsection{\it Outline of the paper}

In Section 2 theoretical description of 2\emph{p} emission will be presented.
Various approaches from effective approximations through microscopic
models to advanced three-body models will be discussed. Section~3 is
devoted to experimental techniques essential for 2\emph{p} decay studies.
Most important aspects of production and identification of ions
beyond the proton drip-line, and detection of their decays will be
reviewed. Discussion of 2\emph{p} emission phenomena will start in Section 4
which is dedicated to democratic decays and 2\emph{p} emission from excited
states. This will be followed by an overview of experiments on
2\emph{p} radioactivity from long lived nuclear ground states in Section 5.
In Section~6 more exotic decay modes, like three-proton or four proton
emission, will be addressed. Related few-body phenomena will be discussed
in Section~7. This will include the relation between proton and neutron
radioactivity, five-body decays, and the question of the tetraneutron.
Finally, a brief summary and outlook for future research will be given
in Section~8.

\renewcommand{\vec}[1]{\mbox{\boldmath $#1$}}
\newcommand{\bra}[1]{\left \langle #1 \right \rvert}
\newcommand{\ket}[1]{\left \rvert #1 \right \rangle}
\newcommand{\E}[1]{\left \langle #1 \right \rangle}
\newcommand{\braket}[2]{\left \langle #1 \middle \rvert #2 \right \rangle}
\newcommand{\braxket}[3]{\left \langle #1 \middle \rvert #2 \middle \rvert #3 \right \rangle}

\section{Theoretical models}
\label{sec:2_theory}

\begin{figure}[tb]
\begin{center}
\begin{minipage}[t]{15 cm}
\includegraphics[width = \columnwidth]{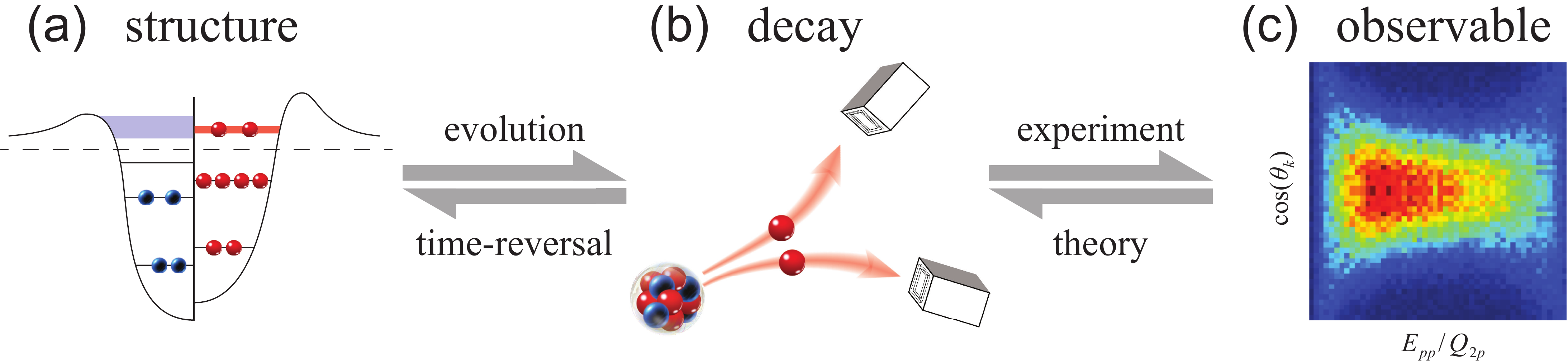}
\end{minipage}
\begin{minipage}[t]{17 cm}
\caption{(Color online) The schematic figure of the connection between (a) inner structure and (c) experimental observables (nucleon-nucleon correlation) through (b) $2p$ decay. The blue and red balls represent the neutron and proton, respectively.}
\label{fig:2_connection}
\end{minipage}
\end{center}
\end{figure}

\begin{figure}[tb]
\begin{center}
\begin{minipage}[t]{15 cm}
\begin{center}
\includegraphics[width = \columnwidth]{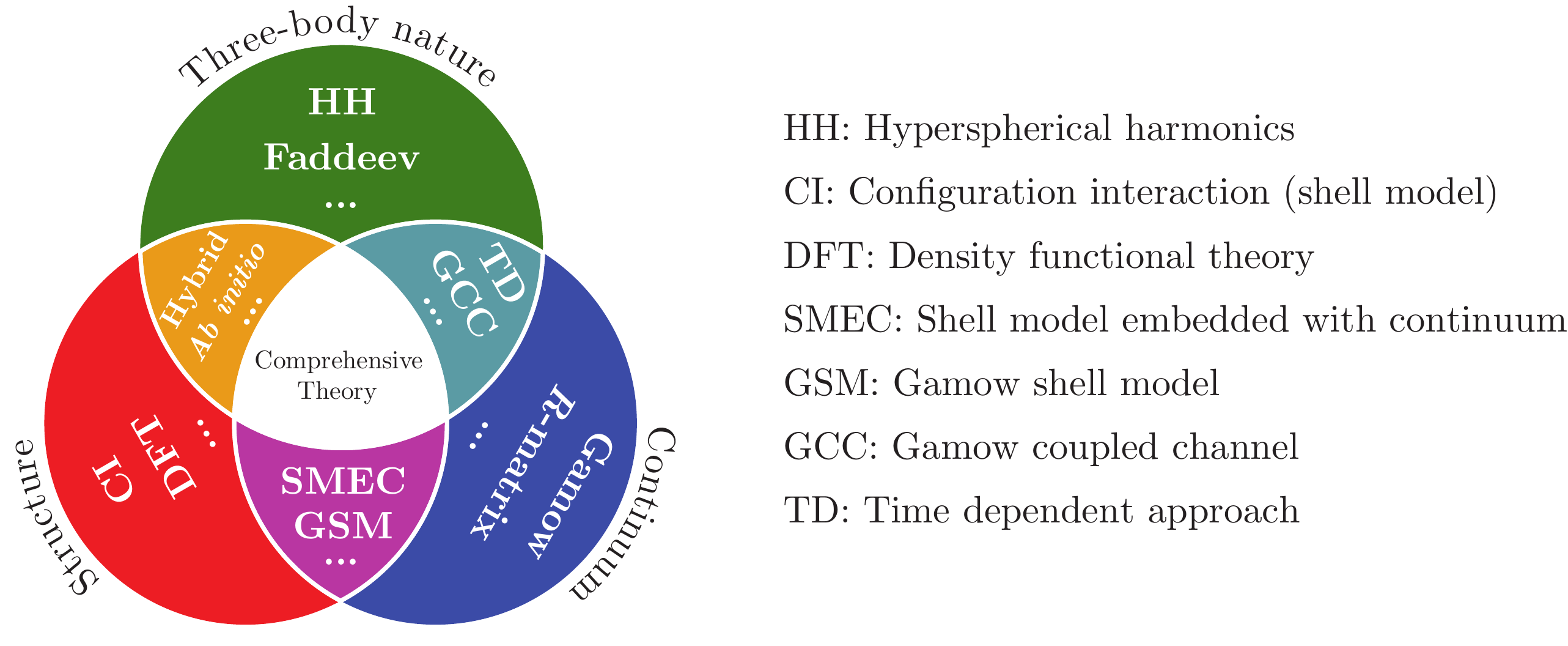}
\end{center}
\end{minipage}
\begin{minipage}[t]{17 cm}
\caption{(Color online) The schematic figure for the categories of the theoretical models applied to $2p$ decay. It contains three different aspects: structure, three-body nature, and continuum effect.}
\label{fig:2_models}
\end{minipage}
\end{center}
\end{figure}

As shown in Section 1, more and more $2p$ emitters have been discovered when going beyond the proton dripline. These nuclei are usually of different properties due to the interplay between the inner structure and open quantum nature. As shown in Fig.\,\ref{fig:2_connection}, in order to decipher the  experimental observables and establish firm connection with structural information, a self-consistent description of the internal and external wave function is needed, which raises a big challenge for nuclear theory. Normally, there are mainly three aspects -- inner structure, three-body nature, and continuum effect -- that are crucial for the $2p$ decay process. The former determines the spectroscopic factor and possible decay channels for the valence protons, while the rest are related to the decay dynamics and asymptotic behavior. Due to the vast difference in decay lifetime and the complexity of the problem, many theoretical frameworks are mainly focused on one or two of these aspects (see Fig.\,\ref{fig:2_models}).  Part of the frameworks has been reviewed in \cite{Pfutzner:2012,Blank:2008c,Grigorenko:2009d,Zhou:2022}, here we will briefly introduce some of approaches published recently. Also presented are some basic theoretical ideas that demonstrate the efforts being made toward a comprehensive description of $2p$ decay.

\subsection{\it Coordinates and degrees of freedom}
\label{sec:2_degree}

To study the $2p$ decay process and the corresponding properties, it is necessary to figure out the degrees of freedom in this system. For the initial state of a $2p$ emitter, the valence protons are located inside the nucleus and coupled with other nucleons mainly through the short-range nuclear interaction. In principle, a fully microscopic framework should start from the nucleon degrees of freedom, and the total wave function $\Psi^{J\pi}$ is a mixing of different configurations. Meanwhile, due to the property of the continuum coupling, it has been found that the structure of the near-threshold resonance tends to be aligned with the corresponding decay channel~\cite{Okolowicz:2020}, which
is in accord with the phenomenon of threshold clustering in various weakly-bound systems~\cite{Ebran:2012,Ikeda:1968,Freer:2018,deGrancey:2016,vonOertzen:1996}. In such a dilute-density environment near the threshold, the correlation caused by the short-range nuclear interaction among the valence nucleons would benefit the cluster formation~\cite{Okolowicz:2012,Wiescher:2017}. This
indicates that, in the case of the $2p$ decay from the low-lying state, the daughter nucleus is possible to be preformed inside the nucleus. Consequently, the degrees of freedom for the nucleons inside the daughter nucleus can be approximately frozen. The interest has also been boosted by exploring the diproton structure, which is likely to occur in $2p$ decay due to the fact of 1) the increasing level density in the presence of the low-lying continuum near the threshold and 2) the combination of channels with different parities. Although the transition from a fully microscopic system to a cluster picture is not totally clear, the three-body assumption is usually taken into account during the emission process. In this case, the corresponding wave function can be described as the relative motion of the daughter nucleus and two valence protons:
\begin{equation}\label{eq:2_WF_3b}
\Psi^{J\pi} = \sum_{J_{p} \pi_{p} j_{c} \pi_{c}}\left[\Phi^{J_{p} \pi_{p}} \otimes \phi^{J_{c} \pi_{c}}\right]^{J \pi},
\end{equation}
where $\Phi ^{J_p\pi_p}$ and $\phi^{J_c\pi_c}$ are the wave functions of the valence part and the core (daughter nucleus), respectively. $J$ and $\pi$ are the corresponding angular momentum and parity. In this case, the final state interaction is mainly dominated by the long-range Coulomb force and the effective interaction among the clusters.

\begin{figure}[tb]
\begin{center}
\begin{minipage}[t]{17 cm}
\begin{center}
\includegraphics[width = \columnwidth]{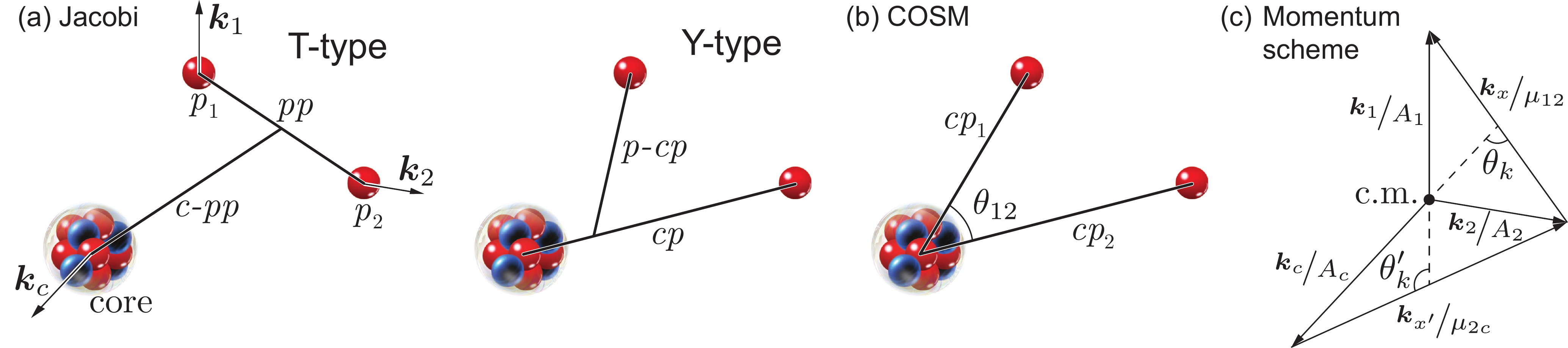}
\end{center}
\end{minipage}
\begin{minipage}[t]{17 cm}
\caption{(Color online) (a) Jacobi and (b) COSM (single-particle) coordinates, and the corresponding momentum scheme.}
\label{fig:2_coordinates}
\end{minipage}
\end{center}
\end{figure}

Based on the three-body nature of the $2p$ decay, it is convenient to introduce the coordinates of the valence protons~\cite{Zhukov:1993}. There are two kinds of coordinates that are commonly used to construct the few-body as well as many-body framework. As shown in  (see Fig.\,\ref{fig:2_coordinates}), one is Jacobi coordinates~\cite{Grigorenko:2002}, and the other is cluster-orbital shell model (COSM)~\cite{Suzuki:1988}. The former is built in the center-of-mass reference frame, also known as the relative coordinates:
\begin{equation}\label{eq:2_Jacobi}
\begin{aligned}
    \vec{x} &= \sqrt{\mu _{x}} (\vec{r}_{i_1} - \vec{r}_{i_2}),\\
    \vec{y} &= \sqrt{\mu _{y}} \left( \frac{A_{i_1}\vec{r}_{i_1} + A_{i_2}\vec{r}_{i_2}}{A_{i_1} + A_{i_2}} -\vec{r}_{i_3} \right),\\
\end{aligned}
\end{equation}
where ${i_1}=p_1, {i_2}=p_2, {i_3}=c$ for T-coordinates and ${i_1}=p_2, {i_2}=c, {i_3}=p_1$ for Y-coordinates, see Fig.\,\ref{fig:2_coordinates}. In Eq.~(\ref{eq:2_Jacobi})
$A_i$ is  the $i$-th cluster mass number, and  $\mu _{x} =  \frac{A_{i_1}A_{i_2}}{A_{i_1}+A_{i_2}}$ and $\mu _{y} = \frac{(A_{i_1}+A_{i_2})A_{i_3}}{A_{i_1}+A_{i_2}+A_{i_3}} $ are the reduced masses associated with $\vec{x}$ and $\vec{y}$, respectively. In this way, the Jacobi coordinates automatically eliminate center-of-mass motion, and allow for the exact treatment of the asymptotic wave functions. Therefore, it is suitable for nuclear reactions and the description of other asymptotic properties. However, the complicated coupling/transformation coefficients and antisymmetrization prevent it from extending to larger fermionic systems, although efforts have been made in recent years with the help of harmonic oscillator basis~\cite{Quaglioni:2013,Quaglioni:2016} or generalized hyperspherical harmonics~\cite{Das:2016}.

The coordinate space is useful to reveal the inner structure of a nucleus, but, in experiments, usually free particles in  momentum space are measured by detectors positioned far away from the source. Therefore, it is also convenient to introduce  the relative momenta:
\begin{equation}
\begin{aligned}
    \vec{k}_x &= \mu _{x} \left(\frac{\vec{k}_{i_1}}{A_{i_1}} - \frac{\vec{k}_{i_2}}{A_{i_2}}\right),\\
    \vec{k}_y &= \mu _{y} \left( \frac{\vec{k}_{i_1} + \vec{k}_{i_2}}{A_{i_1} + A_{i_2}} -\frac{\vec{k}_{i_3}}{A_{i_3}} \right),\\
\end{aligned}
\end{equation}
Where $\vec{k}_i$ is the momentum of the $i$-th cluster. Since there is no c.m. motion, it is easy to notice that $\sum_i \vec{k}_i = 0$, and $\vec{k}_y$ is in the opposite direction of $\vec{k}_{i_3}$. $\theta_k$ and $\theta^\prime_k$ are the opening angles of ($\vec{k}_x$, $\vec{k}_y$) in T- and Y-Jacobi coordinates, respectively (see Fig.\,\ref{fig:2_coordinates}).
The kinetic energy of the relative motion of the emitted nucleons is given by
$E_{pp} = \frac{\hbar^2 k_x^2}{2 \mu_x}$  and $E_{{\rm core}-p}$ is that of the core-nucleon pair, which represents the energy $E_{x} = \frac{\hbar^2 k_x^2}{2 \mu_x}$ associated with $\vec{x}$ in T- and Y-type coordinates, respectively. Similarly, $E_{y} = \frac{\hbar^2 k_y^2}{2 \mu_y} = Q_{2p}-E_{x}$, where $Q_{2p}$ is the  two-nucleon decay energy given by the binding energy difference of  parent and daughter nuclei. Therefore, the T-type ($\theta_k$, $E_{pp}$) and Y-type ($\theta^\prime_k$, $E_{{\rm core}-p}$) distributions reveal the nucleon-nucleon correlation and the structural information about the mother nucleus. Finally, the total momentum $k$ is defined as $\sqrt{{k_x^2}/{\mu_x} + {k_y^2}/{\mu_y}}$, which approaches the limit ${\sqrt{2mQ_{2p}}}/{\hbar}$ at late times of the final state.

On the other side, the COSM is based on the single-particle coordinates ($\vec{r}_n - \vec{r}_c$) with respect to the core or the origin in the laboratory frame. In this framework, all the valence nucleons are treated equally, which would make it easy to calculate matrix elements and extend to the many-body systems by introducing the configuration mixing between valence protons and the daughter nucleus. Hence, it is favorable in many microscopic models such as configuration interaction. However, since it is not in the c.m. coordinate, one needs to be very careful when treating asymptotic observables. For example, a recoil term of $\left(\hat{\boldsymbol{p}}_{1}\cdot\hat{\boldsymbol{p}}_{2} / m_{c}\right)$ is needed to deal with the extra energy introduced by center-of-mass motion~\cite{Suzuki:1988,Michel:2002}, where $ m_{c}$ is the mass number for the core. This also indicates that Jacobi-Y and COSM coordinates become identical with an infinite-mass core.

The benchmarking between COSM and Jacobi coordinates has been done in Ref.\,\cite{Wang:2017b}, in which both weak-bound and unbound systems ($^6$He, $^6$Li, $^6$Be, and $^{26}$O) were investigated using the Berggren ensemble technique~\cite{Berggren:1968}. Consequently, the COSM-based Gamow shell model and the Jacobi-based Gamow coupled-channel method give practically identical results for those structural properties and observables related to the inner wave function of a nucleus~\cite{Wang:2017b}. Also, it has been shown that the Jacobi coordinates capture cluster correlations (such as dineutron and deuteron-type) more efficiently.

\subsection{\it Decay properties and simplified models}
\label{sec:2_decay}

Among the most fundamental properties of an unstable system at the edge of nuclear stability, the first to be established is the decaying mode and corresponding half-life~\cite{Pfutzner:2012,Delion:2010}. In order to capture the decaying properties, the continuum effect needs to be taken into account. Since the spectrum of an observable is real in Hilbert space, the decaying resonance is hidden among the scattering eigenstates of the Hamiltonian, which enhances the difficulty of describing and studying these open quantum systems. To this end, one can utilize Gamow theory by introducing complex-energy eigenstates in the rigged Hilbert space, or $R$-matrix through the flux current of the asymptotic wave function. For the former one, the complex energy can be written as
\begin{equation}
\tilde{E} = E -\mathrm{i} \frac{\Gamma}{2},
\end{equation}
where $\Gamma$ is the decay width. The imaginary part represents the uncertainty of the energy for an unstable state, and is connected with the half-life through the uncertainty principle:
\begin{equation}
T_{1 / 2}=\frac{\hbar \ln 2}{\Gamma}.
\end{equation}
One can also extract this relation by comparing the exponential decay and the temporal part of a resonance, i.e. $\mathrm{e}^{\mathrm{i} E t / \hbar} \mathrm{e}^{-\Gamma t /(2 \hbar)}$. In this way, a decaying system can be treated in a quasi-stationary formalism~\cite{Baz:1969}.

Equivalently, the flux current is another way to extract the decay width or half-life~\cite{Wang:2017b,Grigorenko:2000,Grigorenko:2007}. Due to the three-body character of the final state, the decay width can be obtained with the asymptotic wave function using the expression~\cite{Humblet:1961}:
\begin{equation}
\Gamma=i \frac{\int\left(\Psi^{\dagger} \hat{H} \Psi-\Psi \hat{H} \Psi^{\dagger}\right) d \vec{x} d \vec{y}}{\int|\Psi|^{2} d \vec{x} d \vec{y}},
\end{equation}
where $\hat{H}$ is the Hamiltonian, and ($\vec{x}$, $\vec{y}$) are the Jacobi coordinates. Utilizing the hyperspherical harmonics expansion technique, the Jacobi coordinates can be transferred to the hyperradius $\rho$ and hyperangles $\Omega$, and the asymptotic behavior for the non-interaction three-body system $\hat{H}_0$ can be obtained. However, this is not straightforward for the $2p$ decay, since, in a three-body system, the long-range Coulomb interaction does not commute with $\hat{H}_0$. According to the $R$-matrix theory,  if the contribution from the off-diagonal part of the Coulomb interaction in the asymptotic region is neglected, the hyperradial wave function of the resonance $\Phi_{\xi}(\rho)$ is proportional to the outgoing Coulomb function $H^+(\eta_{\xi},k\rho)$~\cite{Grigorenko:2009b,Descouvemont:2006,Vasilevsky:2001}, where $\eta_{\xi}$ is the parameter related to the effective charge $Z^{\rm eff}_{\xi}$ for the channel $\xi$ and $k$ is the total momentum. For quasi-bound systems, such as medium-mass nuclei, the proton decay widths are usually below the numerical precision of the complex-energy method. One can still safely use the flux current by adopting the expression  $\Phi ^{\prime}/ \Phi = k {H^+}^\prime/H^+$~\cite{Barmore:2000,Kruppa:2004}. In this way, it bypasses the numerical derivative of the small wave function in the asymptotic region that appears in the original current expression and increases numerical precision dramatically~\cite{Esbensen:2000}.

In principle, the two methods mentioned above require a microscopic wave function with relatively high precision. Meanwhile, during the decay process, the valence protons are transiting from a strongly coupled and localized initial stage to a spatially separated but correlated final state, which also increases the complexity of the problem. Therefore, many approximations have been made based on the primary degrees of freedom in this problem.

\subsubsection{\it Direct decay approximation}

Following the picture of Gamow theory and the three-body character of the final state, the daughter nucleus can be assumed to be preformed right before the emission, while the valence protons move inside the mother nucleus bouncing upon the nuclear surface, and eventually penetrate quantum mechanically the surrounding Coulomb and centrifugal barrier. If neglecting the final-state interaction, the two valence protons are emitted independently but under the rules of energy and angular conservation, which is known as {\it direct decay}. Based on the $R$-matrix, the total $2p$ decay width is convolution of two single-proton emissions \cite{Kryger:1995,Azhari:1998,Barker:1999,Barker:2001,Barker:2003,Brown:2003}. The expression can be written in Jacobi-Y coordinates:
\begin{equation}
\begin{aligned}
\Gamma_{2p} &= \theta_{2p}^2 \int_0^{Q_{2p}}\Gamma(E) d E, \\
\Gamma(E) &= \Delta_x(E) \cdot \Gamma_y(Q_{2p}-E)
\end{aligned}
\label{eq:2_direct_decay}
\end{equation}
where $\Gamma_y$ is the decay width of the particle emission associated with $\vec{y}$, and $\Delta_x$ is the normalized level density. $\theta_{2p}^2$ is the spectroscopic factor for the daughter nucleus of $2p$ decay, which, in principle, needs to be evaluated by the microscopic models focused on the structural aspect (see Fig.\,\ref{fig:2_models}). Eq.(\ref{eq:2_direct_decay}) represents the sum over all the decaying channels for particle $p_1$ with energy $E_{p_1} = Q_{2p}-E$. For the {\it direct decay}, $\Delta_x$ is approximately given by a Breit-Wigner type distribution of particle $p_2$:
\begin{equation}
\label{eq:2_level_density}
\Delta_x(E) = \frac{1}{2\pi} \frac{\Gamma_{x}(E)}{(E-E_{p_2})^2+\Gamma_{x}(E)^2/4}.
\end{equation}
Since $\Gamma_{x/y}$ is the single-particle decay width, it can be obtained in the WKB approximation shown in Eq.\,(\ref{eq:1_Gamow}) or the $R$-matrix methodology for single-particle decay~\cite{Lane:1958}:
\begin{equation}
\Gamma_{x/y}(E) =2 P(E, R_{x/y}, Z^{\rm eff}_{x/y}) \gamma^{2}(E, R_{x/y}) = \frac{k\hbar^2}{\mu_{x/y}} \left|\frac{\Phi(E, R_{x/y})}{H^{+}(\eta_{x/y}, k R_{x/y})} \right|^2,
\end{equation}
which is a product between the penetrability $P$ and reduced width squared $\gamma^2$. In the assumption of a multi-step two-body process, the parameter $\eta = { \mu e^2 Z^{\rm eff}}/({\hbar^2}k)$, and the asymptotic behavior of the radial wave function $\Phi$ is proportional to the outgoing Coulomb function $H^+$ for a resonance, that is
$\lim_{R\rightarrow \infty} \Phi(E, R) = C_{\xi} H^{+}\left(\eta, k R\right)$,
where $C_{\xi}$ is the asymptotic normalization coefficient (ANC) for the channel $\xi$. Consequently, the obtained $\Gamma_{x/y}$ does not depend on $R_{x/y}$ at the external region. In practice, $R_{x/y}$ is chosen as the channel radius.

A similar expression for the {\it direct decay} approximation is obtained based on the three-body framework \cite{Galitsky:1964,Grigorenko:2007,Grigorenko:2003},
\begin{equation}
\label{eq:2_direct_decay_3b}
\Gamma_{2p} = 2\pi \E{V_3} \int_0^{Q_{2p}}\Delta_x(E)\cdot\Delta_y(Q_{2p}-E) d E, \\
\end{equation}
where $\E{V_3} = D(Q_{2p}-E_{p_1}-E_{p_2})^2$ and $D$ is a constant depending on details of nuclear structure. Although Eqs.\,(\ref{eq:2_direct_decay}) and (\ref{eq:2_direct_decay_3b}) are different by a factor of about four for the true $2p$ decay from the ground state, they are qualitatively in agreement for the general behavior of decay properties.

The energy-dependent decay width $\Gamma(E)$ in Eqs.\,(\ref{eq:2_direct_decay}) and (\ref{eq:2_direct_decay_3b}) also represents the energy correlation of the corresponding Jacobi coordinates. If the valence protons of the mother nucleus decay sequentially and the intermediate state is relatively narrow compared to $Q_{2p}$ (see Fig.\,\ref{fig:1_EnergyConditions}b), the decay process would favor some specific energy $Q_p$. Consequently, this would result in a sharp peak in the distribution of $\Gamma(E)$. On the contrary, if it is a true three-body decay as shown in Fig.\,\ref{fig:1_EnergyConditions}a, the most optimized decay path would be that the two valence protons emit with roughly equal energies ($E_{p_1}\approx E_{p_2}$ or $E_{x}\approx E_{y}$). In the latter case, neither of the valence protons spends a long time tunneling through the Coulomb and centrifugal barrier, otherwise, the total $2p$ decay width would be significantly suppressed. This results in a bell-shaped $E_{{\rm core-}p}$ energy correlation for the Jacobi-Y coordinates~\cite{Miernik:2007b}. It is interesting to point out that this simple estimation gives results in qualitative agreement with the three-body exact theory.

\subsubsection{\it Diproton decay approximation}
\label{sec:2_diproton}

Based on the fact that the most optimized decay path is under the condition of $E_{p_1}\approx E_{p_2}$ for a true three-body decay, the valence protons can be approximately treated as a cluster (diproton) and decay simultaneously \cite{Goldansky:1961}. After tunneling through the barrier, the diproton tends to separate due to the repulsive Coulomb force. The presumption of two distinct processes is still true. Therefore, one can express Eq.\,(\ref{eq:2_direct_decay}) in Jacobi-T coordinates. Consequently, $\Gamma_{y}(E)$ represents the diproton decay width, which can be evaluated through the WKB approximation or $R$-matrix theory:
\begin{equation}
\Gamma_{y}(E) =2 P(E, R_{c-pp}, 2Z_{c}) \gamma^{2}(E, R_{c-pp}).
\end{equation}
Based on the formula $\Gamma(E) = \Delta_x(E) \cdot \Gamma_y(Q_{2p}-E)$ in Eq.\,(\ref{eq:2_direct_decay}), the level density $\Delta_x$ in Jacobi-T coordinates describes the status of the diproton state ($\ell_x = S = 0$). For a proton-proton system in free space, it is likely to be a threshold resonance~\cite{Kok:1980}, which manifests itself as a scattering feature near the threshold. Therefore, a simple approximation would be $\Delta_x \approx \delta(E-E_{pp})$, where $E_{pp}$ is an average energy between two proton~\cite{Goldansky:1961,Janecke:1965,Brown:1991,Nazarewicz:1996,Ormand:1997}. Later on, Ref.\,\cite{Barker:2001} proposed a level-density form $\Delta_x \approx \sin^2[\delta_s(E)]$ in the spirit of Migdal-Watson approximation, where $\delta_s$ is the $s$-wave phase shift for proton-proton scattering.

In Ref.\,\cite{Delion:2013}, the two-proton decay process was studied within the framework of scattering theory, in which two particles are emitted from a correlated pairing state. By using similar proton-proton interactions, the BCS equations and the external dynamics are treated in a self-consistent way. The results show that it would be incorrect to consider the decaying system as a simple diproton. The decay proceeds neither through a diproton particle nor as an uncorrelated two-proton channel but rather as a configuration in between these two extremes.

Recently, a dynamic dinucleon model was proposed in Refs.\,\cite{Grigorenko:2018,Grigorenko:2020}, which combines a semi-realistic internal structure for the nuclear interior with a nucleon-nucleon interaction solely governing the emission process. The formalism is defined in a three-body framework but with restrictions to illustrate the isolated ``dineutron emission'' aspect of this problem. The results show that a broad variety of ``dineutron'' correlation patterns are possible.

Besides those approximate approaches introduced above, there are some other developments in the phenomenological or quasi-classical methods, including extended Geiger-Nuttall law for $2p$ decay \cite{Sreeja:2019,Delion:2022} and effective liquid drop model \cite{Cui:2020,Santhosh:2021,Cui:2021}. All the approximate approaches in this Section\,\ref{sec:2_decay} are based on different assumptions. Although the real situation of $2p$ decay is a complicated process, these approaches could be useful for systematical studies and demonstrating the main physics process behind some specified $2p$ decays.

\subsection{\it Internal structure and configuration mixing}
\label{sec:2_structure}

In the initial state of a $2p$ emitter, the valence protons could be strongly coupled with the rest of the nucleons inside the mother nucleus. To determine the possibility of the preformation or spectroscopic factor $\theta_{2p}^2$ of the daughter nucleus, one needs to utilize some microscopic theory of the structural aspects, such as density functional theory (DFT) or configuration interaction (CI). The former aims to describe the bulk properties of the nuclei, which is suitable for systematically exploring the candidates of $2p$ emitters (see Section\,8 and Refs\,\cite{Nazarewicz:1996,Fomichev:2011,Olsen:2013,Olsen:2013b,Neufcourt:2020} for details). The latter is focused on the configuration mixing among the valence nucleons of the mother nuclei located near the shell closures~\cite{Brown:2019,Rotureau:2005,Rotureau:2006,Michel:2003}.

\subsubsection{\it Configuration interaction}

As one of the most successful models in nuclear theory, CI is also known as the interacting shell model, and has been widely applied in structural and spectroscopic studies across the nuclear landscape. For a regular CI framework, the many-body Hamiltonian can be expressed in the COSM coordinates with the recoil term:
\begin{equation}
\label{eq:2_CI_Hamiltonian}
\hat{H}=\sum_{i=1}^{n}\left[\frac{p_{i}^{2}}{2 \mu}+U_{i}\right]+\sum_{i<j}^{n}\left[V_{i j}+\frac{\vec{p}_{i}\cdot \vec{p}_{j}}{m_{c}} \right]
\end{equation}
where $\mu$ is the reduced mass of the valence particles, $V_{i j}$ is the two-body interaction. The first and second terms in Eq.\,(\ref{eq:2_CI_Hamiltonian}) represent the one- and two-body operators, respectively. In some models, the latter one could be approximately replaced by an effective interaction or matrix elements. Since all the valence particles are treated in the same framework, one can define the single-particle orbital $q$ with the quantum numbers, and build the total wave function $|\Psi^{J\pi}\rangle = |A\omega J\rangle$ based on the corresponding basis. $A$ is the mass number of a nucleus, and $\omega$ contains the quantum numbers for the system. Consequently, this many-body framework allows us to analyze the preformation factor and structural information of the daughter nucleus  via the two-nucleon decay amplitudes (TNAs). For the removal of two protons from the initial state $|A\omega^\prime J^\prime\rangle$, leaving it in the final state $\langle(A-2)\omega J |$, TNA is given by the reduced matrix element as~\cite{Brown:2003}
\begin{equation}
\label{eq:2_TNA}
\operatorname{TNA}\left(q_{a}, q_{b}\right)=\frac{\left\langle(A-2) \omega J||\left[\tilde{a}_{q_{a}} \otimes \tilde{a}_{q_{b}}\right]^{J_{o}}|| A \omega^{\prime} J^{\prime}\right\rangle}{\sqrt{\left(1+\delta_{q_{a} q_{b}}\right)(2 J+1)}},
\end{equation}
where $\tilde{a}_{q}$ is an operator that destroys a proton in the orbital $q$. The spectroscopic factor $S=\theta^2_{2p}$ is given by a {\it coherent sum} over TNAs. In the case of diproton decay,
\begin{equation}
\label{eq:2_SF}
\theta_{2p}= \sum_{a, b} \sqrt{2J+1}\cdot {\rm TNA}(q_{a}, q_{b}) \cdot \langle LS_{\rm COSM} | j_a j_b \rangle \cdot \langle LS_{\rm T} | LS_{\rm COSM} \rangle,
\end{equation}
where $\langle LS | j_a j_b \rangle$ is the transformation between $LS$ and $jj$ schemes, and $\langle LS_{\rm T} | LS_{\rm COSM} \rangle$ is that between COSM and Jacobi-T coordinates, since the latter is natural for the description of diproton. In a regular CI framework, the wave function is constructed on a harmonic oscillator (HO) basis. In this case, $\langle LS_{\rm T} | LS_{\rm COSM} \rangle$ can be written as the product of a coefficient  $[{A}/({A-2})]^{(N_a+N_b) / 2}$ and Talmi-Moshinsky-Smirnov brackets $\langle {\mathfrak n}_x \ell_x;{\mathfrak n}_y \ell_y | {\mathfrak n}_{a} \ell_{a};{\mathfrak n}_{b} \ell_{b} \rangle_L$ \cite{Rotter:1968,Anyas-Weiss:1974}, where $N$ and ${\mathfrak n}$ are the total and radial numbers of oscillator quanta, respectively. For simplicity, only the matrix elements with $J_o$ = $L$ = $S$ = ${\mathfrak n}_x$ = $\ell_x$ = 0 have been taken into account for the diproton decay. Furthermore, total decay width can be evaluated using the formulae in Section\,\ref{sec:2_diproton}.

\begin{figure}[tb]
\begin{center}
\begin{minipage}[t]{12 cm}
\includegraphics[width = \columnwidth]{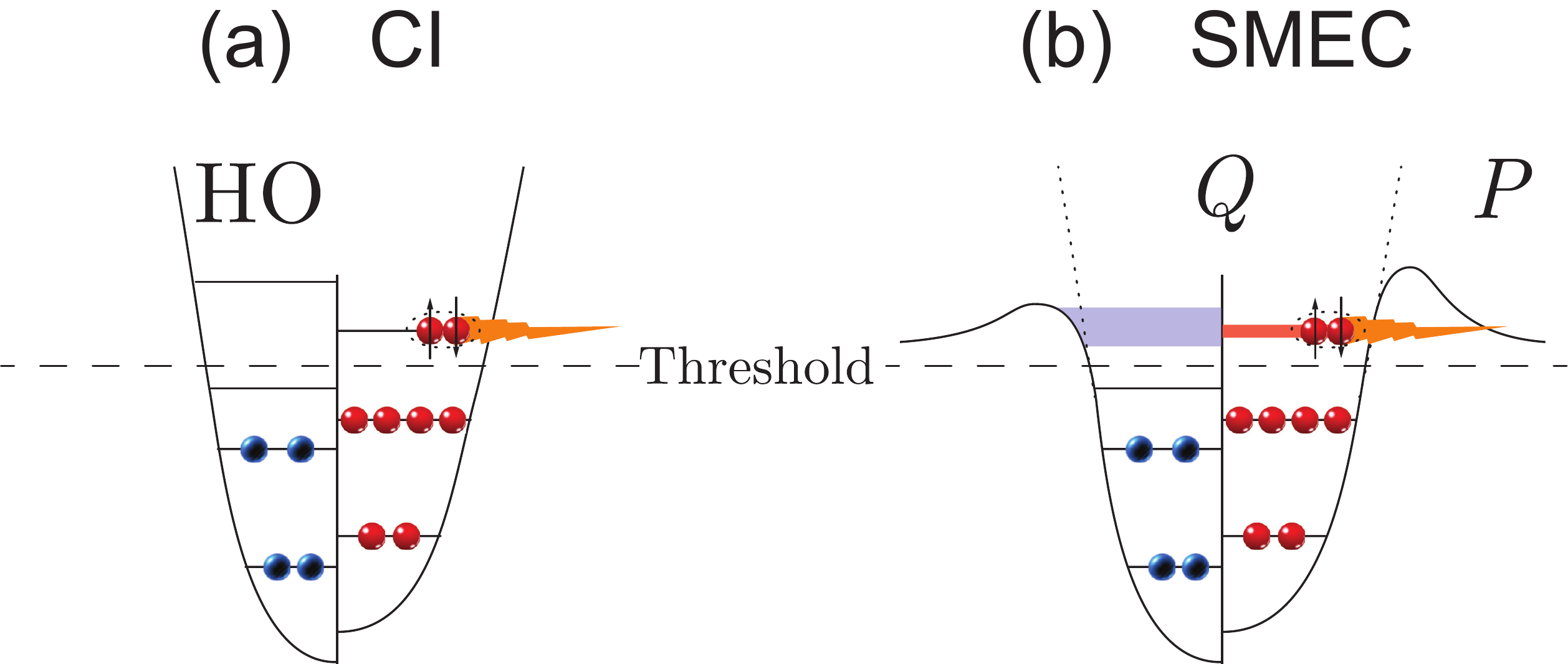}
\end{minipage}
\begin{minipage}[t]{17 cm}
\caption{(Color online) The schematic figure for the comparison between the frameworks of (a) regular configuration interaction and (b) shell model embedded with continuum.}
\label{fig:2_SMEC}
\end{minipage}
\end{center}
\end{figure}

Due to the excellent analytical and numerical properties of the HO basis, the framework described above combines the structural information and decay properties of diproton decay efficiently. However, for a general problem of $2p$ radioactivity, some drawbacks might be critical:
\begin{itemize}
\item
    The structural and decay aspects of the problem are not treated on the same footing. Due to the impact of final-state interaction, the internal and external configurations might be different;
\item
    The diproton decay is an approximation for the $2p$ decay process~\cite{Delion:2013,Grigorenko:2018}, and described as spin-singlet nucleon pair that might not be spatially close in the CI framework. The use of COSM coordinates also makes it hard to capture the proper three-body asymptotic behavior for the $2p$ decay;
\item
    The $2p$ emitter is an unbound system, whose decay process is strongly impacted by the continuum effect. However, the orbitals generated by HO potential are always bound and fully localized as shown in Fig.\,\ref{fig:2_SMEC}. It might fail when dealing with these open quantum systems near the dripline.
\end{itemize}

In order to deal with these problems, new methods have been developed, such as the hybrid model~\cite{Quaglioni:2013,Quaglioni:2016,Brown:2019} and continuum-embedded framework~\cite{Rotureau:2005,Michel:2021}. The former is focused on the mixing between the configuration interaction and three-body decaying dynamics, while the latter is trying to gain insight into the continuum effect in these open quantum systems.

\subsubsection{\it Hybrid model}

The formulae of diproton decay in Section\,\ref{sec:2_diproton} are based on the $R$-matrix theory with that the valence protons are emitted as a cluster, which neglects the three-body dynamics, such as the simultaneous emission with large opening angles. To this end, a hybrid model has been developed~\cite{Brown:2019}, which takes advantage of the detailed structure information from CI and the partial decay width of each channel from the three-body framework~\cite{Grigorenko:2003,Grigorenko:2003b}.

In Ref.\,\cite{Grigorenko:2003b}, a kind of ``three-body penetrability'' was calculated by assuming different dominating internal configurations forming a ``corridor'' of values. Consequently, the total $2p$ decay width $\Gamma_{2p}$ were approximately decomposed into different channels $\ell^2$, as
\begin{equation}
\label{eq:2_hybrid_model}
\Gamma_{2p} \approx \sum_{\ell} S(\ell^2)\Gamma(\ell^2),
\end{equation}
where $S(\ell^2)$ is the weight (or spectroscopic factor if neglecting the preformation factor of the daughter nucleus) of the $\ell^2$ configuration. This analogy could work well if the configuration dominating the internal region is responsible for the decay (with the largest partial decay width). However, the situation becomes complicated when the valence protons are in competitive configurations. In this case, different configurations are connected through the final-state interaction under the decay dynamics of a three-body system, and the valence protons tend to ``penetrate'' from high-$\ell$ to low-$\ell$ orbitals (see Section\,\ref{sec:2_three_body} for details), which would benefit the tunneling process. It indicates that this $\Gamma(\ell^2)$ ideology can not fully take into account the effect of final-state interaction, since the internal structure (in the form of spectroscopic factors) is still disentangled from the decay dynamics in the hybrid model.

Utilizing the estimated partial decay width $\Gamma(\ell^2)$, one can directly adopt the TNA calculated by Eq.\,(\ref{eq:2_TNA}) without transforming to Jacobi-T coordinates. For simplicity, only the case with $J^\prime$ = $J$ = $J_o$ = 0 and $q_a$ = $q_b$ = $q$ has been considered in Ref.\,\cite{Brown:2019}. The total $2p$ decay width can be estimated in two extreme conditions:
\begin{itemize}
\item
    {\it Incoherent sum}.--- Following the assumption described above~\cite{Grigorenko:2003b}, $\Gamma_{2p}$ with structural information embedded can be estimated by replacing $S(\ell^2)$ in Eq.\,(\ref{eq:2_hybrid_model}) with $\left[\mathrm{TNA}\left(q^{2}\right)\right]^{2}$;
\item
    {\it Coherent sum}.--- In this case, all amplitudes combine coherently, as in two-nucleon transfer reactions. This gives
    \begin{equation}
    \Gamma_{2p} \approx \left[\sum_{q} \mathrm{TNA}\left(q^{2}\right)\sqrt{\Gamma\left(\ell^{2}\right)}\right]^2 ,
    \end{equation}
    in which the phase is taken as positive for all terms.
\end{itemize}

\begin{table} [h]
\begin{threeparttable}
\caption{The partial $2p$ half-lives (in ms, ${ }^{19} \mathrm{Mg}$ in picoseconds) calculated by the hybrid model with the {\it incoherent} and {\it coherent} sum of the different amplitudes contributing to the emission process. As a comparison, the experimental $2 p$ emission half-lives (in ms, ${ }^{19} \mathrm{Mg}$ in picoseconds) are also listed. For the four heavier nuclei, the calculations are obtained with and without the contributions from the $s^{2}$ configuration. See Ref.\,\cite{Brown:2019} for details.}
\label{tab:2_hybrid_model}
\vspace{0.5\baselineskip}
\begin{tabular}{ l  c  c  c  c  c }
  \hline
  \hline
  \\[-10pt]
    Nucleus & $T^{2p}_{1/2}$ &  \multicolumn{2}{c}{ $T^{2p}_{1/2}$ without $s^2$ } &  \multicolumn{2}{c}{ $T^{2p}_{1/2}$ with $s^2$ }\\
    $J^\pi$ & Expt. &  Incoherent  & Coherent &  Incoherent  & Coherent\\
    \hline \\[-10pt]
    $^{19}$Mg 1/2$^-$ & 4.0(15) & & & 0.73$^{+1.5}_{-0.17}$ & 0.20$^{+0.40}_{-0.05}$ \\
    $^{45}$Fe 3/2$^+$ & 3.6(4) & 20(8) & 6.6(26) & 5.9(24) & 1.8(7) \\
    $^{48}$Ni 0$^+$ & 3.0(22) & 5.1(29) & 1.8(11) & 1.3(6) & 0.43(22) \\
    $^{54}$Zn 0$^+$ & 1.7(7) & 1.8(8) & 0.9(4) & 1.7(8) & 0.6(3) \\
    $^{67}$Kr 3/2$^-$ & 20(11) & 850(390) & 320(140) & 820(380) & 250(110) \\
    $^{67}$Kr 1/2$^-$ & 20(11) & 904(420) & 290(130) & 940(430) & 360(160) \\
  \hline
  \hline
\end{tabular}
\end{threeparttable}
\end{table}

Within this framework, the half-lives of $2p$ emitters have been evaluated (see Table\,\ref{tab:2_hybrid_model} and Ref.\,\cite{Brown:2019}). The obtained results incorporating three-body dynamics are substantially improved and agree well with the experimental data. Meanwhile, as discussed in Ref.\,\cite{Brown:2019}, it is crucial to include some $s^2$ or other low-$\ell$  components in the decay (owing to their small centrifugal barriers). These components are  often introduced via continuum coupling, three-body dynamics, and core/cross-shell excitations. Meanwhile, the half-lives of  $^{67}$Kr retain a large discrepancy between the experimental data and theoretical predictions \cite{Brown:2019}. As a recently observed $2p$ emitter, the half-life of $^{67}$Kr is systematically over-estimated by various theoretical models \cite{Grigorenko:2003,Goncalves:2017}. It seems that such a short lifetime can only be reproduced by using a large amount of $s$- or $p$-wave components \cite{Brown:2019,Grigorenko:2017}, while Refs.\,\cite{Goigoux:2016,Wang:2018} suggested there might be a deformation effect. This would be interesting for further theoretical or experimental studies.

There are other approaches combining structure and reaction aspects starting from {\it ab initio} framework \cite{Elhatisari:2015,Navratil:2016,Kumar:2017}. In particular, in Refs.\,\cite{Quaglioni:2013,Quaglioni:2016}, each component of the three-body system was calculated using the no-core shell model (NCSM) in Jacobi coordinates. The inter-cluster motion is described using the resonating group method (RGM), which has been widely used in nuclear reactions. Although this framework has not been applied to the study of $2p$ decay yet, it is promising that the method can describe the configuration mixing and three-body dynamics in a robust way.

\subsubsection{\it Continuum-embedded framework}
\label{sec:2_continuum_CI}

The $2p$ emitters are open quantum systems located beyond the proton-dripline, whose decay properties are strongly impacted by the low-lying continuum. In order to properly consider this effect in the CI framework, one has to bypass the localized HO basis. To this end, many continuum-embedded frameworks have been developed~\cite{Michel:2002,Bennaceur:2000,Okolowicz:2003,Volya:2006,Volya:2009,Hagen:2012,Papadimitriou:2013}. Among them, the shell model embedded in the continuum (SMEC)~\cite{Bennaceur:2000,Okolowicz:2003} and Gamow shell model (GSM)~\cite{Michel:2002} have been successfully applied to the study of $2p$ decay~\cite{Rotureau:2005,Rotureau:2006,Michel:2021}.

{\it Shell model embedded in the continuum.}--- In the framework of SMEC~\cite{Rotureau:2005,Rotureau:2006}, the Hilbert space is divided into two orthogonal subspaces: ${\cal Q}\equiv\left\{\ket{\Psi_i^{\rm CI}}\right\}$ and ${\cal P}\equiv\left\{\ket{\xi_E}\right\}$ as shown in Fig.\,\ref{fig:2_SMEC}b. Based on the Feshbach projection technique, $\Psi_i^{\rm CI}$ contains discrete CI states localized inside the nucleus, and $\xi_E$ are scattering states. The corresponding projection operators are
\begin{equation}
\label{eq:2_PQ_space}
\begin{aligned}
    \hat{Q} &=\sum_{i} \ket{\Psi_i^{\rm CI}}\bra{\Psi_i^{\rm CI}},\\
    \hat{P} &=\int_0^\infty dE \ket{\xi_E}\bra{\xi_E}.
\end{aligned}
\end{equation}
Consequently, an open quantum system  description of ${\cal Q}$  includes couplings to the environment of decay channel through the energy-dependent effective Hamiltonian:
\begin{equation}
{\cal H}(E)=\hat{Q} H \hat{Q} +W_{{\cal Q}{\cal Q}}(E) \ ,
\end{equation}
where $\hat{Q} H \hat{Q}$ denotes the standard configuration-interaction (shell-model) Hamiltonian describing the internal dynamics in the closed quantum system approximation, and $W_{{\cal Q}{\cal Q}}(E)$  the energy-dependent continuum coupling term with scattering energy $E$. For the situation with one particle in the scattering continuum,
\begin{equation}
W_{{\cal Q}{\cal Q}}(E)=\hat{Q} H\hat{P_1}\cdot \hat{G}_{{P_1}} \cdot \hat{P_1}H\hat{Q} \ ,
\end{equation}
where $P_1$ is the projection operator for the one-nucleon scattering subspace, and $\hat{G}_{{P_1}}$ is the corresponding one-nucleon Green's function. Following the same methodology, one can obtain the effective Hamiltonian with two particles in the continuum~\cite{Rotureau:2006}. As a result, ${\cal H}(E)$ can take into account all possible emissions for two protons as well as one proton implicitly. Usually, the sequential $2p$ emission may occur either through the resonance or the correlated continuum of an intermediate nucleus, while the one-nucleon scattering subspace is less important for a true $2p$ emission. Therefore, by analyzing the couplings between ${\cal Q}$ and ${\cal P}_{1,2}$ subspaces, one can estimate whether the valence protons of a $2p$ emitter decay sequentially or simultaneously~\cite{Rotureau:2005,Rotureau:2006}.

{\it Gamow shell model.}--- Defined in the rigged Hilbert space, GSM has also been successfully used for studies of  weakly bound and unbound states in dripline nuclei. Similar to SMEC, both frameworks describe the nucleus as a core surrounded by valence nucleons, but they treat the coupling to the unbound  continuum space differently.

\begin{figure}[tb]
\begin{center}
\begin{minipage}[t]{17 cm}
\includegraphics[width = \columnwidth]{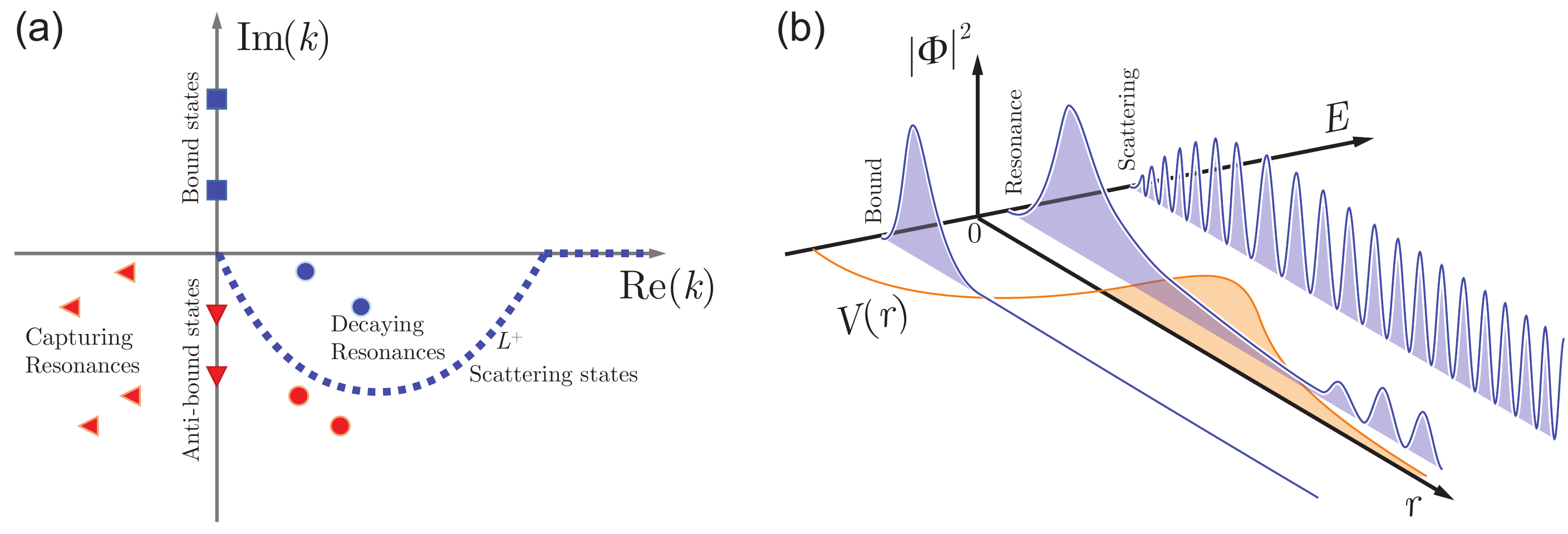}
\end{minipage}
\begin{minipage}[t]{17 cm}
\caption{(Color online) The schematic figure of (a) the $S$-matrix poles with Berggren ensemble in complex momentum plane and (b) the typical wave functions of bound, resonance, and scattering states.}
\label{fig:2_Berggren}
\end{minipage}
\end{center}
\end{figure}

In the GSM, the continuum effects are automatically taken into account by utilizing the Berggren ensemble \cite{Berggren:1968} that contains resonant (bound and decaying) and scattering states (see Fig.\,\ref{fig:2_Berggren}). The completeness relation for the Berggren ensemble can be written as:
\begin{equation}
\label{eq:2_Berggren_ensemble}
\sum_{\mathfrak n} \ket{u_{\mathfrak n}} \bra{u_{\mathfrak n}} + \int_{L^+} \ket{u(k)} \bra{u(k)}~dk = \mathbf{\hat{1}},
\end{equation}
where $\ket{u}$ is a one-body state, and ${\mathfrak n}$ denotes bound states and  decaying resonant (or Gamow)  states lying between the real-$k$ momentum axis in the fourth quadrant of the  complex-$k$ plane. The  $L^+$ contour represents the complex-$k$ scattering continuum, and can be chosen arbitrarily as long as it encompasses the resonances of interest. If the contour $L^+$ is chosen to lie along the real $k$-axis, the Berggren completeness relation reduces to the Newton completeness relation \cite{Newton:1982} involving bound and real-momentum scattering states. The matrix elements of Hamiltonian are calculated using the exterior complex scaling~\cite{Gyarmati:1971} with the Berggren basis. By diagonalizing the complex symmetric Hamiltonian, energies and decay widths are obtained simultaneously as the real and imaginary parts of the complex eigenenergies.

In principle, similar to SMEC, GSM includes all possible emissions for two protons. However, it is not easy to extract the $2p$ decay width from the total one. A convenient way proposed in Ref.\,\cite{Michel:2021} is that the decay mechanism of $2p$ can be roughly estimated by analyzing the total width as a function of the decay energy $Q$ or the related Hamiltonian parameters, since the thresholds of one- and two-proton decay channels vary differently, so are their $G$-factors as shown in Eq.\,(\ref{eq:1_Gamow}). Based on this idea, $^{16}$Ne and $^{18}$Mg have been investigated~\cite{Michel:2021}. The former is a well-known $2p$ emitter studied by many theoretical approaches, while the latter was recently discovered to have a possible $2p$+$2p$ decay mode (see Section\,\ref{sec:6_4p} and Refs.\,\cite{Michel:2021,Jin:2022} for details).

\begin{figure}[tb]
\begin{center}
\begin{minipage}[t]{15 cm}
\includegraphics[width =\columnwidth]{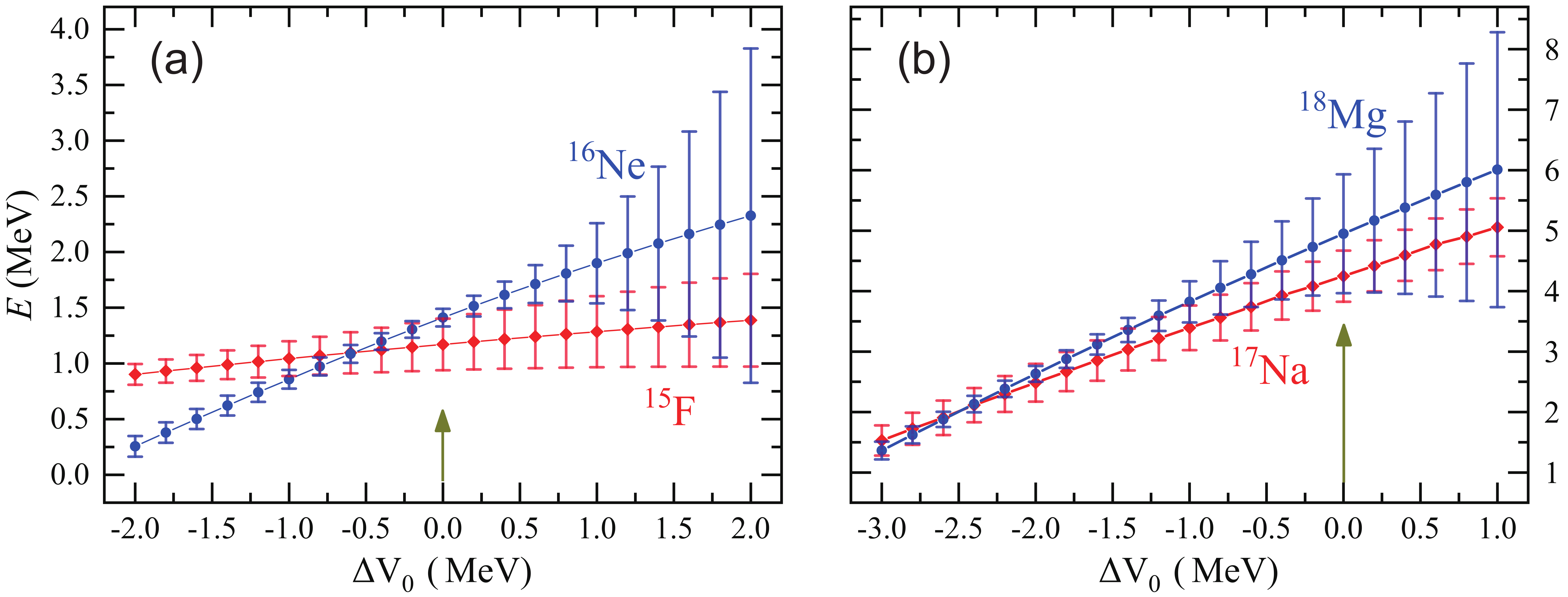}
\end{minipage}
\begin{minipage}[t]{17 cm}
\caption{(Color online) GSM calculated energies and widths (in MeV) of $^{15}$F and $^{16}$Ne (left panel) and of $^{17}$Na and $^{16}$Mg (right panel) as a function of the difference $\Delta V_0 = V_0 - V_0^{\text{(fit)}}$ of the Woods-Saxon central potential depths (see Ref.\,\cite{Michel:2021} for definition). Energies are depicted by blue disks and red diamonds for even and odd nuclei, respectively. Widths are represented by segments centered on disks and diamonds. The widths of $^{16}$Ne and $^{18}$Mg have been multiplied by 20 for readability. Energies are given with respect to the $^{14}$O core. The physical GSM calculation, for which $V_0 = V_0^{\text{(fit)}}$, is indicated by an arrow.
Reprinted with permission from Ref. \cite{Michel:2021}. Copyright (2021) by the American Physical Society.}
\label{fig:2_GSM}
\end{minipage}
\end{center}
\end{figure}

Figure\,\ref{fig:2_GSM} shows the calculated decay energies and widths of the mother and intermediate nuclei as a function of one-body potential depth. As $^{16}$Ne and $^{18}$Mg possess one additional valence proton compared to $^{15}$F and $^{17}$Ne, the binding energy of the former nuclei increases faster with central potential depth than that of the latter. Consequently, $Q_{2p}$ and $Q_{p}$ would change with different speeds. Since the one-proton decay channel usually has a smaller Coulomb barrier compared to the $2p$ one, it is more sensitive to the $Q$ value. It can be noticed that, as $\Delta V_0$ increases, the total decay width has different behaviors. This indicates a transition from true $2p$ decay to the situation where a one-proton decay channel becomes dominated. Since the only requirement is the total decay width, this method is convenient to estimate the decay mechanism, but the behavior of the total decay width needs to be carefully analyzed due to the non-linear property of the $G$-factor.

\subsection{\it Decay mechanism and three-body character}
\label{sec:2_three_body}

Besides the decay width or half-life, another important feature in $2p$ emission is the decay mechanism and dynamics. As discussed above, nuclear decay with three fragments in the final state is a very exotic process. This behavior depends on the status  of the valence protons in the mother nucleus, as well as the property of the neighboring nucleus (see Fig.\,\ref{fig:1_EnergyConditions}). Moreover, it is strongly influenced by an interplay between the internal and external mixing~\cite{Rotureau:2005}, which invalidates an idealized picture of an independent decay mode associated with the pairing field. To gain more sight of this dynamics, the few-body framework could be a proper tool~\cite{Grigorenko:2002,Delion:2013,Braaten:2006,Baye:1994,Hagino:2005,Hagino:2016,Hagino:2016b,Grigorenko:2009}. There is a long history of the developments of the general few-body theory~\cite{TAMURA:1965,Thompson:2012,Fick:1978,Ikeda:1977,Descouvemont:2010}, which has been proven to be very useful for studying the nuclear reaction as well as the decay mechanism. While such models provide a  ``lower resolution'' picture of the nucleus, they can be helpful when interpreting experimental data, providing guidance for future measurements and more microscopic approaches. Here we review some interesting models in the domain that have been  widely used for revealing the valence-proton configuration and the corresponding nucleon-nucleon correlations.

During the decay process, the $2p$ emitter can be treated as a three-body system if neglecting the preformation of the daughter nucleus. Based on these degrees of freedom, the Hamiltonian can be written as:
\begin{equation}
    \hat{H} = \sum^3_{i=1}\frac{ \hat{\vec{p}}^2_i}{2 m_i} +\sum^3_{i>j=1} V_{ij}(\vec{r}_{ij})-\hat{ T}_{\rm c.m.},
\end{equation}
where $V_{ij}$ is the pairwise interaction between clusters $i$ and $j$, including nuclear and Coulomb interactions. In some cases, a phenomenological three-body force $V_3$ is introduced to reproduce the binding energy and spectrum of the system. $\hat{T}_{\rm c.m.}$ stands for the kinetic energy of the center-of-mass, and can be expressed differently in Jacobi~\cite{Grigorenko:2000} and COSM~\cite{Hagino:2005} coordinates. The treatment and properties of COSM coordinates have been discussed in Sections~\ref{sec:2_degree} and \ref{sec:2_structure}. Jacobi coordinates can be expressed as  T- and Y-types, each is associated with a complete basis set. In practice, there are two ways to deal with Jacobi coordinates:
\begin{itemize}
\item
    {\it Faddeev equations} can be defined in momentum~\cite{Faddeev:1960} and coordinate~\cite{Bang:1992} spaces, where the three Faddeev components are expanded on the partial angular momenta related to the Jacobi coordinates. As a result, the valence-proton wave function $\Phi^{J_{p} \pi_{p}}$ in Eq.\,(\ref{eq:2_WF_3b}) can be constructed with three components related to the three sets of Jacobi coordinates~\cite{Nielsen:2001,Thompson:2004}, i.e.
    \begin{equation}
    \label{eq:2_Faddeev}
    \Phi^{J_{p} \pi_{p}} = \sum_{i=1}^3 \Phi_i^{J_{p} \pi_{p}}(\vec{x}_i,\vec{y}_i),
    \end{equation}
    where the components $\Phi_i$ are functions of their own ``natural'' Jacobi coordinate pairs $i$ (see Fig.\,\ref{fig:2_coordinates}), and solutions of the Faddeev coupled equations~\cite{Thompson:2004};
\item
    {\it Hypershperical harmonics (HH) technique} is constructed from five-dimensional hyperangular coordinates $\Omega = \{\arctan(y/x),\Omega_{x},\Omega_{y}\}$ and a hyperradial coordinate $\rho=\sqrt{x^2 + y^2}$~\cite{Garrido:2004,Garrido:2020}. The eigenfunctions of the hyperangular momentum operator are $\mathcal {Y} ^{JM}_{\gamma K} (\Omega)$, where $K$ is the hyperspherical quantum number and $\gamma$ is a set of quantum numbers other than $K$. Utilizing $\mathcal {Y} ^{JM}_{\gamma K} (\Omega)$, the  matrix elements in different sets of Jacobi coordinates can be transformed to one single set through the Raynal-Revai coefficient~\cite{Raynal:1970}, noticing that hyperradius $\rho$ is transformation invariant. Expressed in HH, the total wave-function can be written as \cite{Descouvemont:2003}:
    \begin{equation}
    \label{eq:2_HH}
        \Phi^{J_{p} \pi_{p}} = \rho ^{-5/2} \sum_{\gamma K} \psi ^{J_{p} \pi_{p}}_{\gamma K}(\rho) \cdot \mathcal {Y} ^{J_{p}M}_{\gamma K} (\Omega),
    \end{equation}
    where $\psi ^{J_{p} \pi_{p}}_{\gamma K}(\rho)$ is the hyperradial wave function.
\end{itemize}

The HH framework provides a useful way to analyze the three-body asymptotic behavior. The resulting  Schr\"{o}dinger equation for the hyperradial wave functions can be written as a set of coupled-channel equations:
\begin{equation}
\label{eq:2_Couple_channel}
   \left[ -\frac{\hbar^2}{2m}\left(\frac{d^2}{d\rho^2} - \frac{(K+3/2)(K+5/2)}{\rho^2} \right)-{E} \right] \psi_{\gamma K}(\rho) +\sum_{\xi\equiv \gamma^\prime,K^\prime}\int V^\xi_{\gamma K}(\rho,\rho^\prime)\psi_{\xi}(\rho^\prime)d\rho^\prime=0,
\end{equation}
where $V^\xi_{\gamma K}$ is the effective interaction expanded on the HH basis, and contains local ($\rho$ = $\rho^\prime$) and non-local ($\rho \neq \rho^\prime$) parts. The non-local potential can be generated by the antisymmetrization between core and valence particles. Consequently, some Pauli forbidden states would be introduced, and can be roughly eliminated using the projection operator \cite{Descouvemont:2003,Saito:1969,Kuk:1978} or the supersymmetric transformation method~\cite{Thompson:2004,Sparenberg:1997,Thompson:2000}.

For chargeless particles, $V^\xi_{\gamma K}$ contains only the short-range nuclear force. Therefore, for a decaying system whose three components are spatially apart in the final state, the asymptotic behavior is dominated by the first term of Eq.\,(\ref{eq:2_Couple_channel}), and the asymptotic wave function of a resonance is proportional to the outgoing Hankel function. However, for the charged particles, the presence of the long-range Coulomb potential in a three-body system could mix different channels, which makes the problem theoretically and numerically more complicated~\cite{Grigorenko:2009d,Vasilevsky:2001,Grigorenko:1999}. This influence also manifests itself in the asymptotic nucleon-nucleon correlations, especially for the angular correlation (see the discussion in Section\,\ref{sec:2_correlation} and Refs.\,\cite{Grigorenko:2009b,Wang:2021} for details). In order to extract the correct information from emitted particles, precise three-body solutions at very large distances are required.

From Eq.\,(\ref{eq:2_Couple_channel}), it can be noticed that the centrifugal barrier is determined by the hyperspherical quantum number $K$ in a three-body system. When approaching the threshold, the presence of  the centrifugal barrier is crucial for the tunneling process, and is expected to give raise to  changes in asymptotic correlations. This would rise some interesting phenomena in extreme conditions. For example, when the decay energy $Q_{2p}$ is very small, the $2p$ radioactivity is dominated by the $K=0$ component, which corresponds to the $s$-wave ($\ell_x=\ell_y=0$), while other components are strongly suppressed. Consequently, the energy distribution approaches the universal phase-space limit~\cite{Grigorenko:2003,Grigorenko:2018,Wang:2022}:
\begin{equation}
\label{eq:2_phase_space_limit}
    \frac{d\sigma}{d\varepsilon} \sim \sqrt{\varepsilon(1-\varepsilon)}~~{\rm with}~~\varepsilon=E_{nn}/Q_{2n}.
\end{equation}
At the same time, the angular distribution becomes essentially isotropic. This is a universal property for both $2p$ and $2n$ radioactivity. As the energy of the resonance increases, asymptotic energy and angular correlations quickly start deviating from the phase-space limit shown in Eq.\,(\ref{eq:2_phase_space_limit}). This could provide useful information for testing theoretical frameworks and experimental measurements.

Normally, the numerical results of the three-body method converge fast as the hyperspherical quantum number $K$ increases. However, in some special cases, a large $K$ is required, which will dramatically increase the model space due to the various combinations of angular momentum coupling. Some methods have been proposed to reduce the model space, such as adiabatic approximation~\cite{Nielsen:2001,Lin:1995,Esry:1996,Suzuki:2015} or $G$-matrix~\cite{Grigorenko:2007,Grigorenko:2009}. Both of them treat the kinetic energy and potential adiabatically, and fold the matrix elements into the subspace. A correction with non-adiabatic coupling terms has also been made for the former method~\cite{Suzuki:2015}.

If one considers bound three-body systems, many few-body models have proven to be very useful to analyze the properties of the valence nucleons~\cite{Braaten:2006}, especially models based  on the Lagrange-mesh technique~\cite{Baye:1994} or COSM~\cite{Suzuki:1988}. However, for the description of  resonances and corresponding decay properties, the low-lying continuum should be properly taken into account.  To this end, following the idea introduced in Section\,\ref{sec:2_decay}, many theoretical frameworks have been developed for the few-body system. Here we briefly introduce some of them, which have been successfully applied to studies of two-nucleon decay.

{\it Three-body reaction framework} --- It follows the idea of $R$-matrix theory, and is based on the assumption that the configuration space can be divided into two regions: an internal region where the solution is dominated by nuclear structure aspects, and an external region where the wave function $\Psi^{\rm ext}$ is with a pure outgoing boundary condition for a three-body system. The internal wave function $\Psi^{\rm int}$ can be determined through two kinds of approaches:
\begin{itemize}
\item
    As proposed in Refs.\,\cite{Grigorenko:2000,Grigorenko:2017}, $\Psi^{\rm int}$ can be approximately obtained by
    \begin{equation}
        (\hat{H}-E)\Psi^{\rm int}=0,
    \end{equation}
    and fine-tuned through a kind of perturbative procedure
    \begin{equation}
        (\hat{H}-E)\Psi^{\rm ext}=-i(\Gamma/2)\Psi^{\rm int};
    \end{equation}
\item
    $\Psi^{\rm int}$ can also be approximated by the inhomogeneous Bloch-Schr\"{o}dinger equation~\cite{Descouvemont:2006,Damman:2009,Descouvemont:2010,Lovell:2017}
    \begin{equation}
    \begin{aligned}
    (&\hat{H}+\hat{\mathcal{L}}-E) \Psi^{\rm int}=\hat{\mathcal{L}} \Psi^{\rm ext},\\
    \hat{\mathcal{L}}=\frac{\hbar^{2}}{2 m} &\sum_{\gamma K}\ket{\mathcal{Y}_{\gamma K}^{J M}} \frac{\delta(\rho-R_0)}{\rho^{5 / 2}}\frac{\partial}{\partial \rho} \rho^{5 / 2}\bra{\mathcal{Y}_{\gamma K}^{J M}},
    \end{aligned}
    \end{equation}
    where $R_0$ is the channel radius.
\end{itemize}
Both approaches are complemented with the continuity condition of the wave function $\Psi^{\rm int}(R_m) = \Psi^{\rm ext}(R_m)$ and of its first derivative.

{\it Green's function method} --- Ref.\,\cite{Hagino:2016} shows that the decay energy spectrum $dP/dE$ can be accessed through the Green's function:
\begin{equation}
\frac{d P}{d E}=\frac{1}{\pi} {\rm Im}\braxket{\Phi_{\rm ref}}{\hat{G}}{ \Phi_{\rm ref}},
\end{equation}
where $\hat{G}$ is a two-body Green's function, and $\Phi_{\rm ref}$ is a reference state. $\Phi_{\rm ref}$ can be approximated with the bound analog state of the neighboring isotope (isotone), which captures the main components inside the nucleus for the resonance of interest.

{\it Complex-plane framework} --- As shown in Section\,\ref{sec:2_decay}, in the rigged Hilbert space, one can extract the decay width by the imaginary part of the complex energy. Based on this idea, the complex-scaling method~\cite{Aoyama:2006,Kruppa:2014} is one of the commonly used frameworks. By rotating the Hamiltonian in a complex plane, one can make the wave function $\Psi$ converge at the asymptotic region, which can be used to handle the system with a relatively large width.

Recently, another complex-plane framework Gamow coupled-channel (GCC) method has been developed~\cite{Wang:2017b} utilizing the Berggren ensemble technique (see Section\,\ref{sec:2_continuum_CI}). Since the Berggren basis contains bound, resonant, and scattering states, it allows describing the structure and decays of three-body systems on the same footing. The GCC method has been benchmarked with GSM~\cite{Wang:2017b}, and applied to unraveling the intriguing features of several $2p$ emitters, including extremely unstable $^{11}$O \cite{Wang:2018,Wang:2021,Wang:2022,Webb:2019,Webb:2019a,Webb:2020,Wang:2019}.

Using the methods above, the halflives and other decay properties can be extracted within the three-body framework, which is useful for revealing the status of a $2p$ emitter. The emitted protons carry invaluable information about the internal structure, nucleon-nucleon correlation, and properties of open quantum systems. Since Section\,\ref{sec:2_theory} is focused on the theoretical frameworks, in the following, we will briefly introduce some general observables and properties that can be provided in the three-body method, and leave physics discussions for the other Sections.

\subsubsection{\it Valence-proton configuration}

\begin{figure}[tb]
\begin{center}
\begin{minipage}[t]{15 cm}
\begin{center}
\includegraphics[width =\columnwidth]{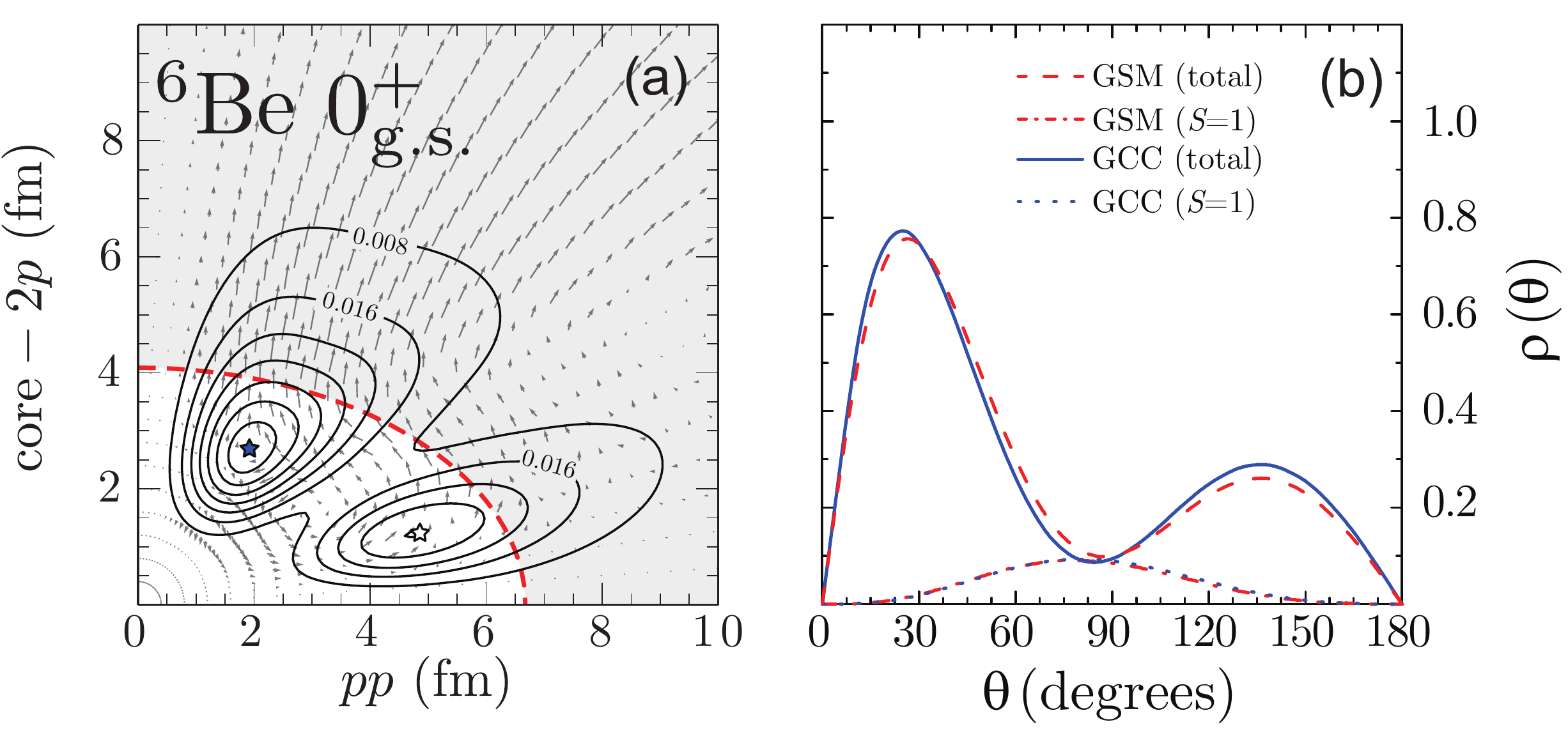}
\end{center}
\end{minipage}
\begin{minipage}[t]{17 cm}
\caption{(Color online) The calculated (a) $2p$ density distribution (marked by contours)  and $2p$ flux (denoted by arrows) in the ground state of $^{6}$Be in Jacobi coordinates $pp$ and ${\rm core}-pp$. The thick dashed line marks the inner turning point of the Coulomb-plus-centrifugal barrier.
The maxima marked by filled and open stars correspond to diproton and cigarlike structures, respectively. (b) Two-nucleon angular densities (total and spin-triplet $S=1$ channels) in the ground-state configurations of $^6$Be, as obtained in GCC and GSM. Part (a) is adopted from Ref. \cite{Wang:2019}
and part (b) from Ref. \cite{Wang:2017b}.}
\label{fig:2_6Be_structure}
\end{minipage}
\end{center}
\end{figure}

The advantage of the three-body models described in this Section\,\ref{sec:2_three_body} is that it can easily capture the configurations and correlations of valence protons~\cite{Hagino:2016,Wang:2019,Casal:2019}. For example, as a typical light-mass $2p$ emitter, $^6$Be has been considered to feature a ``democratic'' decay mode (see Section\,4.1 for details), attributable to the large width of the ground state of its neighboring nucleus $^5$Li ($Q_p$ = 1.97\,MeV, $\Gamma$ = 1.23\,MeV). The density distribution of $^6$Be has been studied in various few-body models; it shows two maxima associated with diproton and cigarlike configurations (see Fig.\,\ref{fig:2_6Be_structure}), which is due to the fact that the valence protons mainly occupy $p$-wave. Meanwhile, the nuclear interaction is attractive at the mid-range, where the diproton configuration is usually more pronounced than the others. The formation of a dinucleon structure also requires the mixing of orbitals with different parities and angular momentum. Therefore, the model space can not be limited to one single shell, and the presence of the low-lying continuum is crucial.

Moreover, the configuration evolution of the internal structure during decay can be analyzed by using the flux current $\vec{j} = {\rm Im}({\Psi}^\dagger \nabla \Psi )\hbar/m$~\cite{Wang:2019}, which shows how the two valence protons evolve within a given state wave function $\Psi$. As we see in the current field of $^{6}$Be (the small arrows in Fig.~\ref{fig:2_6Be_structure}), a competition between diproton and cigarlike configurations occurs inside the inner turning point of the Coulomb-plus-centrifugal barrier associated with the core-proton potential~\cite{Wang:2019}. Near the origin, the dominant diproton configuration tends to evolve toward the cigarlike configuration, owing to the repulsive Coulomb interaction and the Pauli principle. On the other hand, near the surface, the direction of the flux extends from the cigarlike maximum toward the diproton maximum, tunneling through the barrier. Moreover, at the peak of the diproton configuration (located near the barrier), the direction of the flux is almost aligned with the core-$2p$ axis, indicating a clear diproton-like decay~\cite{Wang:2017b,Grigorenko:2009b,Oishi:2014,Oishi:2017}. Meanwhile, the flux current of $^{12}$O indicates that there might be a competition between direct and ``democratic'' $2p$ decay, in which a significant part of the flux from the diproton configuration toward the cigarlike configuration persists up to the potential barrier and beyond~\cite{Wang:2019}.

For the $2p$ emitter with a daughter nucleus well preformed before the decay process, the three-body model can provide accurate information for the valence protons. Usefully, this is true for the weakly bound/unbound systems near the dripline, due to the ``alignment'' effect~\cite{Okolowicz:2020} that the near-threshold resonance would result in a single ``aligned eigenstate'' of the system carrying many characteristics of a nearby decay channel, while the valence protons are weakly coupled to the daughter nucleus~\cite{Wylie:2021}. However, sometimes the daughter nucleus can be exotic and contains mixed configurations. In this case, the lack of structural information regarding the daughter nucleus -- which is usually treated as a frozen core -- becomes one main drawback of the regular three-body model. Therefore, several efforts have been made toward a more microscopic framework on  few-body systems~\cite{Damman:2009,Nunes:1996,Nunes:1996b,Brida:2010,Nesterov:2010,Vasilevsky:2012}. For instance, Ref.\,\cite{Nunes:1996,Nunes:1996b} introduced a deformed core in the three-body method. Later on, microscopic cluster models, such as frameworks based on the generator coordinate method~\cite{Damman:2009} and the stochastic variational method~\cite{Brida:2010}, have been developed to study nuclear structure.

\begin{figure}[tb]
\begin{center}
\begin{minipage}[t]{15 cm}
\begin{center}
\includegraphics[width =\columnwidth]{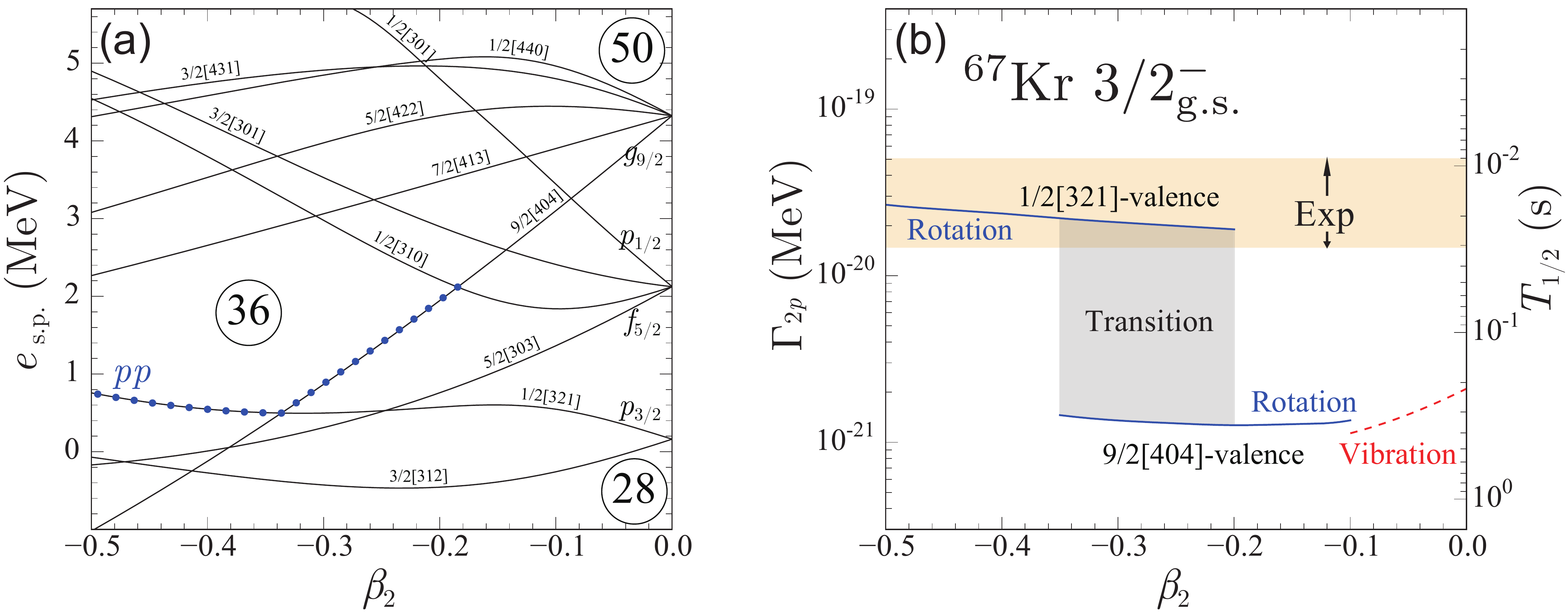}
\end{center}
\end{minipage}
\begin{minipage}[t]{17 cm}
\caption{(Color online) Left: Nilsson levels $\Omega$[Nn$_z\Lambda$] of the deformed core-$p$ potential as functions of the oblate quadrupole deformation $\beta_2$ of the core.
The dotted line indicates the valence level primarily occupied by the two valence protons.
Right: Decay width (half-life) for  the $2p$ ground-state radioactivity of $^{67}$Kr. The solid and dashed lines denote the results within the rotational and vibrational coupling, respectively.
The rotational-coupling calculations were performed by
assuming that the
1/2[321] orbital is either occupied by the core (9/2[404]-valence) or valence (1/2[321]-valence) protons. Figure is taken from Ref.~\cite{Wang:2018}.}
\label{fig:2_67Kr}
\end{minipage}
\end{center}
\end{figure}

For $2p$ emission, the GCC method has been extended to a deformed case~\cite{Wang:2018}, which allows a pair of nucleons to couple to the collective states of the core via nonadiabatic coupling. Consequently, the total wave function of the parent nucleus is the coupling between the valence protons $\Phi ^{J_p\pi_p}$ and the core $\phi ^{J_p\pi_p}$ as shown in Eq.\,(\ref{eq:2_WF_3b}). This method has been used to study the recently observed $2p$ emitter $^{67}$Kr (see Section\,5.4 and Ref.\,\cite{Goigoux:2016} for details), which illustrated how the deformation and exotic structure can impact the decay properties and valence configurations. Figure~\ref{fig:2_67Kr}(a) shows the proton Nilsson levels (labeled with asymptotic quantum numbers $\Omega$[Nn$_z\Lambda$]) of the Woods--Saxon core-$p$ potential. When the deformation of the core increases, a noticeable oblate gap at $Z=36$ opens up, attributable to the down-sloping 9/2[404] Nilsson level originating from the $0g_{9/2}$ shell. The structure of the valence proton orbital changes from the 9/2[404] ($\ell=4$) state at smaller oblate deformations to the 1/2[321] orbital, which has a large $\ell=1$ component. This transition can dramatically change the centrifugal barrier and decay properties of $^{67}$Kr, which results in a dramatic increase in the decay width.

\subsubsection{\it Nucleon-nucleon correlation}
\label{sec:2_correlation}

Another important feature of three-body decay is the nucleon-nucleon correlations. The correlated nucleon-nucleon  pairs  with large relative linear momenta reveal the nature of strong interaction at short distances \cite{Hen:2017}. On the other side of the spectrum, many low-energy properties, such as odd-even staggering of nuclear binding energies, of the atomic nucleus are profoundly affected by pair correlations of its nucleonic (protons and neutrons) constituents \cite{Brink:2005,Broglia:2013,Dean:2003}. The study of $2p$ decay could provide valuable information on the interplay between the short-range nuclear force and long-range electromagnetic interaction~\cite{Pfutzner:2012,Blank:2008,Goldansky:1960}.

At the initial stage of the $2p$ decay, the valence nucleons are highly correlated and spatially compact due to the presence of pairing condensate \cite{Hagino:2005,Pillet:2007,Hagino:2007,Matsuo:2012}. During the emission process, although the initial structure is largely distorted by the Coulomb repulsion and instability of the system, some information leaves an imprint on asymptotic nucleon-nucleon correlations. Of particular importance are the angular and energy correlations between emitted nucleons, which carry information on the interplay between structure and reaction aspects of the nuclear open quantum system~\cite{Wang:2021,Michel:2010,Casal:2021}. To analyze these properties, we briefly introduce these two kinds of nucleon-nucleon correlation, and review the corresponding theoretical developments.

{\it Internal correlation in coordinate space}.--- Similar to the valence-proton configuration, the spatial correlation reveals the structural information mainly inside the nucleus, which is governed by the strong nuclear force. The physics quantity can be accessed in both three-body~\cite{Hagino:2005} and CI~\cite{Papadimitriou:2011} frameworks through a normalized density distribution:
\begin{equation}
\begin{aligned}
\rho\left( r_{cp_1}, r_{cp_2}, \theta \right) &=  \left\langle\Psi\left|\delta\left(r_{cp_1}-r\right) \delta\left(r_{cp_2}-r^{\prime}\right) \delta\left(\theta_{12}-\theta\right)\right| \Psi\right\rangle;\\
\rho\left(\theta \right) &= \int 8\pi r_{cp_1}^2 r_{cp_2}^2\sin\theta\cdot  \rho\left( r_{cp_1}, r_{cp_2}, \theta \right) { d}r { d}r^\prime,\\
\end{aligned}
\end{equation}
where $r_{cp_1}$, $r_{cp_2}$, and $\theta_{12}$ are defined in Fig.~\ref{fig:2_coordinates}(b). It primarily represents the average positions and opening angles of the two valence protons inside the nucleus. Notably, the angular density is defined in coordinate space; it cannot be directly observed in experiments.

The spatial correlation  of the two valence neutrons in the g.s. of  $^{6}$Be is shown in Fig.\,~\ref{fig:2_6Be_structure}b. The distribution $\rho(\theta)$ shows two maxima~\cite{Wang:2017b,Kruppa:2014,Hagino:2016}. In accordance with the density distributions in Jacobi-T coordinates, the higher peak, at a small opening angle, can be associated with a diproton configuration. While, the second maximum, found in the region of large angles, represents the cigarlike configuration. Although Jacobi coordinates are more efficient to describe the diproton configuration, both the three-body model and GSM are able to capture the angular correlations inside the nucleus, which provide practical identical distributions in sufficiently large model space. One can analyze the inner structure of the valence protons by comparing the spatial correlation with analytical expressions. Taking the $(p_{3/2})^2$ or $(p_{1/2})^2$ configuration as an example, $\rho\left(\theta \right)$ is proportional to $\cos^{2} \theta\sin \theta$ and $\sin^{3} \theta$ for spin singlet and triplet states, respectively.

{\it Asymptotic correlation in momentum space}.--- The emitted protons become free particles when far away from the source prior to hitting the detector, which manifest themselves as plane waves in momentum space. As discussed, the asymptotic correlation of the emitted particles is important not only for the experimental accessibility but also for the valuable information it carries. It can be expressed in Dalitz plot and Jacobi coordinates. The former  is mainly focused on the energy correlation ($E_{pp}$, $E_{{\rm core}-p}$), while the latter  contains T-type ($\theta_k$, $E_{pp}$) and Y-type ($\theta^\prime_k$, $E_{{\rm core}-p}$) distributions (see Section\,\ref{sec:2_degree}). To model angular and energy correlations of emitted particles, precise three-body solutions at very large distances are required, and this poses a formidable challenge, especially in the presence of the long-range Coulomb interaction.

\begin{figure}[tb]
\begin{center}
\begin{minipage}[t]{15 cm}
\begin{center}
\includegraphics[width =\columnwidth]{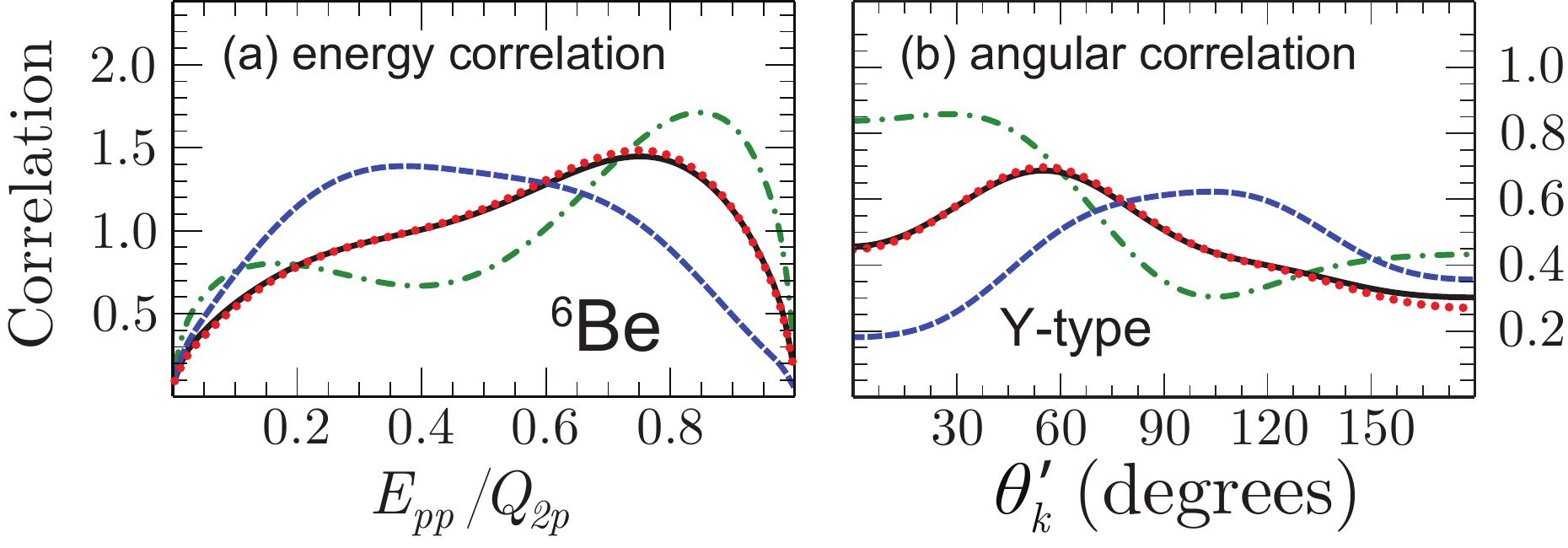}
\end{center}
\end{minipage}
\begin{minipage}[t]{17 cm}
\caption{(Color online) Calculated asymptotic (a) energy and (b) angular correlations of emitted nucleons  from the ground state of $^6$Be calculated at ${t=15}$\,pm/${c}$ with different strengths of  the Minnesota interaction \cite{Thompson:1977}:
standard (solid line), strong (increased by 50\%; dashed line),  and weak (decreased by 50\%; dash-dotted line). Also shown are the benchmarking results obtained within Green's function method (GF; dotted line) using the standard  interaction strength.
$\theta_k^\prime$ is the opening angle between $\vec{k}_{x^\prime}$ and $\vec{k}_1$ in the Jacobi-Y coordinate system, and $E_{pp/nn}$ is the kinetic energy of the relative motion of the emitted nucleons. Figure is taken from Ref.~\cite{Wang:2021}.}
\label{fig:2_Correlation_A6}
\end{minipage}
\end{center}
\end{figure}

Currently, two methods are mainly used to extract the asymptotic correlation for $2p$ decay:
\begin{itemize}
\item
    The first method, as shown in Refs.\,\cite{Grigorenko:2003,Grigorenko:2009}, is to extract the status of the emitted protons by analyzing the flux current at a large distance. In the HH framework, for a three-body decay where all inter-particle distances are large, the flux $\vec{j}$ is independent on the hyperradius $\rho$. In this case, the differential flux at $\rho_{\max }$ becomes identical to the momentum distribution:
    \begin{equation}
        \frac{d \vec{j}\left(\rho_{\max }, \Omega\right)}{d \Omega}=\left.\frac{\hbar^2}{m} \operatorname{Im}\left[\Psi^{\dagger} \rho^{5 / 2} \frac{d}{d \rho} \rho^{5 / 2} \Psi\right]\right|_{\rho=\rho_{\max }} \rightarrow \frac{d \vec{j}\left(\Omega_{k}\right)}{d \Omega_{k}},
    \end{equation}
    where $\Omega_{k} = \{\arctan(k_y/k_x),\Omega_{k_x},\Omega_{k_y}\}$ is the hyperangle in momentum space. To obtain a proper momentum distribution, this method also requires a precise treatment of the asymptotic wave function that has been suppressed and distorted by the long-range Coulomb force.
\item
    An alternative way of tackling the two-nucleon decay is the time-dependent formalism, in which measured inter-particle correlations can be interpreted  in terms of  solutions propagated for long durations. The initial state is a wave packet, and propagated by the time evolution operator (see Section\,\ref{sec:2_time_evolution} for details) which can be expanded on the eigenbasis of the original Hamiltonian~\cite{Oishi:2014,Oishi:2017} or Chebyshev polynomials~\cite{Volya:2009,Loh:2001,Wang:2021}:
    \begin{equation}
        e^{-i \frac{\hat{H}}{\hbar} t}=\sum_{{\mathfrak n}=0}^{\infty}(-i)^{\mathfrak n}\left(2-\delta_{{\mathfrak n} 0}\right) J_{\mathfrak n}(t) T_{\mathfrak n}(\frac{\hat{H}}{\hbar}),
    \end{equation}
    where $J_{\mathfrak n}$ are the Bessel functions of the first kind and $T_{\mathfrak n}$ are the Chebyshev polynomials. Since the evolving wave packet has an implicit cutoff at large distances, the unwanted reflection at the boundary can be avoided, as well as the divergence of the Coulomb interaction in the momentum space~\cite{Wang:2021}. The time-dependent approach is suitable for these short-lived systems, in which decay dynamics can be demonstrated. For those who have relatively long lifetimes, to extract the nucleon-nucleon correlation dominated by the final-state interaction, one needs to propagate the wave function for a long period, and analyze its asymptotic components carefully~\cite{Wang:2022}.
\end{itemize}

The measurement of nucleon-nucleon correlation boosts the interest of $2p$ decay. As shown in Refs.\,\cite{Pfutzner:2012,Grigorenko:2003,Grigorenko:2009}, the asymptotic nucleon-nucleon correlations of $^6$Be, $^{45}$Fe, and several other $2p$ emitters have been reproduced or predicted using the three-body model. Moreover, by comparing with experimental correlation, it indicates that the ground state of $^{45}$Fe contains a moderate amount of the $p$-wave component~\cite{Miernik:2007b}. Similar studies have been done for the light nuclei~\cite{Wang:2022,Webb:2019a,Webb:2020}, among these $psd$-shell $2p$ emitters, the energy correlations of $^{12}$O and its isobaric analog $^{12}$N$_{\rm IAS}$ look similar to that of $^{16}$Ne, but rather different from the energy correlations of $^{6}$Be and $^{8}$B$_{\rm IAS}$. This indicates that the valence protons of $^{12}$O might have more $sd$-shell components, even though, in the naive shell-model picture, these valence protons fully occupy the $p$-wave orbitals. Consequently, these qualitative analyses provide us with useful information about the valence protons and corresponding $2p$ emitters. Moreover, in order to gain deep insight into the connection and extract the structural information from these asymptotic observables, high-quality experimental data and comprehensive theoretical models are anticipated.

The asymptotic nucleon-nucleon correlation also turns out to be useful to reveal the property of nuclear force, and the status of a state. As shown in Fig.\,\ref{fig:2_Correlation_A6}, the asymptotic correlation of $^{6}$Be is sensitive to the nucleon-nucleon interaction strength~\cite{Wang:2021}, which could be used to study the in-medium effect of nuclear force. The states with different spin-parities would result in different patterns of asymptotic correlation. Based on the measured correlation and calculated ones, Ref.\,\cite{Wang:2022} shows that the broad peak observed in $^{11}$O contains multiple states, which are in accordance with the previous energy-spectrum analysis~\cite{Webb:2019,Webb:2020a}.

Moreover, these correlations can help to determine the $2p$ decay mechanism, as well as the corresponding configurations. For example, in the case of $^{45}$Fe, the small-angle emission dominates in the asymptotic region, which may correspond to a diproton decay~\cite{Miernik:2007b}. The Y-type correlation helps to determine the intermediate state of the sequential decay from the 2$^+_2$ state of $^{12}$O~\cite{Webb:2019a}. While, based on the asymptotic correlation, the situation in  $^{6}$Be is considerably more complicated~\cite{Egorova:2012}. However, because the emitted protons are influenced by the centrifugal barrier and long-range Coulomb interactions, one must be careful that these configurations reflecting the circumstances in the asymptotic region might differ from those inside the nucleus.  To better understand decay dynamics on how the inner structure evolves into the asymptotic correlation, the time-dependent framework could be useful (see Section\,\ref{sec:2_time_evolution} for details).

\subsubsection{\it Decay dynamics}
\label{sec:2_time_evolution}

The time-dependent formalism is a useful strategy for tackling the decay process, which allows addressing a broad range of questions -- such as configuration evolution \cite{Volya:2014}, decaying rate \cite{Peshkin:2014}, and fission \cite{Bender:2020}  -- in a  precise and transparent way. Therefore, the complex three-body dynamics of $2p$ decay can be demonstrated within this framework.

\begin{figure}[tb]
\begin{center}
\begin{minipage}[t]{17 cm}
\begin{center}
\includegraphics[width = 1\columnwidth]{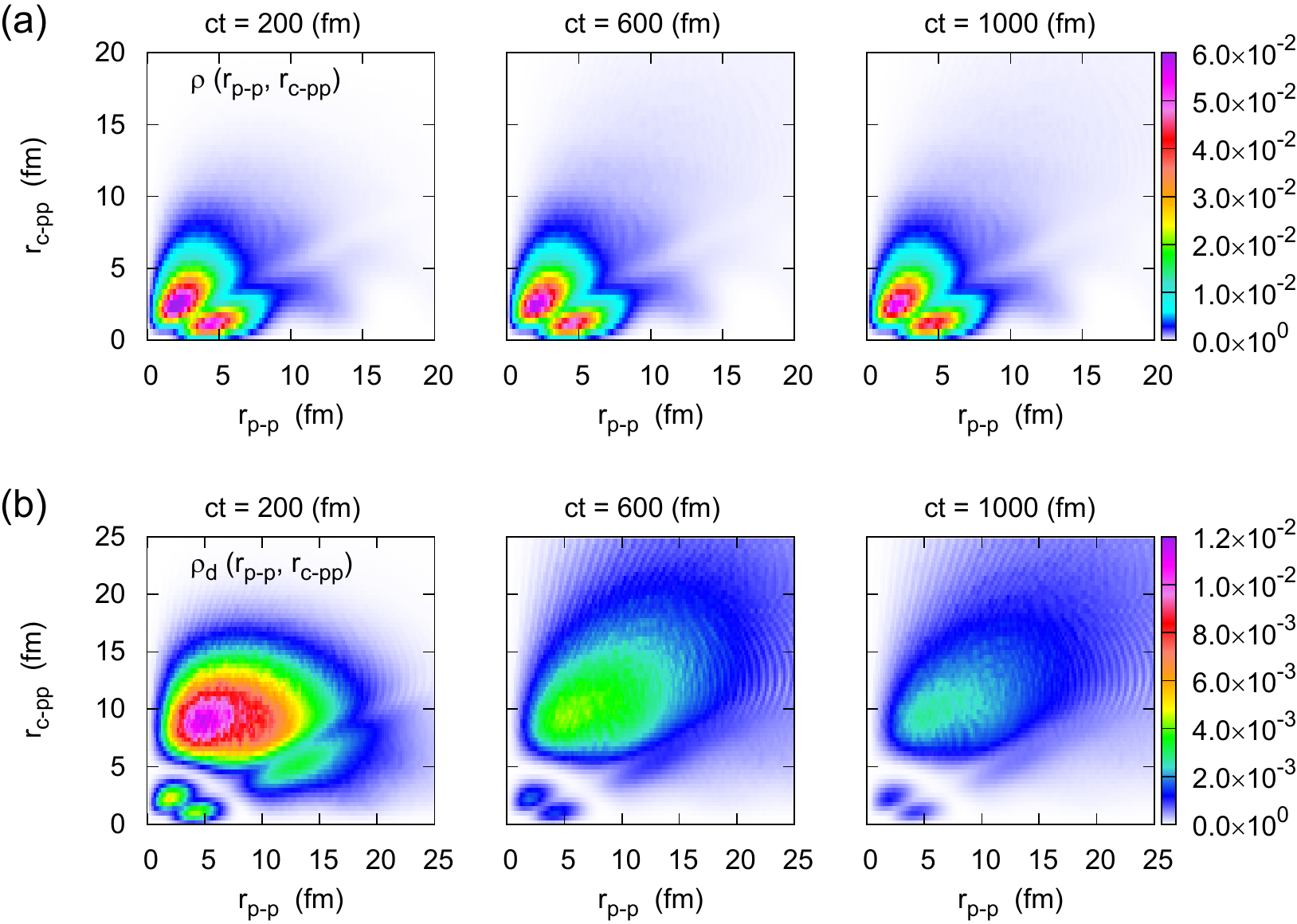}
\end{center}
\end{minipage}
\begin{minipage}[t]{17 cm}
\caption{(Color online) The time evolution for (a) $2p$ density distribution $\rho$ and (b) corresponding decaying density distribution $\rho_{d}$ of the $^6$Be ground state as a function of $r_{p-p}$ and $r_{c-pp}$. Reprinted with permission from Ref. \cite{Oishi:2017}. Copyright (2017) by the American Physical Society.}
\label{fig:2_Density_evolution}
\end{minipage}
\end{center}
\end{figure}

An approximate treatment of $2p$ emission was proposed in Ref.\,\cite{Bertulani:2008}, in which the center-of-mass motion for the two valence protons was described classically. In a more realistic case, the early stage of the $2p$ emission from the ground state of $^{6}$Be was nicely demonstrated using a time-dependent method in Refs.\,\cite{Oishi:2014,Oishi:2017}. A confining potential method was used to make a meta-stable initial  state $\Psi(0)$ be bound by modifying the core-proton potential at the external region. In Ref.\,\cite{Oishi:2017}, the different time snapshots of the density distribution $\rho(r_{pp},r_{c-pp})$ and the corresponding density changes $\rho_{d}$ have been investigated (see Fig.\,\ref{fig:2_Density_evolution}), where $\rho$ and $\rho_{d}$ are defined as:
\begin{equation}
\begin{aligned}
\rho_{(d)}(r_{pp},r_{c-pp}) &=\braxket{\Psi_{(d)}(t) }{\delta(r_x-r_{pp})\delta(r_x-r_{c-pp})}{ \Psi_{(d)}(t) } /\braket{\Psi_{(d)}(t) }{ \Psi_{(d)}(t) };\\
\ket{\Psi_{d}(t)} &= \ket{\Psi(t)}-\ket{\Psi(0)}\braket{\Psi(0) }{ \Psi(t) }.
\end{aligned}
\end{equation}
At the beginning ($t=0$) when the wave function is fairly localized inside the nucleus, the density distribution is in accordance with the static calculations, which shows two maxima for $^6$Be associated with the diproton/cigarlike configuration characterized by small/large relative distance between valence protons. During the early stage of the decay, two strong branches are emitted from the inner nucleus, see Fig.~\ref{fig:2_Density_evolution}. The primary branch corresponds to the protons being emitted at small opening angles, which indicates that a diproton structure is present during the tunneling phase. This can be understood in terms of the nucleonic pairing, which favors low angular momentum amplitudes and  hence lowers the centrifugal barrier and increases the $2p$ tunneling probability  \cite{Grigorenko:2009b,Wang:2019,Oishi:2014,Oishi:2017}. The secondary branch corresponds to protons  emitted in opposite directions. This is in agreement with the flux current analysis shown in Fig.\,\ref{fig:2_6Be_structure}.

In order to  capture the asymptotic dynamics and better understand the role of final-state interaction,  Ref.\,\cite{Wang:2021} has propagated the wave function of $^6$Be to larger distances (over 500\,fm) and longer times (up to 30\,pm/$c$), which allows  analysis of asymptotic observables including nucleon-nucleon correlation. In this framework, the initial state $\Psi_0 ^{J\pi}$ is a complex-momentum eigenstate of the Hamiltonian subject to purely outgoing (decaying) boundary conditions, and can be decomposed into real-momentum scattering states using the Fourier-Bessel series expansion. The resulting wave packet was propagated by the time evolution operator  through the Chebyshev  expansion, and benchmarked with Green's function method, since the time evolution operator can be written as the Fourier transform of $\hat{G} = (E- \hat{H} +i\eta)^{-1}$:
\begin{equation}
	e^{-i \frac{\hat{H}}{\hbar} t}=\frac{e^{\frac{\eta}{\hbar} t}}{2 \pi i} \mathcal{F}\left(\hat{G} , E \rightarrow \frac{t}{2 \pi \hbar}\right).
\end{equation}
The obtained results are practically identical for both methods as shown in Fig.\,\ref{fig:2_Correlation_A6}.

\begin{figure}[tb]
\begin{center}
\begin{minipage}[t]{12 cm}
\begin{center}
\includegraphics[width =\columnwidth]{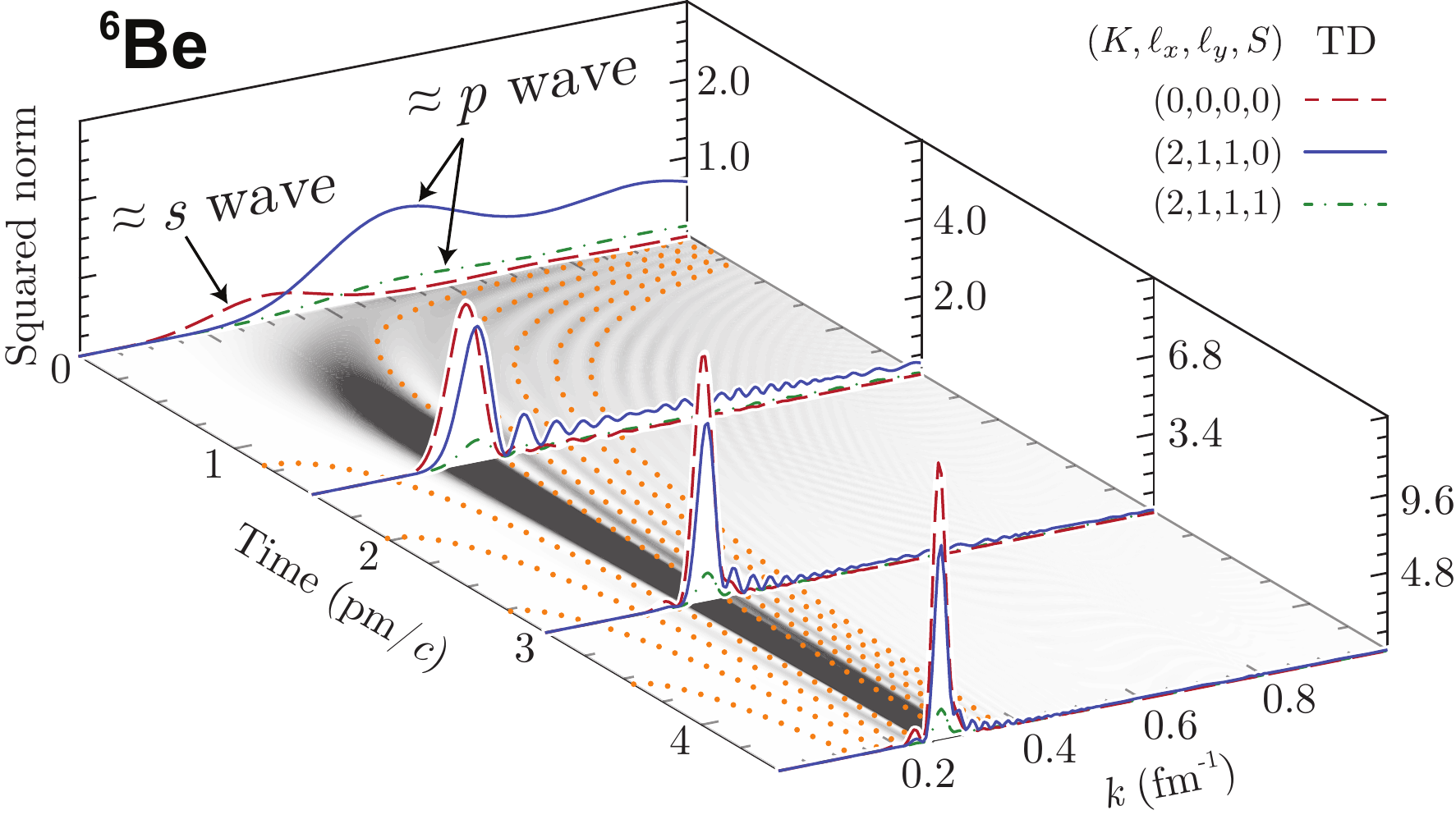}
\end{center}
\end{minipage}
\begin{minipage}[t]{17 cm}
\caption{(Color online) Time evolution of the wave functions of ${^6}$Be. Configurations are labeled as $(K,\ell_x,\ell_y,S)$ in Jacobi-T coordinates. $K$ is the hyperspherical quantum number, $\ell$ is the orbital angular momentum in Jacobi coordinate, and $S$ is the total spin of valence protons. The projected contour map represents the sum of all the configurations in momentum space; the interference frequencies are marked by dotted lines. Figure is taken from Ref.~\cite{Wang:2022}.}
\label{fig:2_Configuration_evolution}
\end{minipage}
\end{center}
\end{figure}

According to Ref.\,\cite{Wang:2021}, after tunneling through the Coulomb barrier, the two emitted protons tend to gradually separate due to  the Coulomb repulsion. Eventually, the $2p$ density becomes spatially diffuse, which is consistent with the broad angular distribution measured in Ref.\,\cite{Egorova:2012}. This corresponds to an opposite trend for the momentum distribution of the wave function, in which the emitted nucleons move  with the well-defined total momentum shown by a narrow resonance peak at long times  \cite{IdBetan:2018}. Figure\,\ref{fig:2_Configuration_evolution} illustrates the dramatic changes in the wave function and configurations of $^6$Be during the decay process. The gradual transition from the broad to narrow momentum distribution exhibits a pronounced interference pattern, which is universal for two-nucleon decays and governed by Fermi's golden rule. The interference frequencies, shown by dotted lines in Fig.\,\ref{fig:2_Configuration_evolution}, can be approximated by $(\frac{\hbar^2}{2m} k^2 - Q_{2p}) t = {\mathfrak n}\pi \hbar$, where ${\mathfrak n}$ = 1, 3, 5 $\cdots$, i.e., they explicitly depend on the $Q_{2p}$ energy.

Moreover, the configuration evolution also reveals a unique feature of three-body decay. As seen in Fig.\,\ref{fig:2_Configuration_evolution}, the initial ground state of $^{6}$Be is dominated by the $p$-wave ($\ell = 1$) components, and the small $s$-wave ($\ell = 0$) component comes from the non-resonant continuum. As the system evolves, the weight of the $s$-wave component -- approximately corresponding to the Jacobi-T coordinate configuration $(K,\ell_x,\ell_y,S)$ = (0,0,0,0) -- gradually increases and eventually dominates because it experiences no centrifugal barrier. Such kind of transition can also be revealed by comparing the internal and external configurations obtained in time-independent calculations \cite{Grigorenko:2003}. This behavior can never happen in the single-nucleon decay due to the conservation of orbital angular momentum but is present for the two-nucleon decay as a correlated dinucleon involves components with different $\ell$-values \cite{Pillet:2007,Catara:1984,Hagino:2014,Fossez:2017}.  In addition,  for $2p$ decays  the Coulomb potential and kinetic energy do not commute  in the asymptotic region \cite{Grigorenko:2009} and this results in additional configuration mixing.

As shown above, the time-dependent approach could reveal the decay dynamics and configuration evolution of the short-lived system. In particular, for $2p$ decay, due to the three-body character, the asymptotic configuration might be different from the inner structure. This indicates that, in order to describe the asymptotic properties correctly, the external wave function is necessary and needs to be treated properly.

So far, only a few 2$p$ emitters have been discovered, and they are located in the vast regions of the nuclear landscape, which have different structures and decay widths. To better understand this exotic phenomenon, impressive progress has been made on the theoretical side  over the past decades. As reviewed in Section\,\ref{sec:2_theory}, these theoretical developments are based on different assumptions and focus on various aspects, such as inner structure, nucleon-nucleon correlation, and decay dynamics. These properties are not isolated but connected to each other, which gives a comprehensive description of $2p$ decay. For example, it has been found that the $2p$ decay width is sensitive to the strength of the pairing interaction \cite{Barker:2003,Wang:2021,Oishi:2017}, since the diproton structure benefits the tunneling process. As shown in Fig.\,\ref{fig:2_Correlation_A6}, the attractive nuclear force is not only responsible for the presence of  correlated dinucleons in the initial state, but it also significantly impacts asymptotic correlations and decay dynamics~\cite{Wang:2021}. This also indicates that, even though the initial-state  correlations are largely lost in the final state, some fingerprints of the dinucleon structure can still manifest themselves in  the asymptotic observables. Although there are a lot of open questions -- such as the $2p$ decay mechanism and the impact of inner structure -- still under debate, further studies with microscopic and self-consistent frameworks would be useful for gaining deep insight into the $2p$ decay process as well as other open quantum systems.


\section{Experimental methods}

\subsection{\it Production of 2p emitters}
\label{sec:3_Production}

As indicated in Section~1.1, the 2\emph{p} emission from
a nuclear ground state can be observed only when this process
is fast enough to compete with $\beta^+$ decay.
Since the latter process is fast for nuclei beyond the
proton drip-line because of large $\beta$ decay energy,
the resulting half-lives of 2\emph{p} decaying nuclei are
shorter than about 10 milliseconds.

\begin{figure}[tb]
\begin{center}
\begin{minipage}[t]{12 cm}
\includegraphics[width = \columnwidth]{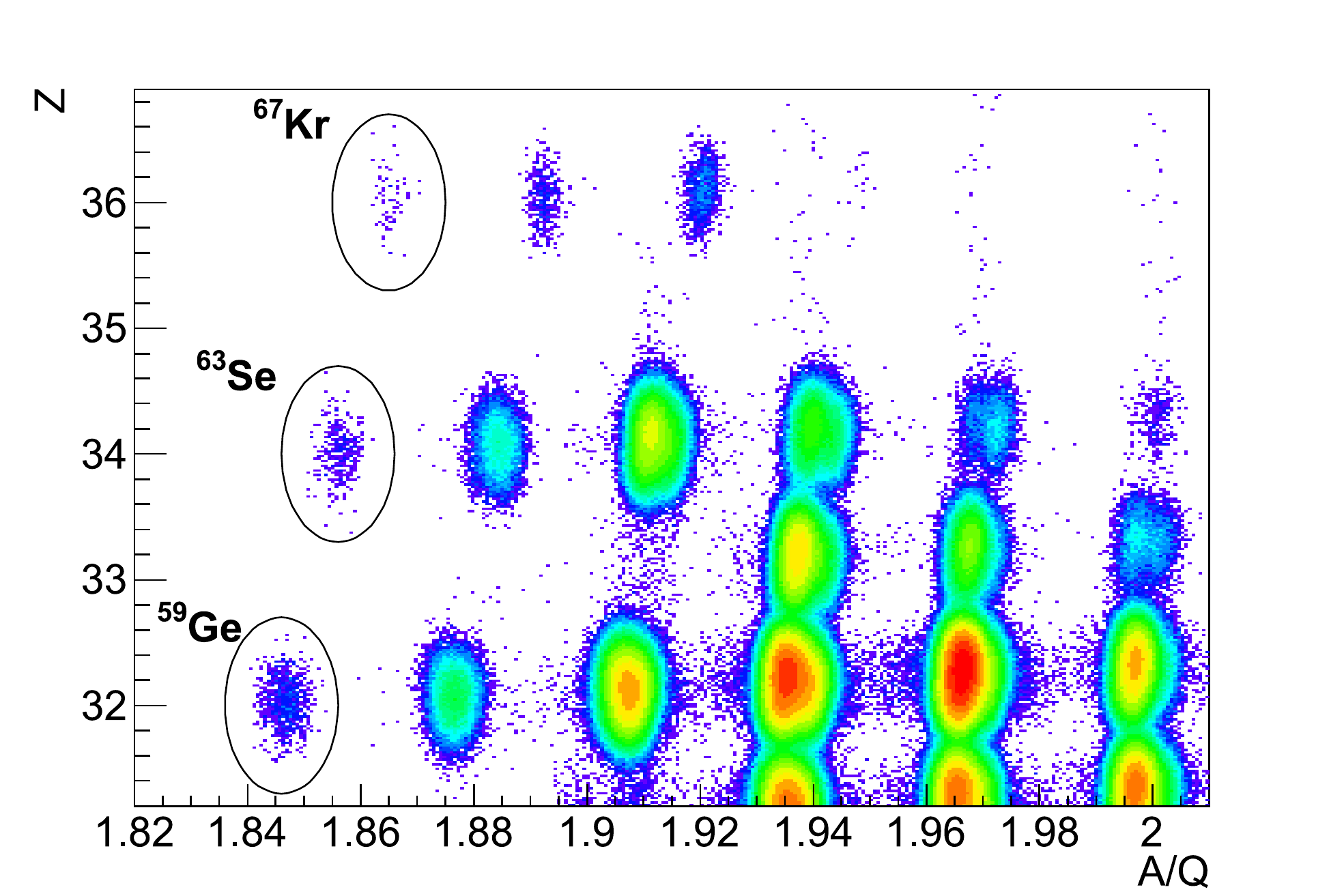}
\end{minipage}
\begin{minipage}[t]{17 cm}
\caption{(Color online) Identification plot of atomic number $Z$ versus the mass-to-charge
 ratio $A/Q$ for ions selected by the BigRIPS separator at the RIKEN Nishina Center.
 The separator setting was optimized for $^{65}$Br.
 Reprinted with permission from Ref.~\cite{Goigoux:2016}. Copyright (2016) by
 the American Physical Society.   }
\label{fig:3_IDPlot}
\end{minipage}
\end{center}
\end{figure}

First we discuss long-lived cases of 2\emph{p} radioactivity with
half-lives longer than a few hundreds of nanoseconds.
The ideal experimental method for study of such cases
is based on projectile fragmentation followed by in-flight
separation and identification \cite{Morrissey:2004}.
In this technique, nuclei of interest are produced by a
high-intensity beam of relativistic ($> 50$~MeV/nucleon) heavy
ions impinging on a target where the reaction of projectile
fragmentation takes place. Selection of very rare ions of interest,
from a huge number of all reaction products is made by means of
a fragment separator. In this device a combination of electromagnetic
elements with wegde-shaped degraders is used to transport
selected ion species to the final focus of the separator
where detection setups are mounted. This technique has a few
key advantages for studies of very exotic nuclei:
\begin{itemize}
  \item The method has no chemical dependence.
  \item The production targets are thick on the absolute scale, of the
        order of 1~g/cm$^2$ which provides \emph{large yields} of products.
        In contrast, a widely used technique to produce proton drip-line
        nuclei, based on fusion-evaporation reaction and the beam energy
        close to the Coulomb barrier, requires targets of about 1~mg/cm$^2$ thick.
  \item The fragmentation targets, however, are thin relative to the range
        of projectiles in the target material. The products emerge from the
        target with \emph{large velocity} (only slightly smaller than that of the
        impinging beam particles) and are transported to the final focus
        within a time of the order of $1 \, \mu$s.
  \item Due to large speed, the products can pass through a series of thin
        detectors, usually located in the second half of the separator. This
        is used for a \emph{full identification} of each single ion arriving to the
        final focus by measurements of its particle rigidity ($B \rho$),
        time-of-flight (TOF), and energy loss ($\Delta E$). The state-of-the-art
        of this procedure is described in Ref.~\cite{Fukuda:2013}.
  \item The beam of fast, selected unstable products can be used as a radioactive
        beam to initiate secondary reactions. The application of this feature
        will be discussed later in this section.
\end{itemize}
The separated and identified in-flight ions are stopped inside the detection setup
where their decays at rest can be observed. This method proved to be very
sensitive and efficient. In fact, all four cases of 2\emph{p} radioactivity shown
in the right part of Fig.~\ref{fig:1_Chart} and in Fig.~\ref{fig:1_S1pS2p}
were discovered with this technique.
An example of the particle identification is presented in Fig.~\ref{fig:3_IDPlot}
showing the first observation of $^{67}$Kr~\cite{Goigoux:2016}.
A list of leading laboratories employing
the projectile fragmentation method is given in Table~\ref{tab:3_Separators}.

\begin{table} [h]
\begin{threeparttable}
\caption{The laboratories where the projectile fragmentation method is used to produce and study
nuclei very far from the stability. In the lower part of the table
          the facilities under construction are listed.}
\label{tab:3_Separators}
\vspace{0.5\baselineskip}
\begin{tabular}{lllcclc}
  \hline
  \hline
  Country & Laboratory  & Driver      &  Beams   & Max beam     & Separator & Reference        \\
          &             & accelerator &          & energy [$A \,$MeV]      &           &     \\
  \hline
  Russia  & FLNR  & cyclotron    & Li--Ar      & 50        &  ACCULINNA-2 & \cite{Fomichev:2018}   \\
  China   & HIRFL & cyclotron    & C--U        & 60        &  RIBLL     & \cite{Sun:2003}\\
  Italy   & LNS   & cyclotron    & H--U        & 80        &  Fribs     & \cite{Russotto:2018} \\
  France  & GANIL & 2 cyclotrons & C--U        & 95        &  LISE      & \cite{Mueller:1991} \\
  USA     & NSCL  & 2 cyclotrons & O--U        & 170       &  A1900\tnote{a} &\cite{Morrissey:2003}\\
  Japan   & RIBF  & 4 cyclotrons & H--U        & 350       &  BigRIPS   & \cite{Kubo:2012}  \\
  Germany & GSI   & synchrotron  & H--U        & 1000      &  FRS       & \cite{Geissel:1992} \\
  \hline
  USA     & FRIB  & linac        & H--U        & 400       &  ARIS      & \cite{Hausmann:2013}  \\
  Germany & FAIR  & synchrotron  & H--U        & 1500      &  Super-FRS & \cite{Winfield:2021} \\
  \hline
  \hline
\end{tabular}
\begin{tablenotes}
\item [a] The A1900 separator was closed in 2020 to be replaced by ARIS
\end{tablenotes}
\end{threeparttable}
\end{table}

The lighter 2\emph{p} emitters, indicated in the left part of Fig.~\ref{fig:1_Chart}
have much shorter half-lives so there is no time to identify them before they
decay. The production of such an emitter and its decay happens essentially
at the same place, in-flight. All recent experiments on such cases employ
the method of radioactive beams by taking advantage of projectile fragmentation.
A fragment separator is used to produce, separate, and identify in-flight
ions of an unstable projectile fragment which play a role of a secondary
radioactive beam. This beam is directed to a secondary production target
where the 2\emph{p} emitting nucleus of interest is produced, mostly by $1n$
or $2n$ knockout reaction. Identification of the 2\emph{p} emitter is done by
detection of the decay products and by the kinematical reconstruction.
Thus the same set of detectors is used to identify the nucleus, its decay
mode, and to study properties of this decay.

\begin{figure}[tb]
\begin{center}
\begin{minipage}[t]{15 cm}
\includegraphics[width = \columnwidth]{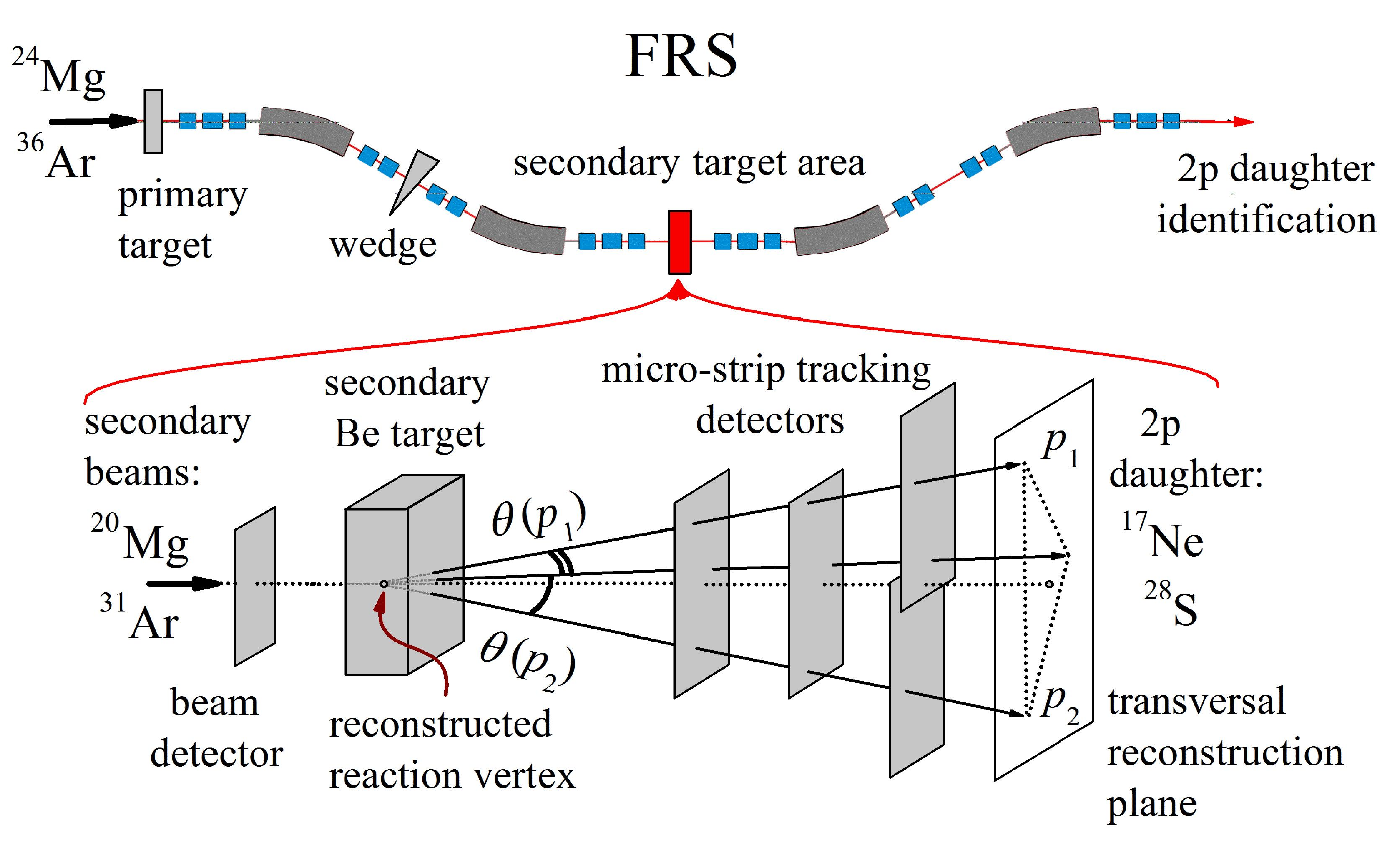}
\end{minipage}
\begin{minipage}[t]{17 cm}
\caption{(Color online) Upper part: A schematic layout of the
FRS fragment separator at GSI Darmstadt. Lower part: Sketch of the detector setup
at the secondary-target area measuring trajectories of the incoming
secondary projectiles and the decay products of the 2\emph{p} decaying
nucleus created in the secondary target. The specified primary and secondary beams,
and 2\emph{p} daughter nuclei, correspond to experiments on the 2\emph{p} decays
of $^{19}$Mg~\cite{Mukha:2007}
and $^{30}$Ar~\cite{Mukha:2015}. See text for more details.}
\label{fig:3_FRSTracking}
\end{minipage}
\end{center}
\end{figure}

An example of such technique is the setup developed at GSI Darmstadt and
used for the first observation of 2\emph{p} emission from $^{19}$Mg \cite{Mukha:2007}.
The idea of the method is illustrated in Fig.~\ref{fig:3_FRSTracking}.
In this experiment, the primary beam of $^{24}$Mg with the energy of
591 MeV/nucleon was bombarding a beryllium target, 4~g/cm$^2$ thick, located
at the entrance of the FRS fragment separator. The first half of the FRS was
used to select ions of $^{20}$Mg which served as the secondary, radioactive
beam. This beam, at about 450 MeV/nucleon, hit the secondary 2~g/cm$^2$ thick
beryllium target at the middle focal plane of the FRS. Here by the 1n knockout
reaction the $^{19}$Mg was produced and promptly decayed by 2\emph{p} emission.
Tracks of both protons were detected by a set of silicon microstrip detectors
and the 2\emph{p} daughter nucleus, $^{17}$Ne, was fully identified in the second half
of the FRS. Triple coincidences of two protons and a $^{17}$Ne ion were used
to define the $^{19}$Mg decay and, in addition, the proton tracks were used
to reconstruct the position of the decay vertex. Later the same method was
used at the FRS to investigate 2\emph{p} emission from $^{16}$Ne~\cite{Mukha:2008},
$^{30}$Ar~\cite{Mukha:2015}, and $^{29}$Ar~\cite{Mukha:2018}.
A very important, and not intuitive,
feature of this technique is that the energy of emitted protons can
be determined solely from the angle of its trajectory with respect to the
beam axis.

A different setup was used at GSI Darmstadt for the first observation of $^{15}$Ne
and its 2\emph{p} emission \cite{Wamers:2014}. Here the full FRS separator was used to
produce the radioactive beam of $^{17}$Ne of 500~MeV/nucleon from the primary
beam of $^{20}$Ne. The fully identified in-flight secondary beam was transported
to the R$^3$B-LAND setup \cite{Langer:2014} on secondary carbon or polyethylene
reaction targets where $^{15}$Ne was created by the 2n knockout reaction.
Its decay products were identified from their trajectories in the magnetic
field of a large-gap dipole magnet (ALADIN) placed behind the target.
The protons and heavy fragments were guided into two separate arms, where
additional energy loss and TOF measurements were performed.

A similar approach was taken at the NSCL/MSU laboratory for a series of
measurements devoted to the 2\emph{p} emission from $^{8}$C~\cite{Charity:2011},
$^{6}$Be~\cite{Egorova:2012}, $^{16}$Ne~\cite{Brown:2014}, $^{12}$O~\cite{Webb:2019},
and $^{11}$O~\cite{Webb:2019a}.
Here, the full A1900 separator was used to prepare the corresponding radioactive beams.
The most recent experiments \cite{Webb:2019,Webb:2019a} could apply
additional electromagnetic time-of-flight filter --- the
Radio Frequency Fragment Separator~\cite{Bazin:2009} --- which
purified the beam to about 80\%. The final secondary beam was impinging
on a Be target and the charged reaction products were detected
further downstream by the High Resolution Array (HiRA)~\cite{Wallace:2007}.
The same setup was used in the study of 2\emph{p}-decay from the first
excited state of $^{16}$Ne~\cite{Brown:2015}.

Decays from excited states populated in $\beta$ decay and from long-lived
isomeric states do not necessarily require fast separation technique.
In such a situation the alternative production method of exotic nuclei,
based on Isotope Separation On Line (ISOL)~\cite{Pfutzner:2012,VanDuppen:2006},
can be applied. Indeed, a large body of information concerning the $\beta 2\emph{p}$
decays was collected with help of the ISOL method \cite{Blank:2008}.

\subsection{\it Detection of 2p decays}
\label{sec:3_Detection}
\subsubsection{\it Silicon detectors}

The most common detection technique in charged-particle
spectroscopy is based on silicon detectors which are characterised
by a good energy resolution and sufficiently good time resolution.
An important advance in this field was the development of
double-sided silicon-strip detectors (DSSSD) \cite{Sellin:1992}.
The granularity achieved by two sets of perpendicular strip
electrodes on the both sides of a DSSSD helps to establish position
correlations between implantation of a heavy ion and its subsequent
decay with emission of charged particles. In this way the detection
sensitivity is enhanced and high counting rates can be accepted.

One method to study 2\emph{p} decay at rest is to implant
a candidate nucleus into a silicon detector. This approach was
applied in the first observation of 2\emph{p} radioactivity of
$^{45}$Fe~\cite{Pfutzner:2002,Giovinazzo:2002}, $^{54}$Zn~\cite{Blank:2005},
and $^{67}$Kr~\cite{Goigoux:2016}. Ions emerging from the fragment
separator are slowed down by means of a dedicated degrader of variable
thickness to be properly stopped in the detector setup.
The products of projectile fragmentation have significant
energy straggling, therefore, to increase the stopping
efficiency, the setup usually
consists of a stack (\emph{a telescope}) of a few Si detectors typically
between 300~$\mu$m and 1~mm thick. Signal generated by the implanted
ion is recorded in coincidence with signals from detectors used
for the in-flight identification. Later, these signals are
correlated with decay signals from the same Si detector and from
the same pixel in case of the DSSSD.
The 2\emph{p} decay energy, $Q_{2p}$, is of the
order of 1~MeV (see Fig.~\ref{fig:1_S1pS2p}) and the range of
1~MeV proton in silicon is about 16~$\mu$m. Since both protons
are emitted simultaneously, they are recorded as a single signal
with the efficiency of almost 100\% and the $Q_{2p}$
energy can be measured with the resolution of about 20~keV.
From the time difference between
implantation and decay signals the half-life can be determined.
Nuclei beyond the proton drip-line undergo $\beta^+$ decay with
large probability of delayed proton emission. Hence, it is
important to distinguish 2\emph{p} decays from $\beta p$ events.
The positron from $\beta^+$ decay leaves a small part of its
energy in the Si detector in which decay occurred and this
energy adds to the energy left by the proton which is emitted
practically at the same time. This effect ($\beta$ \emph{summing})
modifies the proton line shape by producing a high energy tail
which is absent in the lines originating from 2\emph{p} decay.
In addition, the positron has a large probability to be detected
in the neighboring detectors of the Si stack which can be used
to veto the $\beta p$ events. An modern example of such approach
is the the WAS3ABi DSSSD array \cite{Nishimura:2013} which was
used in the 2\emph{p} decay study of $^{67}$Kr \cite{Goigoux:2016}.
WAS3ABi consisted of three 1~mm thick DSSSDs with 60 vertical
and 40 horizontal (Y) strips with a pitch of 1 mm.
A different way to tag $\beta p$ events is to surround
the Si telescope with a high efficiency $\gamma$-ray detector
sensitive to 511~keV photons originating from the positron
annihilation. This method was applied in Ref. \cite{Pfutzner:2002},
by mounting the implantation telescope, consisting of 8
monolithic Si detectors, each 300~$\mu$m thick, inside a
NaI(Tl) barrel composed of 6 crystals. The detection efficiency
of $\beta p$ events in the energy range 0.9--4.0 MeV of this
setup was found to be 93\% \cite{Pfutzner:2002}.

Decay study at rest, after implantation into a silicon detector, has
important advantages of providing large efficiency, close to 100\%
for protons of about 1 MeV, and good energy resolution, usually better
than 30 keV. This method, however, can be applied only to cases
which live long enough to survive the flight from the production target
to the decay station, which is typically a few hundred nanoseconds.
The 2\emph{p} emitters of much shorted half-lives decay in-flight
and thus require a different detection technique.

For the identification of the in-flight 2\emph{p} emission and for
its  kinematical reconstruction, it is necessary to record
trajectories and momenta of all the decay products: two protons and the
2\emph{p} daughter nucleus. Here again silicon detectors prove to be
indispensable. In experiments where 2\emph{p} emitter is created by a
high-energy radioactive beam from a fragment separator, as described
in the previous section, the decay products emerge from the secondary
target also with a high energy due to the Lorentz boost.
As a consequence, they can pass through several detector layers
which facilitates good track reconstruction and identification, and in
addition, it allows to cover large center-of-mass solid angle with
detectors subtending a relatively small solid angle in the laboratory
system. An example is shown in Fig.~\ref{fig:3_FRSTracking}.
Here the ions of the radioactive beam were identified in the
first half of the FRS with help of standard particle identification
detectors and the $B \rho$-TOF-$\Delta E$ method. Their position
on the secondary target was defined by a $6 \times 6$~cm$^2$
DSSSD with $32 \times 32$ strips \cite{Mukha:2007}.
The 2\emph{p} daughter nucleus was characterised (its identity and
the momentum) in the second half of the FRS with the standard
detectors and the $B \rho$-TOF-$\Delta E$ method.
The tracks of the two protons were measured by an array of DSSSD
silicon microstrip detectors \cite{Stanoiu:2008}. It consisted
of four large-area ($7 \times 4$~cm$^2$), 300~$\mu$m thick silicon
detectors with a pitch of 100~$\mu$m. They served for the measurement
of the energy loss and positions of coincident hits of two protons and the
2\emph{p} daughter nucleus. This allowed to reconstruct all fragment
trajectories and derive the coordinates of the decay vertex and
angular correlations between decay products. The achieved transverse
position accuracy by a microstrip detector for protons
was 40~$\mu$m \cite{Mukha:2007}.

In experiments on the lightest 2\emph{p} emitters, performed at the NSCL/MSU laboratory
\cite{Charity:2011,Egorova:2012,Brown:2014,Webb:2019,Webb:2019a}
the secondary beam, delivered by the A1900 fragment separator,
with the energy of about 70~MeV/u impinged on a 1~mm thick beryllium
target. The all 2\emph{p} decay products were detected by the High Resolution
Array (HiRA)~\cite{Wallace:2007} which is a modular and expandable
array of 14 charged-particle telescopes. Each of them,
with an active area of 6.25~cm~$\times$~6.25~cm, consists of a thin
65~$\mu$m single-sided silicon strip-detector with 32 strips, a
1.5~mm double-sided ($32 \times 32$) strip-detector, and four
separate 4~cm long CsI(Tl) crystals, each spanning a quadrant of the
preceding Si detector. The strips on the frontal detectors provide
good definition of the particle position, while the particle identification
is made from the $\Delta E$ vs. $E$ correlations. If the measured
particle stops in the second Si detector, the first, very thin one
delivers the $\Delta E$ information. If the particle range is larger,
then the $\Delta E$ signal is taken from the second, thicker Si detector
and the residual energy $E$ is given by the CsI(Tl) crystal.
This technique was used to identify, and measure energy of
protons, $\alpha$ particles, and isotopes of carbon and oxygen.

In-flight technique is based on recording coincidences between several
detectors which usually have a limited coverage of the solid angle.
Therefore, the total efficiency and energy resolution are worse
than in case of the decay-at-rest spectroscopy. In addition, in contrast
to the latter method which has a lower half-life limit of a few hundreds
nanoseconds, in-flight technique requires that the decay takes place
before the decaying ion passes the first detection array. This imposes
an upper limit for the half-life which depends on geometry of the set-up
and velocity of ions but typically is of the order of a few nanoseconds.
In case when the precise tracking of the decay products is made,
the half-life in the range between 1 ps and 1 ns can be determined from
the distribution of the decay vertex \cite{Mukha:2007}.

\subsubsection{\it Time projection chambers}
\label{sec:3_TPC}

\begin{figure}[h]
\begin{center}
\begin{minipage}[t]{6 cm}
\includegraphics[width = \columnwidth]{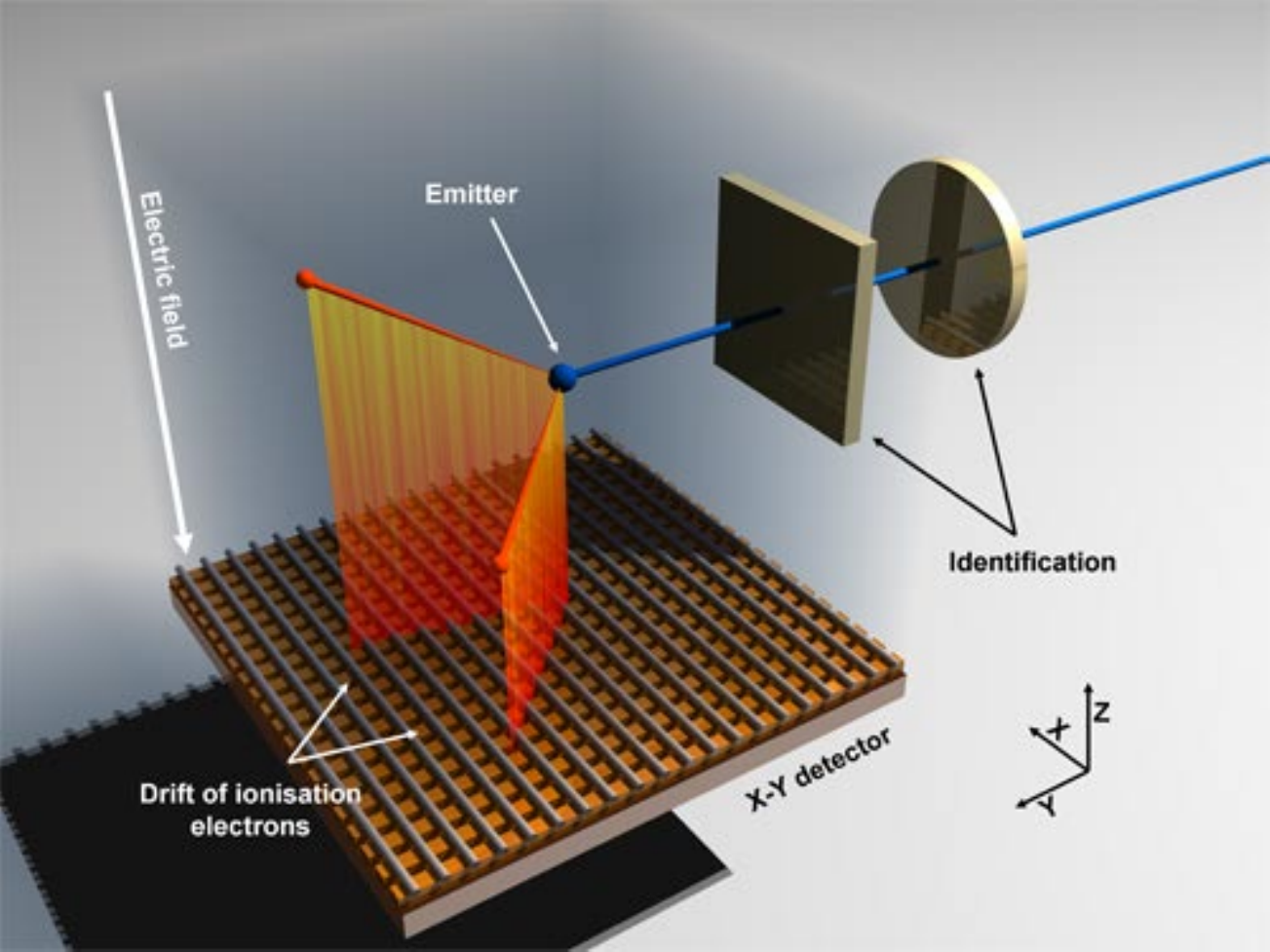}
\end{minipage}
\begin{minipage}[t]{17 cm}
\caption{(Color online) Schematic illustration of the time-projection
chamber operation principle. Reprinted with permission
from Ref.~\cite{Giovinazzo:2007}. Copyright (2007) by the American
Physical Society.}
\label{fig:3_TPCPrinciple}
\end{minipage}
\end{center}
\end{figure}

The method of 2\emph{p} emission study from ions implanted inside a silicon
detector is relatively simple and effective but has an important
drawback that only the total decay energy and the decay time can be determined.
Although both these observables are sufficient to classify the
event as a 2\emph{p} decay, the information on the angle between
protons momenta and on the energy sharing between them is lost.
Since the latter parameters are key to understand the
mechanism of the 2\emph{p} decay, a new experimental approach is
necessary to study this phenomenon in detail.
To address this problem, gaseous detectors based on the
principle of the time projection chamber (TPC) were proposed.
Their main advantage is that such detectors are capable
to record individual tracks of charged particles ejected
within the active volume. The principle of operation is
illustrated in Fig.~\ref{fig:3_TPCPrinciple}. An ion of
interest, identified in-flight and slowed down by a degrader,
stops within the active volume of the chamber filled with a
gas. The stopping ion, as well as charged particles emitted
in the subsequent decay, ionize the gas. The primary ionization
electrons drift in a uniform electric field towards the
anode electrode through a charge amplification section where
they are multiplied. Their pattern on the anode plane represents
the two-dimensional projection of the particles' tracks on that plane.
Since the electrons drift with a constant velocity, the drift
time contains the information on the position along the
electric field direction, perpendicular to the anode plane.

In a TPC detector the decay at rest is observed, so this technique
can be applied only to ions with half-lives longer than the
time-of-flight through the separator (a few hundred nanoseconds),
like in case of implantation into a silicon detector. The efficiency
of observing 2\emph{p} emission is almost 100\% when the ion is
stopped in the active gas volume. The TPC detectors are operated with
the gas at the atmospheric pressure or lower - down to 100 mbar.
As a consequence of fragmentation
reaction mechanism, the range distribution of selected ions in the
gas used in the TPC can be broader than the thickness of the TPC
detector which reduces the overall detection efficiency.
In such case, the special effort is needed to assure the optimal
implantation of ions under study. In addition, because of low
density of the TPC gas, the tracks of particles are much longer
than in a silicon detector. The full kinematical reconstruction of the
decay is possible only when tracks of all decay products are confined
within the active gas volume. Even if this is the case, however, the
energy resolution is worse than provided by a Si detector, typically
above 50 keV. Thus, the two methods of 2\emph{p} decay studies at rest are
complementary and both should be used for the full characterisation
of the decay process.

\begin{figure}[h]
\begin{center}
\begin{minipage}[t]{8 cm}
\includegraphics[width = \columnwidth]{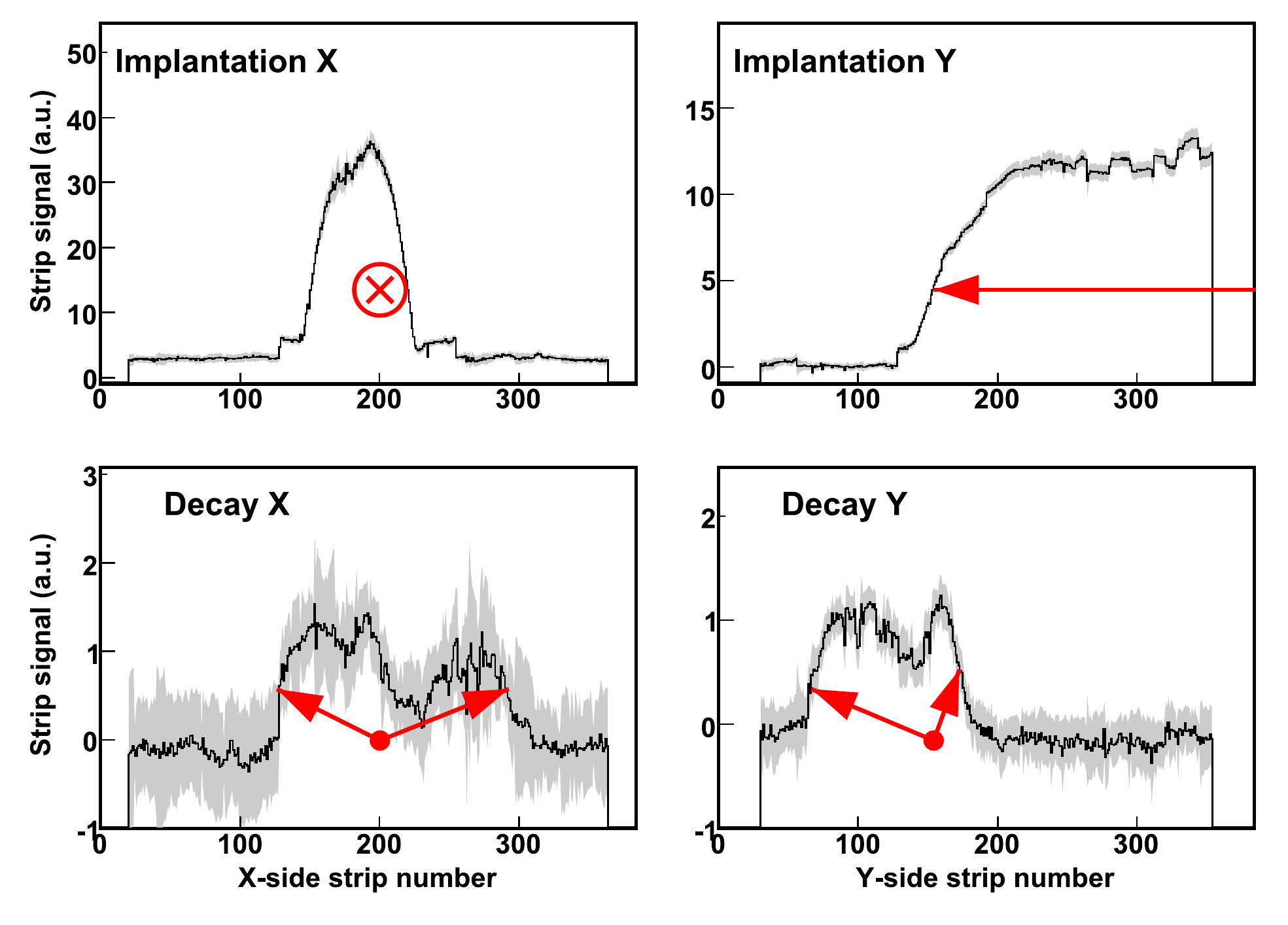}
\end{minipage}
\begin{minipage}[t]{17 cm}
\caption{(Color online) One event of two-proton decay of $^{45}$Fe recorded by
the Bordeaux TPC detector. The top row shows the implantation of a $^{45}$Fe ion.
In the bottom the emission of two protons is seen from the point where the ion
was stopped. Reprinted with permission from Ref.~\cite{Giovinazzo:2007}.
Copyright (2007) by the American Physical Society. }
\label{fig:3_2p45FeBordeauxTPC}
\end{minipage}
\end{center}
\end{figure}

One detector of the TPC type for the study of 2\emph{p} radioactivity was
developed at the CEN Bordeaux-Gradignan \cite{Blank:2010}.
Effectively, it has the active volume of about $14 \times 14 \times 6$~cm$^3$,
filled by the P10 gas (90\% Ar, 10\% CH$_4$). The detector is mounted
inside a tight container which allows to regulate the gas pressure and
thus its density. The typical working pressure is between 500~mbar and 750~mbar.
The charge amplification section is made by a set of four gas electron
multiplier (GEM) foils \cite{Sauli:1997}. The signals are collected
by a double-sided microgroove detector \cite{Bellazzini:1999} equipped with an
application-specific-integrated-circuit (ASIC) read-out.
The detector has two orthogonal sets of 768 strips with the pitch of 200~$\mu$m.
ASIC electronic reads out every second strip which yields 384 channels for each side
of the detector. Other half of the strips is connected in 12 groups of 64
strips and read-out by standard preamplifiers and shapers.
From each of the channels the energy and the timing information
is collected. With this detector the first direct observation of two
protons from the 2\emph{p} decay of $^{45}$Fe was achieved \cite{Giovinazzo:2007}.
An example of such an event is shown in Fig.~\ref{fig:3_2p45FeBordeauxTPC}.
The same TPC detector, with some slight modifications, was used to study
the 2\emph{p} decay of $^{54}$Zn \cite{Ascher:2011}. More advanced
construction of an ACtive TARget and Time Projection Chamber (ACTAR TPC),
to be used also in future 2\emph{p} decay studies, was recently developed
GANIL laboratory~\cite{Mauss:2019,Giovinazzo:2020}.

\begin{figure}[ht]
\begin{center}
\begin{minipage}[t]{8 cm}
\includegraphics[width = \columnwidth]{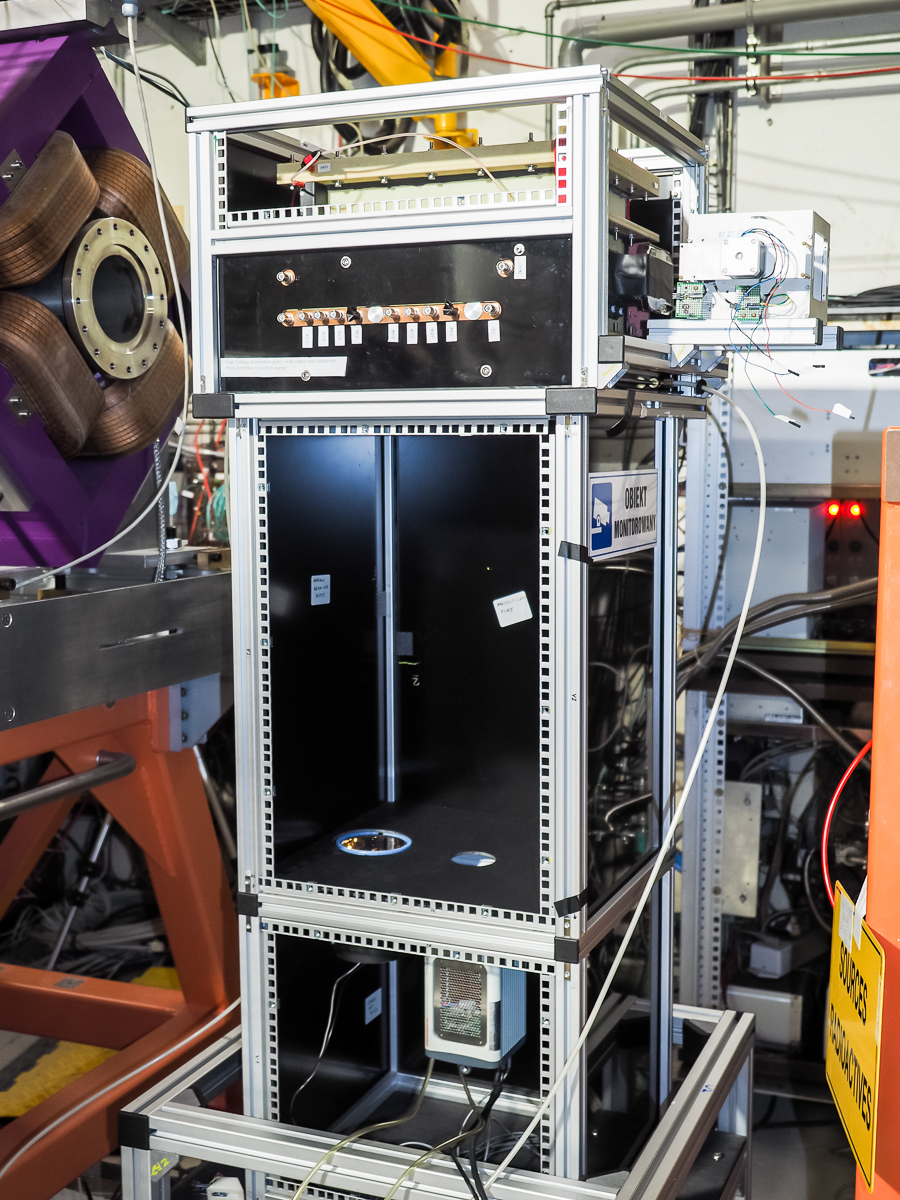}
\end{minipage}
\begin{minipage}[t]{17 cm}
\caption{(Color online) The photograph of the Warsaw OTPC detector being mounted at the
beamline, before installing the light-protecting closure. In the top part
the active gaseous chamber is located. A CCD camera and a PMT can be seen at the
bottom of the lower part. }
\label{fig:3_OTPCPhoto}
\end{minipage}
\end{center}
\end{figure}

A different approach to the signal readout from a TPC detector
was adopted at the University of Warsaw. While the main principle
of operation is the same as described above, the final signal
is not recorded electronically but optically. The key idea,
first applied by Charpak et al.~\cite{Charpak:1988}, is that
the gas atoms and molecules emit light when excited by multiplied
ionization electrons. In the Warsaw Optical Time Projection Chamber
(OTPC) this light is collected by a CCD camera and by a
photomultiplier (PMT) - the first prototype was described
in Ref.~\cite{Miernik:2007}, the later developments
in Ref.~\cite{Pomorski:2014}. In the present version the active
volume amounts to $33 \times 20 \times 21$~cm$^3$ and is filled
by a gas mixture at atmospheric pressure. The mixture is composed
mainly by argon and helium with small admixtures
of CO$_2$, N$_2$ or CF$_4$. The density of the gas is regulated
by changing the proportions of argon and helium. Four GEM foils
are used for charge amplification. The light is generated between
the last GEM foil and the wire mesh anode. The chamber
is closed by a glass plate transparent to visual light.
Below the chamber, in the light-tight box the CCD camera
and the PMT are mounted - see Fig.~\ref{fig:3_OTPCPhoto}.
The acquisition system is triggered by a signal from particle
identification system indicating that an ion of interest
is entering the detector. The CCD camera is opened for a certain
exposure time and the PMT signal is recorded by a digital
oscilloscope. The resulting CCD image shows projection on the
anode plane of tracks of all charged particles traversing
the active volume during the exposure. The PMT waveform
represents the total light intensity as a function of time.
It allows to establish the sequence of events in time during
the exposure and, via the known electron drift velocity,
contains information on the position distribution along
the electric field, perpendicular to the anode plane.
Although the drift time information is contained only in a single
channel (total light intensity), the combination of the CCD
image with the PMT waveform allows to reconstruct tracks
of particles in 3D unambiguously for decays when no more
than two tracks appear at the same time.

\begin{figure}[h]
\begin{center}
\begin{minipage}[t]{7 cm}
\includegraphics[width = \columnwidth]{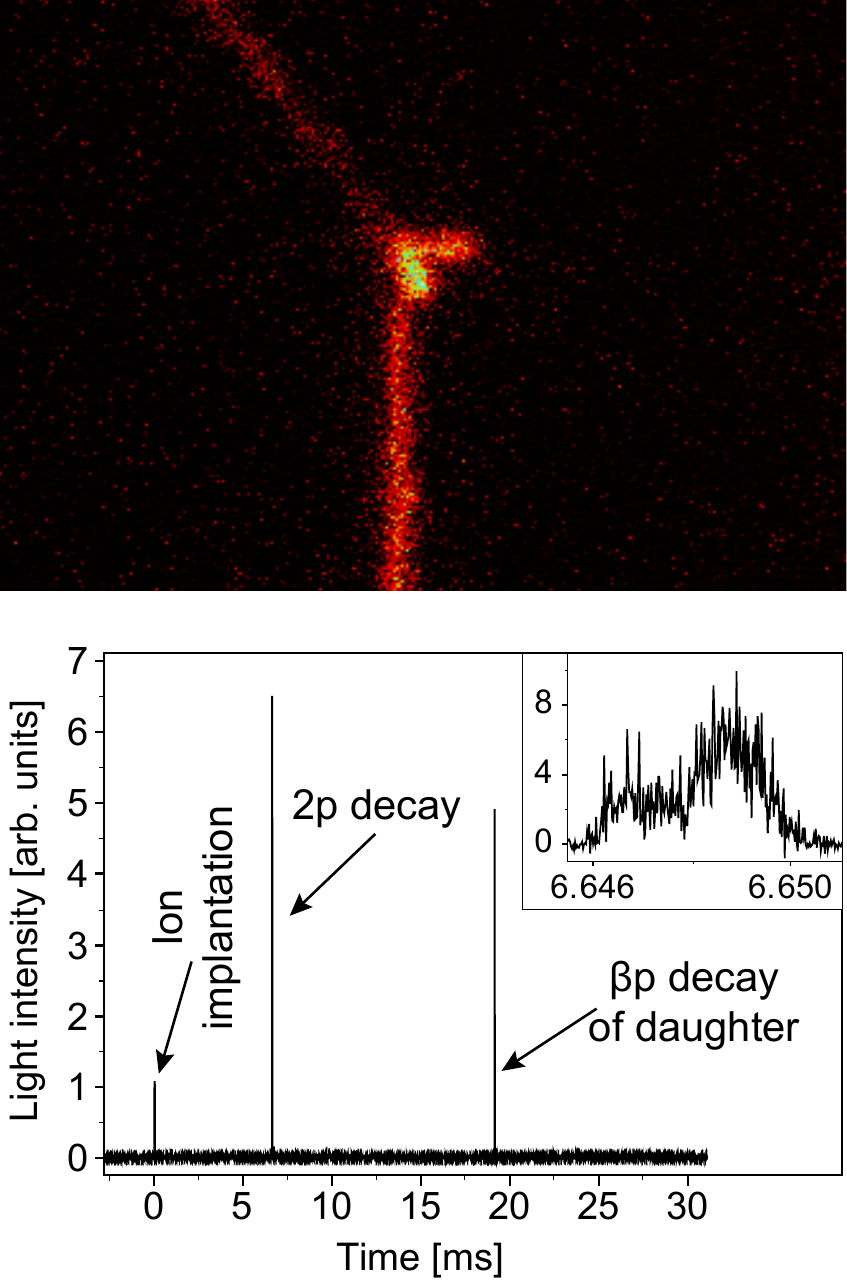}
\end{minipage}
\begin{minipage}[t]{17 cm}
\caption{(Color online) An example event registered by the Warsaw OTPC detector.
Top: the CCD image shows the track of a $^{48}$Ni ion entering the chamber
from below, two short tracks representing the 2\emph{p} decay, and a long track
of the $\beta$-delayed proton emitted by the 2\emph{p} daughter nucleus $^{46}$Fe.
Bottom: the PMT waveform shows the sequence of events: the triggering ion was
implanted at time zero, two protons were emitted about 6.65 ms later, and
the $\beta p$ event occurred about 19 ms after the implantation. The inset shows
a part of the PMT signal zoomed on the 2\emph{p} emission moment, where contributions
from both protons are seen.}
\label{fig:3_OTPCexample}
\end{minipage}
\end{center}
\end{figure}

The OTPC was used in the measurement of the detailed
proton-proton (\emph{p}-\emph{p}) correlations in the
decay of $^{45}$Fe~\cite{Miernik:2007b} and in the discovery
of 2\emph{p} radioactivity of $^{48}$Ni~\cite{Pomorski:2011}.
This detector was also successfully used in studies
of $\beta$-delayed emission of charged particles.
In particular, all cases of $\beta$-delayed emission of
three protons ($\beta 3p$) known to date were discovered
with help of the OTPC~\cite{Miernik:2007c,Pomorski:2011b,Lis:2015,Ciemny:2022}.
An example of a decay event of $^{48}$Ni recorded by the
OTPC is shown in Fig.~\ref{fig:3_OTPCexample}.


\section{2\textit{p} emission from light unbound nuclei}

In this section, we overview 2\textit{p} decays of short-lived nuclear
states, ground and excited ones, which are either faster than classical radioactivity life-times (i.e., shorter than $10^{-14}$  s  according to \cite{Pfutzner:2012}) or radioactive but decay in-flight before reaching experimental detectors (mostly shorter than $10^{-6}$  s ). In both cases, the 2\textit{p}-decay products are detected, which allows for a reconstruction of the energy spectrum and/or half-life value of the 2\textit{p} precursor.

As mentioned in the historical note (Section 1.2), early studies of 2\textit{p} emission were performed on nuclear excited states of bound nuclei populated either in $ \beta $ decays or inelastic nuclear reactions with stable beams. Most of such states decay by a sequential 2\textit{p} emission which can be described as a chain of independent 1\textit{p} emissions via long-lived intermediate states (i.~e., via narrow resonances). Such a decay mechanism is well understood and has been described in a number of approaches, see e.g. the review of R-matrix theory by Lane and Thomas~\cite{Lane:1958}.
Here, we concentrate on reviewing 2\textit{p} emission from light 2\textit{p}-unbound nuclei. These exotic systems located just beyond the proton dripline provide a remarkable playground for studies both of narrow resonance decays and of broad continuum states, where a rich and complex evolution from simultaneous to sequential 2\textit{p} decay is observed with increasing decay energy.

\begin{figure}[h]
	\begin{center}
		\begin{minipage}[t]{18 cm}
\hspace{1. cm}
			\includegraphics[width = 0.9\columnwidth]{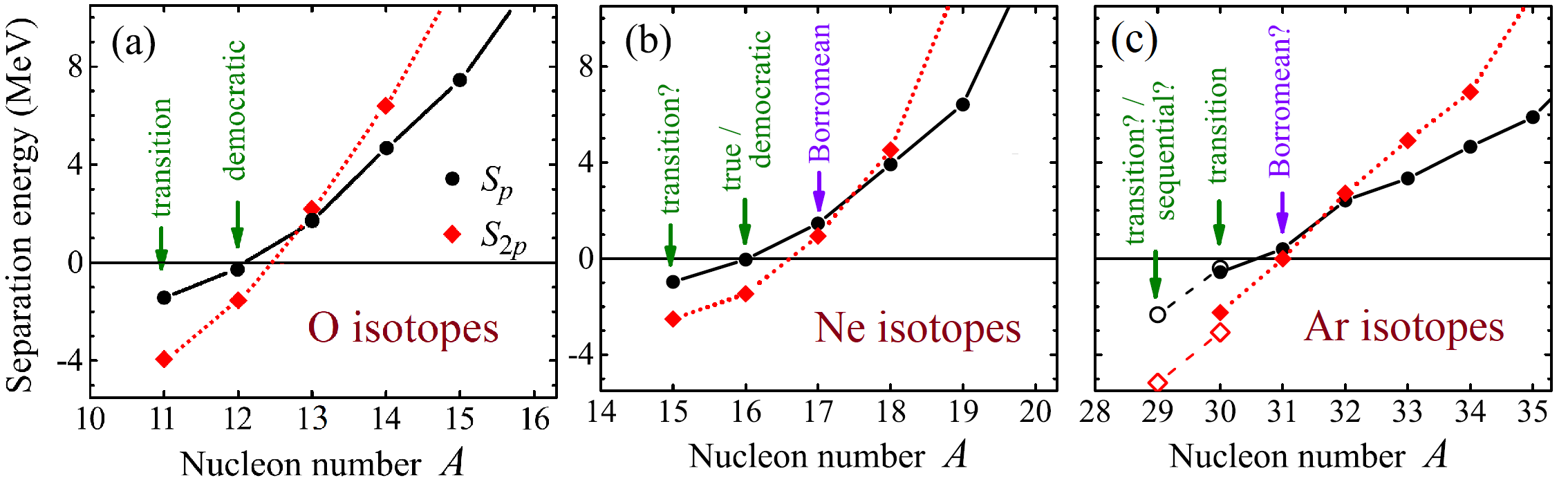}
		\end{minipage}
		\begin{minipage}[t]{16.5 cm}
			\caption{ (Color online) The  \textit{p}- and 2\textit{p}- separation energies ($ S_p $ and $ S_{2p} $, respectively) of oxigen, neon and argon isotopes (circles and diamonds taken from \cite{Wang:2021,Brown:2015,Webb:2019a}) of mass\textit{ A}. The hollow symbols for Ar in panel (c) joined by a dashed line show the mass predictions \cite{Wang:2021}. Isotopes with specific structure and decay properties are highlighted by arrows and the corresponding text legends.}
			\label{fig:4_Sp_O_s_Ne_Ar}
		\end{minipage}
	\end{center}
\end{figure}

As already mentioned, different decay mechanisms (or forms of nuclear dynamics) depend strongly on the ratio of the 1\textit{p}- and 2\textit{p}- separation energies $ S_{p} $ and $ S_{2p} $.
For illustration, the systematics of $ S_{p} $ and $ S_{2p} $ for O, Ne and Ar isotopes near the proton dripline is shown in Fig.~\ref{fig:4_Sp_O_s_Ne_Ar}. The chosen isotope chains represent nuclear structure around the closure of the \textit{p-} shell as well as the lower and the upper parts of the \textit{s-d} shell configurations.
One can see that the \textit{A} dependence of $ S_{p} $ and $ S_{2p} $ values is different, and the corresponding curves intersect near the proton dripline, which causes significant changes of the dynamical properties of nuclides in this region.
Bound nuclei that are closest to the dripline typically exhibit a so-called Borromean structure
(when removal of one out of the three bodies, either one of the protons or the core, leads to disintegration of the whole three-body system) with the condition $ S_p > S_{2p} $.
Unbound nuclei located just beyond the dripline often undergo simultaneous decays, i.e.\ either 2\textit{p} decay by the true or by the democratic mechanism. In addition, a transition situation is possible. Nuclei located further beyond the dripline pass the transition area and occur either in a domain of democratic 2\textit{p} decays (typically for the lightest nuclei) or in a domain of sequential 2\textit{p} decays (rather expected for the heavier, higher-\textit{Z} systems). Thus, the transition dynamics is a quite likely situation in the area between the dominating simultaneous (true or democratic) and sequential 2\textit{p}-decay mechanisms.

Similar evolution of the 2\textit{p}-decay mechanisms might occur in energy spectrum of an individual isotope whose levels decay depending on excitation energy $ E^* $, as $ E_{T}=E^*-S_{2p} $. However, contrary to isotopic chains, the nuclear structure of excited states is usually more complicated than the ground state which may alter the 2\textit{p} decay pattern.

\begin{table} [h]
	\caption{Experimentally studied 2\textit{p}-emitting states in light nuclei which are reviewed in this Chapter. The reactions of population, measured decay energies $ E_{2p} $, widths $ \Gamma_{2p} $ and derived decay mechanisms of the states with spin-parities $ J^\pi $ are collected. The experimental details can be found in the respective sections and references.}
		\label{tab:4_list}
		\vspace{0.5\baselineskip}
\begin{tabular}{|c|c|c|c|c|c|c|}
\hline
\hline
Nuclear & Sec- & Population & $ E_{2p} $,  & $ \Gamma_{2p} $, & Derived mode  & References \\
 state $ J^\pi $ & tion & reactions & MeV & MeV & of 2\textit{p}-decay & \\
\hline
 $ ^6$Be$ \;\; 0^+$ & 4.1 & $^{6}$Li($^{3}$He, \textit{t})$^{6}$Be & 1.37 & 0.09 & 3-body, democratic    & \cite{Geesaman:1977} \cite{Bochkarev:1984} \cite{Bochkarev:1989} \\
 &  & ($^{10}$C, $^{10}$C$^*$)$^{6}$Be+$\alpha$ &  &  & democratic    & \cite{Grigorenko:2009b} \\

 &  & ($^{7}$Be, $^{6}$Be) &  &  & democratic or true    & \cite{Egorova:2012} \\
\hline
 $ ^6$Be$\;\; 2^+$ & 4.1 & $ ^1 $H($^{6}$Li, $^{6}$Be)$ n $ & 3.05 & 1.16 & $ 0^+, 2^+$ interference & \cite{Chudoba:2018} \\
\hline
$ ^6$Be$ ^* $ continuum & 4.1 & $ ^1 $H($^{6}$Li, $^{6}$Be)$ n $ & 5--15 &  & soft dipole mode & \cite{Fomichev:2012} \\
\hline
$ ^{11} $O ($3/2^-$)	& 4.2 & ($ ^{13} $O, $ ^{11} $O) & 4.25 & 2.3  & sequential & \cite{Webb:2020} \\
	\hline
$ ^{12} $O$\;\; 0^+$	& 4.2 &  ($ ^{13} $O, $ ^{12} $O) & 1.74 & <0.07 & democratic & \cite{KeKelis:1978,Kryger:1995,Suzuki:2009,Webb:2019} \\
	\hline
$ ^{15} $Ne ($ 3/2^- $)	& 4.3  & ($ ^{17} $Ne, $ ^{15} $Ne) & 2.52 & 0.59 & transition & \cite{Wamers:2014} \\
	\hline
$ ^{16} $Ne$\;\; 0^+$	& 4.3 & ($ ^{17} $Ne, $ ^{16} $Ne) & 1.46 & <0.08 & democratic & \cite{Mukha:2008,Mukha:2009,Mukha:2010,Brown:2014} \\
	\hline
$ ^{16} $Ne$^*\;\; 2^+$	& 4.3 & ($ ^{17} $Ne, $ ^{16} $Ne) & 3.16 & $ \leq $0.05 &   transition,  & \cite{Brown:2015} \\
& & & & & i.e.~ ``tethered'' 2\textit{p} & \\
	\hline
$^{17}$Ne$^* \; 3/2^-$ & 4.4 & ($ ^{17} $Ne, $ ^{17} $Ne$ ^* $)  & 1.288 & ? & true 2\textit{p} & \cite{Chromik:2002} \\
 &  & $ ^{1} $H($ ^{18} $Ne, $ ^{17} $Ne$ ^* $)\textit{d}  &  &   &  & \cite{Sharov:2017} \\
\hline
$^{17}$Ne$^* \; 5/2^-$ & 4.4 & ($ ^{17} $Ne,$ ^{17} $Ne$ ^* $)  & 1.75 & 0.02 & sequential & \cite{Charity:2018} \\
\hline
$ ^{19} $Mg$\;\; 1/2^- $	& 4.5 & ($ ^{20} $Mg, $ ^{19} $Mg) & 0.75 & 1.1$\cdot10 ^{-10} $ & true 2\textit{p} & \cite{Mukha:2007,Voss:2014,Brown:2017,Xu:2016,Xu:2018} \\
	\hline
$ ^{29} $Ar 	& 4.6 & ($ ^{31} $Ar, $ ^{29} $Ar) & 5.5 & ? & sequential  & \cite{Mukha:2018} \\
	\hline
$ ^{30} $Ar$\;\; 0^+$	& 4.6 & ($ ^{31} $Ar, $ ^{30} $Ar) & 2.25 & <4$\cdot10 ^{-11} $ & transition & \cite{Mukha:2015,Xu:2016, Xu:2018} \\
	\hline
	\hline
\end{tabular}
\end{table}

At the moment, experimental information has been obtained for a number of short-lived 2\textit{p}-emitting nuclei, \textit{i.e.},
$^{6}$Be,
%
$^{11}$O,
$^{12}$O,
%
$^{15}$Ne,
$^{16}$Ne,
%
$^{19}$Mg,
%
%
$^{29}$Ar, and
$^{30}$Ar.
%
Location of the studied nuclei on the nuclear chart is shown on Fig.~\ref{fig:1_Chart}, left panel. Most of them have been populated in reactions with radioactive beams.
In the following subsections we will describe them in more detail
with the focus on the 2\emph{p} decay mechanism and its evolution, as well as on the nuclear-structure information which can be gained in these studies.
In this overview, we include also an interesting case of excited states in $^{17}$Ne.
The summary of the 2\emph{p}-emitting states discussed in this Section is given in Table~\ref{tab:4_list}.

\subsection{\it Decay of $ ^{6}$Be}

The lightest 2\textit{p} unbound system,
$ ^{6} $Be, has been studied in the most detail, since it is relatively easy to be reached experimentally by using reactions with stable beams. Thus, data with very high statistics can be accumulated.
This case may be analysed by assuming a simple
$ \alpha $+\textit{p}+\textit{p} configuration of $ ^{6} $Be, where two valent protons are in a $ p_{3/2} $ shell.

\begin{figure}[h]
	\begin{center}
		\begin{minipage}[t]{16 cm}
\hspace{2. cm}
			\includegraphics[width = 0.7\columnwidth, angle=0.]{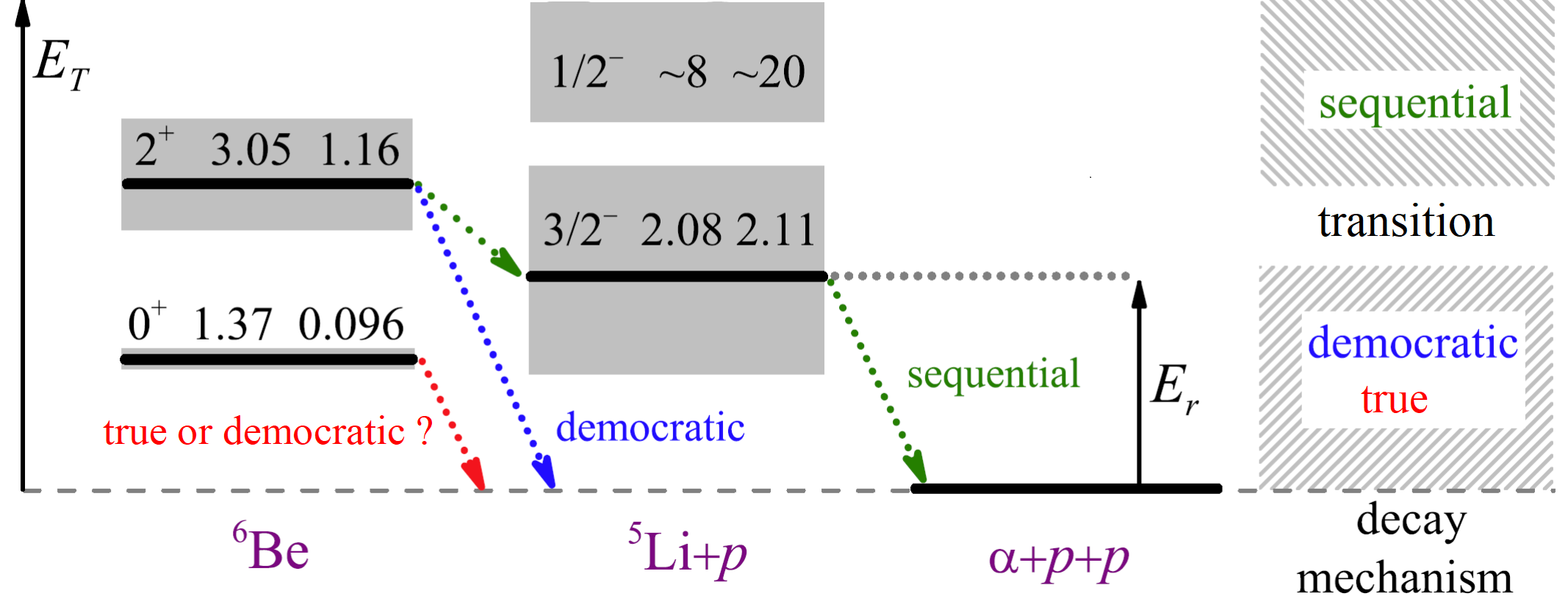}
		\end{minipage}
		\begin{minipage}[t]{16.5 cm}
			\caption{(Color online) Level and decay scheme for $ ^{6} $Be reported in \cite{Egorova:2012} and illustrations of involved decay mechanisms. The continuum
states are labeled with $ J^{\pi}, E_{r}, \Gamma $. Dotted arrows point to 1\textit{p-} and 2\textit{p-} decay transitions. Intermediate resonances $ ^{5} $Li are in the middle. Evolution of decay mechanisms is in the right panel.
			}
			\label{fig:4_6Be_level_scheme}
		\end{minipage}
	\end{center}
\end{figure}

The most recent level scheme of $ ^{6} $Be reported in \cite{Egorova:2012} is shown in Fig.~\ref{fig:4_6Be_level_scheme}. By comparing with the general energy conditions of 2\textit{p} decays in Fig.~\ref{fig:1_EnergyConditions}(b,c), one may expect a democratic decay mode for the ground state of $ ^{6} $Be and a sequential proton emission for its excited states with an intermediate region between these alternative mechanisms.

%

The first investigations of the $^{6}$Be 2\textit{p}
decay started via studies of a charge-exchange reaction $^{6}$Li($^{3}$He,\emph{t})$^{6}$Be \cite{Geesaman:1977, Bochkarev:1984}. %
The measured  \textit{ p-p} correlations from the
ground state 0$ ^{+} $ of $^{6}$Be could not be explained by using simple decay scenarios (phase volume, diproton decay, simultaneous emission of \textit{p}-wave protons),
thus three-body approach was applied in order to understand the measured spectra.
The first kinematically complete experiment on decay of $^{6}$Be (when two decay products, proton and  $\alpha$ particle were measured in coincidence)  demonstrated that much more rich information including three-body  correlations of its fragments can be acquired \cite{Bochkarev:1989}. Though the three-body correlations were measured in a restricted phase volume and with small statistics, they demonstrated presence of specific \textit{p-p} correlation structures  (e.g., \textit{dinucleon}, \textit{cigar} and \textit{helicopter}), which could be described by using the concept of \textit{democratic decay} \cite{Bochkarev:1989}.

\begin{figure}[h]
	\begin{center}
		\begin{minipage}[ht]{8 cm}
			\includegraphics[width = \columnwidth]{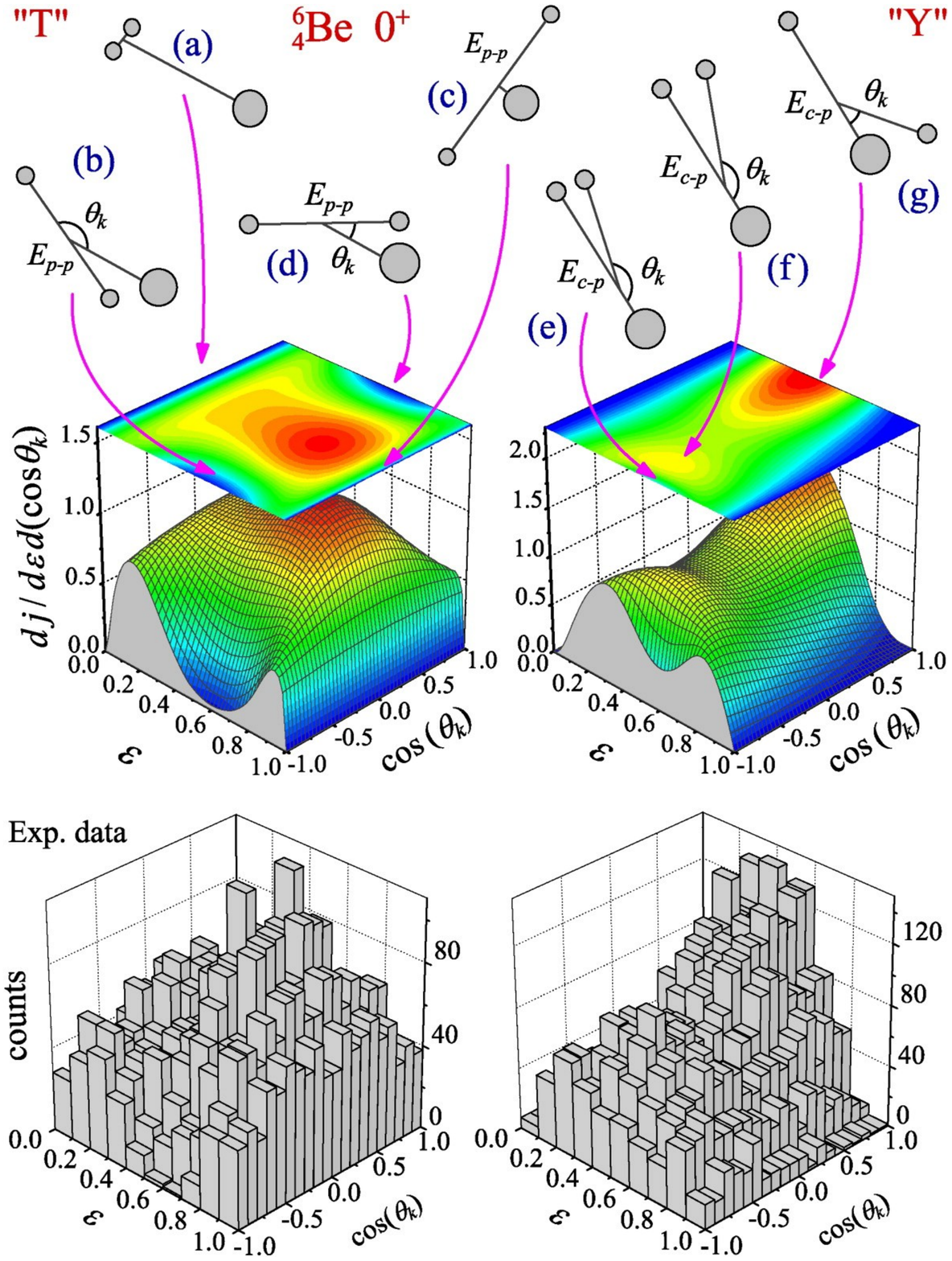}
		\end{minipage}
		\begin{minipage}[t]{16 cm}
			\caption{(Color online) Complete correlation pattern for $ ^{6} $Be g.s.\ decay, presented in the T and Y Jacobi systems (left and right columns, respectively). The upper row is theoretical description with notations of kinematical variables, the
			lower shows the respective experimental data.
			Data and calculations are from \cite{Grigorenko:2009b} and \cite{Grigorenko:2012c}, respectively.
			}
			\label{fig:4_6Be_data_TEXAS}
		\end{minipage}
	\end{center}
\end{figure}

The  first complete-range three-body correlations from decay of the
$ ^{6} $Be ground state were measured  by detecting all decay products, $\alpha$+\textit{p}+\textit{p} \cite{Grigorenko:2009b}.
For the first time, a radioactive beam of $^{10}$C (produced in the reaction
$^{10}$B(\textit{p}, \emph{n})$^{10}$C) was used to populate states in a 2\textit{p} precursor. The secondary-reaction chain was $^{10}$C$ \rightarrow ^{10} $C$^* \rightarrow \alpha + ^{6}$Be$ \rightarrow 2\alpha $+2\textit{p}, and
the recorded coincidences 2$ \alpha $+2\textit{p} were used to derive the three-body $\alpha$+\textit{p}+\textit{p} correlations.
The obtained correlations based on ~1000 decay events of the $^{6}$Be ground state were found to be in a very good agreement with predictions of the theoretical three-body cluster model \cite{Grigorenko:2009b}.
Both experimental and theoretical distributions in the
T and Y Jacobi coordinates are presented in Fig.~\ref{fig:4_6Be_data_TEXAS}.

The measured correlations are sensitive to details of the structure of $ ^{6} $Be ground state.
For example, the energy distribution $\epsilon$ has broader and narrower
profiles in the T and Y systems, respectively, which is an indication of the important
contribution of the $ p^{2} $-configuration in the wave function of $ ^{6} $Be ground state
according to the three-body cluster model \cite{Grigorenko:2012c}.
These data allowed for an analysis of fine details of correlation patterns, which showed
the need for data with higher statistics and with better energy and angular resolution.
\begin{figure}[t]
	\begin{center}
		\begin{minipage}[t]{10 cm}
			\includegraphics[width = \columnwidth]{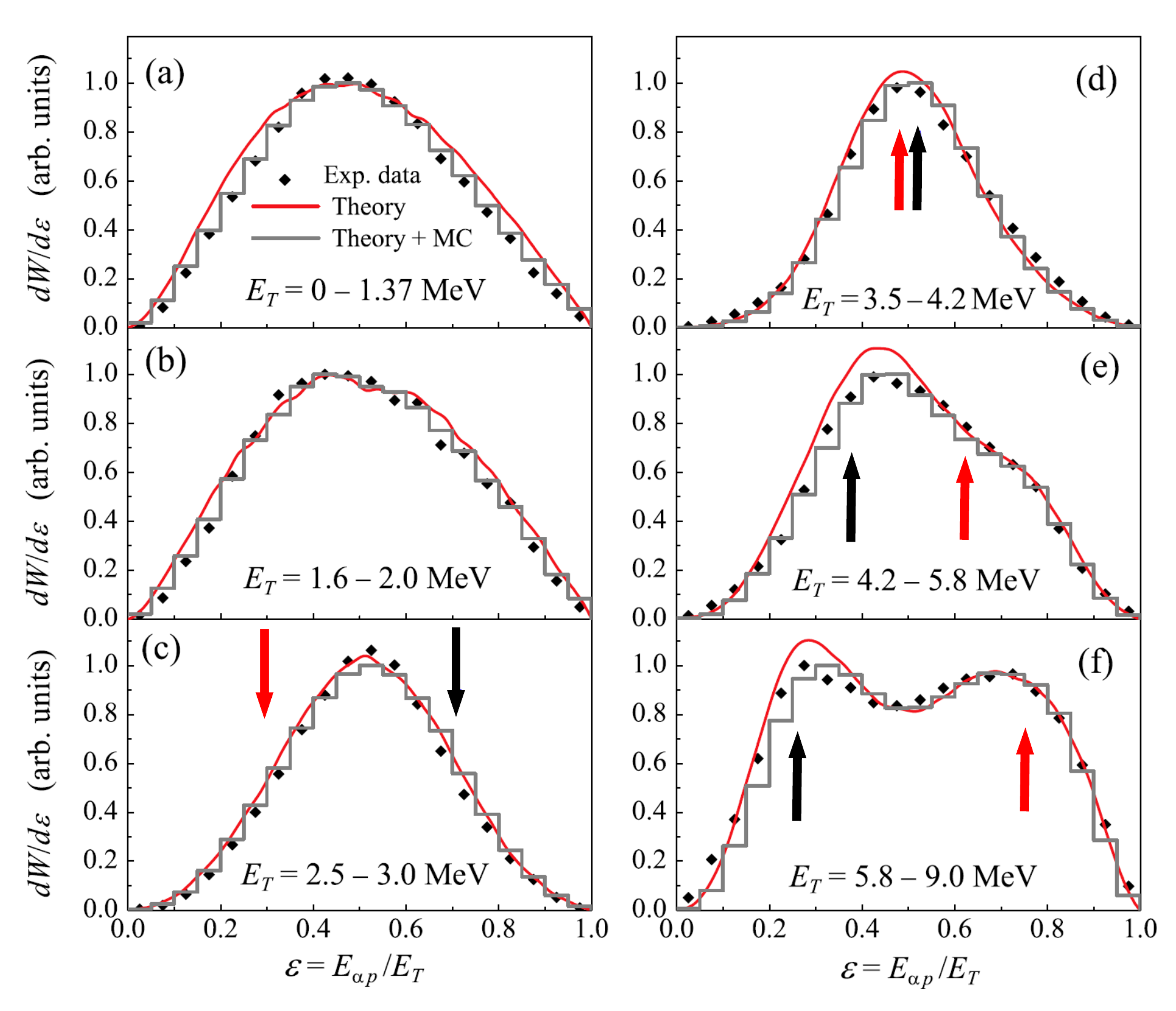}
		\end{minipage}
		\begin{minipage}[t]{16 cm}
	\caption{(Color online) Distributions of relative  energy between the $\alpha$ particle and one	of the protons, $ E_{\alpha p} $  selected with the decay energy $E_T$ of $ ^{6} $Be system, as reported in Ref.~\cite{Egorova:2012}. Evolution of the distribution shapes with $E_T$ reflects a change from the democratic decay mechanism (which dominates in the panels (a,b)) to the sequential proton emission via intermediate $ ^{5}$Li (dominating in the panels (e,f)). The red and black arrows indicate two possible positions of the $ ^{5}$Li ground-state resonance between $\alpha-p_{1}$ and $\alpha-p_{2}$, respectively. Data are taken from \cite{Egorova:2012}.		
		}
			\label{fig:4_6Be_Ycorr_MSU_Egorova}
		\end{minipage}
	\end{center}
\end{figure}

Such a high-statistics experiment on $^{6}$Be was performed by using a 70 \textit{A}MeV $^{7}$Be secondary beam populating levels in $^{6}$Be by neutron-knockout reactions on a Be target
\cite{Egorova:2012}. The three-body decay of the ground and first excited states into the $ \alpha $+\textit{p}+\textit{p} exit channel were detected in coincidence. Precise three-body correlations extracted from the experimental data allowed for insights into the mechanism of the three-body democratic decay and its evolution. This can be illustrated on the example of the
spectrum of relative  energy between the $\alpha$ particle and one
of the protons, $ E_{\alpha p} $. Several such distributions gated by the decay energy of $ ^{6} $Be, $E_T$, are shown in Fig.~\ref{fig:4_6Be_Ycorr_MSU_Egorova}. The data
are in good agreement with a three-body cluster-model calculations, which validate this theoretical approach over a broad energy range.
As seen in Fig.~\ref{fig:4_6Be_Ycorr_MSU_Egorova}(a,b), at low $E_T$ the shape
of the energy distribution has a relatively broad bell-like profile typical for the
true 2\textit{p} decay. However, as $E_T$ increases, the profile first becomes
significantly narrower which happens when the $ ^{5} $Li ground-state resonance enters the decay window, see Fig.~\ref{fig:4_6Be_Ycorr_MSU_Egorova}(c).
For $E_T$ < 2$ E_r $($ ^{5} $Li), the availability of the two-body $\alpha$-\textit{p} resonance
for sequential decay does not lead to a pattern typical for a sequential decay with two peaks. Evidence for such sequential correlations are only observed when
$E_T$ > 2$ E_r $($ ^{5} $Li) + $\Gamma (^{5} $Li), see Fig.~\ref{fig:4_6Be_Ycorr_MSU_Egorova}(f).
This evolution of decay mechanisms is indicated schematically in the right panel of
Fig.~\ref{fig:4_6Be_level_scheme}.

\begin{figure}[t]
\hspace{3. cm}
	\begin{center}
		\begin{minipage}[t]{6 cm}
			\includegraphics[width = \columnwidth]{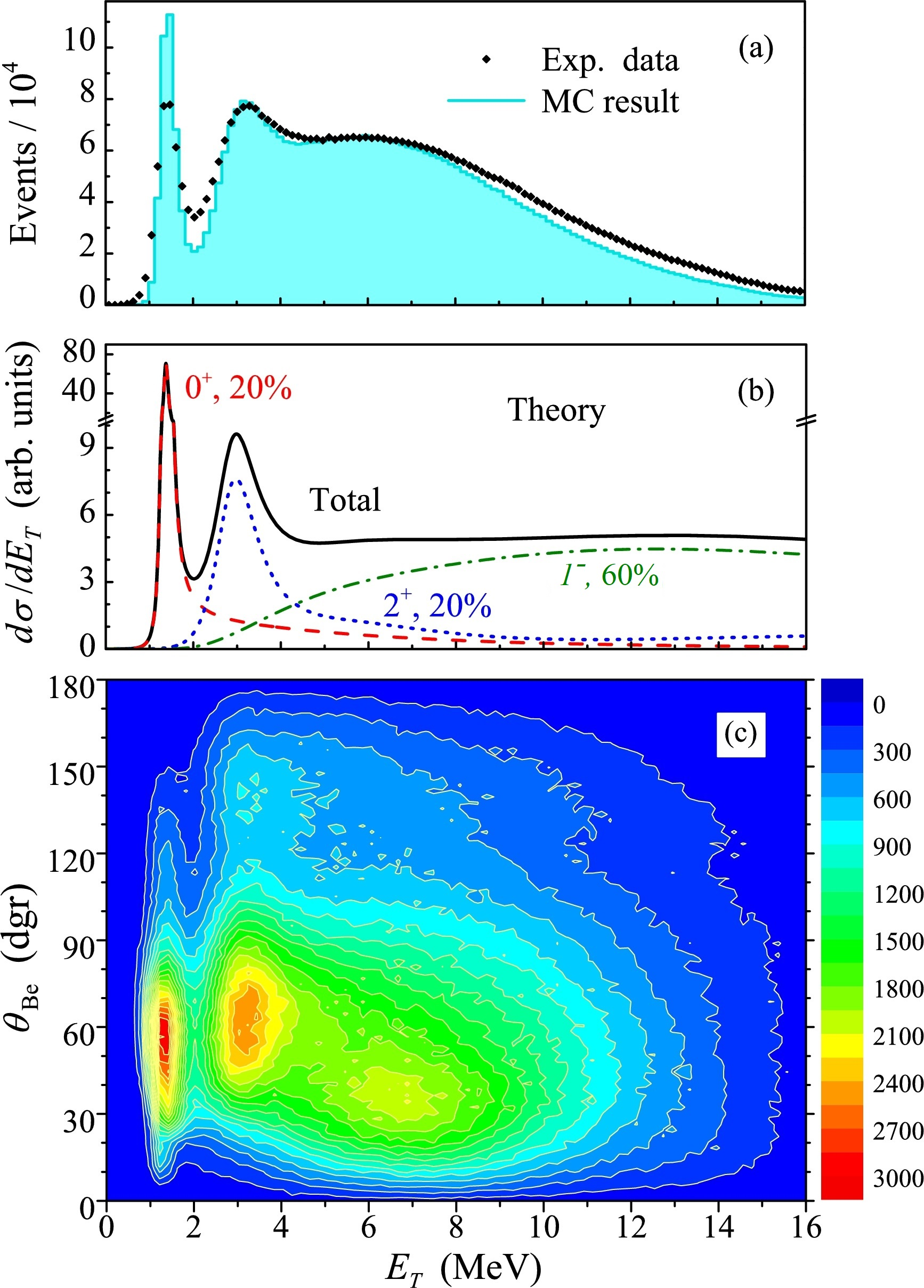}
		\end{minipage}
		\begin{minipage}[t]{16 cm}
\caption{(Color online) (a) Experimental energy spectrum of $ ^{6} $Be (diamonds) and MC simulations (shaded histogram).
 (b) Spectrum obtained  by correcting the data for detection efficiency used in  MC simulation; the contributions of different $ J^{\pi} $ components is indicated.
	(c) Contour plot of the spectrum in the {$ E_{T}, \theta_{\rm Be}$} plane. The $\theta_{\rm Be}$ is the $ ^{6} $Be emission angle in the c.m.\ system of reaction $ ^{1} $H($ ^{6} $Li, $ ^{6} $Be)\textit{n}.
	Data and calculations are from \cite{Fomichev:2012}.
			}
			\label{fig:4_6Be_data_IVSDM}
		\end{minipage}
	\end{center}
\end{figure}

Further progress in studies of $ ^{6}$Be was achieved
by using the $ ^{1} $H($ ^{6} $Li, $ ^{6} $Be)\textit{n} charge-exchange reaction, where population of continuum states in $ ^{6} $Be was observed
up to $ E_{T} $ = 16 MeV  \cite{Fomichev:2012}. In kinematically
complete measurements performed by detecting $\alpha + p + p$ coincidences,
the $ E_{T} $ spectrum of high statistics was obtained.
It provided detailed correlation
information about the  0$ ^{+} $ ground state of $ ^{6} $Be at $ E_{T} $ = 1.37 MeV and its 2$ ^{+} $ state at $ E_{T} $ = 3.05 MeV, see Fig.~\ref{fig:4_6Be_data_IVSDM}.
Moreover, a broad structure extending from 4 to 16 MeV was measured. It contains negative
parity states populated by $L = 1$ angular momentum transfer without other significant contributions. This structure was interpreted as a novel phenomenon, i.e. the isovector soft dipole mode (IVSDM) \cite{Fomichev:2012} associated with the $ ^{6} $Li ground state.
Recently, the same reaction was used to study in detail the $ ^{6}$Be excitation energy region below 3 MeV, where contributions from the ground 0$ ^{+} $ state overlap with the broad first
excited 2$ ^{+} $ state \cite{Chudoba:2018}. Both experiments provide good examples of the
detail investigation of an exotic system via the three-body $\alpha + p + p$ correlations.

\subsection{\it Decays of $ ^{12}$O and $ ^{11}$O}


The two oxygen isotopes beyond the proton drip line, $ ^{12} $O and $ ^{11} $O, may provide information on nuclear structure with a supposedly closed \textit{p}-shell proton configuration and on decay properties of these 2\textit{p}-unbound systems.

\begin{figure}[ht]
	\begin{center}
		\begin{minipage}[t]{12 cm}
			\includegraphics[width = 0.6\columnwidth]{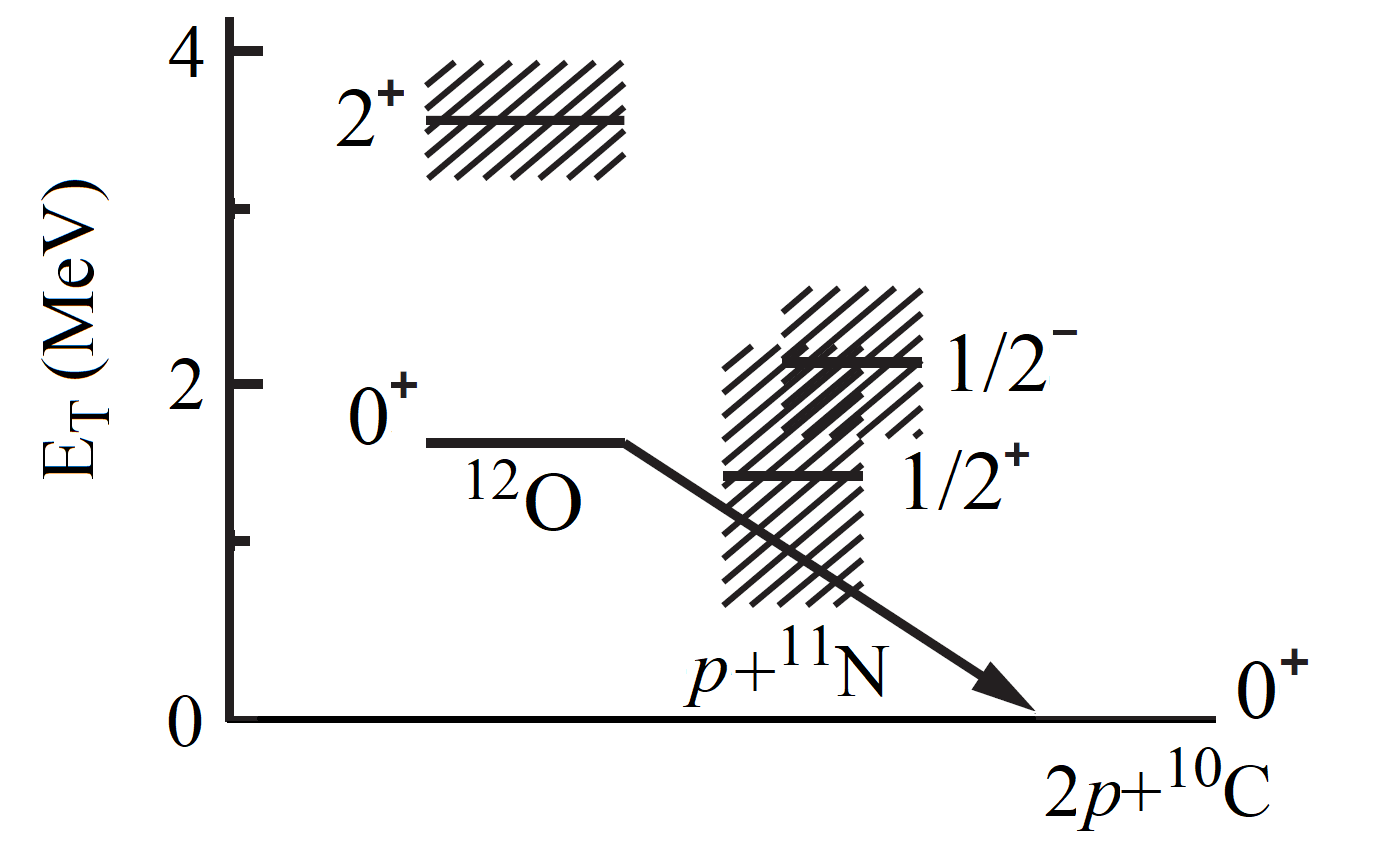}
		\end{minipage}
		\begin{minipage}[t]{16 cm}
	\caption{ Level and decay scheme for $ ^{12} $O reported in \cite{Jager:2012}. The
	$^{12}$O and $^{11}$N	states are labeled  by $ J^{\pi} $. The arrow indicates the
2\textit{p-} decay transition to $^{10}$C.}
			\label{fig:4_12O_level_scheme}
				\end{minipage}
	\end{center}
\end{figure}

The first study of opening angle between protons emitted
from $ ^{12} $O was motivated by the search for a true 2\emph{p} emission \cite{Kryger:1995}.
However, the measured spectrum was explained by the dominating sequential emission via an intermediate $ ^{11} $N state. Later it was found that indeed the ground-state
energy of $ ^{11} $N is below that of $ ^{12} $O \cite{Axelsson:1996,Suzuki:2009}.
The presently-known level and decay scheme of $ ^{12}$O illustrating possible 2\textit{p}-decay mechanism is sketched in Fig.~\ref{fig:4_12O_level_scheme}. The intermediate ground state of $^{11}$N is seen to be very broad, which points to a democratic 2\textit{p}-decay mechanism of the $^{12}$O ground state and to a sequential proton emission for the first excited state $2 ^{+} $.

\begin{figure}[t!]
	\begin{center}
		\begin{minipage}[t]{12 cm}
			\includegraphics[width = 0.7\columnwidth]{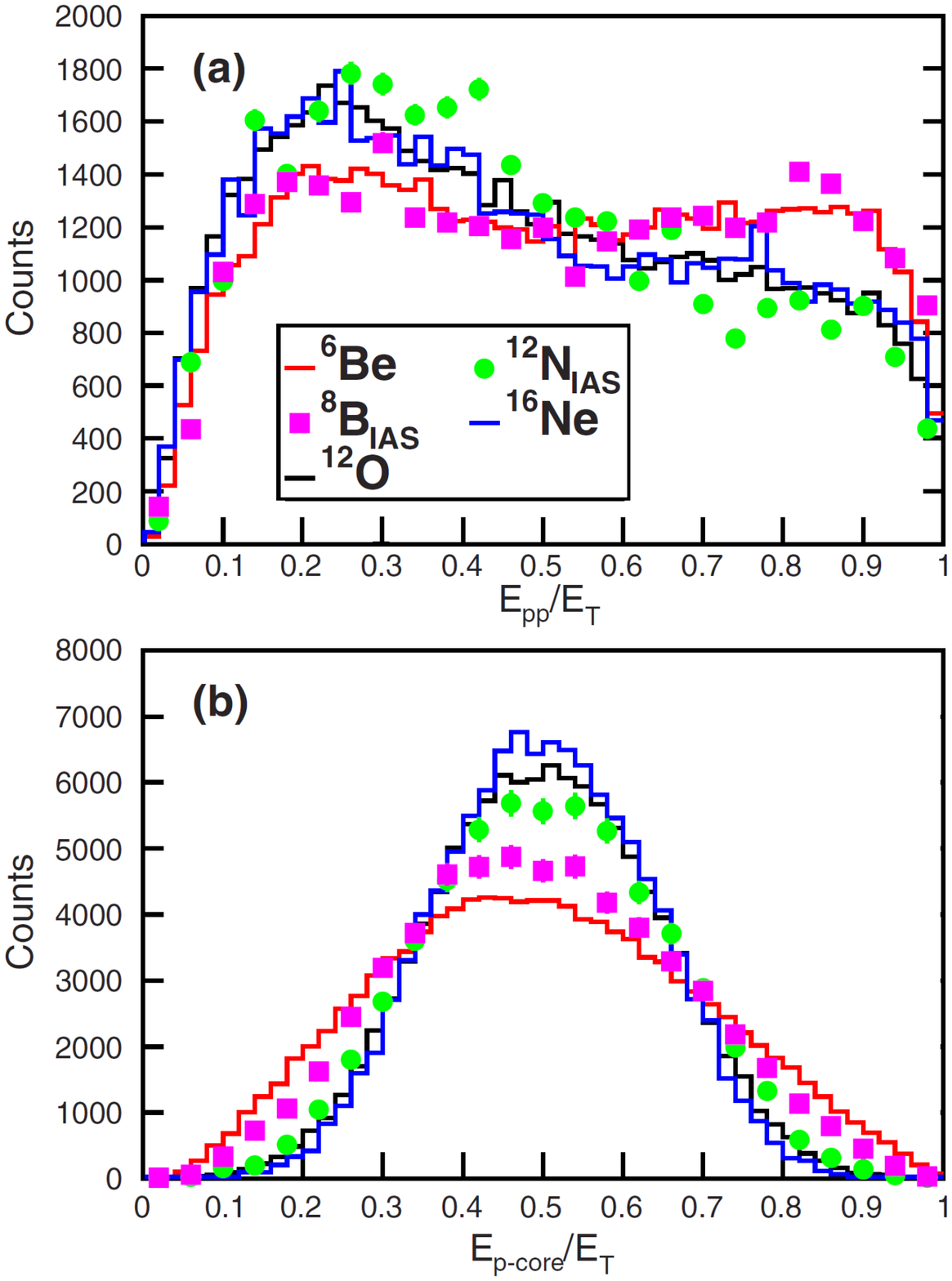}
		\end{minipage}
		\begin{minipage}[t]{16 cm}
			\vspace{-2.9cm}
			\caption{(Color online) Three-body correlations measured in 2\textit{p} decays of $^{12}$O ground state \cite{Webb:2019} compared with those of the $ ^{6} $Be \cite{Egorova:2012} and $ ^{16} $Ne \cite{Brown:2014}. Projections on the relative energies between decay products $ E_{pp} $   and $ E_{p-{\rm core}} $ are shown in the panels  (a) and (b), respectively. Similar spectra are shown for 2\textit{p} decays of isobaric analog states (IAS) of $ ^{ 12} $O in $ ^{12} $N$*$ and  $ ^{ 8} $C in $ ^{8} $B$*$ \cite{Webb:2019}. All energies are normalised to the total decay energy$  E_T $. 	 The term \textit{ core} is for the remnant fragment after 2\textit{p} emission. Reprinted with permission from Ref.~\cite{Webb:2019}. Copyright (2019) by the American Physical Society.
			}
			\label{fig:4_12O_6Be_12N_16Ne_corr}
		\end{minipage}
	\end{center}
\end{figure}


The measured width of the $ ^{12} $O ground state $\Gamma<72$  keV \cite{Jager:2012} was found to be consistent with the theoretical prediction assuming democratic 2\emph{p} decay mechanism \cite{Grigorenko:2002}. The calculations in the three-body cluster model yielded the value of 60~keV for this width, but also pointed to a large difference between the structure of $^{12}$O and its isobaric mirror, with a markedly stronger $s^2$ component in the ground state wave function of $^{12}$O. The mirror asymmetry in nuclei when valence nucleons occupy the $s_{1/2}$ orbital
is known as the Thomas-Ehrman shift (TES) \cite{Ehrman:1951,Thomas:1952}. Its main effect is
the lowering of the $s$-dominated states in the proton-rich partner due to increased radial extent of the weakly-bound valence proton. The finding in $^{12}$O indicated that in addition,
the composition of the wave function may differ between mirror partners. This effect was
called the ``three-body mechanism'' (or ``dynamic'') TES \cite{Grigorenko:2002}.

\begin{figure}[t]
	\begin{center}
		\begin{minipage}[t]{12 cm}
			\includegraphics[width = 0.99\columnwidth]{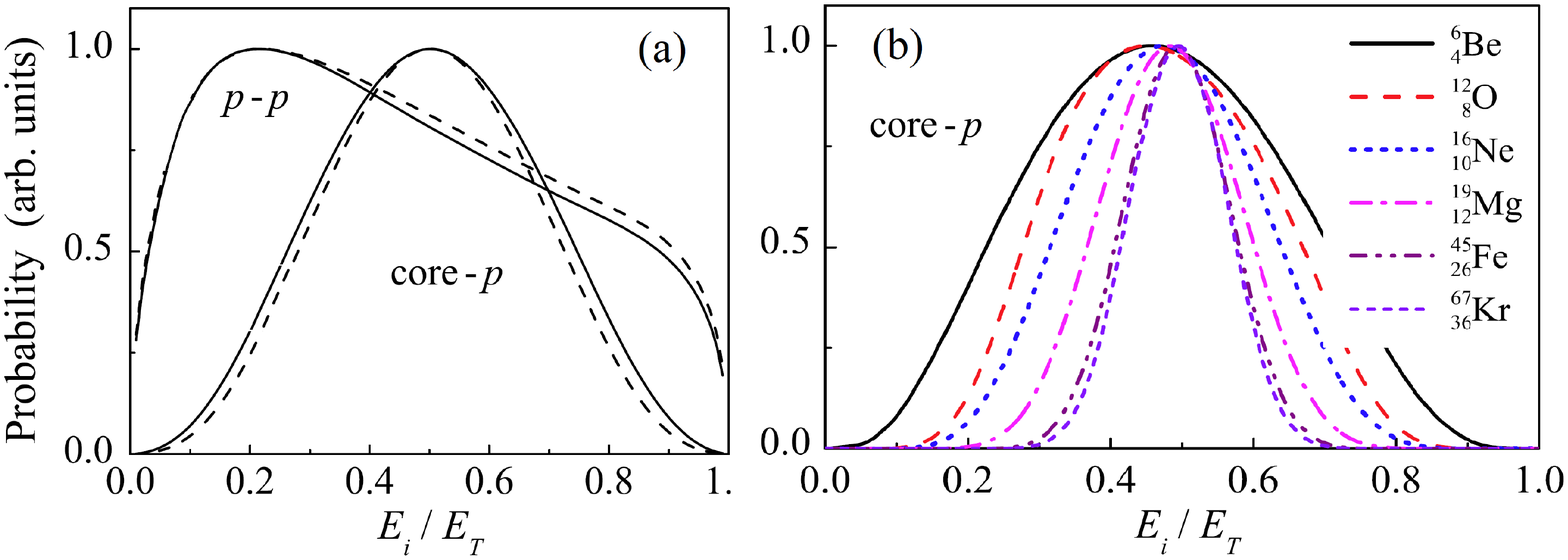}
		\end{minipage}
		\begin{minipage}[t]{16 cm}
			\vspace{-4.5cm}
			\caption{(Color online) Relative-energy ($  E_i $) correlations for \textit{p-p} and core\textit{-p} subsystems in 2\textit{p} decays of: (a) $ ^{ 12} $O and $ ^{16} $Ne (solid and dashed lines, respectively) predicted by the three-body model \cite{Grigorenko:2002}; (b) Illustration of Coulomb repulsion between decay products. Core-\textit{p} correlations of 2\textit{p} emitters  from Be(\textit{Z}=4) to Kr(\textit{Z}=36) calculated by the direct decay model \cite{Grigorenko:2007b}. $  E_i $ is normalised to the total decay energy $ E_T $.}
			\label{fig:4_12O_16Ne_corr_theory}
		\end{minipage}
	\end{center}
\end{figure}

Using one-neutron knock-out reaction with a $^{13}$O secondary beam at the energy of
70~\textit{A}MeV, the ground-state correlations of fragments in 2\textit{p} decay of $ ^{12} $O
were measured \cite{Webb:2019}. The results  are shown in Fig.~\ref{fig:4_12O_6Be_12N_16Ne_corr}
where they are compared with those from the isobaric analog state (IAS) of $ ^{12} $O in $ ^{12} $N$ ^{*} $. In addition, similar spectra are shown for the $ ^{6} $Be \cite{Egorova:2012} where both valence protons are in the $p_{3/2}$ shell configuration.
The measured correlations have basically the same shapes as those predicted by the three-body model \cite{Grigorenko:2002}, see Fig.~\ref{fig:4_12O_16Ne_corr_theory}. The difference is mainly due to Coulomb repulsion in the core-\textit{p} subsystem, see Fig.~\ref{fig:4_12O_6Be_12N_16Ne_corr}(b). Thus, the assumed dominance of a single 2\textit{p} configuration in structure of 2\textit{p} precursor and a three-body decay mechanism are sufficient for description of the data shown.

Recently, similar data on 2\textit{p} decay were reported for the more exotic isotope $^{11}$O \cite{Webb:2019a,Webb:2020a} which is the mirror of the well-known two-neutron halo nucleus $^{11}$Li.
The measured decay pattern of $^{11}$O \cite{Webb:2019a,Webb:2020a} is more complicated for interpretation in comparison with $^{12}$O. The obtained invariant mass
spectrum shows a broad peak of width $ \sim $3.4 MeV, which is difficult to explain by a single component in the 2\textit{p}-configuration  of $^{11}$O by using a $ ^{9} $C+\textit{p}+\textit{p} cluster model. Within the Gamow coupled-channel approach \cite{Wang:2019,Webb:2020a} developed for deformed nuclei, it was concluded that this peak is a multiplet with contributions from the four lowest  $^{11}$O resonant states with $ J^{\pi} $=3/2$^{-}_1$, 5/2$ ^{+}_1$, 3/2$ ^{-}_2$, 5/2$ ^{+}_2$.
Interestingly, only a moderate isospin asymmetry between ground states of $^{11}$O and $^{11}$Li was observed.

\subsection{\it Decays of $ ^{16} $Ne and $ ^{15} $Ne}

The systematics of the measured proton separation energies (see  Fig.~\ref{fig:4_Sp_O_s_Ne_Ar}) indicates, that the ground states of two unbound neon isotopes, $ ^{16} $Ne and $ ^{15} $Ne, decay by emission of two protons. The democratic or true 2\textit{p} decay mechanism is plausible for $ ^{16} $Ne, and the  2\textit{p} emission  of $ ^{15} $Ne occurs in a transition region between direct and sequential mechanisms.

The decay of $ ^{16} $Ne was investigated  experimentally in Refs.~\cite{Mukha:2008,Mukha:2009,Mukha:2010,Brown:2014,Brown:2015}.
It was produced by one-neutron knock-out reactions from the secondary
beam of $ ^{17} $Ne, which was produced first in the fragmentation reaction
of the stable $^{20}$Ne.
\begin{figure}[h]
	\begin{center}
		\begin{minipage}[t]{16 cm}
\hspace {3. cm}
			\includegraphics[width = 0.5\columnwidth]{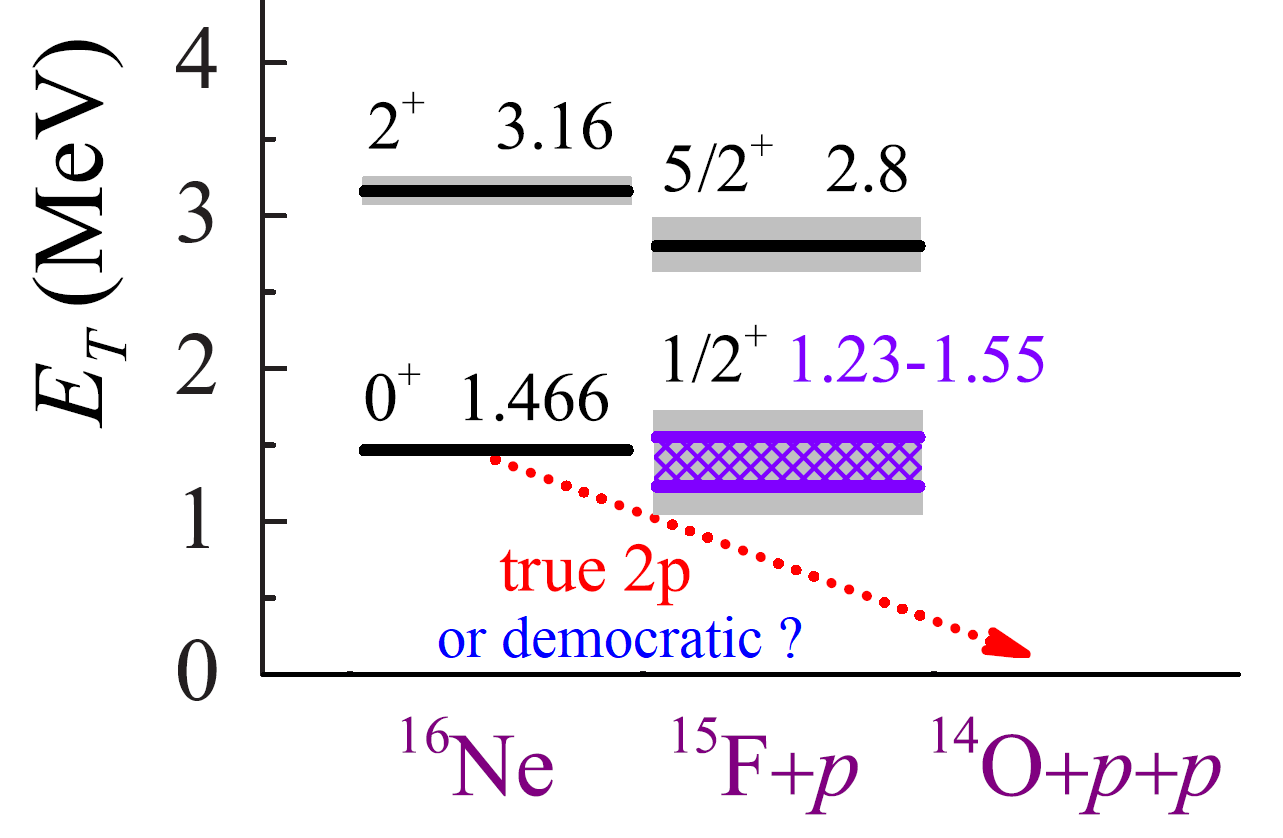}
		\end{minipage}
		\begin{minipage}[t]{16.5 cm}
			\caption{(Color online) Level and decay scheme for $ ^{16} $Ne \cite{Brown:2014}. The	$^{16}$Ne and intermediate $^{15}$F	states are labeled  by $ J^{\pi}$ and $E_T $. The ground state of the intermediate $ ^{15}$F was determined to be in the range 1.23--1.55 MeV. The arrow indicates the 2\textit{p} decay transition to $^{14}$O. }
			\label{fig:4_16Ne_level_scheme}
		\end{minipage}
	\end{center}
\end{figure}

The level scheme of  $ ^{16} $Ne with the ground and first excited states as well as intermediate $^{15}$F states are shown in Fig.~\ref{fig:4_16Ne_level_scheme}. In the first approximation, structure of the ground $0 ^{+} $ state  has a dominating $ [s^{2}] $-wave proton configuration, and the first excited 2$ ^{+} $ state is mainly a $ [d^{2}] $-wave of valence protons. Similar structure may be expected for the $^{15}$Ne ground $3/2 ^{+} $  and first-excited $5/2 ^{+} $ states ($ E_T $ of 2.52 and 4.42 MeV, respectively).


%
\begin{figure}[h!]
	\begin{center}
		\begin{minipage}[t]{16 cm}
			\hspace{1.5 cm}
			\includegraphics[width = 0.8\columnwidth, angle=0.]{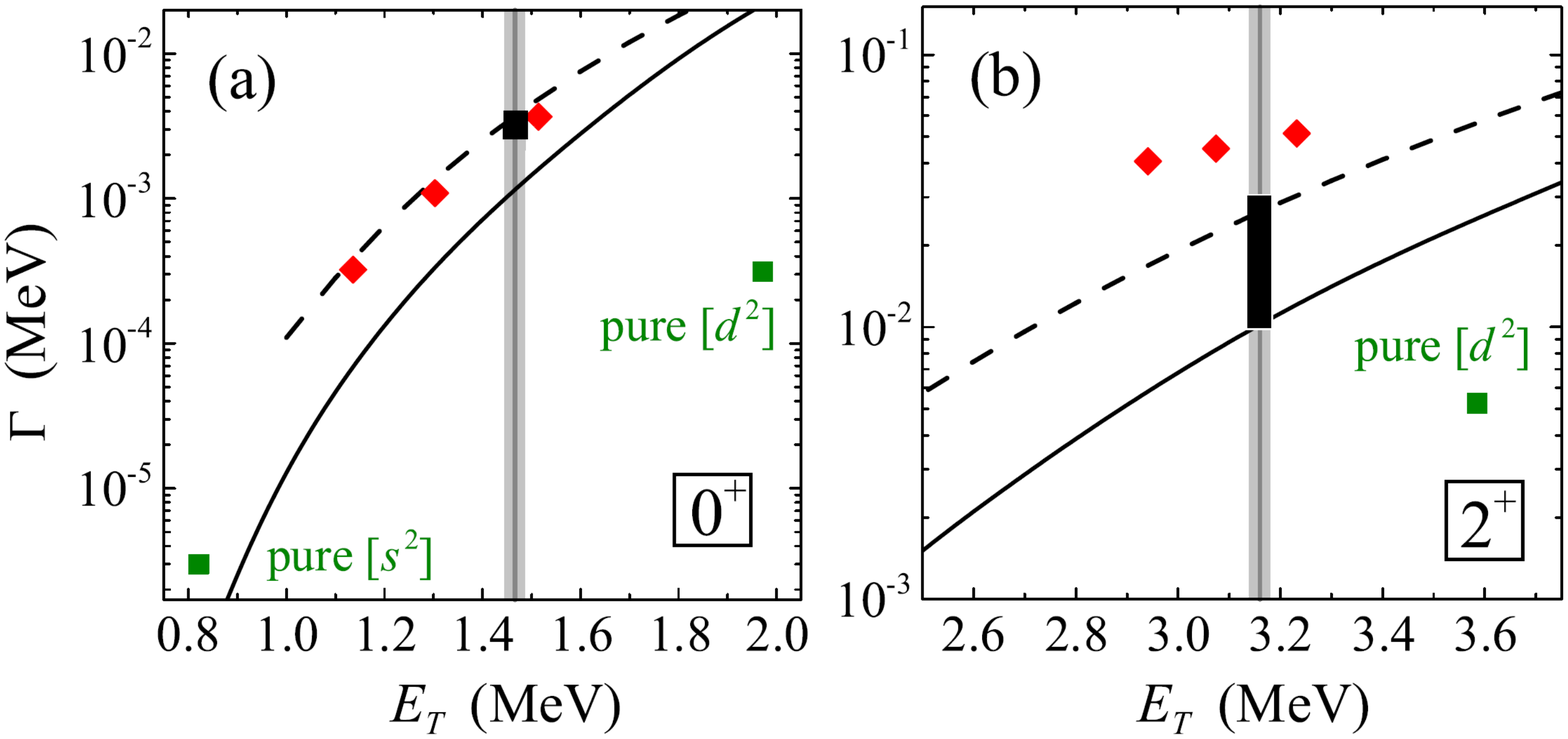}
		\end{minipage}
		\begin{minipage}[t]{16.5 cm}
			\vspace{-2.9cm}
	\caption{(Color online) Widths $ \Gamma $ of the $ ^{16} $Ne 0$ ^{+} $ and 2$ ^{+} $ states (the panels (a) and (b), respectively). The measured widths and decay energies $ E_T $ \cite{Brown:2014} (black rectangles whose dimensions correspond to their uncertainties) are compared with theory. Solid and dashed curves are the  three-body model predictions and the diproton model estimates, respectively, from the early work \cite{Grigorenko:2002}. Red diamonds are the results of the three-body cluster model predicting nuclear \textit{sd}-configuration mixture \cite{Grigorenko:2015} with the three assumed energies
of the $ ^{14} $O-\textit{p} intermediate resonance 1/2$ ^{+} $. Green squares show the results of calculations with limiting cases of nuclear structure: pure 	[$ s^{2} $] or pure [$ d^{2} $] configurations.
	Vertical gray lines and dashed areas show the experimental $ E_T $  with their uncertainties.
			}
			\label{fig:4_16Ne_widths_Grig2015}
		\end{minipage}
	\end{center}
\end{figure}

\begin{figure}[h!]
	\begin{center}
		\begin{minipage}[t]{14 cm}
			\includegraphics[width = 0.99\columnwidth, angle=0.]{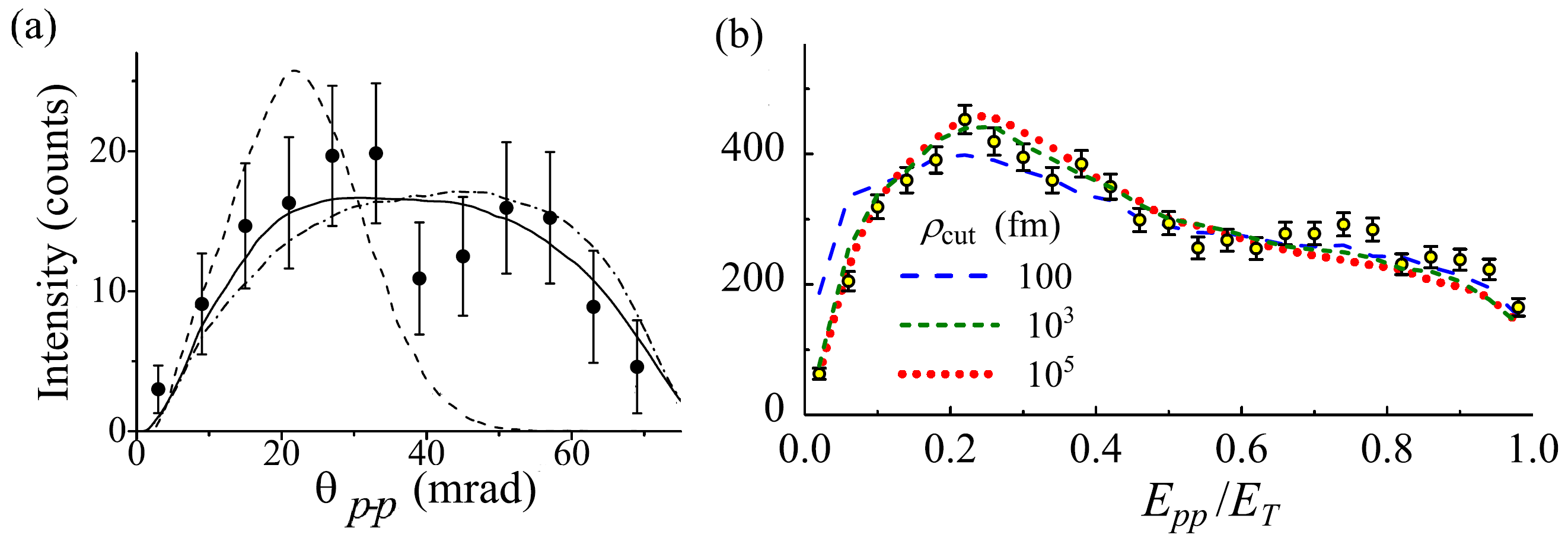}
		\end{minipage}
		\begin{minipage}[t]{16 cm}
\caption{ (a) Angular \textit{p-p} correlations from the 2\textit{p} decay of the ground state of $ ^{16} $Ne  (dots with statistical uncertainties) obtained from the measured $ ^{14} $O+\textit{p}+\textit{p} coincidences in Ref.~\cite{Mukha:2008}.
The	solid curve shows the three-body model calculations with the assumed 54\% of \textit{d}-wave configuration in the ground state. The dashed curve is the diproton model prediction, and the dash-dotted curve is the phase-space simulation of the 2\textit{p} decay assuming an isotropic proton emission.  (b) Relative-energy \textit{p-p} correlations $ E_{pp} $ from the 2\textit{p} decay of the $ ^{16} $Ne ground state measured in Ref.~\cite{Brown:2014} ($ E_{pp} $ is normalized to the decay energy $ E_T $). The theoretical description by three-body model with different parameter $\rho_{\rm cut}$ values reflecting ranges of considered Coulomb interaction are
compared to the experimental data (the detector response is included via the MC simulations).
			}
			\label{fig:4_16Ne_pp_ang}
		\end{minipage}
	\end{center}
\end{figure}

The  measured widths and decay energies of the $ ^{16} $Ne states \cite{Brown:2014} are compared with theoretical predictions \cite{Grigorenko:2015} in Fig.~\ref{fig:4_16Ne_widths_Grig2015}(a,b). One may see, that decays of both ground $0 ^{+} $ and excited $2 ^{+} $ states can not be reproduced by the diproton emission or the three-body 2\textit{p} decay of $^{16} $Ne consisting of a pure $ [s^{2}] $ or $ [d^{2}] $ configuration.
However, calculations of the three-body cluster model with a strong nuclear \textit{sd}-configuration mixture can reproduce the observed width values.
This observation provides the second example, after $^{12}$O discussed above, of the
the ``three-body mechanism'' of the TES \cite{Grigorenko:2002}. It was found that
in the $ ^{16} $Ne-$ ^{16} $C mirror pair, the ``dynamic'' component of the TES
is responsible for about half of the whole TES effect. The strong \textit{sd}-configuration mixing in the ground state of $^{16}$Ne is also supported by the correlations of its 2\textit{p} decay products. Fig.~\ref{fig:4_16Ne_pp_ang}(a) shows the first-measured \textit{p-p} angular correlations from the $ ^{16} $Ne ground state \cite{Mukha:2008}. The solid line represents the three-body cluster model assuming the 54\% $d$-wave contribution.


%

In another study, an evidence for the long-range Coulomb effect in three-body
correlations of fragments was found \cite{Brown:2014}. As illustrated in
Fig.\ref{fig:4_16Ne_pp_ang}(b), a good description of the \emph{p-p} relative energy
in the 2\emph{p} decay of the $^{16}$Ne ground state requires tracing
the three body Coulomb interaction to distances of 1000~fm and beyond.
This observation is important by pointing out that the analysis of the
2\emph{p} decays in heavier systems, where the Coulomb interaction is stronger,
needs a special care.
\begin{figure}[h!]
	\begin{center}
		\begin{minipage}[t]{14 cm}
			\includegraphics[width = 0.99\columnwidth, angle=0.]{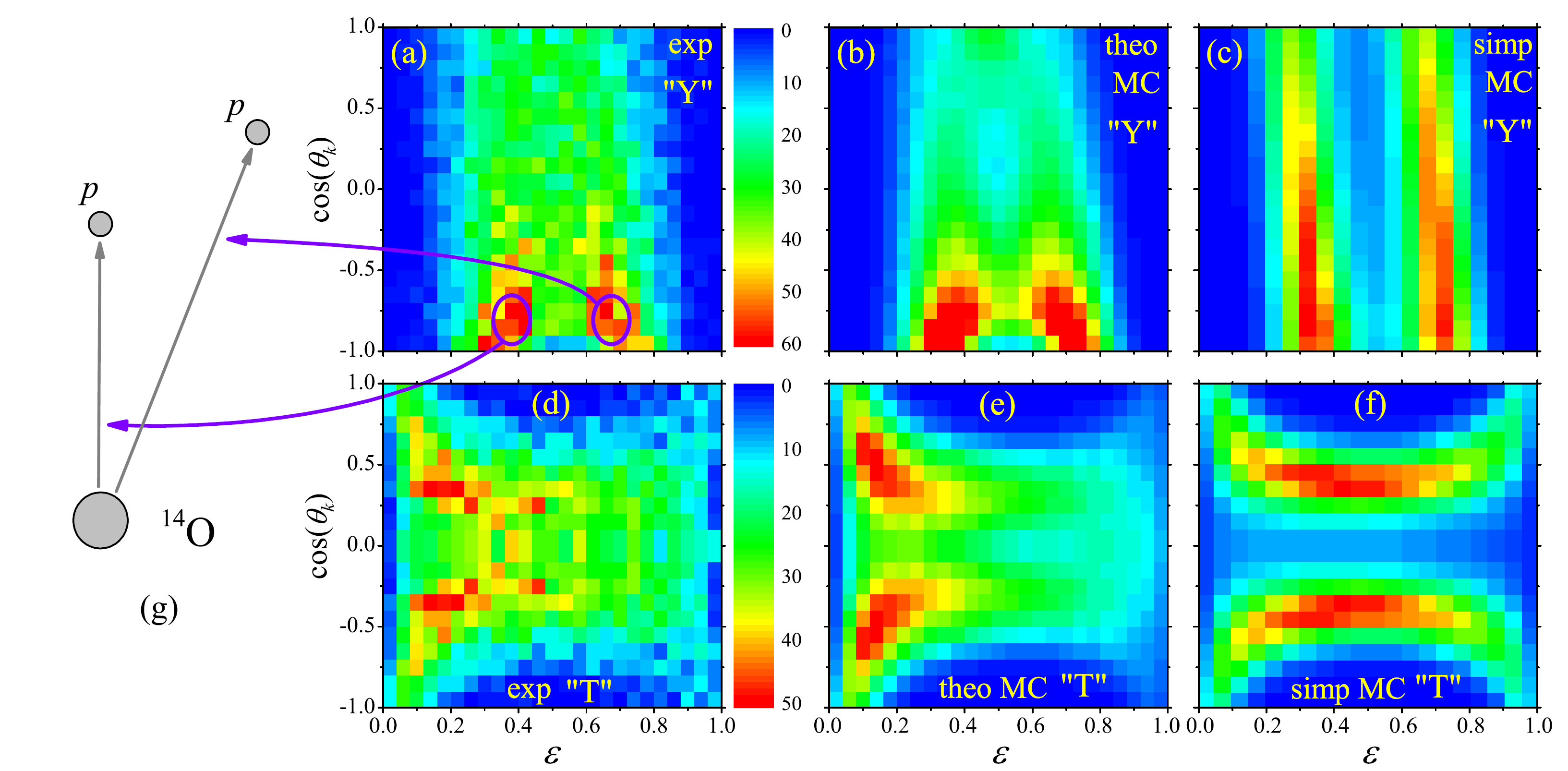}
		\end{minipage}
		\begin{minipage}[t]{16 cm}
\caption{(Color online) Three-body correlations of the 2\textit{p}-decay products of the first excited state 2$ ^{+} $ in $ ^{16} $Ne  \cite{Brown:2015}, presented in the T (top row) and Y (bottom row) Jacobi systems. The $\epsilon $ is either
$ E_{p-^{14}{\rm O}} $ (in system Y) or $ E_{p-p} $ (in system T) relative-energy normalized to the decay energy $ E_T $.
The experimental results are shown in panels (a,d).
The three-body model calculations are in panels (b,e), and those from a sequential decay simulation are in (c,f). The theoretical distributions have been folded with the detector response via MC simulations. (g) Sketch illustrating the relative\textit{ p-p} vectors for the peak regions indicated by blue circles in panel (a). Figure is taken from \cite{Brown:2015}.
			}
			\label{fig:4_16Ne_exc_corr}
		\end{minipage}
	\end{center}
\end{figure}

An interesting interplay of the true and sequential 2\textit{p}-decay mechanisms in the decay of the first excited 2$ ^{+}$ state in $ ^{16} $Ne was reported in Ref.~\cite{Brown:2015}. The results are shown in Fig.~\ref{fig:4_16Ne_exc_corr}(a,d). The $E_{p-^{14}O}$-related distribution in panel (a) shows the presence of two peaks which are expected for the first- and second- emitted protons in the sequential decay mechanism,  however, only when the two protons have low relative energy, which is characteristic for the diproton-like correlation.
Thus, features of two distinct decay mechanisms seem to be observed.
A careful investigation within the three-body model indicated that, indeed, both the
sequential and the simultaneous decay paths do occur here and they interfere resulting
in the observed, strange correlation pattern. It was qualitatively described
as a ``tethered'' decay mechanism \cite{Brown:2015}.

The only experimental work on $^{15}$Ne to date, was reported in Ref.~\cite{Wamers:2014}.
The first observation of this exotic nucleus was achieved in two-neutron knock-out
reaction from a beam of $^{17}$Ne. The decay, proceeding directly to $^{13}$O with
emission of two protons, with the total 2\emph{p} decay energy of 2.522(66)~MeV,
was observed. The energy correlations between the decay products indicate a
democratic decay mechanism. The structure of the $^{15}$Ne ground state was
considered as a coupling of the $^{13}$O core to the valence protons in a mixture
of $[s_{1/2}^2]$ and $[d_{5/2}^2]$ configurations. The contribution of the s-wave
component was determined to be 63(5)\%


%

\subsection{\it Excited states in $ ^{17} $Ne}

\begin{figure}[h!]
	\begin{center}
		\begin{minipage}[t]{9 cm}
			\includegraphics[width = \columnwidth, angle=0.]{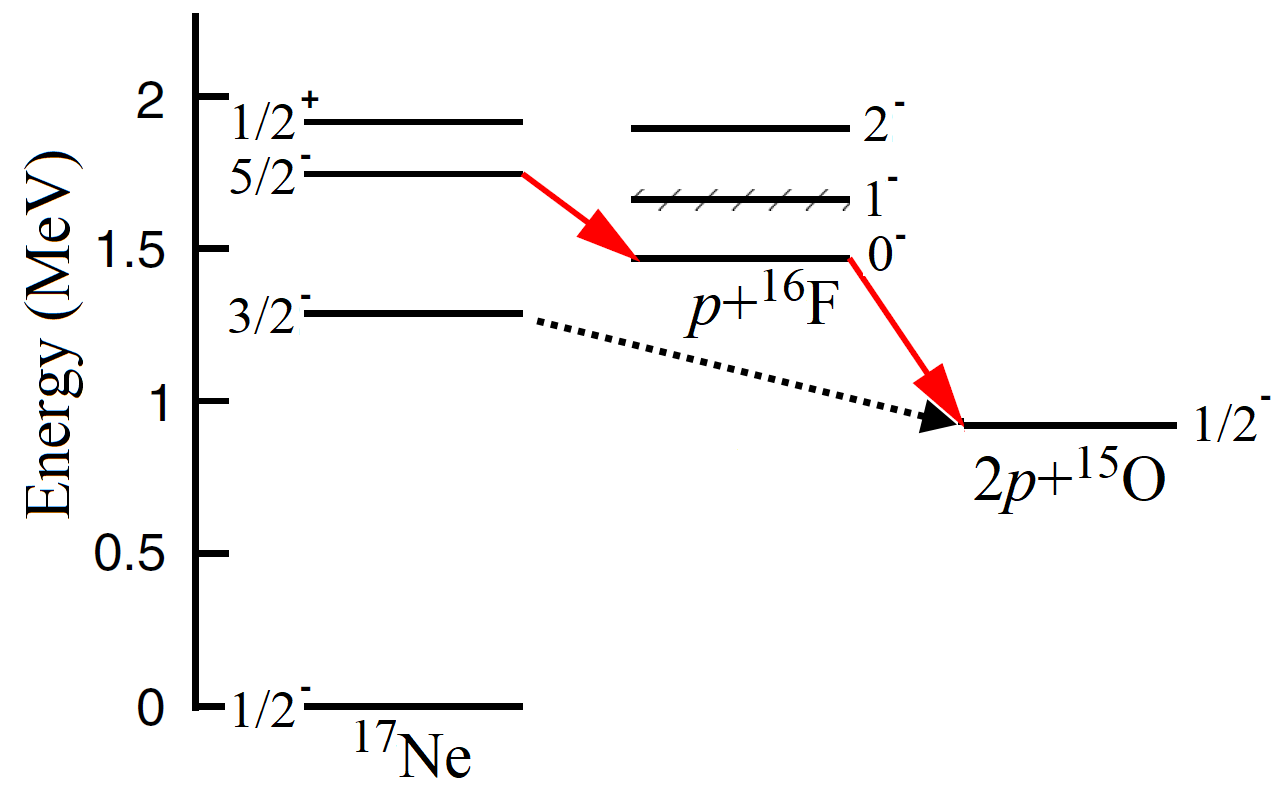}
		\end{minipage}
		\begin{minipage}[t]{16 cm}
		\caption{(Color online)  Level and decay scheme for $ ^{17} $Ne \cite{Charity:2018}. The $^{17}$Ne and intermediate $^{16}$F	states are labeled  by $ J^{\pi}$.  The red arrows show the known sequential proton emission of the second excited state 5/2$ ^{-} $ to $^{15}$O g.s.\ via $^{16}$F, and the dotted arrow indicates the undetected yet true 2\textit{p} decay of the first excited state $3/2 ^{-} $ in $ ^{17} $Ne.
			}
			\label{fig:4_17Ne_pp_scheme}
		\end{minipage}
	\end{center}
\end{figure}

In is interesting to mention in context of this review the case of
$^{17}$Ne, as its first excited state was one of the early candidates
for observation of the true 2\textit{p} decay. The ground state of $^{ 17} $Ne
is bound, but the first excited state ($3/2 ^{-} $ at 1.288 MeV) is
unbound by 355~keV relative to the 2\textit{p} emission and
bound by 181 keV with respect to the 1\textit{p}-decay threshold \cite{Charity:2018},
see the level scheme in Fig.~\ref{fig:4_17Ne_pp_scheme}. Thus, this state fulfils
the energy criterion for the true 2\textit{p} emission. Unfortunately, it was
found to decay predominantly by $\gamma$ emission \cite{Chromik:2002}.
The interest in the 2\emph{p}-decay branch of the $3/2 ^{-} $ state
has an additional motivation from nuclear astrophysics.
The $ ^{15} $O nucleus is known to be a ``waiting point'' in
the \textit{rp}-process \cite{Gorres:1995}. The radiative capture of two
protons by $ ^{15} $O via $^{ 17} $Ne(3/2$ ^{-}$) could be be a possible
bypath for this waiting point.

Recently, a dedicated search for the 2\textit{p} decay branch
of the 3/2$ ^{-} $ state in $ ^{17} $Ne has been performed by
applying the original ``combined mass'' method to reconstruct
the $ ^{17} $Ne excitation spectrum \cite{Sharov:2017}.
In this work a new upper limit for $\Gamma_{2p}$/$\Gamma_{\gamma} \leq 1.6(3) \times 10^{-4}$
was obtained. This value, however, is still far from the theoretical
estimate which is in the order of $10^{-6}$~\cite{Grigorenko:2007}.

Higher lying excited states of $ ^{17} $Ne are expected to decay by a
sequential 2\textit{p} emission through the levels of $^{16}$F, as indicated
in Fig.~\ref{fig:4_17Ne_pp_scheme}. Indeed, the second excited state, $5/2 ^{-} $,
was reported to decay sequentially already in Ref.~\cite{Chromik:2002}.
The 2\emph{p} decay of this $5/2 ^{-} $ state was studied in much more detail
by Charity et al.~\cite{Charity:2018}. The width of the ground state of
$^{16}$F was determined to be about 21~keV. The analysis of this width in the Shell
Model Embedded in Continuum (SMEC, see Section~\ref{sec:2_continuum_CI})
allowed to constrain the coupling interaction between the shell-model
states and the scattering continuum. A strong reduction of the
continuum couplings between the unbound proton and the well-bound neutron in
$^{16}$F was found~\cite{Charity:2018}. This is a nice illustration of how
the 2\emph{p} emission study may shed light on the effective
nucleon-nucleon interactions for the open-quantum system.

\subsection{\it Decay of $ ^{19} $Mg}

\begin{figure}[htb]
	\begin{center}
		\begin{minipage}[t]{14 cm}
			\includegraphics[width = 0.99\columnwidth, angle=0.]{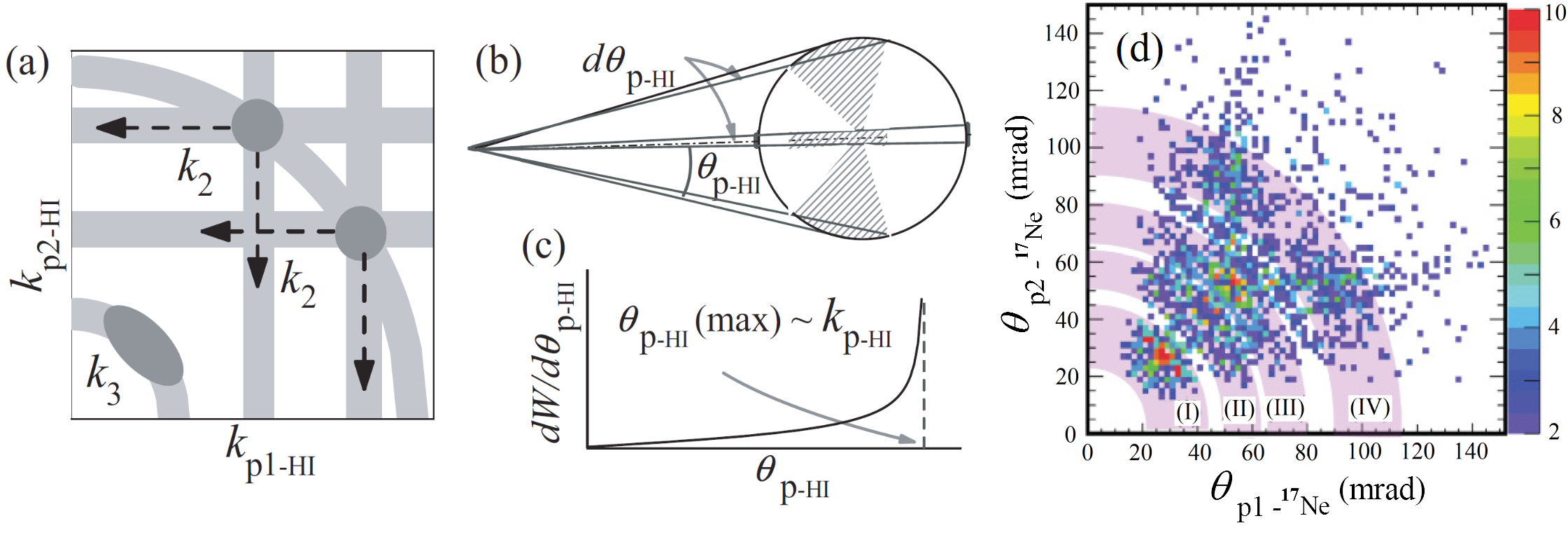}
		\end{minipage}
		\begin{minipage}[t]{16.5 cm}
	\caption{ (Color online) Principles of the in-flight decay tracking method and the data for the decay of $^{19}$Mg.
(a) Two types of transverse momentum correlations $ k_{p1-HI}-k_{p2-HI} $ typical for a direct three-body ($ k_3 $ area) and sequential ($ k_2 $ areas) 2\textit{p}-decay mechanisms.
Arrows show directions of the peak tails.
(b) Kinematical enhancement of angular \textit{p}-HI correlations at the maximum
possible angle for a fixed $ k_{p-HI} $.
(c) The corresponding angular \textit{p}-HI distribution.	
(d)	Measured angular ($ p_1-^{17} $Ne)-($ p_2-^{17} $Ne) correlations (color boxes with
	a scale on the right-hand side) \cite{Mukha:2012}. The shadowed arc areas (I$-$IV)
	indicate locations of simultaneous or sequential
	2\textit{p} decays of the mostly populated states in $ ^{19}$Mg.
			}
			\label{fig:4_pNe_kinem_corr}
		\end{minipage}
	\end{center}
\end{figure}

The ground and excited states in the unbound $ ^{19} $Mg decay by 2\textit{p} emission. They were studied experimentally by measuring angular correlations of the in-flight decay products, $ ^{17} $Ne+\textit{p}+\textit{p} \cite{Mukha:2007,Mukha:2008,Mukha:2012,Xu:2016}.
In these experiments, the in-flight tracking technique was applied, see Section \ref{sec:3_Production} and Figure~\ref{fig:3_FRSTracking}.
The main principle of the method is explained schematically in Figure~\ref{fig:4_pNe_kinem_corr}. Panel a) shows two types of \emph{p-p}
momenta correlations: a sequential emission (loci marked as $k_2$), and
a simultaneous emission (region $k_3$). Panels b) and c) explain how, and why,
the opening angle of a proton trajectory reflects the value of its momentum.
Essentially, a clear peak in the angular distribution may be directly
related to the energy taken by the proton.
Finally, the panel (d) shows the measured angular ($ p_1 $-$ ^{17} $Ne)--($ {p_2} $-$ ^{17} $Ne) correlations  \cite{Mukha:2012}. The shadowed arc areas (I--IV) indicate locations of either simultaneous or sequential 2\textit{p} decays of the most intensively
populated states in $ ^{19} $Mg. The decay events with the smallest \textit{p}-$ ^{17} $Ne angles, which are located within the arc area (I) in Fig.~\ref{fig:4_pNe_kinem_corr}(d), were assigned to the ground state of $ ^{19}$Mg, while the events in the arcs (II--IV) correspond to decays of the excited states of $ ^{19}$Mg \cite{Mukha:2007,Mukha:2012,Xu:2016}.

%
\begin{figure}[h!tb]
	\begin{center}
		\begin{minipage}[t]{7 cm}
			\includegraphics[width = \columnwidth, angle=0.]{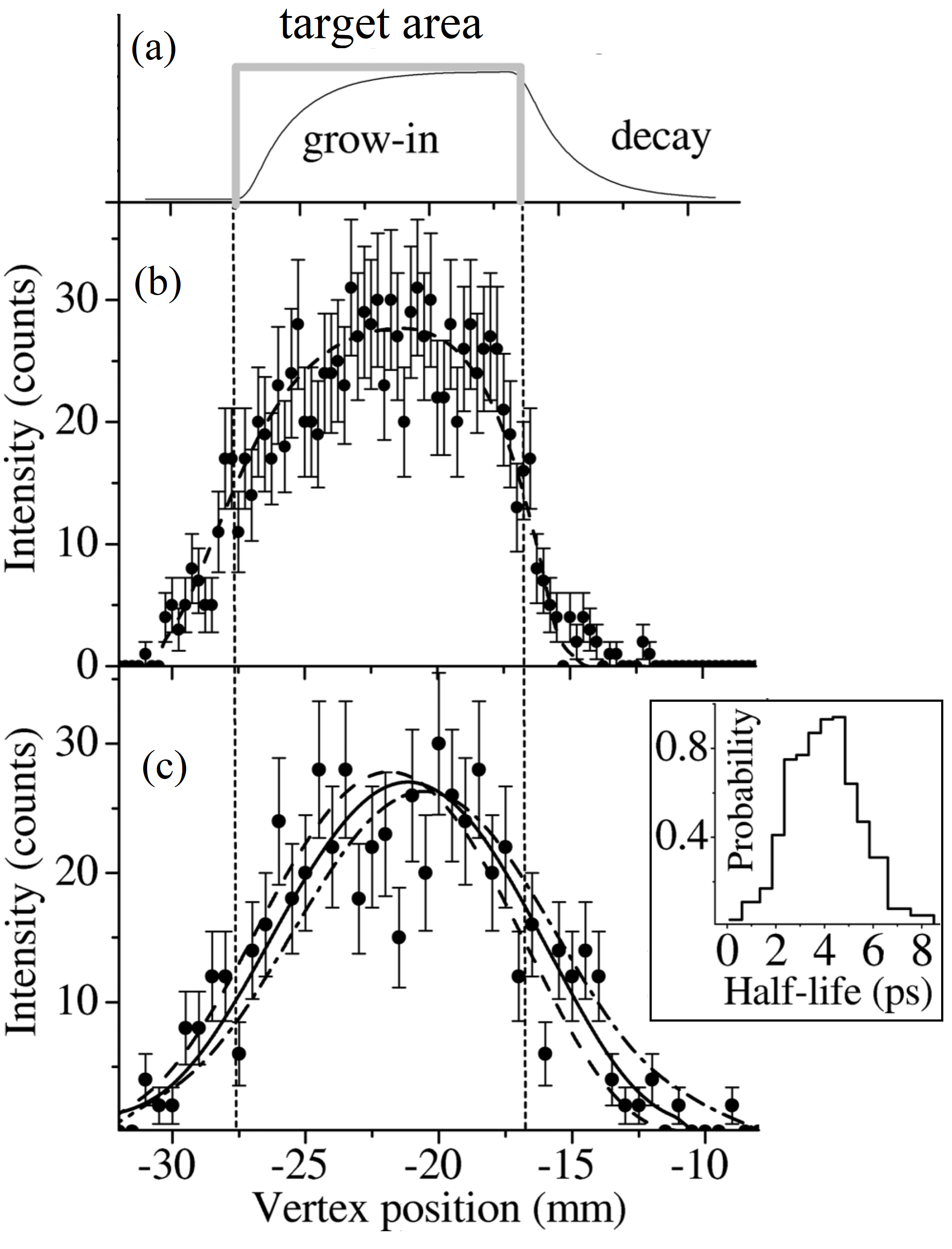}
		\end{minipage}
		\begin{minipage}[t]{16.5 cm}
\caption{Profiles of the decay vertices along the beam direction
from the target to the closest microstrip detector in the $ ^{19}$Mg experiment \cite{Mukha:2007}. (a) Ideal
profiles of prompt (gray curve) and delayed (black curve) decays
expected in a thick target. (b) Vertex distribution of $ ^{17} $Ne+\textit{p}+\textit{p} events gated by large \textit{p}-$ ^{17} $Ne angles (and thus large \textit{ p}-$ ^{17} $Ne relative energies), which corresponds to short-lived excited
states in $ ^{19} $Mg (full circles with statistical uncertainties). The
dashed curve shows the MC simulations of the detector
response for the 2\textit{p}-decay $ ^{19}$Mg$\rightarrow^{17}$Ne$+p+p $ with $ T_{1/2} \sim$0.1 ps. (c) The same as (b) but gated by small \textit{p}-$ ^{17} $Ne angles
corresponding to the ground-state group (I) in Fig.~\ref{fig:4_pNe_kinem_corr}(c). The
dashed curve is a simulation of the 2\textit{p}-decay component
with $ T_{1/2} \sim$0. The solid and dash-dotted curves are fits to the
data assuming
the radioactivity of $ ^{19}$Mg with $ T_{1/2}$ values of 4 and
8 ps, respectively. The inset in (c) shows the probability  that the simulations match the data as a function
of the assumed half-lives.			
			}
			\label{fig:4_19Mg_vertex_2007}
		\end{minipage}
	\end{center}
\end{figure}

The measured distribution of the decay vertex, along the beam direction is shown in  Fig.~\ref{fig:4_19Mg_vertex_2007}. The vertex profile corresponding
to excited states (b) is consistent with the prompt decay when the limited
detector resolution and the angular straggling is taken into account.
In contrast, the profile representing the ground state decay (c) is broader
and shifted downstream, clearly indicating the delayed activity.
The fitting of the Monte Carlo simulated profiles yielded the
ground-state half-life of $^{19}$Mg to be $ T_{1/2} = 4.0(15)$ ps as
illustrated in the inset in Fig.~\ref{fig:4_19Mg_vertex_2007}(c) \cite{Mukha:2007}.
An independent measurement, using a different technique, yielded
the lifetime of $ ^{19} $Mg in the range from $ 1.75^{+0.43}_{-0.42} $ to $ 6.4^{+2.4}_{-2.7} $ ps \cite{Voss:2014}, which is consistent with the result of Ref.~\cite{Mukha:2007}.

\begin{figure}[h!]
	\begin{center}
		\begin{minipage}[t]{12 cm}
			\includegraphics[width = 0.99\columnwidth, angle=0.]{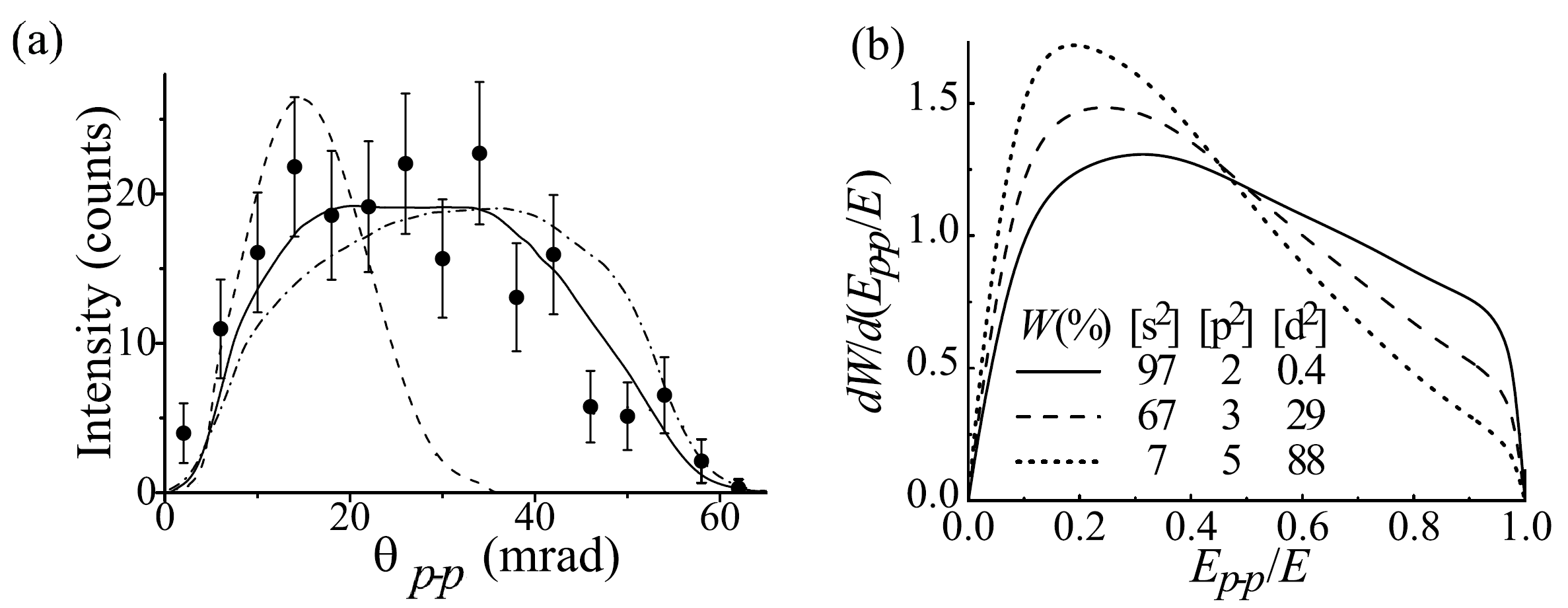}
		\end{minipage}
		\begin{minipage}[t]{16 cm}
	\caption{ (a) Angular \textit{p-p} correlations from the 2\textit{p} decay of the ground state of $ ^{19} $Mg  (dots with statistical uncertainties) obtained from the measured $ ^{17} $Ne+\textit{p}+\textit{p} coincidences in Ref.~\cite{Mukha:2008}.
The	solid curve shows the three-body model calculations with the best fit by assuming an 88\% of \textit{d}-wave configuration in structure of the $ ^{19} $Mg g.s. The dashed curve is the diproton model prediction, and the dash-dotted curve is the phase-space simulation of the 2\textit{p} decay assuming an isotropic proton emission.  (b) Relative-energy \textit{p-p} correlations $ E_{p-p} $
	from the  $ ^{19} $Mg g.s.\ calculated for different weights \textit{W} of its 	\textit{spd}-shell configurations \cite{Grigorenko:2003}.
			}
			\label{fig:4_19Mg_pp_ang}
		\end{minipage}
	\end{center}
\end{figure}

The angular \emph{p-p} correlations corresponding to the ground-state decay of
$ ^{19} $Mg are shown in Fig.~\ref{fig:4_19Mg_pp_ang}(a) \cite{Mukha:2008}.
The data are well described by the three-body model which is sensitive to
the structure of the decaying nucleus, as illustrated in
Fig.~\ref{fig:4_19Mg_pp_ang}(b)~\cite{Grigorenko:2003}.
The best fit of the model was found for the $ ^{19} $Mg ground-state wave function
having the dominant 88\% \emph{d}-shell configuration, which is also consistent
with the half-life information \cite{Mukha:2007}.

%
\begin{figure}[h!tb]
	\begin{center}
		\begin{minipage}[t]{12 cm}
	\hspace{1.cm}
			\includegraphics[width = 0.9\columnwidth, angle=0.]{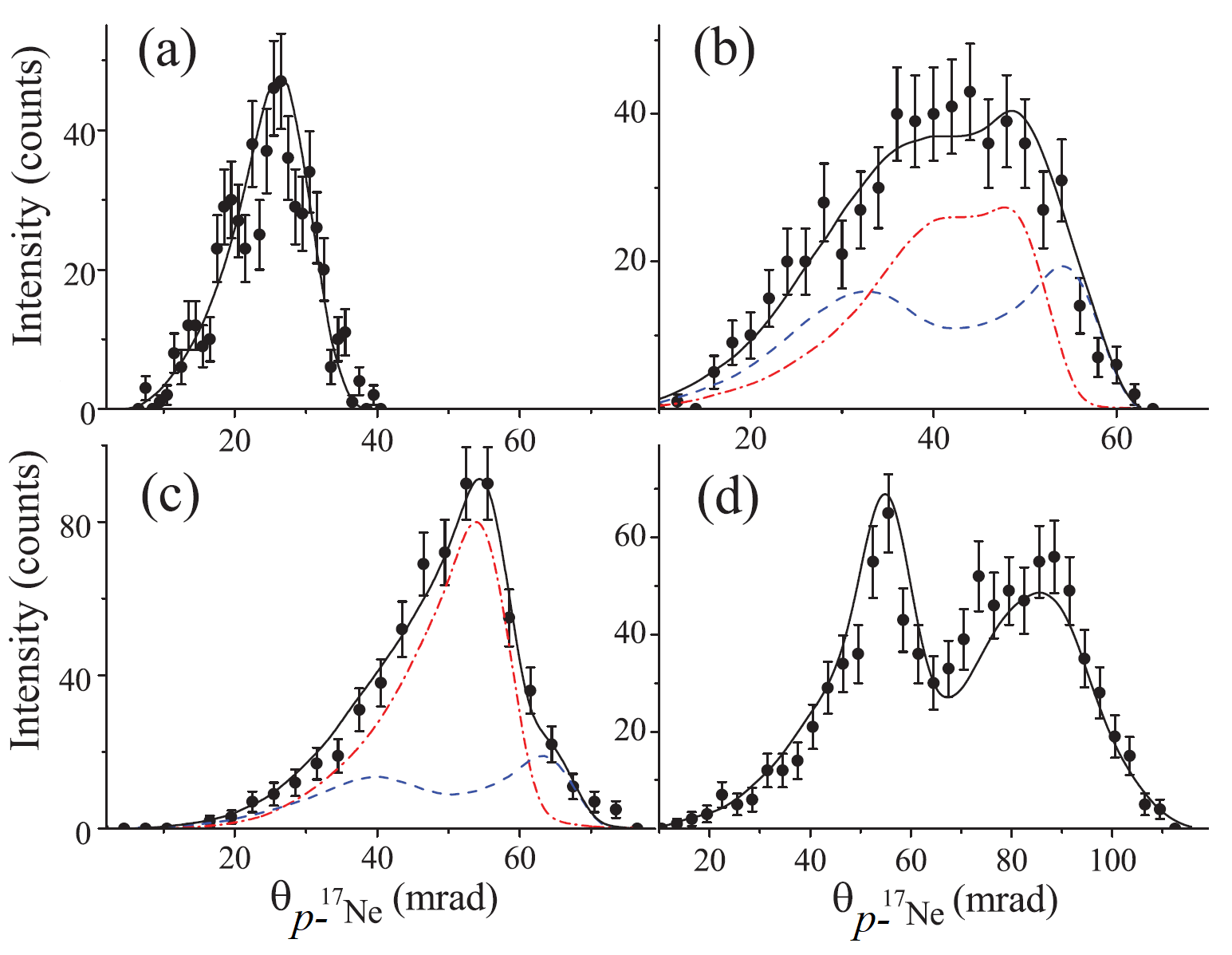}
		\end{minipage}
		\begin{minipage}[t]{16 cm}
\caption{(Color online) Angular \textit{p}-$ ^{17}$Ne correlations (dots
with statistical uncertainties) selected from the $ ^{17}$Ne+\textit{p}+\textit{p} data of the  $ ^{19}$Mg experiment \cite{Mukha:2012} by choosing the arc gates (I-IV) shown in Fig.~\ref{fig:4_pNe_kinem_corr}(d). (a) The 2\textit{p} decay of
the $ ^{19} $Mg g.s.\ selected by the gate (I). The solid curve shows the best-fit simulation of the three-body model using a 2\textit{p}-decay energy of $ E_T $ = 0.76(6) MeV. (b) The decay of the ''first-excited'' state $ ^{19} $Mg selected by the gate (II). The solid curve displays the simulation of the sequential 2\textit{p} decay of the state at 2.14 MeV via two intermediate states in $ ^{18} $Na, the g.s.\ at 1.23 MeV (dash-dotted line), and the 2$ ^{-} $ state at 1.55 MeV (dashed line). (c) The decay of the ``second-excited'' state in $ ^{19} $Mg selected by the gate (III). The solid line is the best-fit simulation of the sequential 2\textit{p} decay of the $ ^{19} $Mg state at 2.9 MeV via the 2$ ^{-} $ and 3$^{-} $ states in $ ^{18} $Na (the dash-dotted and dashed curve, respectively). (d) The decay of the suggested high-lying state in $ ^{19} $Mg at $ E_T $=5.5 MeV selected by the gate (IV). The solid curve represents the best-fit simulation of the decay of this state by sequential emission via $ ^{18} $Na$ ^{*} $(2$ ^{-} $).			
             }
           \label{fig:4_19Mg_pNe_2012}
		\end{minipage}
	\end{center}
\end{figure}


\begin{figure}[h!]
	\begin{center}
		\begin{minipage}[t]{10 cm}
			\includegraphics[width = \columnwidth, angle=0.]{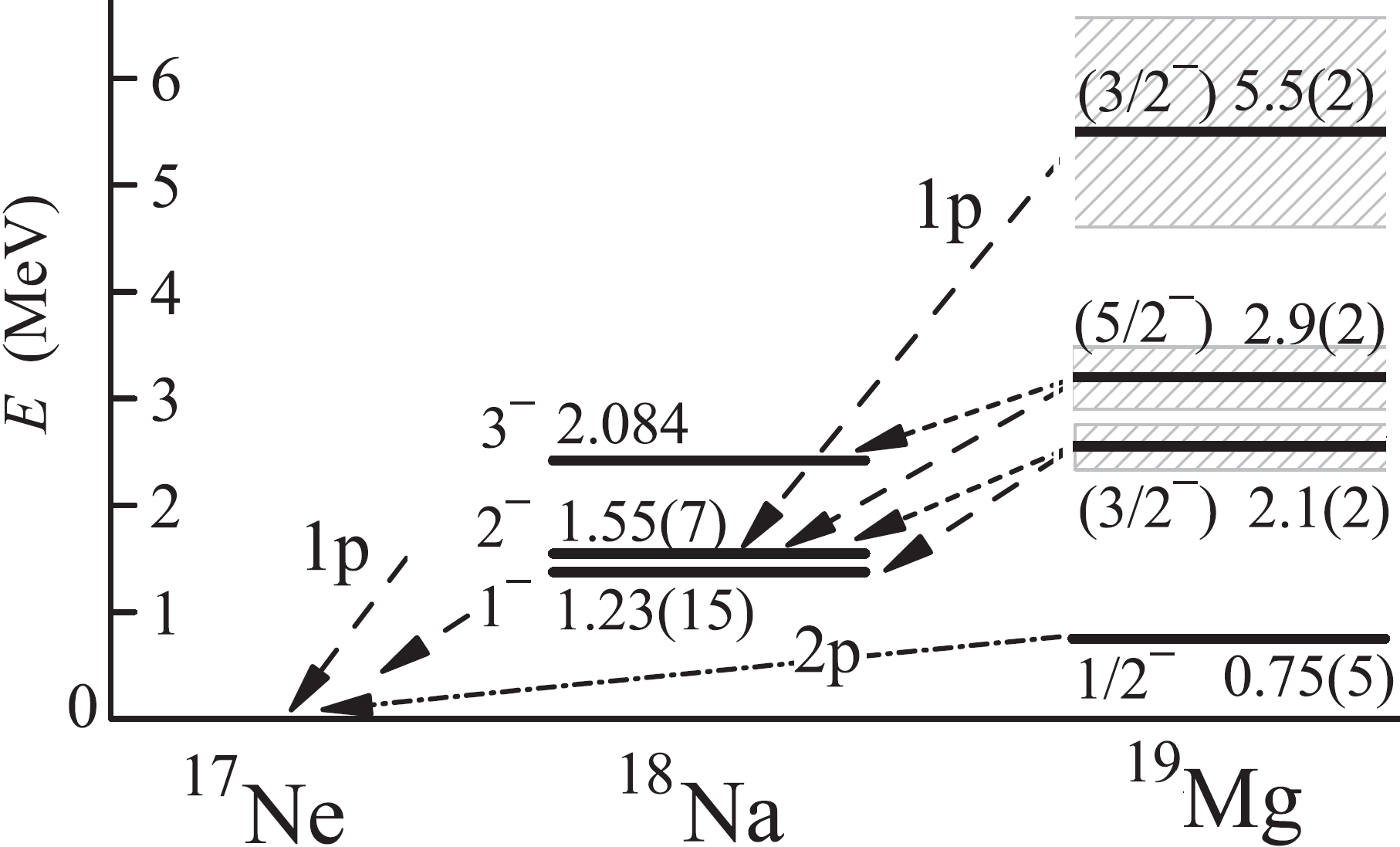}
		\end{minipage}
		\begin{minipage}[t]{16 cm}
	\caption{ Low-excitation part  of the $ ^{19} $Mg spectrum reported in \cite{Mukha:2008,Mukha:2012}, the decay scheme and illustrations of involved 2\textit{p} decay transitions. The derived energies of
	states are  complemented by tentative ($ J^{\pi}$) assignments. Dash-dotted arrow points to the true 2\textit{p} decay, and dashed (short-dashed) arrows indicate sequential proton emissions. Intermediate resonances $ ^{18} $Na are in the middle.
			}
			\label{fig:4_19Mg_scheme_2012}
		\end{minipage}
	\end{center}
\end{figure}

The angular correlations $ ^{17}$Ne+\textit{p}+\textit{p} from the decay of $ ^{19}$Mg
allowed also for the determination of the energies of the most intensively-populated
excited states in $ ^{19} $Mg \cite{Mukha:2012}.
For this purpose, angular \textit{p}-$ ^{17}$Ne correlations were selected by choosing
the arc gates (I-IV) in Fig.~\ref{fig:4_pNe_kinem_corr}(d). The results are shown in Fig.~\ref{fig:4_19Mg_pNe_2012}. It can be seen that the patterns corresponding to
the excited states are well described by sequential proton emission
via intermediate states in $ ^{18} $Na. This analysis allowed to propose
the level scheme of $^{19}$Mg, shown in Fig.~\ref{fig:4_19Mg_scheme_2012}.
\begin{figure}[t!]
	\begin{center}
		\begin{minipage}[ht]{13 cm}
			\includegraphics[width = \columnwidth, angle=0.]{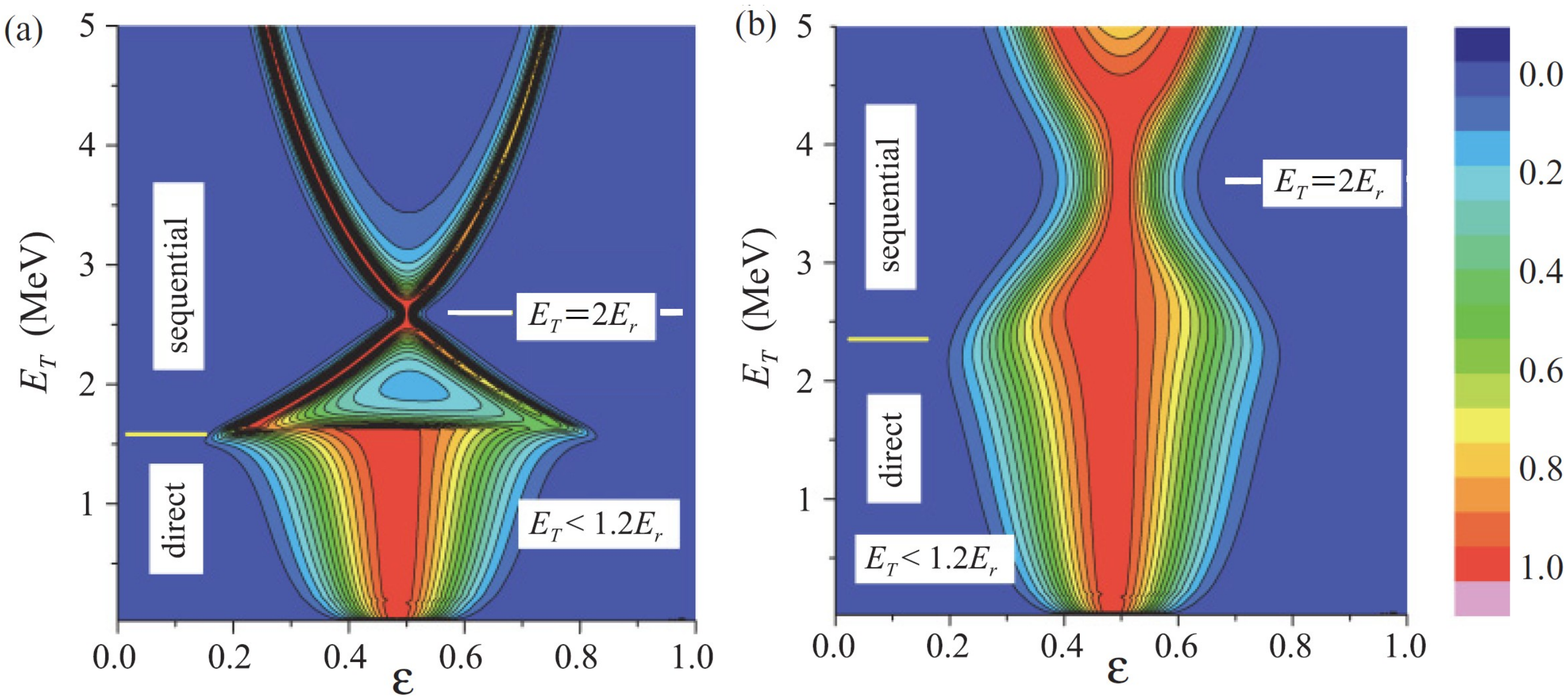}
		\end{minipage}
		\begin{minipage}[t]{16 cm}
			\vspace{-3.cm}
	\caption{ (Color online) Intensity distributions of 2\textit{p} decays from $ ^{19} $Mg continuum as a function of the 2\textit{p}-decay energy $ E_T $ and the parameter  $ \epsilon \sim E_{p-^{17}{\rm Ne}}/E_T $.  The distributions are calculated with a
	three-body model \cite{Mukha:2012} assuming a single resonance in the intermediate nucleus $ ^{18} $Na with energy $ E_r $ and width $ \Gamma_r $. (a) Decays by taking
	into account the narrow g.s.\ of $ ^{18} $Na at $ E_r $=1.23 MeV with $ \Gamma_r $=0.1 MeV. (b) Decays via the broad $1^{-}_2 $
	state in $ ^{18} $Na at $ E_r $=2.03 MeV and $ \Gamma_r $=0.9 MeV.		
			}
			\label{fig:4_Mg19_exc_Ep_calc}
		\end{minipage}
	\end{center}
\end{figure}

Since the three-body model was so successful in description of all
$^{19}$Mg decay features, an interesting exercise was made
to investigate how the predicted energy sharing between two protons
would depend on the 2\emph{p} decay energy $E_T$ \cite{Mukha:2012}.
Results of this simulations are shown in Fig.~\ref{fig:4_Mg19_exc_Ep_calc},
where on the left the decay through the narrow $^{18}$Ne ground state is assumed,
while on the right the decay through a single, broad $1^{-}_2$ resonance
in $^{18}$Ne is considered. In the former case, at low energies, as expected,
the protons share the decay energy $E_T $ in a broad peak centered around
$ \epsilon$=0.5, which is a characteristic
feature of the direct, simultaneous 2\textit{p} emission.
Only when the decay energy reaches the value $ E_T \simeq 1.2 \cdot E_r $
a sudden change in the proton-energy distribution occurs. Two narrow
peaks appear which represent the sequential 2\textit{p} emission via
the narrow $^{18}$Na state. With the further increase of $E_T $
these peaks approach each other until the point when $ E_T  $=2$ \cdot E_r$
i.e. energies of both protons are equal. In Fig. \ref{fig:4_Mg19_exc_Ep_calc}(b),
a very similar pattern is visible, although the distribution is smeared out
due to the broad intermediate state. An important conclusion from this
simulation is, that the sequential decay pattern takes over the simultaneous one,
only when the decay energy $E_T$ is larger that $1.2 \cdot E_r$ (compare Figure~\ref{fig:1_EnergyConditions}).

\subsection{\it Transition mechanism of 2p decay of $ ^{30} $Ar and decay of $^{29}$Ar}


Two unbound isotopes of argon, $ ^{30} $Ar and $ ^{29} $Ar were investigated  experimentally in Refs.~\cite{Mukha:2015,Xu:2016,Xu:2018} and \cite{Mukha:2018}, respectively. These isotopes were produced in either -1\textit{n} or -2\textit{n}  neutron knock-out reactions of the secondary beam of $ ^{31} $Ar which was produced in fragmentation reaction of a primary beam of $ ^{36} $Ar at high energy, see Figure~\ref{fig:3_FRSTracking}. The method of angular correlations of decay-in-flight products was applied, and the respective level energies and decay mechanisms were assigned in an analogous manner to the case of $^{19}$Mg described above.

\begin{figure}[htb]
	\begin{center}
		\begin{minipage}[htb]{18 cm}
			\includegraphics[width = 0.99\columnwidth, angle=0.]{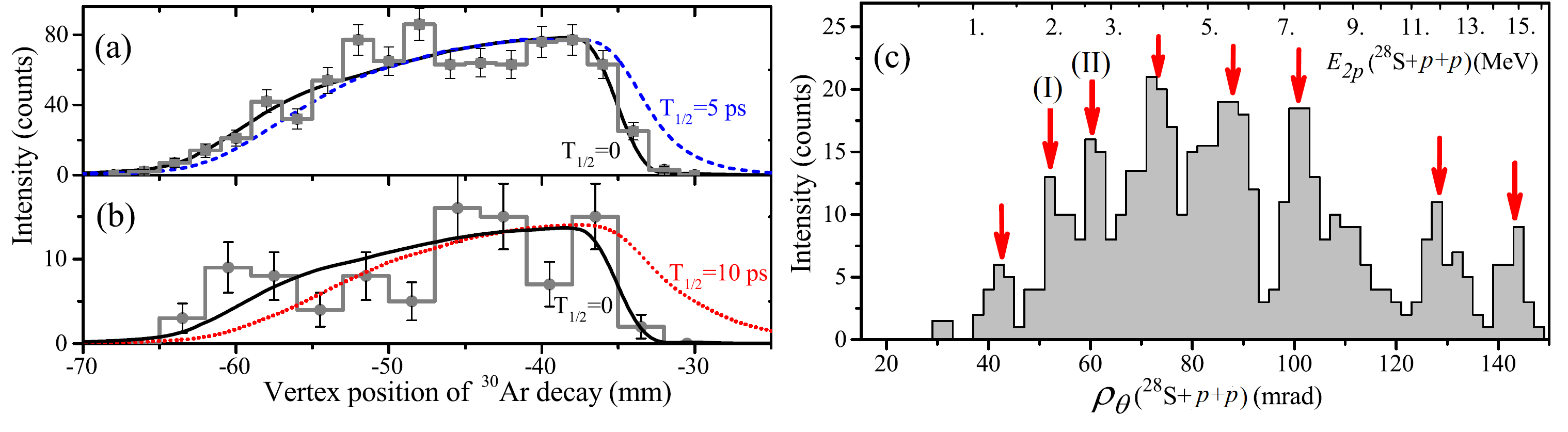}
		\end{minipage}
		\begin{minipage}[t]{16.5 cm}
\caption{ (Color online) (a),(b) Profiles of the
$ ^{30} $Ar$\rightarrow ^{28}$S+\textit{p}+\textit{p}
decay vertices measured along the beam direction (histograms with statistical uncertainties) \cite{Mukha:2015}.
(a) The data gated by large $\rho_\theta$ angles shown in panel (c), which
corresponds to short-lived excited states in $ ^{30} $Ar. (b) The data
gated by the (I) peak in panel (c) where the g.s.\ of $ ^{30} $Ar is
established. Solid, dashed, and dotted curves show the MC
simulations of the detector response for the $ ^{30} $Ar 2\textit{p} decays with
half-life $ T_{1/2} $ of 0, 5, and 10 ps, respectively. (c) Angular
correlations $\rho_\theta$=$\sqrt{\theta^2(p_1\mathrm{-}^{28}\!S)+\theta^2(p_2\mathrm{-}^{28}\!S)}$  of the measured $^{28}$S+\textit{p}+\textit{p} coincidences (filled
histogram), which reflect the excitation spectrum of $ ^{30} $Ar. The pointed peaks suggest $ ^{30} $Ar states whose 2\textit{p}-decay energies are shown in the upper axis. The  peaks labeled (I) and (II) correspond to the assigned ground and first excited states in $ ^{30} $Ar, respectively.
			}
			\label{fig:4_30Ar_vertex_rho}
		\end{minipage}
	\end{center}
\end{figure}

%
\begin{figure}[h!]
	\begin{center}
		\begin{minipage}[t]{16 cm}
			\hspace{3. cm}
			\includegraphics[width = 0.6\columnwidth, angle=0.]{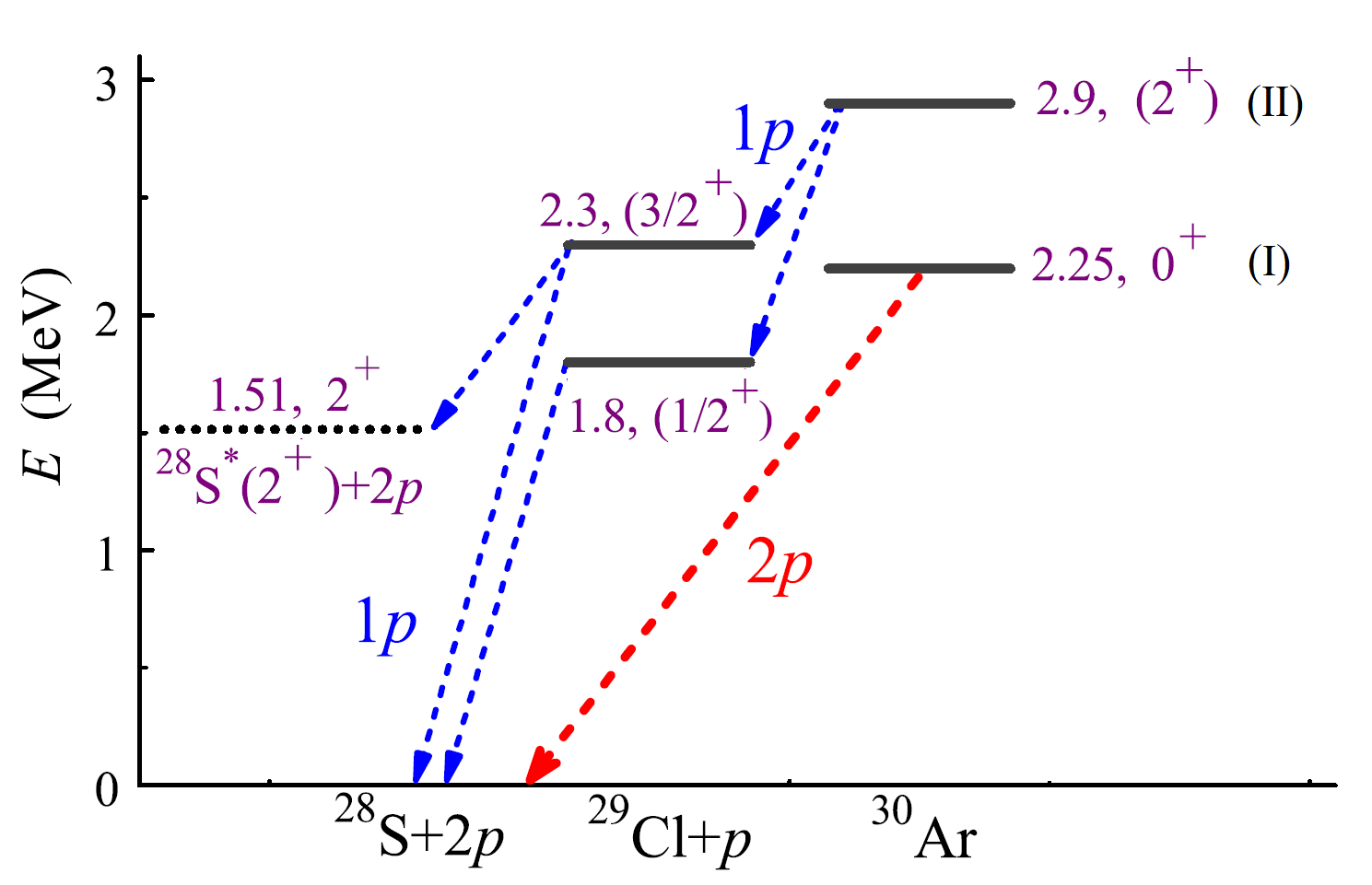}
		\end{minipage}
		\begin{minipage}[t]{16.5 cm}
\caption{ (Color online) Low-lying states and decay scheme of $ ^{30} $Ar derived in \cite{Mukha:2015}. The assigned ground and first-excited states correspond to the peaks (I) and (II) in Fig.~\ref{fig:4_30Ar_vertex_rho}(c), respectively.
The derived energies of
states and tentative ($ J^{\pi}$) assignments are shown. Dashed arrow points to the g.s.\ 2\textit{p} decay, and short-dashed arrows indicate the assigned sequential proton emissions. The established intermediate resonances $ ^{29} $Cl are in the middle.
			}
			\label{fig:4_30Ar_scheme}
		\end{minipage}
	\end{center}
\end{figure}

The results for $^{30}$Ar are shown in Figure~\ref{fig:4_30Ar_vertex_rho}.
In the right panel (c) the excitation spectrum of $ ^{30} $Ar obtained as a function of the generalized correlation angle $\rho_\theta$=$\sqrt{\theta^2(p_1\mathrm{-}^{28}\!S)+\theta^2(p_2\mathrm{-}^{28}\!S)}$
derived from the angular $^{28} $S+$ p_1 $+$ p_2 $ correlations.
The decay vertex profiles are displayed in the left part
of Figure~\ref{fig:4_30Ar_vertex_rho}. The panels a) and b) correspond to the events
assigned to the decay of the first excited and the ground state state of $^{30}$Ar,
contained in the peak marked as (II) and (I) in the panel c), respectively.
In both cases, only an upper limit of 5 ps for the half-life could be established.
The low-lying states and their decay scheme, derived from this work are
presented in Figure~\ref{fig:4_30Ar_scheme} \cite{Mukha:2015}.
Interestingly, the ground state of the intermediate $^{29} $Cl is located below
the ground state of $ ^{30} $Ar, what has a consequence for the 2\emph{p} decay mechanism.

\begin{figure}[h!tb]
	\begin{center}
		\begin{minipage}[t]{8 cm}
			\includegraphics[width = \columnwidth, angle=0.]{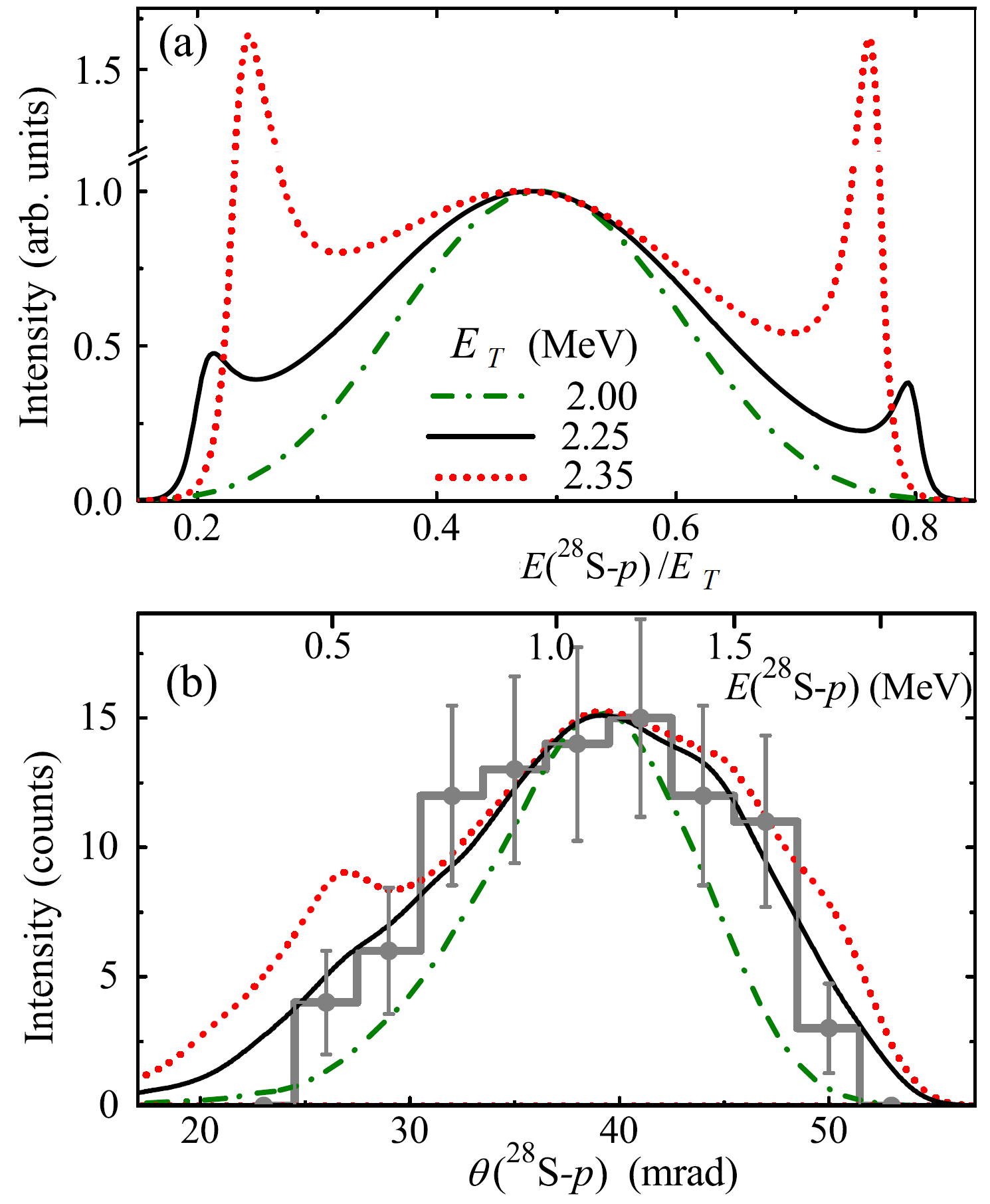}
		\end{minipage}
		\begin{minipage}[t]{16.5 cm}
\caption{(Color online)  (a) Transition from the direct 2\textit{p} decay to
the sequential 1$ p $-emission mechanism: the proton spectra
calculated by varying the 2\textit{p}-decay energy $ E_T $ of
$ ^{30} $Ar. (b) The measured angular $ ^{28} $S-\textit{p} correlations (histogram	with statistical uncertainties) from the $ ^{30}$Ar ground state selected by choosing the gate (I) in Fig.~\ref{fig:4_30Ar_vertex_rho}(c)~\cite{Mukha:2015}. The data are compared with the
respective MC simulations of the proton spectra in panel (a).			
			}
			\label{fig:4_30Ar_pS_corr}
		\end{minipage}
	\end{center}
\end{figure}

The proton energy distribution from 2\textit{p} decays of the $ ^{30} $Ar
ground state, together with the results of model calculations, is shown in Fig.~\ref{fig:4_30Ar_pS_corr}. It appears broader than the one predicted
by the simultaneous decay mechanism, but it does not exhibit a clear double-peak
feature expected from a sequential 2\textit{p} emission. What is observed,
represents an interplay, or a transition, between the two distinct decay scenarios.

The first observation, and the only one to date, of $^{29}$Ar was reported in Ref.~\cite{Mukha:2018}. The evidence was provided by the detection of the
$^{27}$S+\textit{p}+\textit{p} correlations. Despite the limited statistics,
the observed correlations were interpreted as representing a sequential
2\emph{p} decay of a state in $^{29}$Ar through levels in $^{28}$Cl.
The total 2\emph{p} decay energy was measured as 5.50(18) MeV.
It is not clear whether this energy corresponds to the ground state of $^{29}$Ar
or to an excited state. Further studies are necessary to clarify this case.

\subsubsection{\it Ground state of $ ^{31} $Ar deduced from 2p decays of its excited states }

\begin{figure}[h!tb]
	\begin{center}
		\begin{minipage}[t]{10 cm}
			\hspace{1. cm}
			\includegraphics[width = \columnwidth, angle=0.]{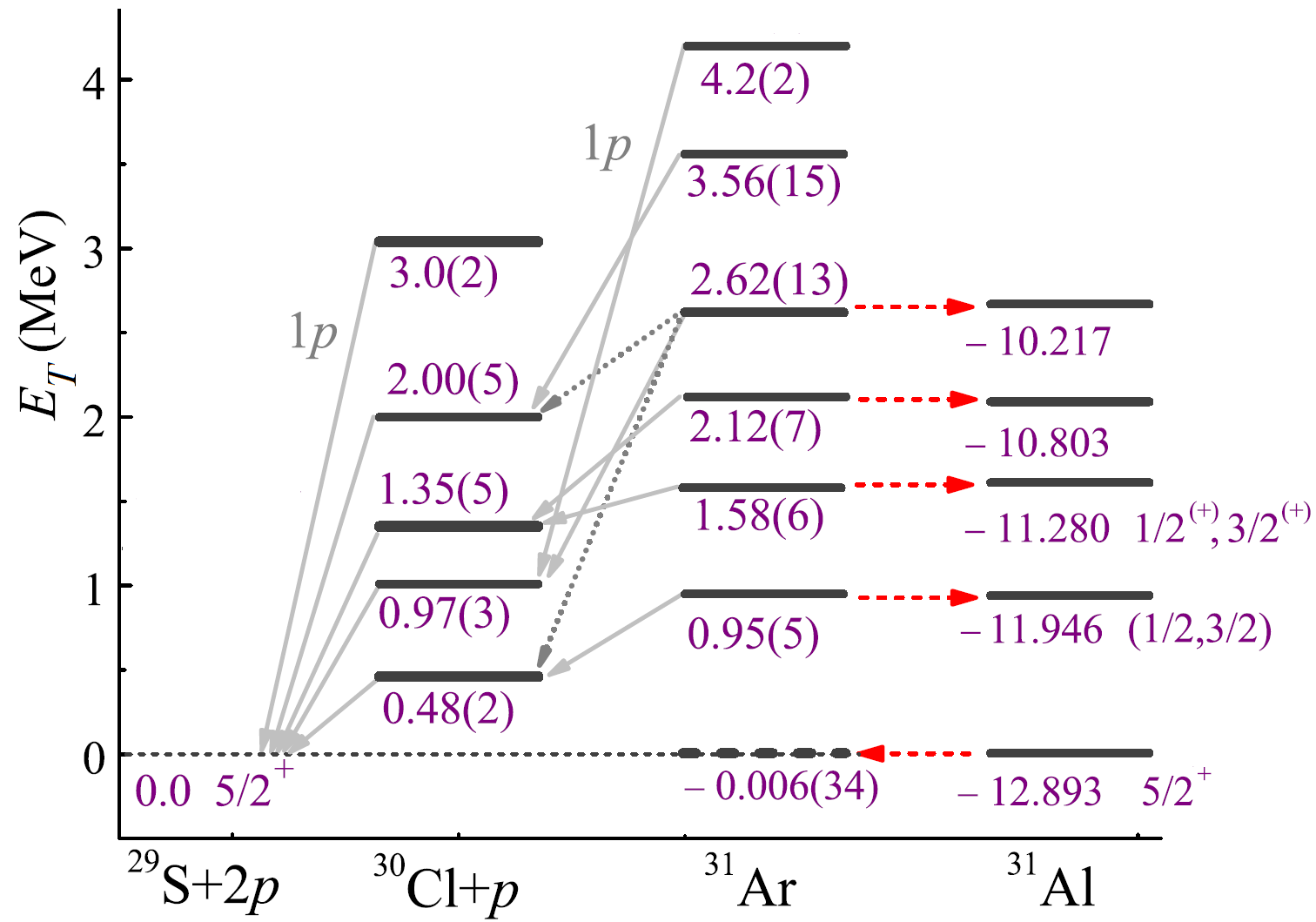}
		\end{minipage}
		\begin{minipage}[t]{16.5 cm}
		\caption{(Color online) The decay and level schemes of $ ^{31} $Ar and $ ^{30} $Cl isotopes derived in Ref.~\cite{Mukha:2018}. The assigned 1\textit{p} transitions are shown by the light-gray solid arrows. The dotted arrows show two undistinguished
decay branches of the 2.62-MeV state in $ ^{31} $Ar. The vertical axis shows the decay energies of the $ ^{30} $Cl) and $ ^{31} $Ar states. The four lowest excited states of isobaric mirror partner	$ ^{31} $Al are aligned with corresponding observed states of $ ^{31} $Ar (the
		correspondence of the levels is indicated by red dashed arrows) and the
		$ ^{31} $Ar g.s.\ energy is inferred based on isobaric symmetry assumptions.
		The legends for $ ^{31} $Al levels show energies relative to the 2\textit{p}-breakup
		threshold and the spin-parity $ J^\pi $ of the state. The scheme is taken from \cite{Mukha:2018}.			
			}
			\label{fig:4_31Ar_scheme}
		\end{minipage}
	\end{center}
\end{figure}

As a byproduct of experiments aimed at $^{29,30}$Ar \cite{Mukha:2015,Xu:2018},
interesting information was obtained for the excited states of $ ^{31} $Ar \cite{Mukha:2018}.
The ground state of $ ^{31} $Ar is known to be located very close the 2\textit{p}-decay threshold, but all excited states should decay by 2\textit{p} emission.
From the measured $ ^{29} $S+\textit{p}+\textit{p} coincidences, the level scheme
of $ ^{31} $Ar was derived, as shown in Figure~\ref{fig:4_31Ar_scheme}.
It can be seen, that the observed excitation spectrum of $ ^{31} $Ar
matches well the excitation spectrum of its isobaric mirror $ ^{31} $Al.
Based on this symmetry, the position of the $ ^{31} $Ar ground state
was determined by calculating the weighted mean of energy differences
between four pairs of the respective levels in $ ^{31} $Ar and $ ^{31} $Al.
The result of this procedure is $S_{2p} = +0.006(34)$ MeV for the
ground state of $ ^{31} $Ar. One should note that the claimed uncertainty of 34~keV
does not account for possible systematic errors which may arise from a number of
assumptions on spin-parity of the considered states in $ ^{31} $Ar.
In particular, it is nor clear what is the role of the Thomas-Ehrman shift
in this case. The evaluated decay energy is consistent with the previous
$ S_{2p}$ value of $-0.003$(110)~MeV obtained in $ \beta $-decay studies
of $ ^{31} $Ar \cite{Axelsson:1998}.

The near-zero value of $ S_{2p} $ of $ ^{31} $Ar allows speculations
about the possible near-threshold phenomena like a 2\textit{p} halo, which may
be examined by further, more precise studies of this isotope.

\subsection{\it Summary of 2p decays of light nuclei}

 \begin{figure}[h!tb]
 	\begin{center}
 			\includegraphics[width = 0.6\columnwidth, angle=0.]{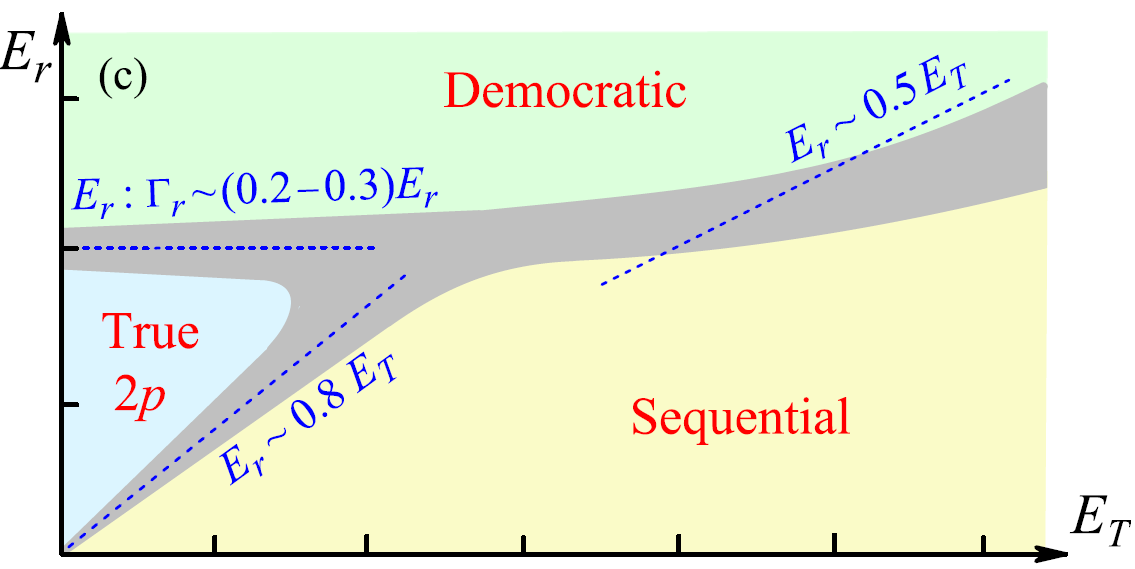}
 		\begin{minipage}[t]{16.5 cm}
 			\caption{ (Color online)   Overall view of evolution of three types of 2\textit{p} decay mechanisms in dependence on 2\textit{p}-  and 1\textit{p}-  decay energies  ($E_T$ and $E_r$, respectively) \cite{Golubkova:2016}, see also Figure~\ref{fig:1_EnergyConditions}.
 Three main areas of dominating decay mechanisms are marked  as ``True'' 2\textit{p}-decay (also referred as 2\textit{p}-radioactivity for long-lived states), as ``Sequential'' decay via narrow intermediate states in the sub-system, and as ``Democratic'' decay either directly to the daughter nucleus or via broad intermediate states.
 The dashed lines delineating the boarders between the areas correspond to the established ratios of the $E_r$ to $E_T$ values. The transition regions where these regimes interplay are shown in gray.	The figure is adopted from Ref.~ \cite{Golubkova:2016}.
 				}
 				\label{fig:4_30Ar_Golub2016}
 			\end{minipage}
 		\end{center}
 	\end{figure}

Nearly half of the neutron-deficient \textit{p}- and \textit{sd}-shell nuclei located 1-2
neutrons beyond the proton dripline decay by 2\textit{p} emission. In the last decade, extensive information on 2\textit{p} emitters in this light, $ A\leq $30, nuclear-mass region was obtained.

A general investigation of different mechanisms of 2\textit{p} decay and
of the governing physical conditions was attempted in Ref.~\cite{Golubkova:2016}.
For this survey, an improved three-body model was constructed which was found to
describe adequately all qualitative features of 2\emph{p}-decay distributions in
the whole kinematical space. In particular, it provided a good description of three-body correlations in 2\textit{p} decays of low-lying states in $^{16}$Ne \cite{Brown:2015}.
The main factors driving the 2\emph{p} decay mechanism were identified as ratios of three parameters, the 2\textit{p}-decay energy $E_T$, the 1\textit{p}-decay energy of an intermediate proton resonance $ E_r $, and its width $ \Gamma_r $. A schematic map identifying
the regions of dominance of various mechanisms, and the regions of interplay
between them, is shown in Fig.~\ref{fig:4_30Ar_Golub2016}.
One conclusion of this study is, that the change from one decay mechanism to another, which is reflected by momentum distributions of the decay products, occurs within rather narrow transition areas, i.e. within small changes of the relevant decay parameters.
In this transition areas, interplay of two decay mechanisms results in a significant change of three-body correlations, like for example in the ``tethered'' 2\textit{p}-decays of $ ^{16} $Ne$ ^{*} $(2$ ^{+} $).

Studies of 2\emph{p} emission from light nuclei demonstrated that this decay process
is a source of information on the structure of nuclei beyond the proton dripline,
which would be very hard, if not impossible, to access otherwise.
There is substantial evidence that the width of the 2\emph{p} unstable state,
as well as correlation pattern between emitted protons, do depend on the
details of the wave function of the initial nucleus.
This in turn can be used to probe the isospin symmetry
between 2\textit{n}-bound and 2\textit{p}-unbound isobaric mirrors.
An example of such an approach is the study of $ ^{12} $O--$ ^{12} $Be
and $ ^{16} $Ne--$ ^{16} $C pairs \cite{Grigorenko:2002}. The wave function
composition determined from the 2\emph{p} decay width was found to differ from
that of the mirror partner. This effect was named the ``dynamic'', or ``three-body''
mechanism of the Thomas-Ehrman shift.
Another example is the $ ^{6} $Be low-energy spectrum where negative-parity (0$ ^{-} $, 1$ ^{-} $, 2$ ^{-} $) continuum states dominate in charge-exchange reactions. Such a domination is  interpreted as a novel phenomenon of the low-energy  mode of giant dipole resonance of isovector type (or isovector soft dipole mode) in nuclei with losely-bound nucleon pairs \cite{Fomichev:2012}.
Further experiments on such excitations may offer new opportunities in studies of the few-body nuclear structure.

\section{Two-proton radioactivity}

Here, we focus on cases of ground-state 2\emph{p} radioactivity which live
long enough to study their decays at rest. Another characteristic feature
of this category is the fact that 2\emph{p} decay competes
with $\beta^+$ decay, so in addition to the half-life of such an emitter
also the branching ratio for 2\emph{p} decay, $b_{2p}$ must be measured in order
to determine the partial half-life for this decay channel, $T^{2p}_{1/2}$.
Below, we discuss each of the four cases known separately.

\subsection{\it The decay of $^{45}$Fe}

The most neutron-deficient isotope of iron to date, $^{45}$Fe, was observed
for the first time at the FRS facility in GSI Darmstadt \cite{Blank:1996}.
Three ions of $^{45}$Fe were identified among products of projectile
fragmentation of a $^{58}$Ni beam at 600 MeV/nucleon on a 4~g/cm$^2$ beryllium target.
Although no decay information was collected at that time, this experiment
paved the way to more detailed further studies.

\begin{figure}[h]
\begin{center}
\begin{minipage}[t]{12 cm}
\includegraphics[width = \columnwidth]{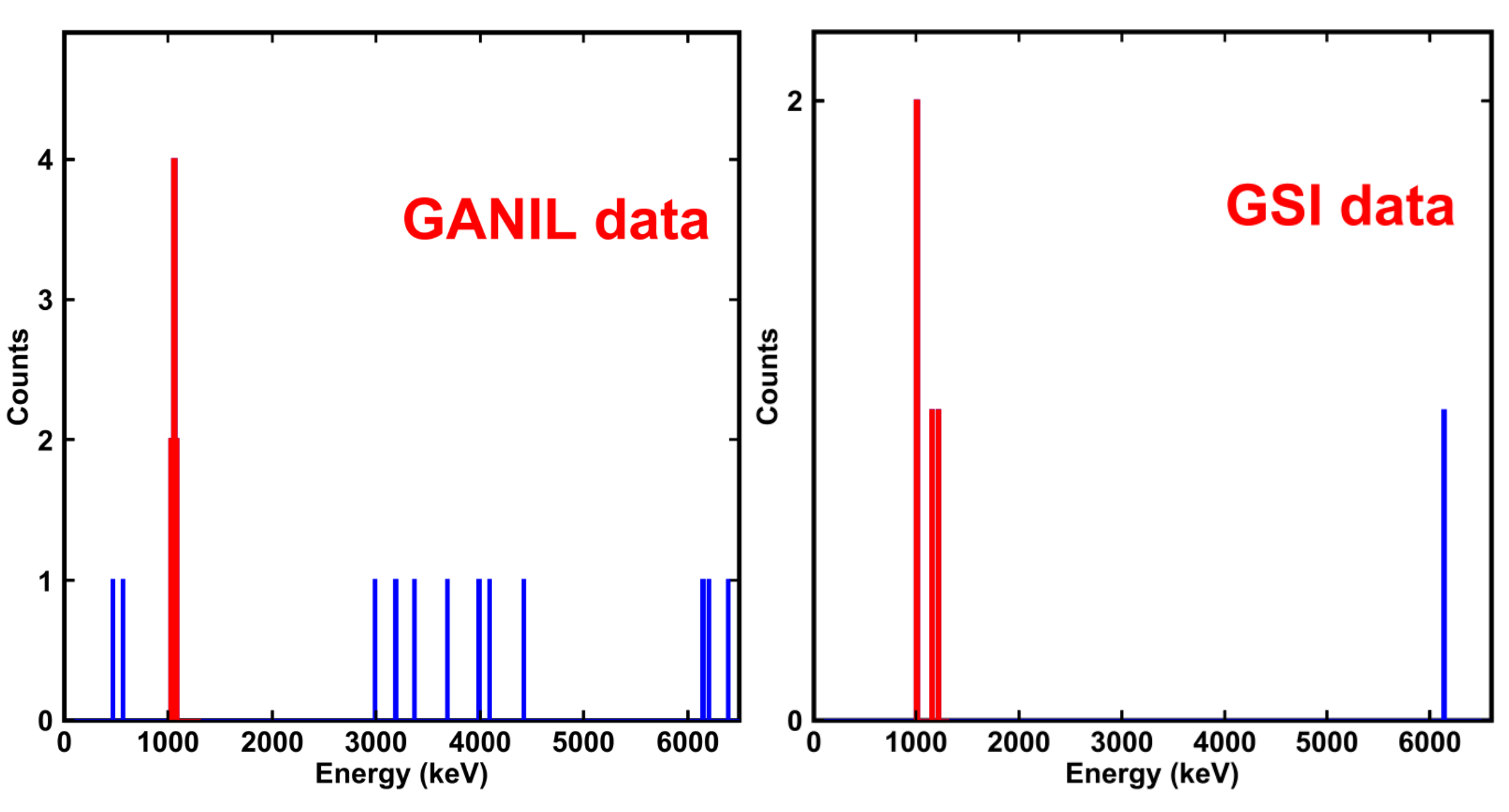}
\end{minipage}
\begin{minipage}[t]{17 cm}
\caption{(Color online) Energy spectra correlated with implantation of $^{45}$Fe ions representing
 the first evidence for the 2\emph{p} decay of this nucleus.
Right: spectrum from the FRS at GSI~\cite{Pfutzner:2002}. Left: spectrum from the
LISE3 at GANIL~\cite{Giovinazzo:2002}. Copyright IOP Publishing.
Reproduced with permission from Ref.~\cite{Blank:2008c}. All rights reserved. }
\label{fig:5_2p45FeDiscovery}
\end{minipage}
\end{center}
\end{figure}

The first data on decay of $^{45}$Fe were obtained in two experiments
performed at the LISE3 separator at GANIL \cite{Giovinazzo:2002} and
at the FRS in GSI \cite{Pfutzner:2002}. In both cases the beam of $^{58}$Ni
was used, at 75 MeV/nucleon on a natural nickel target, and at 650 MeV/nucleon
on a beryllium target, respectively. Selected ions were implanted into
silicon detectors but two setup differed in a method to detect $\beta$
decay events. At GANIL a silicon detector mounted next to the
implantation detector could directly register positrons. At GSI a NaI barrel
surrounding the silicon implantation telescope was used to detect 511~keV
annihilation photons. In addition, at GSI a special effort was taken to
detect decays as soon as a few microseconds after the implantation of
a heavy ion to achieve sensitivity for very fast decays \cite{Pfutzner:2002b}.
These precautions turned to be not necessary --- the half-life happened
to be in a range of milliseconds. At GSI out of six implanted ions of $^{45}$Fe,
decays of five were recorded, yielding the half-life of
$3.2^{+2.6}_{-1.0}$~ms. Four decays were consistent with the decay energy
of 1.1(1)~MeV. At GANIL 22 ions of $^{45}$Fe were implanted and 12 decays
were found in a narrow peak of 1.14(5)~MeV, indicating the half-life of
$4.7^{+3.4}_{-1.4}$~ms. None of the events in the 1 MeV region seen in
both experiments was coincident with emission of a positron. In addition,
the peak observed at GANIL did not show any distortion due to $\beta$ summing.
All these findings could be explained only by the hypothesis of the 2\emph{p}
emission from the ground state of $^{45}$Fe and this was the first
case of the new radioactive decay mode. Energy spectra from the two
experiments are shown in Fig.~\ref{fig:5_2p45FeDiscovery} and the
decay data are collected in Table~\ref{tab:5_2p45FeData}.

\begin{table} [h]
\begin{threeparttable}
\caption{Number of 2\emph{p} decay events observed ($N_{2p}$) and decay energies ($Q_{2p}$), half-lives ($T_{1/2}$), 2\emph{p} branching ratios, and the 2\emph{p} partial half-lives ($T_{1/2}^{2p}$) determined in experiments on decay of $^{45}$Fe.}
\label{tab:5_2p45FeData}
\vspace{0.5\baselineskip}
\begin{tabular}{lccccc}
  \hline
  \hline
  Reference             & $N_{2p}$    & $Q_{2p}$ (MeV)  &  $T_{1/2}$ (ms) & $b_{2p}$   &  $T_{1/2}^{2p}$ (ms) \\
  \hline
  Pf\"utzner \emph{et al}. \cite{Pfutzner:2002}  &  4&$1.1 \pm 0.1$     &$3.2^{+2.6}_{-1.0}$& $0.80 \pm 0.15$  &$4.0^{+3.3}_{-1.8}$ \\
  Giovinazzo \emph{et al}. \cite{Giovinazzo:2002}& 12&$1.140 \pm 0.040$ &$4.7^{+3.4}_{-1.4}$& $0.55 \pm 0.12$  &$8.5^{+6.4}_{-3.2}$ \\
  Dossat \emph{et al}. \cite{Dossat:2005}        & 17&$1.154 \pm 0.016$ &$1.6^{+0.5}_{-0.3}$& $0.57 \pm 0.10$  &$2.8^{+1.0}_{-0.7}$ \\
  Miernik \emph{et al}. \cite{Miernik:2007b}     & 87&       -          &$2.6 \pm 0.2$      & $0.70 \pm 0.4$   &$3.7 \pm 0.4$       \\
  Audirac\emph{ et al}. \cite{Audirac:2012}      &  7&$1.21 \pm 0.05$   &$3.6^{+1.6}_{-0.8}$& $0.78^{+0.14}_{-0.22}$&           \\
  \hline
  \hline
\end{tabular}
\end{threeparttable}
\end{table}

Second experiment at GANIL \cite{Dossat:2005}, which used a similar setup as reported
in \cite{Giovinazzo:2002}, succeeded in implantation of 30 ions of $^{45}$Fe into a
DSSSD detector. The 17 decay events were observed in the peak representing
the 2\emph{p} radioactivity. This measurement resulted in the most precise
determination of the 2\emph{p} decay energy for $^{45}$Fe to date,
$Q_{2p} = 1.154 \pm 0.016$~MeV, see Table~\ref{tab:5_2p45FeData}.
This value agrees very well with theoretical predictions, as shown in Fig.~\ref{fig:1_S1pS2p}.

\begin{figure}[h]
\begin{center}
\begin{minipage}[t]{10 cm}
\includegraphics[width = \columnwidth]{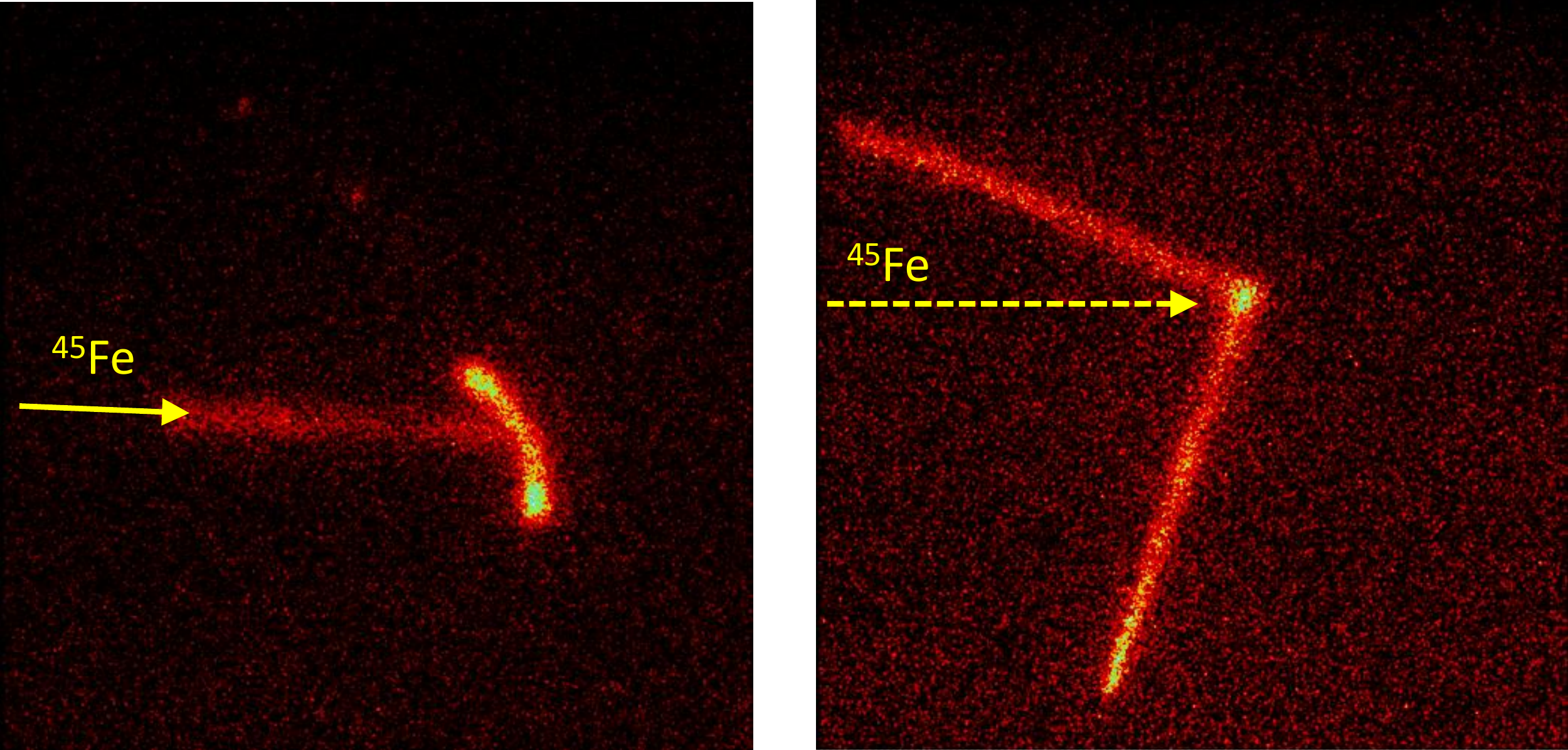}
\end{minipage}
\begin{minipage}[t]{17 cm}
\caption{(Color online) Example CCD images of decays of $^{45}$Fe recorded by the OTPC
detector in the experiment described in Refs.~\cite{Miernik:2007b,Miernik:2007c}.
Left: 2\emph{p} decay event, the track of a $^{45}$Fe ion entering
the chamber from left is seen. Right: $\beta 2p$ decay event, the track of the
entering ion is not seen because the exposure of the camera was started
just after implantation. Two long tracks represent delayed protons which
escape the active detector volume.}
\label{fig:5_2pb2p45Fe}
\end{minipage}
\end{center}
\end{figure}

New chapter in the studies of $^{45}$Fe was opened by application of TPC detectors.
First, Giovinazzo et al. \cite{Giovinazzo:2007} published the first direct observation
of the two protons ejected from an implanted ion of $^{45}$Fe using the Bordeaux TPC
detector, see Section~\ref{sec:3_TPC} and Fig.~\ref{fig:3_2p45FeBordeauxTPC}.
Detailed analysis of this experiment, published a few years later, presented the
full reconstruction of seven 2\emph{p} events \cite{Audirac:2012}.

\begin{figure}[h]
\begin{center}
\begin{minipage}[t]{10 cm}
\includegraphics[width = \columnwidth]{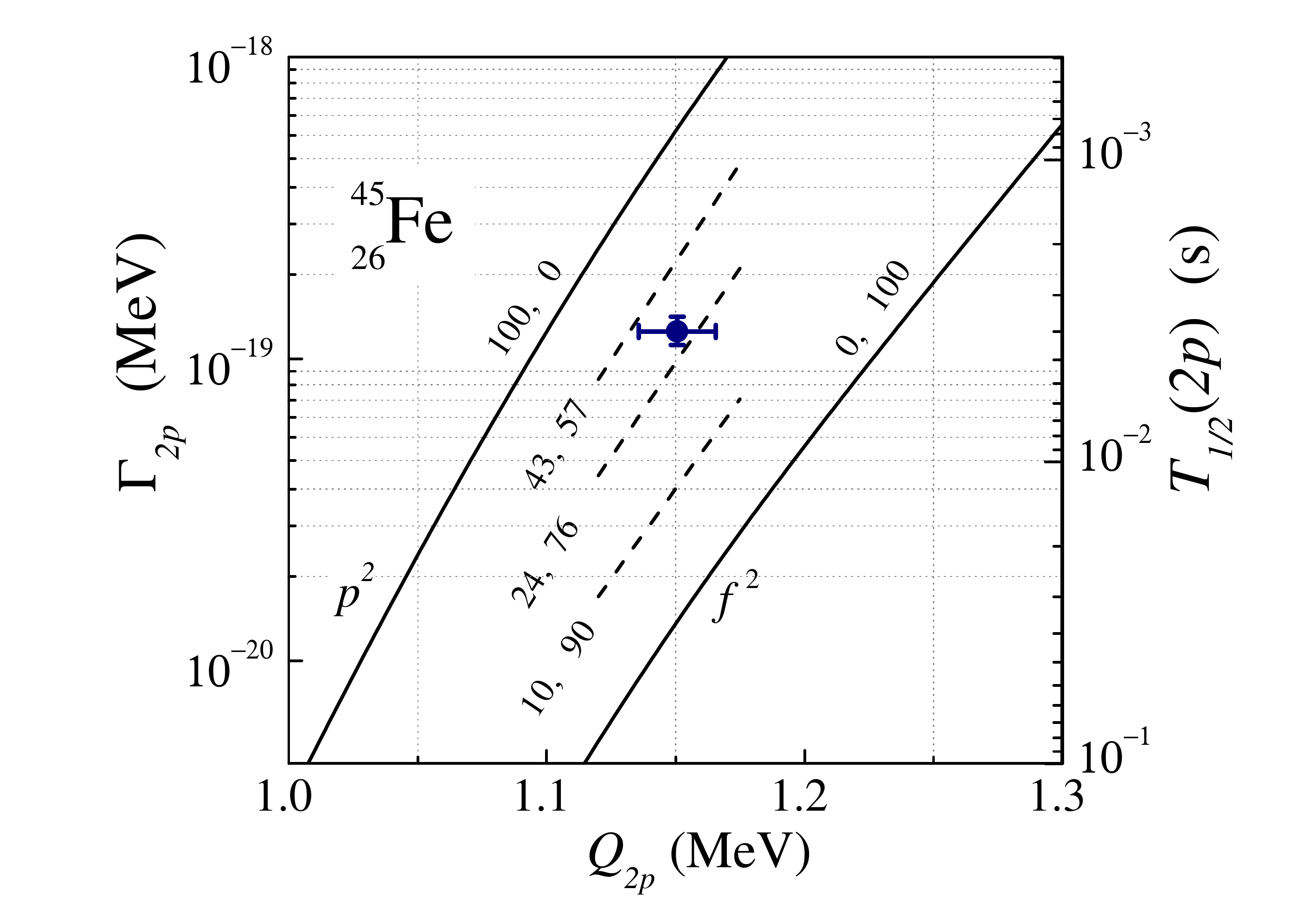}
\end{minipage}
\begin{minipage}[t]{17 cm}
\caption{The partial 2\emph{p} half-life of $^{45}$Fe as a function of the 2\emph{p} decay energy. The experimental result is shown together with predictions of the
3-body model \cite{Grigorenko:2003}. The decay width is taken from Ref.~\cite{Miernik:2007b} and the decay energy from Ref.~\cite{Dossat:2005}.
The numerical labels indicate the relative weights of the $p^2$ and $f^2$
configurations, respectively. Reprinted with permission from Ref.~\cite{Miernik:2007b}. Copyright (2007) by the American Physical Society.  }
\label{fig:5_45FePRL2}
\end{minipage}
\end{center}
\end{figure}

Miernik at al. \cite{Miernik:2007b,Miernik:2007c} applied the Warsaw OTPC detector
(see Section~\ref{sec:3_TPC} and Fig.~\ref{fig:3_OTPCPhoto}) to study the decay
of $^{45}$Fe at A1900 separator at the NSCL/MSU laboratory. In total, 125 decays
of $^{45}$Fe were observed, 87 of them showed a clear 2\emph{p} decay pattern
and 38 represented $\beta$ decay followed by emission of protons. The latter channel
is easy to identify since $\beta$-delayed protons have much larger energies
than protons from 2\emph{p} decay. Examples of 2\emph{p} and $\beta 2p$ decays
of $^{45}$Fe observed with the OTPC are shown in Fig.~\ref{fig:5_2pb2p45Fe}.
The relatively high statistics (the largest to date!)
allowed the determination the 2\emph{p} branching ratio and the half-life of $^{45}$Fe
with the highest accuracy, see Table~\ref{tab:5_2p45FeData}. The resulting partial
2\emph{p} half-life is compared in Fig.~\ref{fig:5_45FePRL2} with predictions
of the 3-body model \cite{Grigorenko:2003}. The theoretical lines are labeled
with the relative weights of the $p^2$ and $f^2$ initial proton configurations.
Note that both the experimental decay energy and the decay width are
best described by the model with the $p^2$ contribution of 24\%.

\begin{figure}[h]
\begin{center}
\begin{minipage}[t]{16 cm}
\includegraphics[width = \columnwidth]{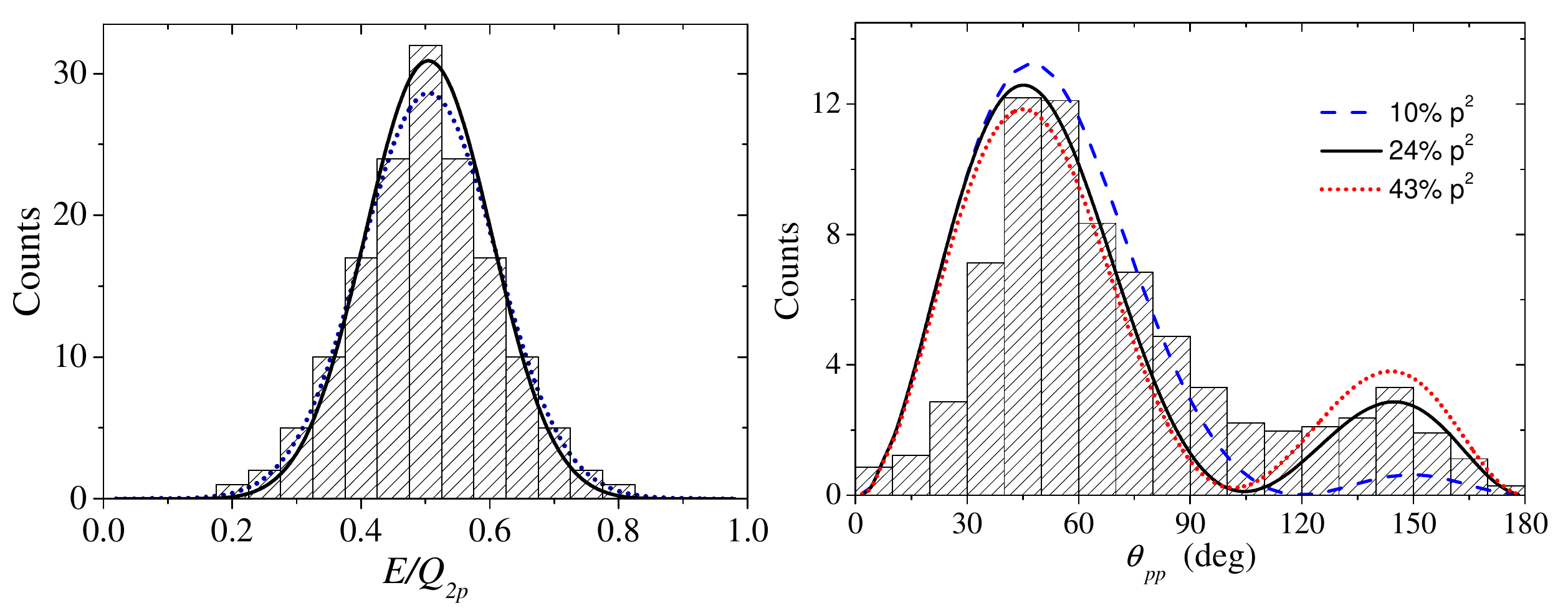}
\end{minipage}
\begin{minipage}[t]{17 cm}
\caption{Left: The energy distribution of individual protons emitted in
the 2\emph{p} decay of $^{45}$Fe (histogram). The energy is given in
units of the total decay energy $Q_{2p} = 1.15$~MeV. The solid line
indicates the prediction of the 3-body model Ref.~\cite{Grigorenko:2003}
while the dotted line shows this theoretical prediction folded with a
Gaussian function representing the detector energy resolution of 20\%.
Right: The measured distribution of the opening angle between two protons
emitted in the decay of $^{45}$Fe (histogram). Lines show the predictions
of the 3-body model for the indicated contribution of $p^2$ configuration.
Reprinted with permission from Ref.~\cite{Miernik:2007b}. Copyright (2007) by the American Physical Society.}
\label{fig:5_45FePRL34}
\end{minipage}
\end{center}
\end{figure}

An interesting byproduct of this experiment was the
first observation of $\beta 3p$ decay channel
for $^{45}$Fe \cite{Miernik:2007c} and for $^{43}$Cr~\cite{Pomorski:2011b}.
The most important result, however, reported in Ref.~\cite{Miernik:2007b} was the
reconstruction of 2\emph{p} decay events in 3D yielding the full \emph{p-p}
correlation pattern, see Fig.~\ref{fig:5_45FePRL34}. As expected, the most
probable emission occurs when each protons takes half of the available
decay energy. More interesting is the distribution of the opening
angle between proton momenta (Fig.~\ref{fig:5_45FePRL34}b). The visible
two-bump structure corresponds well to the prediction of the
3-body model \cite{Grigorenko:2003} and contradicts models
which assume two-body, diproton-type of decays or uncorrelated,
isotropic emission. Moreover, the detailed shape of the correlation
depends on the initial configuration of protons and the experimental
data are best described by the $p^2$ contribution of 24\%.
Thus, the same assumption on the initial wave function
consistently describes the 2\emph{p} decay width and the angular
correlation between protons. This observation indicates
that 2\emph{p} radioactivity may become a unique and sensitive
tool for probing the nuclear structure of these very exotic nuclei.

\subsection{\it The decay of $^{48}$Ni}

The $^{48}$Ni was observed for the first time at the LISE3 separator
at GANIL \cite{Blank:2000b}. Four events of this nucleus were found
among products of fragmentation of a $^{58}$Ni beam at 75~MeV/nucleon
on a natural nickel target. No decay data were recorded. With the isospin
projection $T_z = -4$ the supposedly doubly magic $^{48}$Ni is the most
neutron-deficient nucleus ever identified.

The first information on decay of $^{48}$Ni was obtained at GANIL
in the same experiment which provided the most accurate measurement
of the $Q_{2p}$ energy for $^{45}$Fe~\cite{Dossat:2005}. Again, only
four ions of $^{48}$Ni were identified and implanted into a DSSSD.
The four decay energies were found to be scattered between 1.3~MeV
and 4.5 MeV. One of them, at 1.35(2)~MeV, happened in the region where
2\emph{p} decay of $^{48}$Ni was expected and was not coincident with
a $\beta$ particle. This event could have come from 2\emph{p} decay but
such a single energy value is not sufficient for a conclusive evidence.
Three other events with larger energies were interpreted as resulting
from $\beta$ delayed emission of protons. The half-live of $^{48}$Ni
determined from these four events was $T_{1/2}=2.1^{+2.1}_{-0.7}$~ms.

\begin{figure}[h]
\begin{center}
\begin{minipage}[t]{6 cm}
\includegraphics[width = \columnwidth]{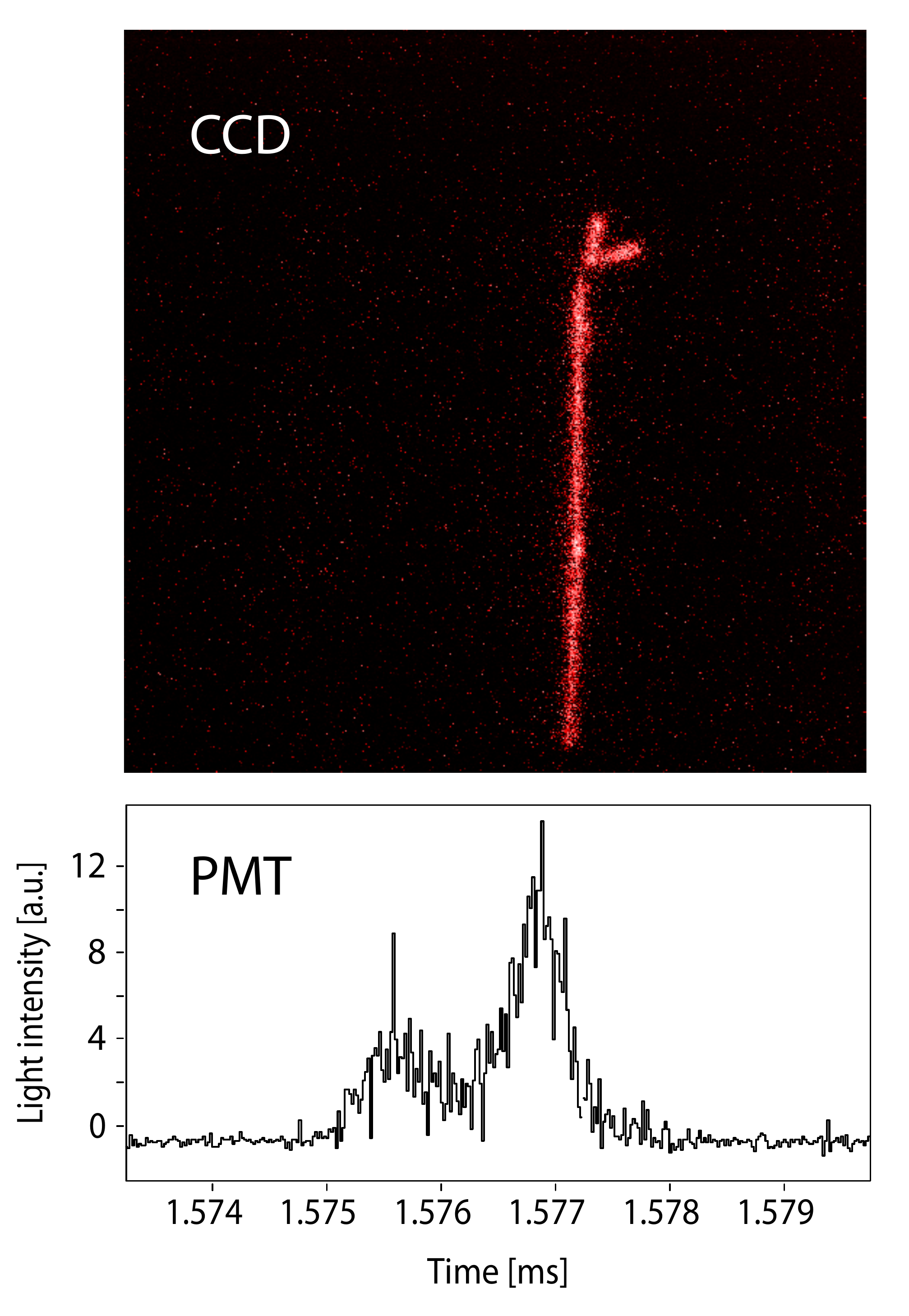}
\end{minipage}
\begin{minipage}[t]{17 cm}
\caption{(Color online) An example of a 2\emph{p} decay event of $^{48}$Ni
recorded by the Warsaw OTPC detector.
Top: the image recorded by the CCD camera. A track of the $^{48}$Ni ion entering
the chamber from below is seen. The two short tracks are protons emitted
1.576~ms after the implantation.
Bottom: a part of the time profile of the total light intensity measured
by the PMT showing in detail the 2\emph{p} emission.}
\label{fig:5_2p48Ni}
\end{minipage}
\end{center}
\end{figure}

The first unambiguous observation of the 2\emph{p} decay of $^{48}$Ni
was made at A1900 separator at NSCL with help of the OTPC
detector \cite{Pomorski:2011,Pomorski:2014}. The primary beam of $^{58}$Ni
at 160~MeV/nucleon and a natural nickel target were used. Ten ions
of $^{48}$Ni were identified by the acquisition system and decays
of six of them was recorded by the OTPC detector. Two of them were
consistent with an emission of a proton of large energy escaping
the detector, thus representing $\beta p$ decay channel. However,
the other four events clearly showed the emission of two protons
of low energy. One of these events was shown in Fig.~\ref{fig:3_OTPCexample}
the second one is presented in Fig.~\ref{fig:5_2p48Ni}.
The analysis of the six decay events yielded the half-life
for $^{48}$Ni of $T_{1/2}=2.1^{+1.4}_{-0.6}$~ms which agrees
very well with the result reported in Ref.~\cite{Dossat:2005}.
However, the branching for the 2\emph{p} decay appeared
to be $b_{2p}=0.7 \pm 0.2$ which differs from what is suggested
in Ref.~\cite{Dossat:2005}. With this branching the partial
2\emph{p} decay half-life of $^{48}$Ni is $T_{1/2}^{2p}=3.0^{+2.2}_{-1.2}$~ms.
The full reconstruction of 2\emph{p} events provided the
decay energy $Q_{2p}=1.29 \pm 0.04$~MeV~\cite{Pomorski:2014}
which is in good agreement with theoretical predictions,
see Fig.~\ref{fig:1_S1pS2p}. The momentum correlations between
protons obtained from the four events do not, unfortunately,
allow for meaningful conclusions. Future experiments are clearly
needed to provide data with larger statistics.

\subsection{\it The decay of $^{54}$Zn}

\begin{figure}[h]
\begin{center}
\begin{minipage}[t]{10 cm}
\includegraphics[width = \columnwidth]{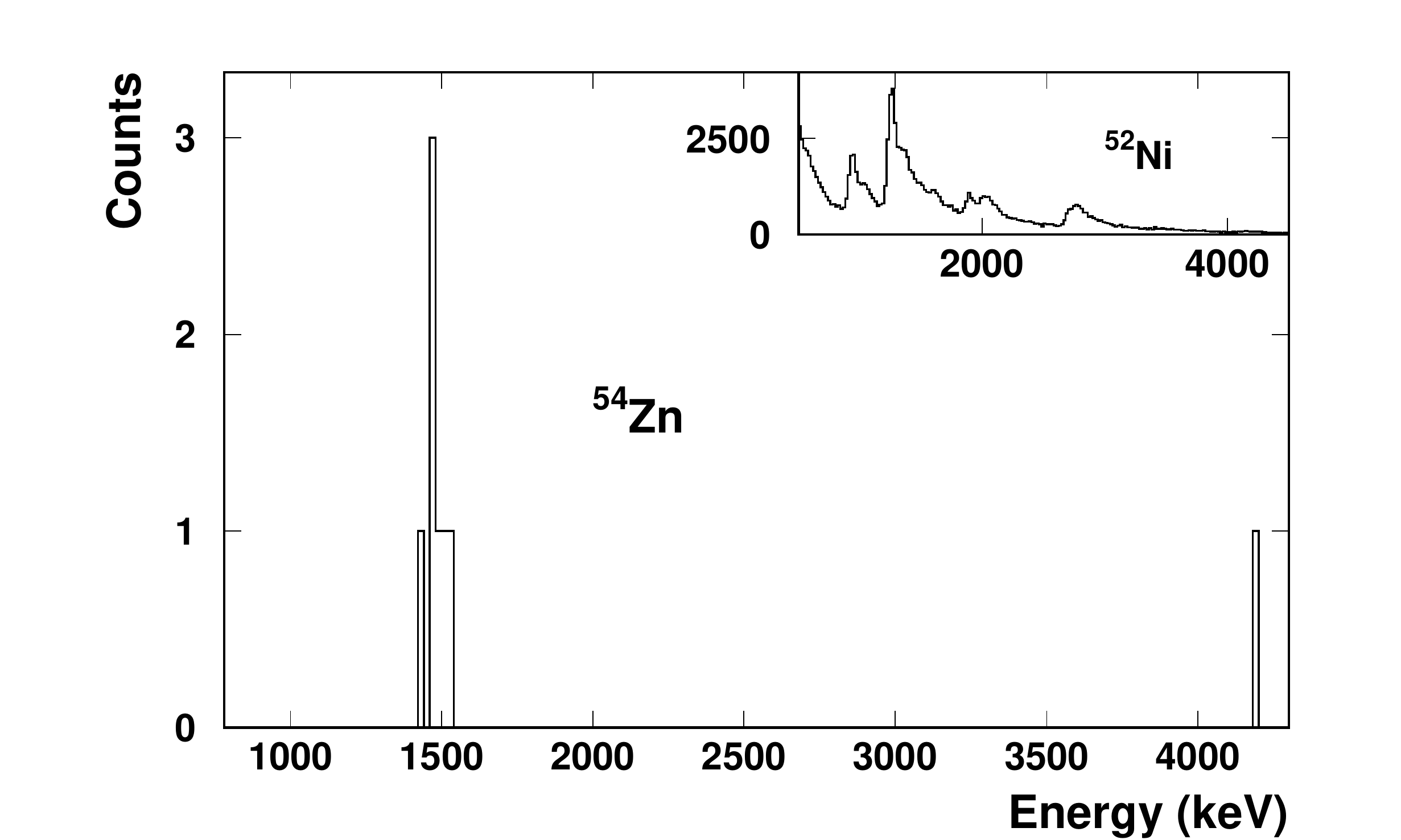}
\end{minipage}
\begin{minipage}[t]{17 cm}
\caption{Decay energy spectrum measured for $^{54}$Zn. The peak at 1.48(2)~MeV
represents 2\emph{p} decay of this nucleus. The inset shows the decay energy spectrum
for $^{52}$Ni dominated by $\beta p$ decays. Reprinted with permission
from Ref.~\cite{Blank:2005}. Copyright (2005) by the American Physical Society.  }
\label{fig:5_2p54ZnDiscovery}
\end{minipage}
\end{center}
\end{figure}

The $^{54}$Zn was identified for the first time at the LISE3 at GANIL
in an experiment with $^{58}$Ni beam at 75~MeV/nucleon on a natural
nickel target \cite{Blank:2005}. Eight ions of $^{54}$Zn were
implanted into a DSSSD detector and their decays were recorded.
The extracted half-life amounted to $3.2^{+1.8}_{-0.8}$~ms.
The decay energy spectrum is shown in Fig.~\ref{fig:5_2p54ZnDiscovery}.
Seven events can be seen located in a narrow energy peak
corresponding to the decay energy of $1.48 \pm 0.02$~MeV.
None of these events was coincident with a positron signal
in adjacent detectors, and in addition, the peak does not
show a broadening due to $\beta$ summing.
For comparison, the inset in Fig.~\ref{fig:5_2p54ZnDiscovery}
shows the spectrum of $\beta$-delayed protons from the
decay of $^{52}$Ni where such broadening is clearly visible.
Therefore, the peak at 1.48(2)~MeV was interpreted as originating
from 2\emph{p} decay of $^{54}$Zn which proceeds with the
branching ratio of $87^{+10}_{-17}$\%.
The one event at about 4.2~MeV can be explained as a result
of $\beta p$ emission. Theoretical predictions for the
2\emph{p} decay energy of $^{54}$Zn, shown in Fig.~\ref{fig:1_S1pS2p},
have a larger spread than for other 2\emph{p} emitters, but the experimental
value of 1.48(2)~MeV is close to the average of these predictions.

\begin{figure}[h]
\begin{center}
\begin{minipage}[t]{10 cm}
\includegraphics[width = \columnwidth]{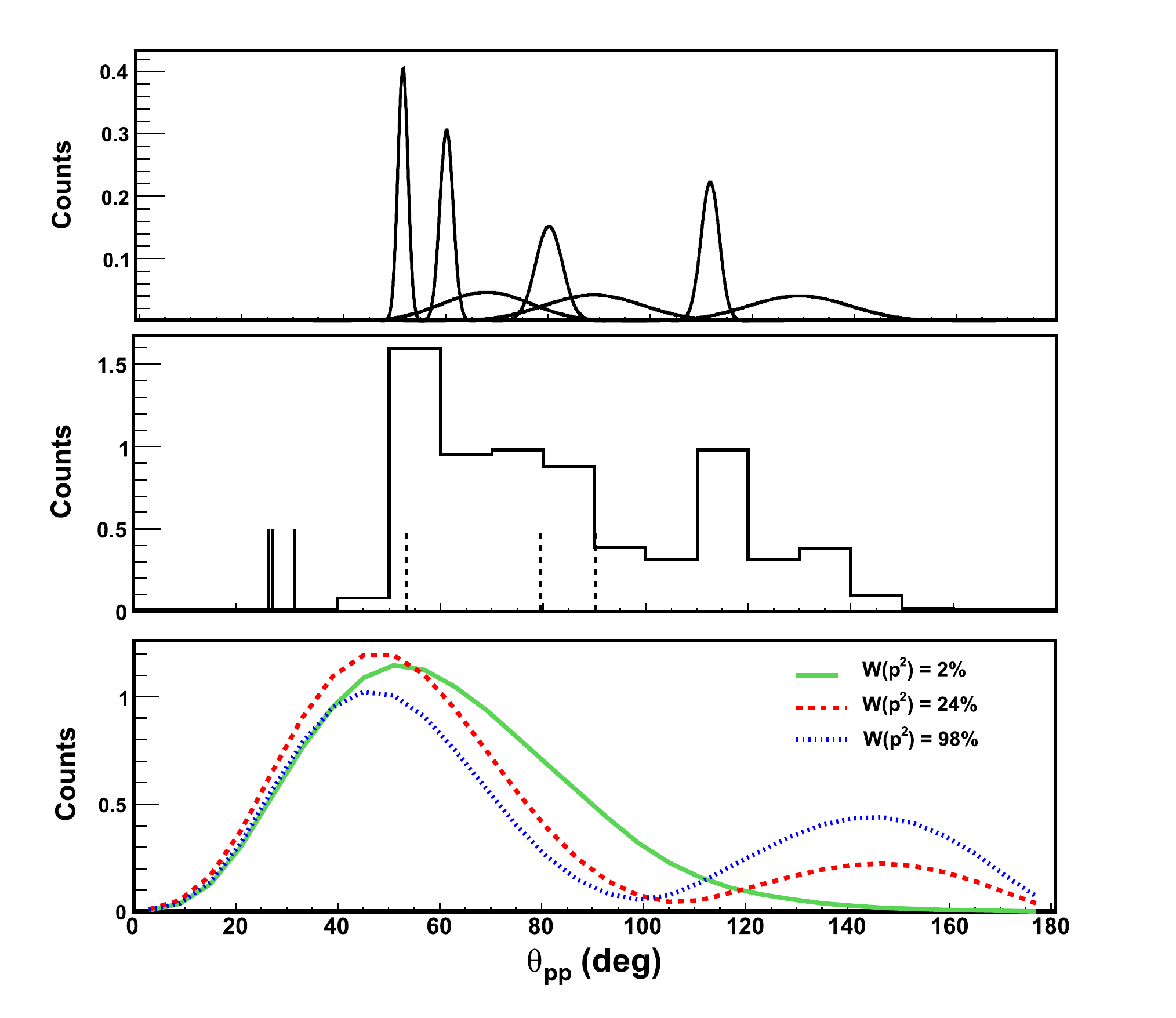}
\end{minipage}
\begin{minipage}[t]{17 cm}
\caption{(Color online) The opening angle between protons emitted in 2\emph{p} decay
of $^{54}$Zn. Top: experimental values for seven events analysed. Each event is
convoluted with a Gaussian representing the angular resolution.
Middle: Sum of seven events shown in the top represented as a histogram. The
dashed and full lines represent two possible angles for three
not fully reconstructed events.
Bottom: Theoretical predictions by the three-body model for three values of
the $p^2$ configuration in the initial wave function.
Reprinted with permission from Ref.~\cite{Ascher:2011}.
Copyright (2011) by the American Physical Society.}
\label{fig:5_54ZnCorrelations}
\end{minipage}
\end{center}
\end{figure}

In the next experiment at GANIL, using the same reaction, decay
of $^{54}$Zn was investigated by means of the Bordeaux TPC
detector and provided the first direct observation of 2\emph{p}
decay for this nucleus \cite{Ascher:2011}.
This time 18 events of $^{54}$Zn were identified and for 10
of them decay information could be extracted. The analysis
yielded new values for the half-life, $T_{1/2}=1.59^{+0.60}_{-0.35}$~ms,
and the 2\emph{p} branching ratio, $b_{2p}=92^{+6}_{-13}$\%.
For the 2\emph{p} decay energy the value of $1.28 \pm 0.21$~MeV
was found which is consistent with the energy reported
in Ref.~\cite{Blank:2005} but less precise.
Seven 2\emph{p} decay events could be fully reconstructed in 3D.
The distribution of the opening angle between emitted protons
is shown in Fig.~\ref{fig:5_54ZnCorrelations}. Experimental
data do indicate a double-hump structure, as predicted by the
three-body model and also observed for $^{45}$Fe. However,
much larger statistics is clearly needed before meaningful
conclusions could be drawn from this case.

We note that the reaction used in both GANIL experiments could
not be the pure projectile fragmentation which leads only to removal
of nucleons from the projectile. The pick-up of two
protons was necessary to make zinc from nickel. This means
that other reaction channels, like a multinucleon transfer,
contributed to the reaction process.
The cross section for the production of $^{54}$Zn in this
reaction was estimated in Ref.~\cite{Blank:2005} to be about
100~fb.

A different approach to the production of $^{54}$Zn was undertaken
at RIKEN where the fragmentation of $^{78}$Kr beam at 345 MeV/nucleon
on a beryllium target was used. The production
cross section for $^{54}$Zn was determined to be
$3.5 \pm 0.7_{\rm (stat)} \pm 1.2_{\rm(syst)}$~fb \cite{Kubiela:2021}.
This value is smaller by a factor of about 30 from the cross
section obtained at GANIL with $^{58}$Ni beam at 75~MeV/nucleon
on a natural nickel target. However, the available intensity of $^{78}$Kr
beam at RIKEN is larger than the intensity of $^{58}$Ni beam at GANIL
and, in addition, the higher energy of $^{78}$Kr allows using a
thicker target. Both these factors lead to a much larger
luminosity which overcompensates the smaller cross section.
In consequence, the final production yield of $^{54}$Zn ions at the
end of the fragment separator at RIKEN may reach 70 ions per day, thus
by a factor of about 3 lager than the corresponding maximal yield at GANIL \cite{Kubiela:2021}.

\subsection{\it The decay of $^{67}$Kr}

\begin{figure}[h]
\begin{center}
\begin{minipage}[t]{15 cm}
\includegraphics[width = \columnwidth]{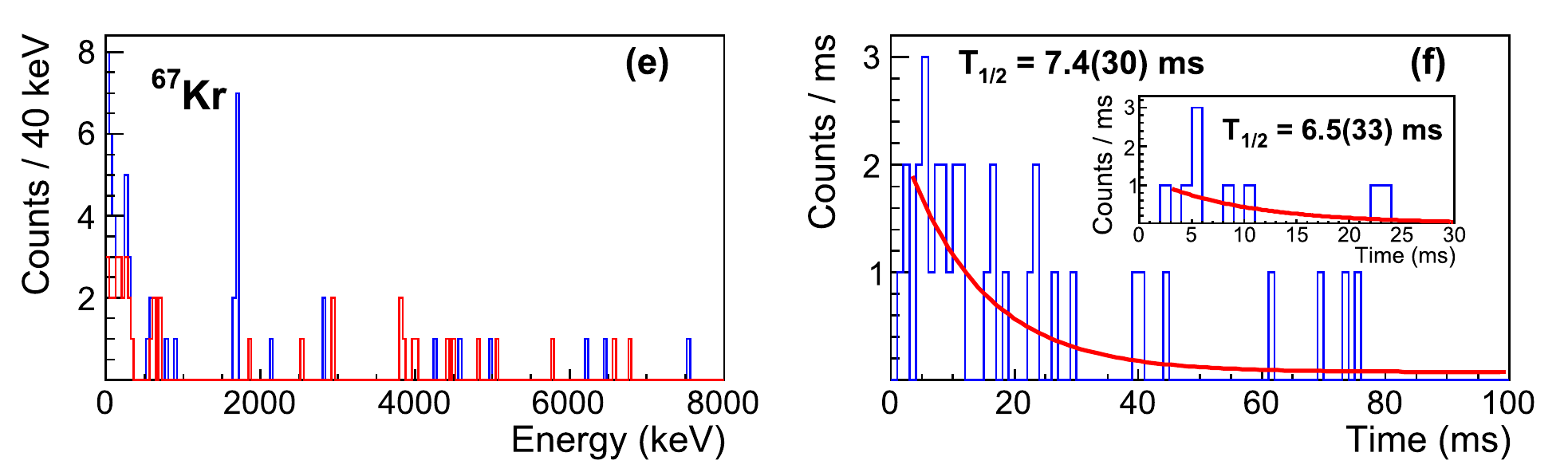}
\end{minipage}
\begin{minipage}[t]{17 cm}
\caption{(Color online) Left: decay energy spectrum of $^{67}$Kr.
The narrow peak at 1.69~MeV is due to 2\emph{p} radioactivity.
Events shown in red were in coincidence with positrons detected
in neighboring detectors. Right: Decay time distribution for $^{67}$Kr.
In the inset the decay time for nine events in the 1.69~MeV peak is shown.
Reprinted with permission from Ref.~\cite{Goigoux:2016}.
Copyright (2016) by the American Physical Society.}
\label{fig:5_2p67KrDiscovery}
\end{minipage}
\end{center}
\end{figure}

The heaviest 2\emph{p} emitter known to date, $^{67}$Kr, was discovered
at the RIKEN Nishina Center in an experiment at the BigRIPS
fragment separator \cite{Goigoux:2016}. The primary beam of $^{67}$Kr
at energy of 345~MeV/nucleon was impinging on a beryllium production target.
The identification plot of ions coming to the final detection setup
is shown in Fig.~\ref{fig:3_IDPlot}. These ions were implanted into
the WAS3ABi DSSSD array \cite{Nishimura:2013}. Out of 82 ions of $^{67}$Kr
identified at the end of BigRIPS, 36 were successfully implanted.
The measured energy spectrum is shown in Fig.~\ref{fig:5_2p67KrDiscovery} (left part).
A narrow peak of nine events is seen at the energy of $1.690 \pm 0.017$~keV.
None of these events was found in coincidence with $\beta$ particles,
thus they are interpreted as an evidence for 2\emph{p} radioactivity.
The 2\emph{p} decay energy of 1.690(17)~MeV is in a very good agreement
with theoretical predictions, see Fig.~\ref{fig:1_S1pS2p}.
Other events in the spectrum are attributed to $\beta$ delayed particle
channels. The 2\emph{p} branching ratio was found to be 37(14)\%.
In the right part of Fig.~\ref{fig:5_2p67KrDiscovery} the time
distribution for decay events of $^{67}$Kr is shown with the best fitted
decay curve corresponding to the half-life $T_{1/2}=7.4 \pm 3.0$~ms.
Taking into account the 2\emph{p} branching ratio,
the partial 2\emph{p} half-life of $^{67}$Kr is $T_{1/2}^{2p}=20 \pm 11$~ms.
Further experiments with a TPC detector are planned to observe
directly two protons emitted in the decay of $^{67}$Kr and to
study their correlation pattern.

The authors of Ref.~\cite{Goigoux:2016} noted the surprisingly short
partial 2\emph{p} half-life of $^{67}$Kr. Predictions of the three-body
model for this nucleus, published in Ref.~\cite{Grigorenko:2003},
depend on the relative weight of $p^2$ and $f^2$ configurations
of the initial state. The fastest 2\emph{p} decay, for the pure
$p^2$ configuration, is predicted to proceed with the half-life
of 0.28~s, thus by an order of magnitude longer than determined
in the experiment. It was concluded that this could be due to
configuration mixing effects and/or nuclear deformation.
The short half-life of $^{67}$Kr motivated Grigorenko and
others \cite{Grigorenko:2017} to reexamine the decay mechanism
of this nucleus. They noted that it is plausible that the
decay of $^{67}$Kr is at the borderline between true three-body
\emph{2p} and the sequential two-body channels.

The possible influence of deformation on the decay of $^{67}$Kr was
addressed by Wang and Nazarewicz \cite{Wang:2018} who applied their recently
developed model of nuclear three-body decays --- the Gamow coupled
channel (GCC) method in Jacobi coordinates \cite{Wang:2017b}.
They were able to show that the deformation effects
strongly affect the 2\emph{p} decay of $^{67}$Kr.
For the oblate deformation of the $^{65}$Se core
at $\beta_2 \approx -0.3$, the Nilsson orbital 1/2[321]
with large $l=1$ amplitude becomes available to valence
protons what significantly reduces the 2\emph{p} decay half-life.
The GCC model yielded for this case $T^{2p}_{1/2}=24^{+10}_{-7}$~ms
in good agreement with experiment (see also discussion in Section 2.4.1
and Fig.~\ref{fig:2_67Kr}).

It is expected that the planned proton-proton correlation studies will
shed more light on the \emph{2p} decay of $^{67}$Kr and clarify what
mechanism is responsible for the half-life reduction in this case.


\section{Beyond 2p emission. Multi-particle nuclear decays}

The most exotic nuclei located in the very remote outskirts of the nuclear landscape become unbound in respect of new decay channels. Such exotic decay modes play increasingly important role with a precursor decay energy growing. In particular, there are a few nuclei established as the ground-state 3\textit{p}-emitters,  $^{7}$B \cite{Charity:2011}, $^{17}$Na \cite{Brown:2017},$^{31}$K \cite{Kostyleva:2019}, and $^{13}$F \cite{Charity:2021}. The measured  decay patterns in all cases include  2\textit{p} emission as part of a sequential 3\textit{p}-decay mechanism.
Furthermore, two ground-state 4\textit{p} emittersc, $^{8}$C \cite{Charity:2011}
and $^{18}$Mg \cite{Jin:2022}, decay by a sequential 2\textit{p}-2\textit{p} emission of two \textit{pp} pairs.
Besides, there is a number of nuclear exited states decaying via 3\textit{p}, 4\textit{p} emissions and even more exotic exit channels.

In this section, the experimental information relevant to the mentioned exotic decays will be reviewed as well as  applications of the established  2\textit{p} decay mechanisms to the measured cases. In addition, few examples of three-body nuclear decays without \textit{pp} or \textit{nn} pairs in exit channel will be discussed.

\subsection{\it Three proton emission}

In most known nuclei, three-proton decay thresholds are located higher than the respective 1\textit{p}, 2\textit{p} decay thresholds. Therefore, such 3\textit{p}-decay channels can be explored more easily by populating excited nuclear states.

Beta decay offers one way to populate such states as the $\beta$-decay energies ($Q_{\beta}$) for
nuclei far from stability are large and highly excited states in daughter nuclei may be fed.
Indeed, beta-delayed one- ($\beta p$) and two-proton ($\beta 2p$) emissions are quite common
and studied since decades \cite{Pfutzner:2012,Blank:2008}. The first instance of $\beta$-delayed
three-proton emission ($\beta 3p$) was discovered for $^{45}$Fe \cite{Miernik:2007c} using the
OTPC detector (see section 3.2.2). In the same experiment, $^{43}$Cr was also identified as the
$\beta 3p$ emitter \cite{Pomorski:2011b}. Later, using the same detection technique, the delayed
\emph{3p} emission was observed also for $^{31}$Ar \cite{Lis:2015} and for $^{23}$Si \cite{Ciemny:2022}.
In all these measurements, however, the full decay kinematics
could not be reconstructed, as at least some protons escaped the detector volume.
The case of $^{31}$Ar was studied in more detail by Koldste et al. \cite{Koldste:2014,Koldste:2014b}.
Using an array of DSSSD detectors it was possible to measure energies of individual protons and to
establish the $Q_{3p}$ decay energy. It was found that about half of $\beta 3p$ decays stem from
the Isobaric Analog State (IAS) in the $\beta$-daughter $^{31}$Cl while the rest follows
the Gamow-Teller (GT) transitions to $^{31}$Cl levels above the IAS. Furthermore, it was
estimated that the latter represent about 30\% of the observed GT strength in the decay
of $^{31}$Ar \cite{Koldste:2014b}. Noting that the total branching for $\beta 3p$ decay
channel is only $7(2) \times 10^{-4}$ \cite{Lis:2015}, this vividly illustrates the importance
of delayed multi-particle emission for $\beta$ decay spectroscopy. The mechanism of such
decays is not fully resolved. The two protons from $\beta 2p$ decay of $^{31}$Ar are emitted
sequentially \cite{Fynbo:2000} and the same is assumed for the $\beta 3p$ channel.
Future measurements of momentum correlations between emitted protons will shed light
on this question.

A number of excited states were reported to be 3\textit{p} emitters  in studies dedicated to 2\textit{p} precursors. For example, high-lying continuum states in $ ^{16} $Ne \cite{Mukha:2009} decay by a cascade three-step proton emission $ ^{16} $Ne$ ^{*}\!\rightarrow \:$\textit{p}+$ ^{15} $F$ ^{*}\!\rightarrow \:$2\textit{p}+$ ^{14} $O$ ^{*}\!\rightarrow \:$3\textit{p}+$ ^{13} $N via narrow intermediate states in
$ ^{15} $F and $ ^{14} $O \cite{Mukha:2009}. This observation was reported as a by-product of the measurements of 2\textit{p} decays of $ ^{16} $Ne \cite{Mukha:2008}.
Another example  is the high-excited state in $ ^{12} $N observed in the
experiment dedicated to $ ^{12} $O \cite{Webb:2019,Webb:2020}.
The $ \sim $22 MeV state decays as $ ^{12} $N$ ^{*}\! \rightarrow \:$3$ {p} $+$ ^{9} $Be, and the proton emission is sequential via intermediate states in $ ^{10} $B.

\begin{figure}[htb]
	\begin{center}
		\begin{minipage}[t]{16 cm}
			\includegraphics[width = 0.99\columnwidth, angle=0.]{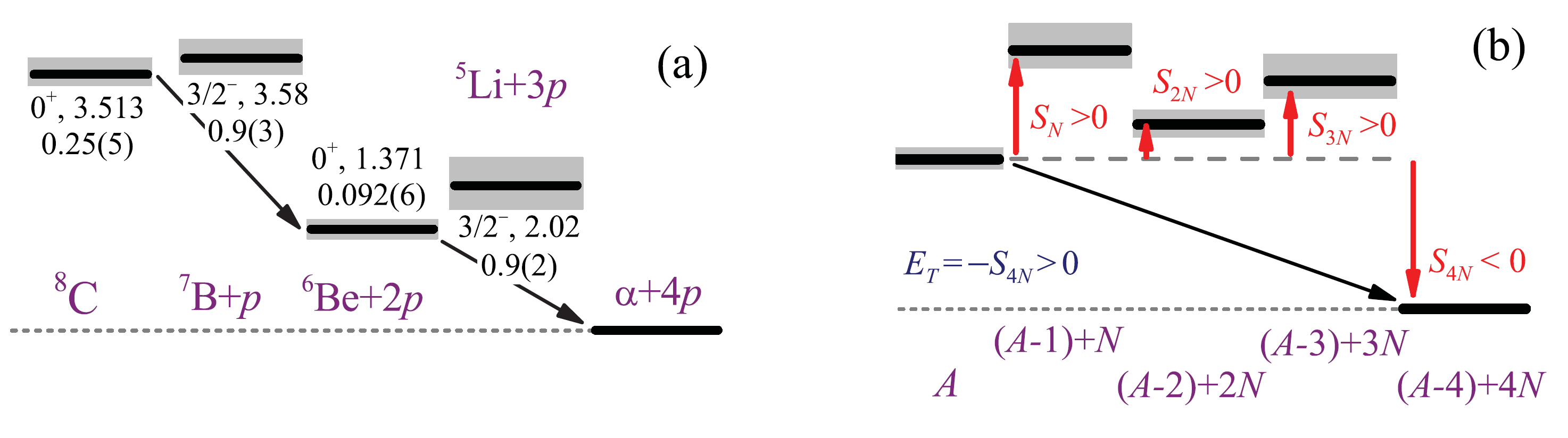}
		\end{minipage}
		\begin{minipage}[t]{16.5 cm}
			\caption{(Color online)  (a) The 4\textit{p}-decay scheme of $ ^{8} $C with intermediate resonances $ ^{7} $B, $ ^{6} $Be and $ ^{5} $Li  reported in \cite{Charity:2011}.  The level spin-	parities $ J^\pi $, decay energies $ E_T $, and widths (in MeV) are indicated. Arrows illustrate sequential 2\textit{p}-2\textit{p} decay mechanism of $ ^{8} $C. (b) General energy conditions required for the true four-nucleon, 4\textit{N}, decay (black arrow) of isotope with mass \textit{A}. Red arrows illustrate the separation energies $ S_{(N,2N,3N)} $ of the respective intermediate nuclei with masses (${A-1}$), (${A-2}$), (${A-3}$).	
			}
			\label{fig:6_sheme_4p}
		\end{minipage}
	\end{center}
\end{figure}

Exotic isotopes located by several atomic-mass units beyond the proton drip line, may be unbound in respect to 3\textit{p} emission. The lightest ground-state 3\textit{p} precursor $ ^{7} $B was produced in fragmentation reactions of a radioactive beam $ ^{9} $C at intermediate energy \cite{Charity:2011}. The ground state resonance was measured by  applying  invariant-mass method to the detected  $\alpha$+3\textit{p} decay products.
The derived 3\textit{p}-decay energy of 3.58 MeV of the $ ^{7} $B ground state and \textit{p}-$ ^{6} $Be$ _{\rm g.s.} $ correlations are consistent with its sequential \textit{p}-2\textit{p} decay via intermediate g.s.\ of $ ^{6} $Be, see the level and decay scheme in Fig.~\ref{fig:6_sheme_4p}(a).


%

Recently a very exotic  3\textit{p} resonance, $ ^{13} $F was observed following a charge-exchange reaction of an intermediate-energy $ ^{13} $O beam \cite{Charity:2021}. The resonance was found in the invariant-mass distribution of $ ^{10}$C+3\textit{p} events and has tentatively been assigned as a 5/2$ ^{+} $ excited state. The ground state was also expected to be populated, but was not  resolved from the background. The observed level decays via initial proton emissions to both the ground and the
first 2$ ^{+} $ state of $ ^{12} $O, which subsequently undergo 2\textit{p} decay.

The ground-state 3\textit{p} precursor $ ^{17} $Na was observed in a charge-exchange reaction of a radioactive beam $ ^{17} $Ne at intermediate energy \cite{Brown:2017}. Its spectrum has been derived from the measured $ ^{14} $O+3\textit{p} coincidences by the invariant-mass method. The $ ^{17} $Na decay-energy spectrum showed a broad bump at $ \simeq $4.8 MeV with the ground and excited states unresolved. Then an upper limit of the 3\textit{p}-decay energy of 4.85(6) MeV  was reported for the ground state of $ ^{17} $Na.

The most heavy isotope located beyond the proton dripline by 4 mass units, $ ^{31} $K  is also unbound with respect to 3\textit{p} emission. It was produced in a charge-exchange reaction of the $ ^{31} $Ar secondary beam at high energy, and its decays were detected in flight by
tracking the trajectories of all decay products using microstrip detectors \cite{Kostyleva:2019}. The 3\textit{p} emission processes were studied by the means of angular correlations of the decay products $ ^{28}$S+3\textit{p}  as well as the respective decay vertices. The energies of the ground and excited states of $ ^{31} $K have been determined. This provided its 3\textit{p} separation energy
$S_{3p} =4.6(2)$~MeV. The upper half-life limit of 10~ps for the observed $ ^{31} $K states have been derived from the distributions of the decay vertices. The reported states undergo a sequential \textit{p}-(2\textit{p}) decay $ ^{31} $K$ \rightarrow ^{30}$Ar+\textit{p}$\rightarrow ^{28}$S+3\textit{p} via intermediate 2\textit{p}-resonance  $ ^{30} $Ar \cite{Kostyleva:2019}.

For all cases of 3\textit{p} emission mentioned above, a dominant sequential decay mechanism
(e.g., 1\textit{p}-2\textit{p}) was concluded. In the analysis of these cases, the application of the established  2\textit{p} decay mechanisms is essential.

\subsection{\it Four proton emission}
\label{sec:6_4p}
The first-measured  ground-state 4\textit{p} emitter is $^{8}$C. Its decay mechanism was studied in the experiment where $^{8}$C was produced in a neutron knockout reaction from a $ ^{9} $C beam \cite{Charity:2011}. The $^{8}$C  decay products, 4\textit{p}+$ \alpha $  were measured, and the invariant-mass method was applied.
The five-body decay of the $ ^{8} $C ground state was found to proceed in two steps by a sequential 2\textit{p}-2\textit{p} emission of two successive \textit{pp} pairs via the $ ^{6} $Be$ _{\rm g.s.} $ intermediate state.
The isobaric analog of the $ ^{ 8} $C ground state in $ ^{8} $B was also found to undergo 2\textit{p} decay to the isobaric analog of $ ^{6} $Be$ _{\rm g.s.} $ in $ ^{6} $Li. Such conclusions were reached on the basis of the corresponding five-body correlations of the decay products, which were projected on the respective three-body systems like $ ^{6} $Be. The derived 4\textit{p}-decay scheme with the related intermediate states is shown in Fig.~\ref{fig:6_sheme_4p}.

Very recently, another 4\textit{p}-emitter $ ^{18} $Mg was observed by the invariant-mass reconstruction of $ ^{14} $O+4\textit{p} events \cite{Jin:2022}. The derived
ground state decay energy and the width are $ E_T $=4.87(3) MeV and $\Gamma$=115(100) keV, respectively. The observed momentum correlations between the five particles are reported to be consistent with two sequential steps of prompt 2\textit{p} decay passing through the intermediate
ground state of $ ^{16} $Ne.

Four-proton decay channel was also observed in a more complicated 6-particle
decay of a  high-excitation energy state in $ ^{12} $O into a
4\textit{p}+2$ \alpha $ channel.
In a neutron knock-out reaction from a $ ^{13} $O beam, a state at $ \simeq $8 MeV was observed in $ ^{12} $O by detecting the $4 p + 2 \alpha $ decay products using invariant-mass method \cite{Webb:2019}.
It was shown that this $4 p + 2 \alpha $ emitter decays by a two-step sequential process via an intermediate  $ ^{6} $Be + $ ^{6} $Be system where each $ ^{6} $Be subsystem is either in ground or excited states whose 2\textit{p}-decay patterns are well studied.
Analogs to these \textit{T}=2 states in $ ^{12} $O were found in $ ^{12} $N in the 2\textit{p}+$ ^{10} $B and 2\textit{p}+$ \alpha $+$ ^{6} $Li channels.

The available data on experiments with a 4\textit{p}-decay channel  revealed  only a sequential (e.g., 2\textit{p}-2\textit{p}) decay mechanism. The established  2\textit{p}-decay mechanisms were applied to decays of subsystems like $ ^{6} $Be, which facilitates identification of 4\textit{p}-decay modes.
However, in analogy to true 2\textit{p} decay, a direct simultaneous  4$ {p} $ emission, called the ``true 4\textit{p}'' decay mode could be also possible \cite{Grigorenko:2011}.
General separation energy conditions for the true four-nucleon emission are shown in Fig.~\ref{fig:6_sheme_4p}(b). They require that $S_{4N}$<0 and {$ S_N $, $ S_{2N} $, $ S_{3N} $}>0, in analogy with the true two-nucleon decay.

For proton-unbound nuclei, it is unlikely that the
energy conditions for the true 4\textit{p} emission are fulfilled in reality.
This statement given in Ref.~\cite{Grigorenko:2011} was illustrated by the example of two isobaric mirror pairs, $ ^{6} $He-$ ^{6} $Be and $ ^{8} $He-$ ^{8} $C. Their energies were  estimated in the
independent particle model as
\[
	E_{^6\mathrm{He}} = 2E_{^5\mathrm{He}} + E_{nn}(^6\mathrm{He}),
\]
\[	
 E_{^8\mathrm{He}} = 2E_{^7\mathrm{He}} + E_{nn}(^8\mathrm{He}).
\]
The phenomenological pairing energy value
$ E_{nn}$($ ^{6} $He)$\simeq2.8$ MeV in $ ^{6} $He was found to be very close to that
one $ E_{nn}$($ ^{8} $He)$\simeq 3.1$~MeV  in $ ^{8} $He. Within the same approximation,
the respective energies of the $ ^{6,8} $He isobar mirrors in $ ^{6} $Be, $ ^{8} $C are
\[
	E_{^6\mathrm{Be}} = 2E_{^5\mathrm{He}} + E_{nn}(^6\mathrm{He}) + 2V ^{coul}_{\alpha-p} + V ^{coul}_{ p-p},
\]
\[
	 E_{^8\mathrm{C}} = 2E_{^7\mathrm{He}} + E_{nn}(^8\mathrm{He}) + 4V ^{coul}_{\alpha-p} + 6V ^{coul}_{ p-p} .
\]

The Coulomb energy contributions are growing much faster than the corresponding nuclear
contributions with increasing number of valence nucleons.
These simple estimates demonstrate that the decay-energy conditions for true 4\textit{p} emitters are rather unfavourable, and that the
4\textit{p} decays should have the mechanism of  sequential 2\textit{p}-2\textit{p}
emission. Indeed, the energy conditions for the true 2\textit{p} decay
are realized in $ ^{6} $Be but not in $ ^{8} $C (see Fig.~\ref{fig:6_sheme_4p}), and the sequential 2\textit{p}-2\textit{p} emission is the dominating decay mode \cite{Charity:2011}.
However one may not exclude, that some special nuclear configuration
may be realized beyond the proton drip line, which can make the true 4\textit{p} emission possible.

The unbound isotopes with large decay energy are open to channels with even larger number of protons.
For example, the unobserved yet isotope $^{20}$Si is open   into the 6\textit{p}+$^{14}$O exit channel with the expected sequential decay chain
$^{20}$Si$ \rightarrow 2p$+$^{18}$Mg$ \rightarrow 4p$+$^{16}$Ne$ \rightarrow 6p$+$^{14}$O.
Such isotopes can be populated in a 2\textit{n} knock-out reaction ($^{22}$Si,$^{20}$Si), which
may be feasible in large-scale RIB facilities like RIBF (Japan), FRIB (USA) or FAIR (Germany).

\subsection{\it Three-body decays without \textit{pp} pairs in exit channel}

Three-body approaches developed for 2\textit{p} decays can be applied
also to other 3-particle decay channels. First such attempt was dedicated to decays of the ground state of $ ^{9} $B into $ \alpha $+$ \alpha $+\textit{p}  and its IAS in $ ^{9} $Be$^*\rightarrow \alpha $+$ \alpha $+\textit{n} \cite{Bochkarev:1990a}, where a democratic decay channel was identified.

Emission of 3$ \alpha $ from $ ^{12} $C excited states, and in particular from the astrophysically-important Hoyle state was studied in a number of experiments, see the Ref.~\cite{Itoh:2014} and  references therein.
Special three-body treatment of three identical $ \alpha $-particles in the exit channel
is required there, and a sequential decay via intermediate $ ^{8} $Be is established.

An asymmetric fragment emission, $ ^{8} $Li$^*$(4$ ^{+} $)$\rightarrow \alpha $+\textit{t}+\textit{n}  was reported in Ref.~\cite{Grigorenko:2002a}, and its
inspection has revealed a democratic decay mechanism.
However, the decay mechanism of its mirror state  $ ^{8} $Be$^*\rightarrow \alpha $+\textit{t}+\textit{p}  was interpreted as a sequential emission via the intermediate state $ ^{7} $Li$ ^{*} $(7/2$ ^{-} $) \cite{Charity:2011}.

The unbound nuclei with large decay energy are open to a number of channels with multi-particle emission. Investigations of four-, five- and even  six-particle decay channels of high-lying levels in light nuclei were reported, where the nuclei of interest were populated by using reactions with a $ ^{9} $C beam at intermediate energy \cite{Charity:2011}.

\subsection{\it  Towards limits of  existence of nuclear structure of proton-rich isotopes}

One of the fundamental goals of nuclear science is to establish the limits of existence of
nuclear structure (i.e., individual states) in nuclear systems.
In particular, there is a number of yet unobserved discrete proton resonances which
may be distinguished from the continuum. Such resonances are usually identified as
individual nuclei. A general trend of ground-state properties along an isotope chain
is that their widths become broader with increasing imbalance between proton and neutron numbers,
which makes the identification of such states (and thus isotope assignment) increasingly difficult.
One may expect a limit of existence of such states (\textit{i.e.}, nuclear structure)
far beyond the proton drip line.

To shed light on this question, a simple quasi-classical approach was applied in Ref.~\cite{Grigorenko:2018}.
A nuclear (\textit{Z,A}) configuration was assumed to have an individual structure,
with at least one distinctive state (typically the ground state), if the orbiting valence
protons of the system are reflected from the corresponding
nuclear barrier at least one time. Then, the nuclear half-life value (or width of the state)
may be used as a criterion of such an existence. The very long-lived particle-emitting states
may be considered as quasistationary. For example, the known heavy 2\textit{p} radioactivity
precursors (\textit{i.e.}, $ ^{45} $Fe, $ ^{48} $Ni, $ ^{54} $Zn, and $ ^{67} $Kr)
have half-lives of few milliseconds. For the long-lived states, modifications
of nuclear structure by the coupling with continuum are negligible.
In contrast, the continuum coupling becomes increasingly important for the  very short-lived
unbound ground states, which can be regarded as a transition to continuum
dynamics of the open nuclear system.
In Ref.~\cite{Grigorenko:2018}, the evolution of the ground state widths was
considered for the argon and chlorine isotope chains. The isotopes $ ^{26} $Ar and $ ^{25} $Cl
were predicted as the most remote nuclear configurations
with identified ground states having widths smaller than 3--5 MeV.

Similar estimates can be applied to other isotope chains.
The width estimates for argon and chlorine isotope chains \cite{Grigorenko:2018} were recently complemented by a global prediction of decay energies for proton unbound nuclei \cite{Neufcourt:2020}.
Such estimates allow to expect a number of previously unknown unbound isotopes located
within a relatively broad (2--5 neutrons) area along the proton drip line.
For more exotic nuclear systems, beyond such a domain, no ground states, and thus no new
isotope identification, are expected. Then a new borderline indicating the limits
of existence of isotopes on the nuclear chart, and the transition to chaotic nucleon
matter may be established.

The most-remote identified isotopes like $ ^{17} $Na or $ ^{31} $K  (located 4 neutrons
beyond the proton dripline) were observed in charge exchange reactions with radioactive beams.
In future, similar charge exchange reactions with the most exotic
beams like $ ^{48} $Ni or $ ^{67} $Kr may be used in order to gain access to
nuclear systems located up to 7 mass units beyond the proton drip
line. In such nuclear systems, the validity of the basic concepts, like the mean-field
and the Pauli principle, may be tested \cite{Kostyleva:2019}.

\section{Related few-body phenomena}

\subsection{\it Two-neutron vs. two-proton emission }
\label{sec:7_2n}

%
%

In analogy to true, democratic and sequential 2\textit{p} decays, similar mechanisms may be considered for 2\textit{n} emission by using the same energy criteria shown in Fig.~\ref{fig:1_EnergyConditions}. The true 2\textit{n} decay is the direct analog of true 2\textit{p} decay across the isobar.
With the progress in reaching experimentally the neutron drip-line, there is possibility that
neutron(s) emission may even take the form of 2\textit{n} radioactivity.

\begin{figure}[h!tb]
	\begin{center}
		\begin{minipage}[t]{16.9 cm}
			\includegraphics[width = 0.99\columnwidth, angle=0.]{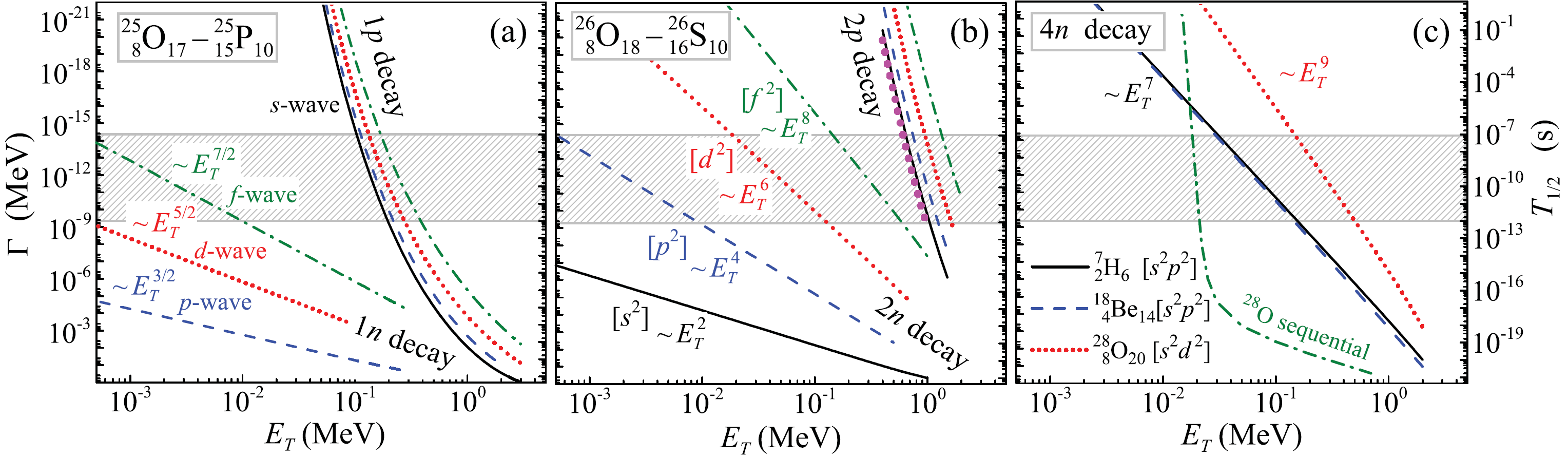}
		\end{minipage}
		\begin{minipage}[t]{16.9 cm}
			\caption{(Color online)  Widths $\Gamma$ and respective half-lifes $ T_{1/2} $ for (a) one-nucleon, (b) true two-nucleon, (c) true 4\textit{n} emission calculated for different orbital
			configurations as functions of total decay energy $ E_T $ \cite{Grigorenko:2011}. The estimates use the integral formulas in a simplified semi-analytical approach \cite{Grigorenko:2007}. In the case of neutron emission, the low-energy behavior of the widths has an asymptotic dependence $ E^\alpha_T $, as is indicated next to the corresponding curves. The hatched area shows the lifetime range accessible by decay-in-flight techniques. The results for $ ^{26} $S from
			Ref.~\cite{Fomichev:2011} are represented in panel (b) by the thick dots. The dash-dotted curve in panel (c) provides the estimate of the sequential (2\textit{n})--(2\textit{n})
			decay of $ ^{28} $O via $ ^{26} $O g.s., whose decay energy is assumed to be 20 keV.
			The figure is adopted from \cite{Grigorenko:2011}.		
			}
			\label{fig:7_neurad_T_Grigor_2011}
		\end{minipage}
	\end{center}
\end{figure}

A simple theoretical model to estimate the widths of true multi-nucleon decays was formulated in Ref.~\cite{Grigorenko:2011} on basis of the integral formulas developed for 2\textit{p} radioactivity by using a simplified semi-analytical approach \cite{Grigorenko:2007}. A number of candidate nuclei
emitting 1\textit{n}, 2\textit{n}, 4\textit{p} and 4\textit{n} in \textit{s}-, \textit{p}-, \textit{d}-, \textit{f}-waves were considered.  The calculated widths and respective half-lives are shown in Fig.~\ref{fig:7_neurad_T_Grigor_2011} as functions of the total decay energy which was unknown for the most cases. In this section, we discuss only true 2\textit{n} decays.  The prospects of a true four-nucleon decay  will be discussed in Section~\ref{sec:7_5body}.

For the reference case of 1\textit{p}/1\textit{n} emission,
simple width estimates obtained by the standard R-matrix
expression are shown in Fig.~\ref{fig:7_neurad_T_Grigor_2011}(a).
For example,  the decay
energy window for proton radioactivity (with half-life exceeding 1 ps)
ranges from 50 to 200 keV for $ ^{25} $P. Then its isobar mirror $ ^{25} $O
may undergo neutron radioactivity if the decay
energy $ E_T \leq$1 keV  (for assumed \textit{d}-wave). One may conclude, that the realistic chance to find 1\textit{n} radioactivity
may appear only for \textit{f} wave and higher $L$ states in the heavier
neutron drip-line nuclei which are un-observed yet.

Width and half-life calculations for the true
two-nucleon decays are given in Fig.~\ref{fig:7_neurad_T_Grigor_2011}(b).
The example of isobaric pair $ ^{26} $S-$ ^{26} $O is presented for
\textit{s}-, \textit{p}-, \textit{d}-, \textit{f}-waves.
The possible 2\textit{n} radioactivity has a few important differences
in comparison to the 1\textit{n} radioactivity \cite{Grigorenko:2011}.
\begin{itemize}
\item
	Low-energy \textit{s}-wave neutron emission could take place in
	the form of a virtual state, which cannot be interpreted in terms
	of width, and thus the neutron \textit{s}-wave curve is missing
	in Fig.~\ref{fig:7_neurad_T_Grigor_2011}(a).
	In contrast, the true 2\textit{n } emission has an additional effective
	centrifugal barrier, and a narrow resonance state is formed even
	for the decay of a [$ s^2 $] configuration.
\item
	The decay-energy window for 2\textit{p} radioactivity of
	$ ^{26}$S ranges from 0.5 to 1.7 MeV, thus, is about 10 times
	broader than for the proton radioactivity of $^{ 25 }$P. The energy
	window for the true 2n decay is much broader than for
	one-neutron decay: for the [$ d^2 $] and the [$ f^2 $] configurations,
	the decays would be ascribed as radioactive for decay energies
	ranging up to 200 and 600 keV, respectively. Such energy
	ranges make the existence of true 2\textit{n} radioactivity much more
	probable.
\item
	At variance to the 1\textit{n} situation, the 2\textit{n} estimates in
	Fig.~\ref{fig:7_neurad_T_Grigor_2011}(b) should be interpreted
	as lifetime limits due to the
	possibility of configuration mixing. The [$ s^2 $] and [$ p^2 $] curves
	are likely to provide lower lifetime limits for \textit{s-d} and \textit{p-f}
	configurations, respectively. The [$ d^2 $] and [$ f^2 $] curves provide
	upper lifetime limits for them.
\end{itemize}
The prospects of experimental search for 1\textit{n}/2\textit{n} radioactivity concluded in \cite{Grigorenko:2011} are:
(i) The observation of neutron radioactivity in \textit{s-d}
shell nuclei seems unrealistic, but sufficiently long lifetimes
may occur in decays of heavier (\textit{p-f} shell) systems.
(ii) The estimated  lifetimes for true 2\textit{n} emission are much
longer compared to the lifetimes of 1\textit{n} emitters with
the same energy. For that reason, the existence of 2\textit{n}
 radioactivity is plausible, since the energy windows
corresponding to the radioactive timescale is reasonably broad.

The experiments on neutron unbound nuclei provided data in qualitative
agreement with these general theoretical predictions.
Below, we present two examples of 2\textit{n} decays from the ground
states of $ ^{16} $Be and $ ^{26} $O.

\subsubsection{\it 2n decay of the $ ^{16} $Be}

\begin{figure}[h!tb]
	\begin{center}
		\begin{minipage}[t]{16 cm}
			\hspace{2.5 cm}
			\includegraphics[width = 0.6\columnwidth, angle=0.]{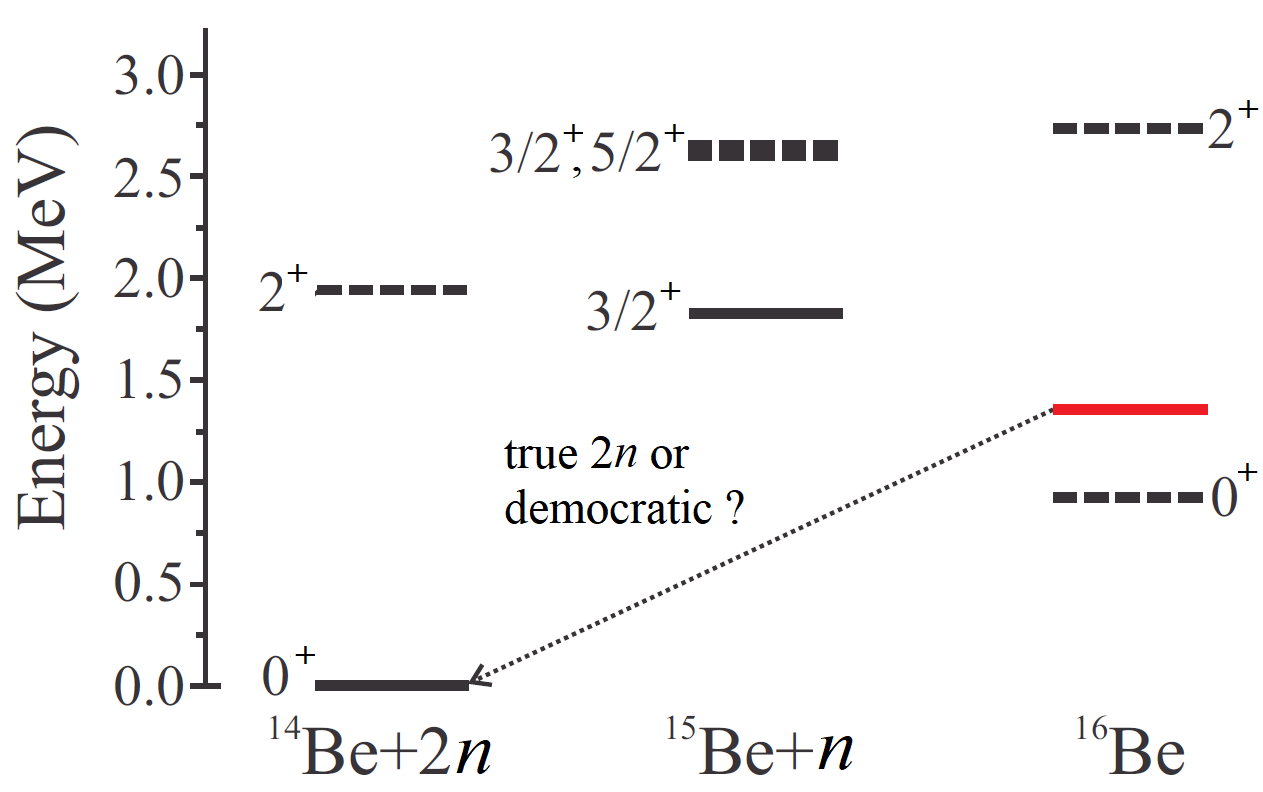}
		\end{minipage}
		\begin{minipage}[t]{16.5 cm}
		\caption{(Color online) The decay and level schemes of $ ^{16} $Be and $ ^{15} $Be isotopes derived in Ref.~\cite{Spirou:2012,Snyder:2013}. The measured 2\textit{n} decay of $ ^{16} $Be ground state (red line) is shown by the dotted arrow.  The vertical axis shows the decay energies with respect to the $ ^{14} $Be+2\textit{n} threshold. The dashed black lines represent shell model calculations within the
		\textit{s-p-sd-pf} model space \cite{Spirou:2012}. 		
			}
			\label{fig:7_16Be_scheme}
		\end{minipage}
	\end{center}
\end{figure}

The 2\textit{n} emission from the $ ^{16} $Be ground state
was observed following the 1\textit{p} removal reaction from a $ ^{17} $B beam
at energy of 53 \textit{A}MeV. The neutrons were detected with the
Modular Neutron Array MoNA in coincidence with charged fragments.
The three-body decay energy, as well as the $ ^{14} $Be\textit{-n-n} correlations
were measured \cite{Spirou:2012}. The ground state energies of $ ^{16} $Be and $ ^{15} $Be
were determined to be 1.35(10)~MeV and 1.8(1)~MeV, respectively, see Fig.~\ref{fig:7_16Be_scheme} \cite{Snyder:2013}.

The reported widths of the ground states of $ ^{15} $Be and $ ^{16} $Be are 0.58(20) and 0.8(2) MeV, respectively. These values suggest that both true and democratic 2\textit{n}-decay mechanisms are possible, and a transition mechanism can not be excluded as well. The measured $ ^{14} $Be\textit{-n-n} events demonstrate strong \textit{n-n} correlations of a di-neutron type, which called for further studies with three-body approaches.
The dominating di-neutron configuration and the computed width of the $0 ^{+} $ ground state \cite{Casal:2019} are consistent with the data, as well as with the R-matrix
calculations for the true three-body continuum \cite{Lovell:2017}. Both three-body models predict
the $0 ^{+} $ width of 0.17 MeV which underestimates the measured value of 0.8(2) MeV. If the theoretical predictions are correct, the true 2\textit{n} decay should occur, while the experimental data suggests rather a democratic mechanism. Therefore, final conclusions on the 2\textit{n}-decay mechanism require more precise measurements and calculations.

\subsubsection{\it 2n decay of $ ^{26} $O. Is 2n radioactivity observed? }

\begin{figure}[h!tb]
	\begin{center}
		\begin{minipage}[t]{16 cm}
			\hspace{2.9 cm}
			\includegraphics[width = 0.6\columnwidth, angle=0.]{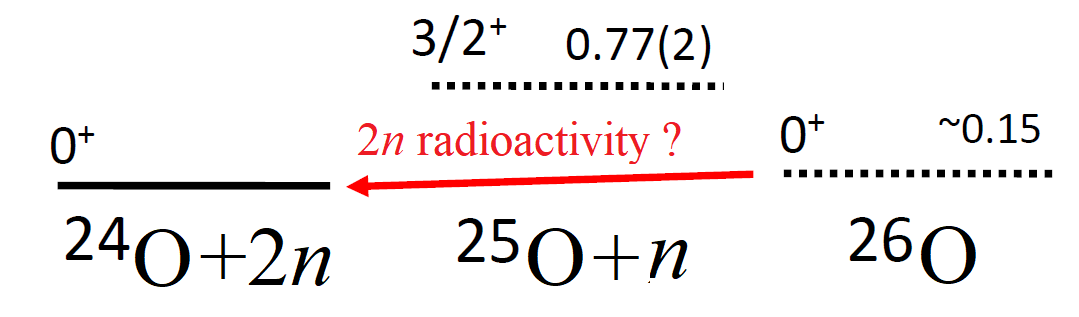}
		\end{minipage}
		\begin{minipage}[t]{16.5 cm}
			\caption{(Color online) The decay and level schemes of $ ^{26} $O and $ ^{25} $O isotopes (dashed lines with $J^{\pi}$ and energies with respect to the $ ^{24} $O+2\textit{n} threshold) derived in Ref.~\cite{Lundenberg:2012}. The measured 2\textit{n} decay of $ ^{26} $O g.s.\  is shown by the red arrow.  		
			}
			\label{fig:7_26O_scheme}
		\end{minipage}
	\end{center}
\end{figure}

Striking  results were obtained is studies of 2\textit{n} decay of $ ^{26} $O.
The definite proof that $ ^{26} $O is indeed unbound was established in an invariant mass measurement
at NSCL/MSU \cite{Lundenberg:2012}. The setup was similar to the previously discussed $ ^{16} $Be
experiment. Ions of $ ^{26} $O were produced by the 1\textit{p} removal reaction from
a secondary $ ^{27} $F beam. The three-body decay energy spectrum obtained from the measured $ ^{24} $O-\textit{n}-\textit{n} correlations showed a clear peak near the threshold. The extracted decay energy was $150 ^{+50}_{-150} $ keV \cite{Lundenberg:2012}. The observed width was dominated by the experimental resolution so that the expected very narrow width of the ground state could not be determined. This limit was confirmed by the GSI-LAND group which reported the $ ^{26} $O ground state to be unbound by less than 120 keV \cite{Caesar:2013}.
The decay scheme of $^{26} $O is shown in Fig.~\ref{fig:7_26O_scheme}. The ground state of the intermediate, unbound $ ^{25} $O is located about 600~keV above
the $ ^{26} $O ground state and this speaks in
favour of the true 2\textit{n} decay mechanism.

\begin{figure}[h!tb]
	\begin{center}
		\begin{minipage}[t]{16 cm}
\hspace{1.2 cm}
			\includegraphics[width = 0.8\columnwidth, angle=0.]{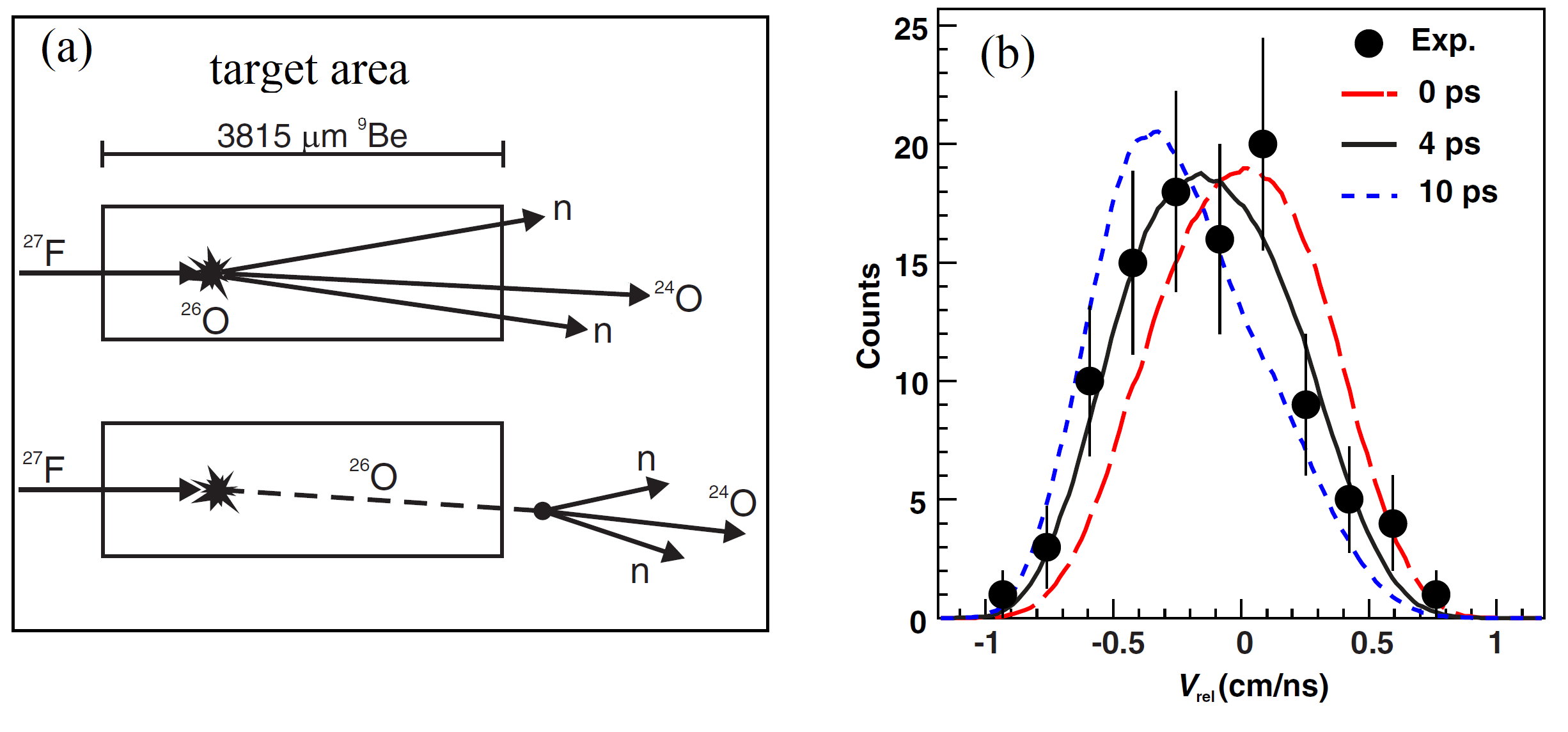}
		\end{minipage}
		\begin{minipage}[t]{16.5 cm}
			\caption{(Color online) (a) Scheme of the in-flight decay of  $ ^{26} $O within the thick $ ^{9} $Be target measured in \cite{Kohley:2013}, which is illustrated
			for two cases: (on top) very short lifetime corresponding to an
			immediate decay and (on bottom) a lifetime $\geq$30 ps which
			allows the $ ^{26} $O to exit the target before decaying.
			(b) Experimental  distribution of relative $ ^{24} $O--\textit{n} velocity $V_{rel}$ from	the decay of $ ^{26} $O compared to the MC simulations where
			the $ ^{26} $O half-life value is set as 0, 4, and 10 ps according to Ref.~\cite{Kohley:2013}. The figures (a,b) are 	
			reprinted with permission from Ref.~\cite{Kohley:2013}. Copyright (2013) by
			the American Physical Society.	
			}
			\label{fig:7_26O_Kohley_2013}
		\end{minipage}
	\end{center}
\end{figure}
In a following experiment at NSCL/MSU, an original technique developed to measure the lifetimes of neutron unbound nuclei was applied to $ ^{26} $O \cite{Kohley:2013}. The decay of $ ^{26} $O$\rightarrow^{24}$O+\textit{n}+\textit{n} was examined in the picosecond time range.
The scheme illustrating half-life measurements of the in-flight decay of  $ ^{26} $O within the thick target is shown in Fig.~\ref{fig:7_26O_Kohley_2013}(a). Due to energy loss of $ ^{24} $O fragment in the thick target, an average value of relative $ ^{24} $O--\textit{n} velocities $V_{rel}$ should be either negative or zero depending where the decay occurs. The experimental distribution of $V_{rel}$ is shown in Fig.~\ref{fig:7_26O_Kohley_2013}(b) where it is compared with MC simulations of delayed decays.
The half-life of $ ^{26} $O extracted by fitting the $V_{rel}$ distribution
was 4.5$ ^{+1.2}_{-1.5} $(stat)$\pm 3$(syst) ps. This corresponded to $ ^{26} $O having a finite lifetime at an 82\% confidence level and, thus, suggested the possibility of 2\textit{n} radioactivity.
To verify this conclusion, a detailed study of long lived
(radioactive) true 2\textit{n} emitters was performed \cite{Grigorenko:2013}.
 Using a dedicated three-body $ ^{24} $O-\textit{n}-\textit{n} model,
it was found that the evidence for 2\textit{n}
radioactivity of $ ^{26} $O with the reported lifetime should correspond
to extremely low decay energy of $\leq$1 keV. Or, reversely,  the reported decay energy
should result in much shorter lifetime. Such a contradiction between the measured decay energy and width calls for more accurate studies of this phenomenon by using finer experimental techniques. Additionally, it has been demonstrated that the three-body force is crucial in describing the ground states of dripline systems \cite{Stroberg:2021,Maris:2022,Zhang:2022}. The interplay between the three-body force and the low-lying continuum is essential in ensuring the proper binding of $^{26}$O, as has been established in several studies \cite{Otsuka:2010,Hagen:2012,Bogner:2014,Jansen:2014,Holt:2013,Stroberg:2017,Hu:2019,Ma:2020,Stroberg:2021}. However, despite these findings, there has been little research into the impact of the three-body force on the multi-nucleon decay mechanism. This is an area that warrants further investigation and could lead to interesting discoveries.

The next measurement of three-body $ ^{24} $O-\textit{n}-\textit{n} correlations  \cite{Kohley:2015} following decays of $ ^{26} $O  was not sensitive to the decay mechanism due to the experimental resolutions. However, the three-body correlations were found to be sensitive to the resonance energy
of $ ^{26} $O. An upper limit of 53 keV was extracted for the decay energy of the $ ^{26} $O ground state.
Finally, the best up-to-date invariant mass spectroscopy of $ ^{26} $O,  following one-proton
removal from $ ^{27} $F, has been performed at RIKEN \cite{Kondo:2016}. The ground state was found to lie
only $18 \pm 3$(stat)$\pm4$(syst) keV above the 2\textit{n}
decay threshold. According to Ref. \cite{Grigorenko:2013}, such an energy favors a half-life much
shorter ($\sim10^{-15}$ s) than that reported in Ref.~\cite{Kohley:2013}.
Thus, the intriguing race for the discovery of 2\textit{n} radioactivity is still going on.

\subsection{\it  Five-body decays: 4n vs. 4p emission}
	\label{sec:7_5body}

Similarly to 4\textit{p} decays which are considered in Section~\ref{sec:6_4p},
true 4\textit{n} emission is also possible for the case when two more neutrons added on top
of a true 2\textit{n} precursor constitute a system that has
only a true 4\textit{n}-decay branch, which is illustrated in Fig.~\ref{fig:6_sheme_4p}(b)  where the necessary binding energies of all subsystems of a 4\textit{n} precursor are present.

The important feature of the true few-body emission is the existence
of effective few-body centrifugal barriers which grow
rapidly as the number of emitted particles increases. The predictions of half-lives of true 4\textit{n} emitters, given by
the integral formula method \cite{Grigorenko:2011}, are presented in Fig.~\ref{fig:7_neurad_T_Grigor_2011}(c) by examples of [$ s^2p^2 $] and [$ s^2d^2 $] valence-neutron configurations. One may see that the estimated  half-lives for true 4\textit{n} emission are much
longer compared to the half-lives of 2\textit{n} emitters with
the same energy (compare the panels (b) and (c) in Fig.~\ref{fig:7_neurad_T_Grigor_2011}). The given examples of $ ^{7} $H (neutron [$ s^2p^2 $] configuration) and $ ^{28} $O ([$ s^2d^2 $] configuration) demostrate a dramatic increase  of the estimated half-life by the factor of 1000, which allows for searches in the broader range of 4\textit{n}-decay energies,  up $\sim$1 MeV. Then 4\textit{n} radioactivity is plausible, since the energy window
corresponding to the radioactive timescale is estimated to be
reasonably broad. Therefore,
the prospects to search  for 4\textit{n} radioactivity could
be also promising like the searches for 2\textit{n} radioactivity.

\begin{figure}[h!tb]
	\begin{center}
		\begin{minipage}[t]{16 cm}
			\hspace{1.5 cm}
			\includegraphics[width = 0.7\columnwidth, angle=0.]{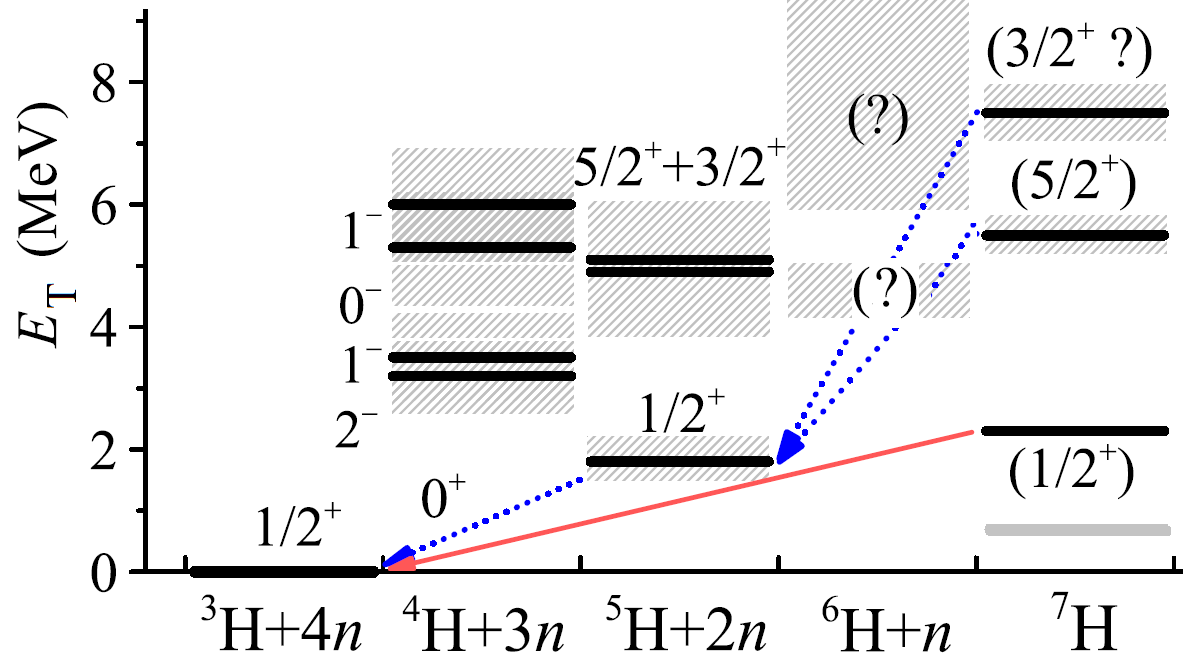}
		\end{minipage}
		\begin{minipage}[t]{16.5 cm}
			\caption{(Color online) The decay and level schemes of $ ^{7} $H isotope and its
subsystems $ ^{4} $H, $ ^{5} $H \cite{Korsheninnikov:2001,Golovkov:2005,Wuosmaa:2017}, and $ ^{6} $H \cite{Caamano:2008,Nikolskii:2021a}. The solid red arrow illustrates the decay mechanism of the $ ^{7} $H ground state which is expected to be true 4\textit{n} emission \cite{Bezbakh:2020}. The dotted blue arrows illustrate the decay mechanism of the higher excitations in $ ^{7} $H \cite{Muzalevskii:2021}, which are expected to be sequential 2\textit{n}+2\textit{n} and \textit{n}+\textit{n}+2\textit{n} emissions
via the $ ^{5} $H and $ ^{6} $H excited states, respectively. The grey line shows the
$ ^{7} $H g.s.\ assigned in Ref.~\cite{Caamano:2007}.
The figure is adopted from \cite{Muzalevskii:2021}.
			}
			\label{fig:7_7H_scheme_2020}
		\end{minipage}
	\end{center}
\end{figure}

There were several experiments searching for a $ ^{ 7} $H resonance
by using a reaction of proton removal from $ ^{8} $He secondary beam. The
$ ^{1} $H($^8 $He,$ ^{2} $He) reaction was studied in Ref.~\cite{Korsheninnikov:2003}, and evidence for the population of the $ ^{7} $H spectrum right above the $ ^{3} $H+4\textit{n}
threshold was demonstrated. Then the search for a long-living  $ ^{7} $H produced in the $ ^{2} $H($ ^{8} $He, $ ^{3} $He) reaction
was made in Ref.~\cite{Golovkov:2004a} resulting in the ascertainment of the
lower decay-energy limit $ E_T $>50--100 keV for the $ ^{7} $H ground state.
The observation of a low-lying $ ^{7} $H resonant state  populated
in the $ ^{12} $C($ ^{8} $He, $ ^{7} $H)$ ^{13} $N reaction was declared in Ref.~\cite{Caamano:2007}. Despite the difficulty of the reaction-channel
identification in the active target MAYA, the ground state was claimed to be at $ E_T $=0.57$^{+0.42}_{-0.21}$ MeV. A similar measurement by using  reaction $ ^{19} $F($ ^{8} $He, $ ^{7} $H)$ ^{20} $Ne provided another estimate $ E_T $=0.73$^{+0.58}_{-0.47}$ MeV \cite{Caamano:2020}. The next attempt to discover $ ^{7} $H was made using the $ ^{2} $H($ ^{8} $He, $ ^{3} $He) reaction \cite{Nikolskii:2010}, and an indication for a state at $\sim$2 MeV was obtained.
The recent most accurate data on
the $ ^{7} $H system were obtained in two subsequent experiments using the  reaction $ ^{2} $H($ ^{8} $He, $ ^{3} $He)$ ^{7} $H at energy of 26 \textit{A}MeV \cite{Bezbakh:2020,Muzalevskii:2021}.
The $ ^{7} $H missing
mass energy spectrum, the $ ^{3} $H energy, and the angular distributions in the $ ^{7} $H decay frame were reconstructed, which allowed for ascribing low-energy states of $ ^{7} $H, and in particular its ground state at 2.2(5) MeV. The derived level and decay scheme of $ ^{7} $H is shown in Fig.~\ref{fig:7_7H_scheme_2020} together with its subsystems $ ^{4-6} $H. Results of these experiments suggest the true 4\textit{n} decay mechanism for the ground state of $ ^{7} $H,
although unknown states of the least studied system, $ ^{6} $H, could affect this conclusion.

In order to clarify a 4\textit{n} decay mechanism of $ ^{7} $H, its subsystem $ ^{6} $H was studied in the direct $ ^{2} $H($ ^{8} $He, $ ^{4} $He)$ ^{6} $H transfer reaction
with a  secondary $ ^{8} $He beam at energy of 26 \textit{A}MeV \cite{Nikolskii:2021a}.
The measured missing mass spectrum showed a broad bump
at 4--8 MeV energy relative to the $ ^{3} $H+3\textit{n} decay threshold. This bump was interpreted as a
broad resonant state in $ ^{6} $H at 6.8(5) MeV. The obtained missing mass spectrum  allowed also to derive the lower limit for the
possible resonant state energy in $ ^{6} $H at 4.5(3) MeV, and there were no events registered below 3 MeV which contradicts the results of Ref.~\cite{Caamano:2008}. According to the paring energy estimates in \cite{Nikolskii:2021a}, such a 4.5(3) MeV resonance was mentioned as a candidate for the $ ^{6} $H ground state, see Fig.~\ref{fig:7_7H_scheme_2020}.
This would confirm that the decay mechanism of the $ ^{7} $H g.s.\ (located at 2.2 MeV above the $ ^{3} $H+4\textit{n} threshold) is the true 4\textit{n} emission.

One should note that the half-life of the true 4\textit{n}-precursor $ ^{7} $H ground state at 2.2 MeV is estimated to be shorter than 10$ ^{-19} $ s according to the calculations shown in Fig.~\ref{fig:7_neurad_T_Grigor_2011}(c), which excludes 4\textit{n} radioactivity here. However, if the ground state of $ ^{7} $H is at 0.57$^{+0.42}_{-0.21}$ MeV, as assigned in Ref.~\cite{Caamano:2007}, then its decay would correspond to the radioactivity time range.

\subsection{\it Tetraneutron}
\label{sec:7_Tetraneutron}

The study of multi-nucleon emission can shed light on the four-neutron system --- the tetraneutron ---  which bridges the finite systems with nuclear matter and provides valuable insights into the short-range nuclear interaction in the presence of the low-lying continuum. Due to its extreme proton-neutron ratio, the tetraneutron is situated at the edge of nuclear stability. The question of the tetraneutron is a complex one, and this manifests itself in the controversies concerning experimental studies as well as theoretical predictions.

The experimental searches for a four-neutron system have been performed since 1963 \cite{Schiffer:1963}, resulting in only a few indications of its existence so far \cite{Marques:2002,Kisamori:2016,Faestermann:2022,Duer:2022}.

To figure out the true nature of an observed resonant state, one needs to distinguish the genuine resonance from a final-state effect (scattering feature). A similar situation arises in the case of neutron-neutron or proton-proton dimers. The former is unable to form neither as a bound state nor as a resonance. It gives rise to a virtual (antibound) state, which is characterized by the negative energy and zero decay width, but has a different asymptotic behavior than the bound state \cite{Ohanian:1974}.
In case of the diproton, the presence of the Coulomb barrier gives rise to a subthreshold resonance that has negative energy but a large decay width. While such exotic resonant states cannot be directly observed, they can be identified by enhanced scattering strength just above the threshold and strong final-state effects, making them more akin to scattering features than true resonances.

Regarding the tetraneutron, it has been established that the attractive nuclear interaction alone is insufficient to produce a bound state \cite{Sofianos:1997, Pieper:2003}. However, the question of whether the system can exist as a resonance under the interplay among pairing, many-body correlation, and continuum effects remains an open and ongoing area of research \cite{Sobotka:2022}.

The status of the tetraneutron was investigated using the No-Core Shell Model (NCSM) based on a realistic JISP16 interaction \cite{Shirokov:2016, Shirokov:2018}. As the continuum was absent, the single-state harmonic oscillator representation of scattering equations (SS HORSE) technique was used to calculate the $S$-matrix resonant parameters. The results predicted a resonance near the threshold, with an energy of 0.8 MeV and a width $\Gamma$ of about 1.4 MeV, which was found to be insensitive to the choice of the nucleon-nucleon interaction \cite{Shirokov:2016, Lashko:2008}. The Green Function Monte Carlo (GFMC) method also supported the existence of the tetraneutron resonance \cite{Pieper:2003, Gandolfi:2017}. However, it has been debated that such extrapolation might not be accurate due to the discontinuity at the threshold \cite{Deltuva:2019,Gandolfi:2019}, also known as the Wigner cusp \cite{Wigner:1948}. The No-Core Gamow Shell Model (NCGSM) was developed to explicitly include the continuum \cite{Fossez:2017_2, Li:2019}, which is crucial for an open quantum system such as the tetraneutron. By utilizing the Berggren basis, the bound, Gamow, and scattering states were described on the same footing, and the pole obtained in Ref.\,\cite{Fossez:2017_2} was found to represent a scattering feature rather than a genuine nuclear state. A later study by Ref.\,\cite{Li:2019}, using a larger model space, indicated a resonance at 2.64 MeV, with a width of 2.38 MeV, which was short enough to influence the formation of a nucleus and make it close to a threshold resonance.

On the other side, the system was investigated in the few-body frameworks, such as the Gaussian expansion method and the Faddeev-Yakubovsky formalisms \cite{Hiyama:2016,Lazauskas:2017}. These studies utilized a four-body Jacobi coordinate to give a proper asymptotic behavior and the complex-scaling method to obtain the resonance with the outgoing boundary condition. Consequently, it was found that the results are sensitive to the isospin $T$ = 3/2 three-body force. Ref.\,\cite{Hiyama:2016}, based on the parameters constrained by the low-lying states of ${ }^4 \mathrm{H}$, ${ }^4 \mathrm{He}$, and ${ }^4 \mathrm{Li}$ and the ${ }^3 \mathrm{H}+n$ scattering, did not obtain the tetraneutron resonance and suggested that the tetraneutron signal could possibly come from some unknown dynamical phenomena. A similar conclusion was drawn by Ref.\,\cite{Deltuva:2018}, in which the problem was solved by Alt, Grassberger, and Sandhas (AGS) equations. Within an adiabatic hyperspherical framework \cite{Higgins:2020,Higgins:2021}, it has been demonstrated that, due to the long-range universal physics analogous to the Efimov effect, an enhancement of the density of states, or of the Wigner-Smith time delay, would appear near the threshold of the tetraneutron even without forming a resonance. Most recently, Ref.\,\cite{Lazauskas:2023} showed through a reaction model that the sharp low-energy peak observed in the experiment \cite{Duer:2022} could also be a consequence of dineutron-dineutron correlations and is expected in the decay of other systems containing four-neutron halos.

As discussed above, distinguishing between a scattering feature and a resonance is challenging, and further experiments with other observables \cite{Wang:2023} or crosschecks from different reactions may be necessary. Multi-nucleon emission also provides a way to study the nucleon-nucleon/dinucleon-dinucleon correlations, as well as the in-medium effect for the tetraneutron system.


\section{Outlook and Conclusions}
\label{sec:8_conclusions}
\subsection{\it Global predictions}

\begin{figure}[h]
\begin{center}
\begin{minipage}[t]{11 cm}
\includegraphics[width = \columnwidth]{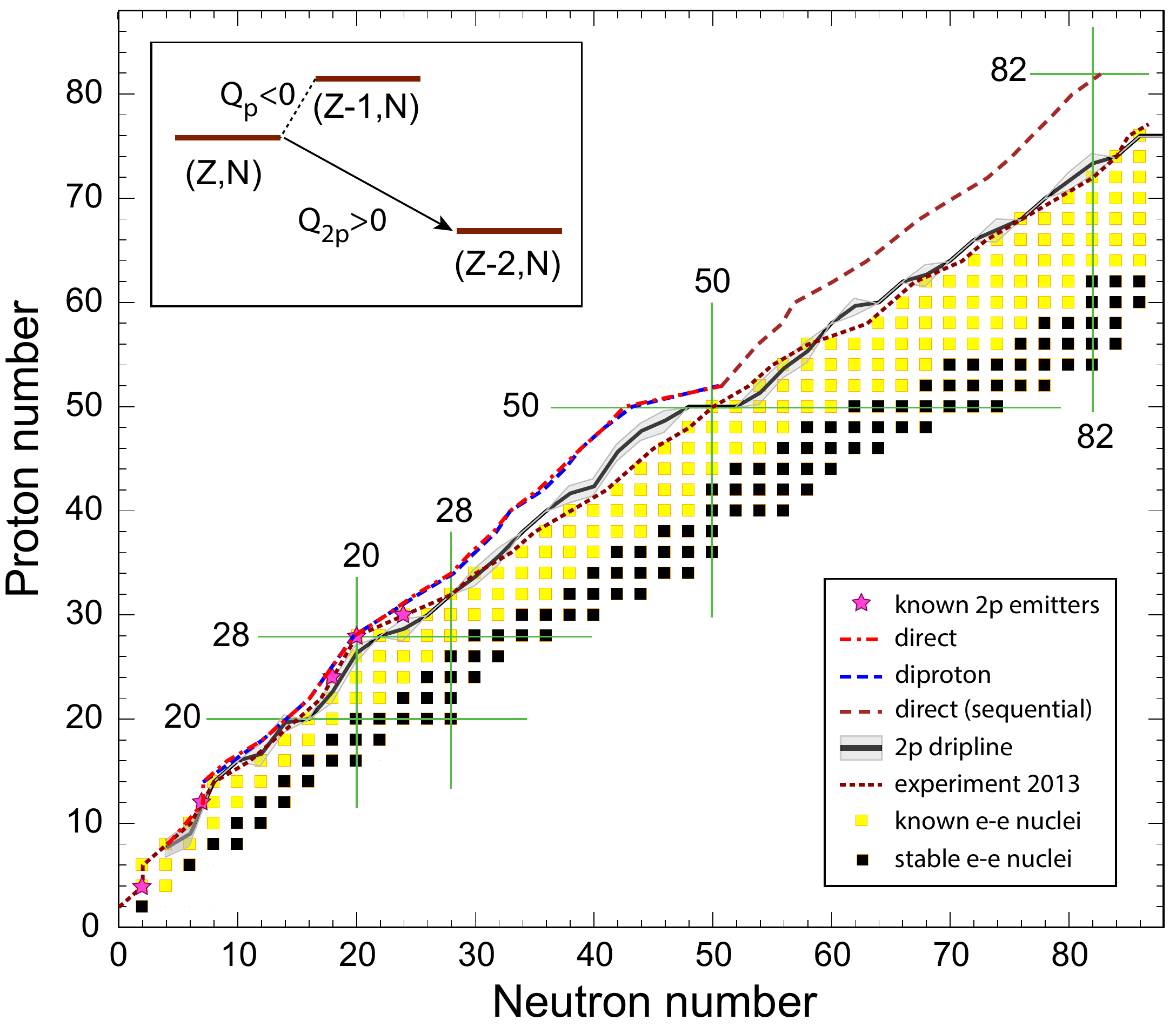}
\end{minipage}
\begin{minipage}[t]{17 cm}
\caption{(Color online) The landscape of ground-state 2\emph{p} radioactivity.
The known proton-rich even-even nuclei are marked by yellow squares,
stable even-even nuclei by black squares, and 2\emph{p} emitters known in 2013 by stars.
The dotted line shows the experimental reach at 2013. Other lines represent
averages of predictions by six interaction models considered. The solid line
shows the 2\emph{p} drip-line, the dashed line indicates the average position
of the true 2\emph{p} (blue) and sequential \emph{pp} (brown) decay candidates from
the direct decay model.
The dash-dotted (red) line indicates the position of the true 2\emph{p}
candidates from the diproton model. Figure from Ref.~\cite{Olsen:2013b}.}
\label{fig:8_OlsenPrediction}
\end{minipage}
\end{center}
\end{figure}

Presently the landscape of 2\emph{p} decay studies is limited to nuclei
with atomic number up to $Z=36$ (krypton). Natural question arises whether
this phenomenon does occur, and can be experimentally studied, in heavier
regions of the nuclidic chart. The first, global investigation of the
ground-state 2\emph{p} emission along the whole chart of nuclei was carried
out by Olsen et al.~\cite{Olsen:2013,Olsen:2013b}. This work was based on
large-scale mass table calculations which used the state-of-the-art
nuclear energy density functional (EDF) theory with six different effective Skyrme
interaction models \cite{Erler:2012}. These calculations were used
to predict all relevant separation energies. The results were not as
accurate as in the method using the Coulomb displacement energies, but
they allowed for a global, qualitative survey. To estimate
half-lives, two simplified models of 2\emph{p} emission were used:
a direct-decay model and a diproton model \cite{Olsen:2013} (see Section 2.2).
In addition, a competition from $\alpha$ decay was considered
by employing a global phenomenological formula for the $\alpha$-decay
half-life from Ref.~\cite{Koura:2012}.
In search for true 2\emph{p} decay candidates the energy criterion
$Q_{2p}>0, \, Q_p < 0.2 Q_{2p}$ was applied. It follows the observation
(see Section 4.5 and Fig.~\ref{fig:4_30Ar_Golub2016}) that if the
decay energy for the single-proton emission
is positive but small enough, the simultaneous 2\emph{p} emission dominates
the decay. In order to assure that the decay can compete with $\beta^+$
decay and with $\alpha$ decay, but is slow enough for the projectile
fragmentation technique, the additional time criteria were adopted:
$100 \, {\rm ns} < T_{1/2}^{2p} < 100 \, {\rm ms}$,
and $T_{1/2}^{2p} < 10 T_{1/2}^{\alpha}$, where
$T_{1/2}^{\alpha}$ is the partial half-life for $\alpha$ decay.
Nuclei fulfilling these
criteria were found for all elements between argon ($Z=18$) and
tellurium ($Z=52$). The position of these candidates, averaged over
six mass models considered, is shown in Fig.~\ref{fig:8_OlsenPrediction}
with dashed and dashed-dotted lines for the direct and diproton
model, respectively. No such candidates were found above
tellurium \cite{Olsen:2013b}.
For the region between tellurium and lead ($52<Z\leq 82$), however,
the candidates were found which fulfilled, in addition to the
time criteria mentioned above, the energy
condition $Q_{2p}>0, \, Q_p > 0.2 Q_{2p}$, which corresponds to the
sequential emission of two protons (\emph{pp}) \cite{Olsen:2013b},
see Fig.~\ref{fig:8_OlsenPrediction}.
This is a consequence of the high Coulomb barrier in these
nuclei. To meet the adopted half-life conditions, the system has
to be so far beyond the drip-line, that the single-proton
emission can proceed unsuppressed. That also means that
beyond the proton drip-line there exist a large territory
of $\beta$-decaying nuclei.
Finally, above $Z=82$ the $\alpha$ decay was found to dominate totally.
Comparison of the predicted location of the most probable 2\emph{p} decay
candidates with known experimental 2\emph{p} emitters indicated
that the uncertainty of the predictions presented in
Refs.~\cite{Olsen:2013,Olsen:2013b} is about one mass unit for
a given element. A very interesting finding was that for two
candidates, $^{103}$Te and $^{145}$Hf, a possible competition of
$\alpha$ decay with 2\emph{p} and \emph{pp} emission, respectively,
may be expected.

\begin{figure}[h]
\begin{center}
\begin{minipage}[t]{16 cm}
\includegraphics[width = \columnwidth]{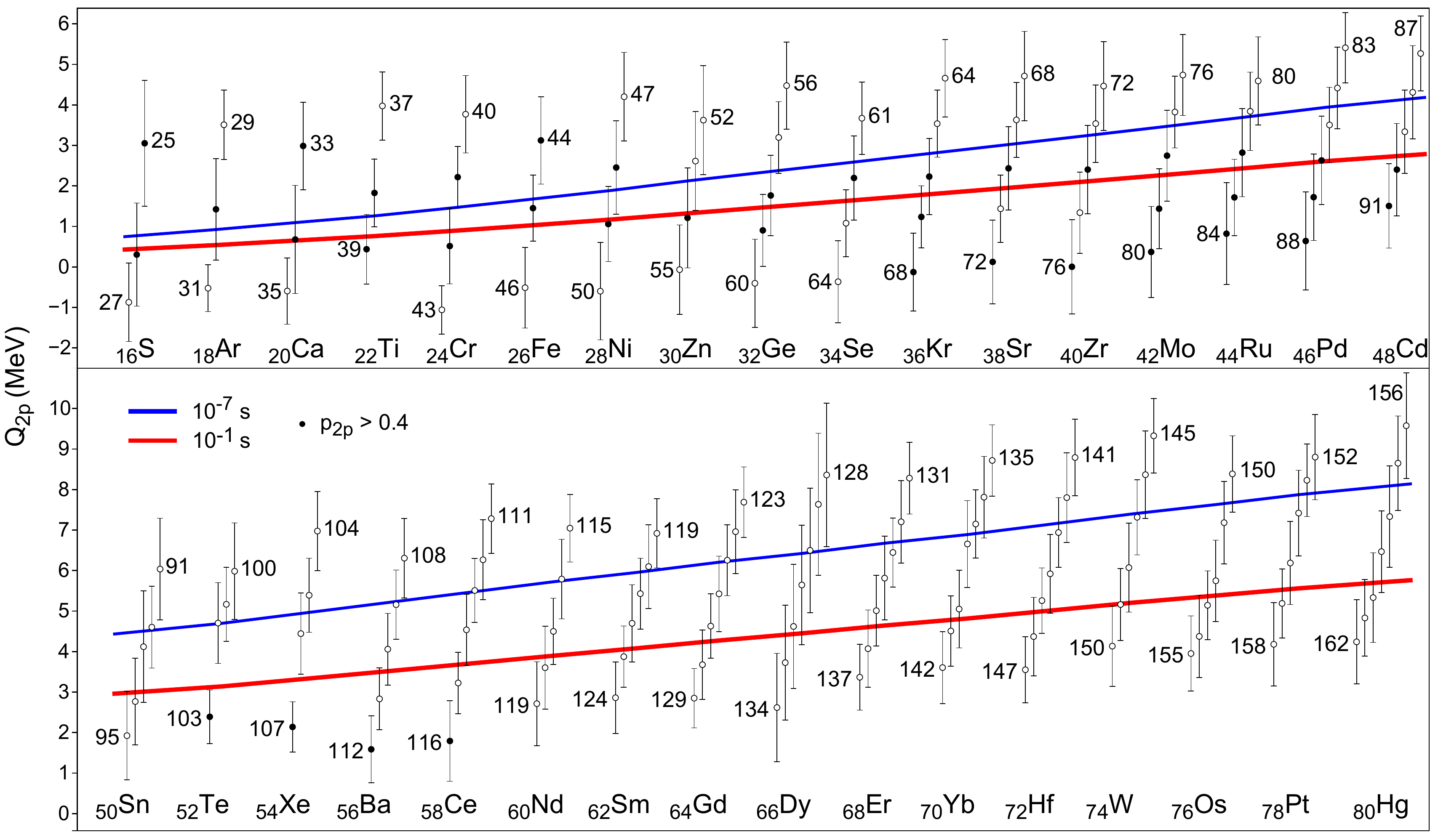}
\end{minipage}
\begin{minipage}[t]{17 cm}
\caption{(Color online) $Q_{2p}$ values predicted by the Bayesian Model Averaging
method for even-even isotopes beyond the proton drip-line between sulfur ($Z=16$)
and mercury ($Z=80$). The solid lines (red and blue) mark the adopted half-life
limits. The mass numbers of selected isotopes are indicated. The nuclei with the
2\emph{p} decay probability larger than 0.4 are marked by filled circles.
Reprinted with permission from Ref.~\cite{Neufcourt:2020}.
Copyright (2020) by the American Physical Society.}
\label{fig:8_BayesianPrediction}
\end{minipage}
\end{center}
\end{figure}

Recently, a new approach to the global prediction of the ground-state 2\emph{p}
emission was undertaken, based on the Bayesian Model Averaging (BMA)
analysis of nuclear masses \cite{Neufcourt:2020}. In this method
the quality of mass predictions is improved by including the current
experimental information through machine learning techniques.
Patterns of systematical deviations between predicted and measured
masses are quantified using statistical Bayesian Gaussian process
techniques~\cite{Neufcourt:2018} and then averaged over different
mass models taken into account. In this way, the ``collective'' prediction
of maximized accuracy, rooted in available experimental knowledge,
is obtained. This approach was presented in more details and
applied to the neutron drip-line predictions in Ref.~\cite{Neufcourt:2019}.
In the proton drip-line analysis, density functional theory with
nine different EDFs, and a few different Bayesian averaging methods
were considered. Using the condition for the true 2\emph{p} decay
$Q_{2p}>0, \, Q_p < 0$, the probability for the 2\emph{p} decay was
determined. The 2p-decay half-lives were estimated with the diproton
model described in Ref.~\cite{Olsen:2013} and the practical range
of the half-lives: $100 \, {\rm ns} < T_{1/2}^{2p} < 100 \, {\rm ms}$
was imposed in addition. The resulting prediction for 2\emph{p} radioactivity
is shown in Fig.~\ref{fig:8_BayesianPrediction}. In general, they
are consistent with the current experimental status of the 2\emph{p} decay and
with the predictions of Refs.~\cite{Olsen:2013,Olsen:2013b}. The promising
candidates for the 2\emph{p} radioactivity are found in all even-$Z$
elements under consideration until tellurium ($Z=52$). For heavier
elements ($Z>52$) the Coulomb barrier is too high for the true 2\emph{p}
decay, that is why the sequential \emph{pp} emission is expected
in this region.

\subsection{\it Summary}

The program of 2\emph{p} emission studies got a new impetus about 20 years ago when the
ground-state 2\emph{p} radioactivity of $^{45}$Fe was discovered \cite{Pfutzner:2002,Giovinazzo:2002}.
The progress in this field, especially on experimental side, has been continuous,
albeit slower than desired. The main impeding factor is the difficulty to produce
the nuclei of interest which are beyond the proton dripline. Short half-lives of 2\emph{p}
radioactive nuclei require high-energy in-flight production techniques. Only a few
fragmentation facilities worldwide have been providing sufficient conditions for
this task, as shown in Table 2. And still among the four cases of medium-mass 2\emph{p} radioactivity,
shown in the right panel of Fig.~\ref{fig:1_Chart}, the record of statistics -- about 100 decays observed --
was achieved for $^{45}$Fe in 2007 \cite{Miernik:2007b}. The further advances in this research
field, especially in view of predictions discussed above, are expected when new facilities
will start operation. The FRIB laboratory, at the Michigan State University (USA) just
started running, initially with lower beam intensities. In the next few years, when the
full power will be reached, the significant progress in study of known 2\emph{p} emitters and
discoveries on new 2\emph{p}-emitting nuclei, are expected. Similarly, the FAIR facility
near Darmstadt (Germany), in spite of delays in construction, is hoped to
provide excellent conditions for investigations of proton emitting heavier nuclei in the next decade.

On the other hand, a more substantial progress was achieved in studies of short-lived
2\emph{p}-emitting states in lighter nuclei, shown in the left part of Fig.~\ref{fig:1_Chart}.
These democratic decays from ground states and excited resonances are
investigated using techniques developed for reaction studies, and involve
beams of light nuclei, which are easier attainable with sufficient intensities.
The state-of-the-art in this field may be represented by a detailed study
of $^{6}$Be decay, where the transition from the democratic 2\emph{p} decay
of the ground-state to the sequential emission of protons from excited resonances
was documented and analysed  \cite{Egorova:2012}.
Among new interesting phenomena observed is a complex interplay between democratic
and sequential 2\emph{p} emission from the first excited $2^+$ state of $^{16}$Ne,
described as a ``tethered'' decay \cite{Brown:2015}. A superposition of
both 2\emph{p} decay mechanisms was also established for the ground-state decay
of $^{30}$Ar shedding light on the separation criterion between simultaneous
and sequential emission of protons \cite{Mukha:2015}. Understanding of
the 2\emph{p}-emission mechanism is still far from complete, so this type of
studies are going to be intensively continued in the future.

The interest in 2\emph{p} emission naturally extends to processes where more
protons are emitted, and to decays with emission of other particles, like two neutrons.
Experimental data on 3\emph{p}- and 4\emph{p}-emission are scarce but all of them
suggest that they proceed sequentially by 1\emph{p}-2\emph{p}
and 2\emph{p}-2\emph{p} decay mechanism, respectively. Although it seems unlikely,
it is not yet excluded that there exist a nuclear system fulfilling the energy
criterion for the simultaneous true 4\emph{p} emission. Studies of 2\emph{n} emissions
are very hard because of experimental difficulties to reach beyond the neutron
dripline. A few cases of ground-state emission of two neutrons
were observed but for one of them, $^{26}$O, the half-life of a few picoseconds
was reported, and thus classifying this decay as a radioactive one.
Subsequently, different results were obtained for this nucleus, rendering the
search for the first case of 2\emph{n} radioactivity still open.

The 2\emph{p}-decay studies provide knowledge and tools useful for investigations of
other nuclear processes where three-body aspects play a not negligible role.
An example is the study of isospin symmetry breaking in particle-unbound
nuclei, in particular the Thomas-Ehrman shift. Another such topic is the
soft dipole mode, representing a collective motion of two valence nucleons
against the core. Such an excitation is predicted to occur in 2\emph{p}-unbound
systems. It can play a significant role in the two-proton capture reaction
which is of relevance in astrophysical context. The impact of this reaction,
the reverse of 2\emph{p} emission, on nucleosynthesis modeling is still
a matter of study.

Finally, the theoretical description of 2\emph{p} emission also has seen a significant progress
in the past two decades. The initial simplified approaches, based on two-body
approximations gave way to models taking explicitly into account the three-body
nature of the process. The latter, in turn, evolve from schematic descriptions
of the initial nuclear state, to more advanced ones, including different nuclear degrees
of freedom, nucleon-nucleon correlations, and decay dynamics. Despite of impressive
advances in this field, however, we are still far from the comprehensive
description of the 2\emph{p} decay, which would integrate the main aspects of
this phenomenon. The efforts to provide a unifying picture of the three-body
2\emph{p} emission, with the realistic description of the initial nucleus,
and taking properly into account Coulomb asymptotic and continuum effects,
will remain in focus of theoretical activities in the near future.
One may hope that these efforts will go hand-in-hand with accomplishments
of experimentalists providing a stream of new data on two-proton emission
and related phenomena.

\section*{Acknowledgments}

We are grateful to Leonid Grigorenko for fruitful discussions.
This work was partly supported by the National Science Center, Poland, under Contract No. 2019/33/B/ST2/02908 and by the National Natural Science Foundation
of China under Contract No.\,12147101.

\providecommand{\noopsort}[1]{}\providecommand{\singleletter}[1]{#1}

\end{document}